\documentclass[namedreferences]{solarphysics}
\usepackage{amsmath}
\usepackage[hyperref,optionalrh,solaromanenum]{spr-sola-addons} 
\usepackage{graphicx}                    
\usepackage{color}                       
\usepackage{epstopdf}
\AppendGraphicsExtensions{.tif}
\usepackage{hyphenat}

\usepackage{solphys_macros}

\usepackage{xcolor}
\usepackage{soul}

\begin{document}

\begin{article}

\begin{opening}

\title{The Fibre Resolved opticAl and Near-ultraviolet Czerny-Turner Imaging Spectropolarimeter ({\sc{francis}})}

%
\author[addressref={aff1,aff2},corref,email={d.jess@qub.ac.uk}]
{\inits{D.B.~}
\fnm{David~B.~}
\lnm{Jess}}

\author[addressref={aff1},email={samuel.grant@qub.ac.uk}]
{\inits{S.D.T.~}
\fnm{Samuel~D.T.~}
\lnm{Grant}}

\author[addressref={aff1},email={wbate02@qub.ac.uk}]
{\inits{W.~}
\fnm{William~}
\lnm{Bate}}

\author[addressref={aff3,aff4},email={jiajialiu@ustc.edu.cn}]
{\inits{J.~}
\fnm{Jiajia~}
\lnm{Liu}}

\author[addressref={aff5,aff6},email={shahin.jafarzadeh@mps.mpg.de}]
{\inits{S.~}
\fnm{Shahin~}
\lnm{Jafarzadeh}}

\author[addressref={aff1},email={p.keys@qub.ac.uk}]
{\inits{P.H.~}
\fnm{Peter~H.~}
\lnm{Keys}}

\author[addressref={aff7},email={luis.vieira@inpe.br}]
{\inits{L.E.A.~}
\fnm{Lu{\'{i}}s~E.~A.~}
\lnm{Vieira}}

\author[addressref={aff7},email={alisson.dallago@inpe.br}]
{\inits{A.~}
\fnm{Alisson~}
\lnm{Dal~Lago}}

\author[addressref={aff7},email={fernando.guarnieri@inpe.br}]
{\inits{F.L.~}
\fnm{Fernando~L.~}
\lnm{Guarnieri}}

\author[addressref={aff2},email={damian.christian@csun.edu}]
{\inits{D.J.~}
\fnm{Damian~J.~}
\lnm{Christian}}

\author[addressref={aff8,aff9},email={douglas.gilliam@njit.edu}]
{\inits{D.~}
\fnm{Doug~}
\lnm{Gilliam}}

\author[addressref={aff10},email={directoraries@aries.res.in}]
{\inits{D.~}
\fnm{Dipankar~}
\lnm{Banerjee}}

%
\runningauthor{D.B. Jess et al.}
\runningtitle{{\sc{francis}}: a fibre-fed, near-UV solar IFU prototype}

\address[id={aff1}]{Astrophysics Research Centre, School of Mathematics and Physics, Queen's University Belfast, Belfast, BT7 1NN, UK}
\address[id={aff2}]{Department of Physics and Astronomy, California State University Northridge, Northridge, CA 91330, USA}
\address[id={aff3}]{Deep Space Exploration Lab/School of Earth and Space Sciences, University of Science and Technology of China, Hefei, 230026, China}
\address[id={aff4}]{CAS Key Laboratory of Geospace Environment, Department of Geophysics and Planetary Sciences, University of Science and Technology of China, Hefei, 230026, China}
\address[id={aff5}]{Max Planck Institute for Solar System Research, Justus-von-Liebig-Weg 3, 37077 G\"{o}ttingen, Germany}
\address[id={aff6}]{Rosseland Centre for Solar  Physics, University of Oslo, P.O. Box 1029 Blindern, 0315 Oslo, Norway}
\address[id={aff7}]{National Institute for Space Research (INPE/Brazil), S{\~{a}}o Jos{\'{e}} dos Campos, S{\~{a}}o Paulo 12227-010, Brazil}
\address[id={aff8}]{Big Bear Solar Observatory, 40386 North Shore Lane, Big Bear City, CA 92314, USA}
\address[id={aff9}]{National Solar Observatory, University of Colorado Boulder, 3665 Discovery Drive, Boulder, CO 80303, USA}
\address[id={aff10}]{Aryabhatta Research Institute of Observational Sciences, Nainital, 263000, Uttarakhand, India}

\begin{abstract}
The solar physics community is entering a golden era that is ripe with next-generation ground- and space-based facilities, advanced spectral inversion techniques, and realistic simulations that are becoming more computationally streamlined and efficient. With ever-increasing resolving power stemming from the newest observational telescopes, it becomes more challenging to obtain (near-)simultaneous measurements at high spatial, temporal and spectral resolutions, while operating at the diffraction limit of these new facilities. Hence, in recent years there has been increased interest in the capabilities integral field units (IFUs) offer towards obtaining the trifecta of high spatial, temporal and spectral resolutions contemporaneously. To date, IFUs developed for solar physics research have focused on mid-optical and infrared measurements. Here, we present an IFU prototype that has been designed for operation within the near-ultraviolet to mid-optical wavelength range, which enables key spectral lines (e.g., Ca~{\sc{ii}}~H/K, H$\beta$, Sr~{\sc{ii}}, Na~{\sc{i}}~D$_{1}$/D$_{2}$, etc.) to be studied, hence providing additional spectral coverage to the instrument suites developed to date. The IFU was constructed as a low-budget proof-of-concept for the upcoming $2${\,}m class Indian National Large Solar Telescope and employs circular cross-section fibres to guide light into a Czerny-Turner configuration spectrograph, with the resulting spectra captured using a high quantum efficiency scientific CMOS camera. Mapping of each input fibre allows for the reconstruction of two-dimensional spectral images, with frame rates exceeding $20$~s$^{-1}$ possible while operating in a non-polarimetric configuration. Initial commissioning of the instrument was performed at the Dunn Solar Telescope, USA, during August~2022. The science verification data presented here highlights the suitability of fibre-fed IFUs operating at near-ultraviolet wavelengths for solar physics research. Importantly, the successful demonstration of this type of instrument paves the way for further technological developments to make a future variant suitable for upcoming ground-based and space-borne telescope facilities.   
\end{abstract}

%

\end{opening}

%
\section{Introduction}
\label{sec:introduction} 

The solar physics community has recently benefited from a number of improved telescopes and instruments that enable the dynamics of the Sun's tenuous atmosphere to be studied with unprecedented spatial, temporal, and spectral resolutions, while also sampling the different polarisation states of the incident light \citep[see, e.g., the recent review by][]{JessLRSP2023}. Such facilities include the CRISP/CHROMIS instruments on the Swedish Solar Telescope \citep[SST;][]{2003SPIE.4853..341S, 2008ApJ...689L..69S, 2017psio.confE..85S}, ROSA/HARDcam imagers on the Dunn Solar Telescope \citep[DST;][]{Dunn1969, 2010SoPh..261..363J, 2012ApJ...757..160J}, balloon-borne experiments such as {\sc{Sunrise}} \citep{2010ApJ...723L.127S, 2011SoPh..268....1B, 2011SoPh..268..103B}, space-based telescopes including the Interface Region Imaging Spectrograph \citep[IRIS;][]{2014SoPh..289.2733D}, Solar Orbiter \citep{2013SoPh..285...25M, 2020A&A...642A...1M}, and Parker Solar Probe \citep{2016SSRv..204....7F}, and of course pioneering ground-based telescopes such as the Daniel K. Inouye Solar Telescope \citep[DKIST;][]{2016AN....337.1064T, 2020SoPh..295..172R, 2021SoPh..296...70R}.

Irrespective of whether a space-based, balloon-borne or ground-based facility is employed, it has always been a goal of the instrument teams to maximise the trifecta of (near-)simultaneous spatial and spectral information obtained across different polarisation states at high temporal cadences. Traditional instruments, which include Fabry-P{\'{e}}rot interferometers \citep[e.g., see the recent review by][]{Bailen2023} and scanning slit-based spectrographs, are unable to obtain simultaneity between spatial information and the associated spectra, either through the need to scan in wavelength space (e.g., Fabry-P{\'{e}}rots) or raster across a field of view (e.g., slit-based spectrographs). Of course, with constantly improving detector sensitivities and more efficient optical configurations becoming commonplace in instrument designs over the last decade-or-so \citep[e.g.,][]{2010MmSAI..81..763J, 2011A&A...535A..14K, 2012SPIE.8446E..6XD, 2012AN....333..880P, 2013OptEn..52h1606P, 2020SoPh..295..172R, 2022SoPh..297...22D, 2022SoPh..297..137J}, the temporal restrictions are now significantly less of a challenge with regards to scientific analyses of the data. Furthermore, multi-slit variations of slit-based spectrographs can be considered an in-between solution that offers better temporal resolution over their single slit counterparts. However, due to the fundamental design of these instruments, they are inherently unable to obtain strict simultaneity of their wavelength and spatial samplings.

In recent times, there have been a number of significant developments in the construction of integral field units (IFUs) suitable for solar observing. As described by \citet{2019OptEn..58h2417I}, IFUs currently come in three distinct formats: image slicers, microlens arrays, and fibre-fed bundles, each with their own technical processes to capture two-dimensional (2D) images and disperse the incident light into simultaneous cubes of [$x$, $y$, $\lambda$], where $x$ and $y$ are the spatial dimensions and $\lambda$ is the wavelength coverage sampled. Thus, it is possible to simultaneously acquire spectral information across a 2D field-of-view, and when performed at high cadences, enables dynamic solar events to be examined with high degrees of precision. Recently, the Diffraction Limited Near Infrared Spectropolarimeter \citep[DL-NIRSP;][]{2022SoPh..297..137J}, the Multi-Slit Image Slicer based on collimator-Camera \citep[MuSICa;][]{2013JAI.....250009C}, the Microlens Hyperspectral Imager \citep[Mi-Hi;][]{2022A&A...668A.149V}, the GREGOR Infrared Spectrograph \citep[GRIS;][]{2012AN....333..872C, 2022JAI....1150014D}, and microlens-etalon coupled spectrographs \citep[e.g.,][]{2018ApJ...867...77R} have all been prototyped, constructed, and tested with great success. Indeed, there are a number of conceptual IFU designs that have been put forward as first-light instruments on the planned European Solar Telescope \citep[EST;][]{2022A&A...666A..21Q}, which is a European led, 4.2~m on-axis facility due to be operational within the next decade. While IFUs for solar science are in their relative infancy, more advancements have been made in the night-time community \citep[see the overview provided by][]{2009arXiv0912.0201L}, with the James Webb Space Telescope employing the Mid-Infrared Instrument \cite[MIRI;][]{2015PASP..127..646W} and the Near-Infrared Spectrograph \citep[NIRSpec;][]{2022A&A...661A..80J} instrument, which employ long-slit configurations similar to an image slicer arrangement \citep{2022SPIE12188E..57R}. Of course, the need for IFUs to place both spatial and spectral domains on a single detector chip naturally results in a reduced field-of-view size, which unfortunately is a limiting factor in any instrument that attempts to observe a three-dimensional data space with two-dimensional detectors. This effect can be minimised during the design phase by using larger format detectors and smaller diameter fibres to better match the diameter of each fibre to the pixel size of the imaging camera.

Fabry-P{\'{e}}rot interferometers, slit-based spectrographs, and IFU spectrographs are all valuable tools in the field of astrophysics, each with its own set of advantages and disadvantages. The choice of which spectrograph to use depends on the specific research objectives and observational conditions. The asynchronous nature of the data taken by Fabry-P{\'{e}}rot interferometers and slit-based spectrographs may result in undesired effects due to changing atmospheric conditions between each exposure, or through the failure to capture dynamic solar features that evolve on timescales less than the typical scan/raster duration \citep{2006A&A...447.1111S, 2023A&A...669A..78S}. IFUs attempt to mitigate this challenge by acquiring spectral and spatial information simultaneously, although they are typically forced to utilise a smaller field-of-view size and/or a reduced spatial resolution \citep{2019OptEn..58h2417I}. Due to IFUs attempting to sample three data dimensions ([$x,y,\lambda$]) with two-dimensional detectors, they become naturally less efficient in the use of imaging detector pixels than their Fabry-P{\'{e}}rot interferometer and slit-based spectrograph counterparts. This means that even with full detector pixel usage efficiency, the field of view would remain smaller than for a comparable Fabry-P{\'{e}}rot instrument at the diffraction limit \citep{2006NewAR..50..244A,2013OptEn..52i0901H}. However, for cases where simultaneous spatial and spectral sampling is paramount, the benefits of IFUs can outweigh the drawbacks of a reduced field of view and/or reduced spatial resolution. For example, it has been shown that the study of impulsive shock waves in sunspot umbral atmospheres can be negatively influenced by the time required for scanning Fabry-P{\'{e}}rot instruments to construct spectral profiles \citep{2018A&A...614A..73F, 2020ApJ...896..150Y}, hence making an IFU the ideal choice of instrument.

To date, many IFUs have been developed to work primarily in the (near-)infrared portion of the electromagnetic spectrum, including DL-NIRSP, the SpectroPolarimetric Imager for the Energetic Sun \citep[SPIES;][]{2012SPIE.8446E..1DL, 2019ApJ...882..161A}, and GRIS. However, the blue/green portion of the electromagnetic spectrum offers numerous spectral lines for examination of the lower solar atmosphere, including recent studies involving Ca~{\sc{ii}}~H/K \citep[e.g.,][]{2013A&A...549A.116J, 2015ApJ...806..132G, 2017ApJS..229...11J, 2022A&A...668A.153M, 2022A&A...664A...8M}, H$\beta$ \citep[e.g.,][]{2020A&A...642A..52A, 2021A&A...648A..54R, 2022ApJ...928..190K, 2022ApJ...937...56K}, Na~{\sc{i}}~D$_{1}$/D$_{2}$ \citep[e.g.,][]{2010ApJ...719L.134J, 2016ApJ...832..147K, 2017ApJ...836...18C}, as well as other magnetically-sensitive lines that have not yet been fully exploited for solar research \citep[e.g., Ca~{\sc{i}} 422.7~nm;][]{2020A&A...641A..63C}, none of which are interspersed with telluric lines. Furthermore, it has been shown that the Ca~{\sc{ii}}~H \& K lines, with effective Land{\'{e}} $g$-factors of $g_{\mathrm{H}} = 1.333$ and $g_{\mathrm{K}} = 1.167$, respectively, are highly suited for circular polarisation studies \citep{1990ApJ...361L..81M}, especially since due to the atomic transition of the Ca~{\sc{ii}}~H line \citep[$j=\frac{1}{2}$, $j'=\frac{1}{2}$;][]{1980A&A....84...60S, 1984SoPh...91....1L}, no linear polarisation is produced in the core of this line due to resonance scattering. Motivated by such cutting-edge studies, it now appears timely to pursue the construction of an IFU dedicated to the blue/green portions of the electromagnetic spectrum. Furthermore, the lower wavelengths associated with these regions, when compared to (near-)infrared alternatives, allows for thinner cladding thicknesses around each fibre core \citep{Goto:15} that helps to promote a better observational filling factor of the 2D fibre array without needing to employ microlenses, which (if used) come at an additional financial cost.

Here, we report on the design, construction, commissioning and science verification of a new fibre-fed IFU named the Fibre Resolved opticAl and Near-ultraviolet Czerny-Turner Imaging Spectropolarimeter ({\sc{francis}}), which is initially put forward as a prototype instrument for the proposed Indian National Large Solar Telescope \citep[NLST;][]{2010AN....331..628H, 2014MNRAS.437.2092D}. The NLST will be a state-of-the-art 2m-class telescope, built close to the Chinese border in Merak, designed for high-resolution studies of the solar atmosphere. Merak, which is in the Himalayan mountain region at an altitude greater than 4000~m, has an extremely low water vapour content, resulting in excellent conditions for optical and infrared observations. Furthermore, the NLST will be the second largest solar telescope in the world when it sees first light in the latter stages of the decade and will be ideally situated 15.5 hours ahead of the world's largest solar telescope (DKIST) in Hawaii, offering the unique ability to provide unrivalled 24/7 ground-based observations of evolving solar phenomena when combined with other world-class optical facilities, such as the those available in North America \citep[e.g., the DST and Big Bear Solar Observatory, BBSO;][]{2010AN....331..620G}, the Canary Islands \citep[e.g., the SST and GREGOR;][]{2012AN....333..796S}, and China \citep[e.g., the New Vacuum Solar Telescope, NVST, and the Chinese Large Solar Telescope, CLST;][]{2014RAA....14..705L, 2015JATIS...1b4001R, 2020SCPMA..6309631R}.

\begin{figure} 
\centerline{\includegraphics[width=0.9\textwidth,clip=]{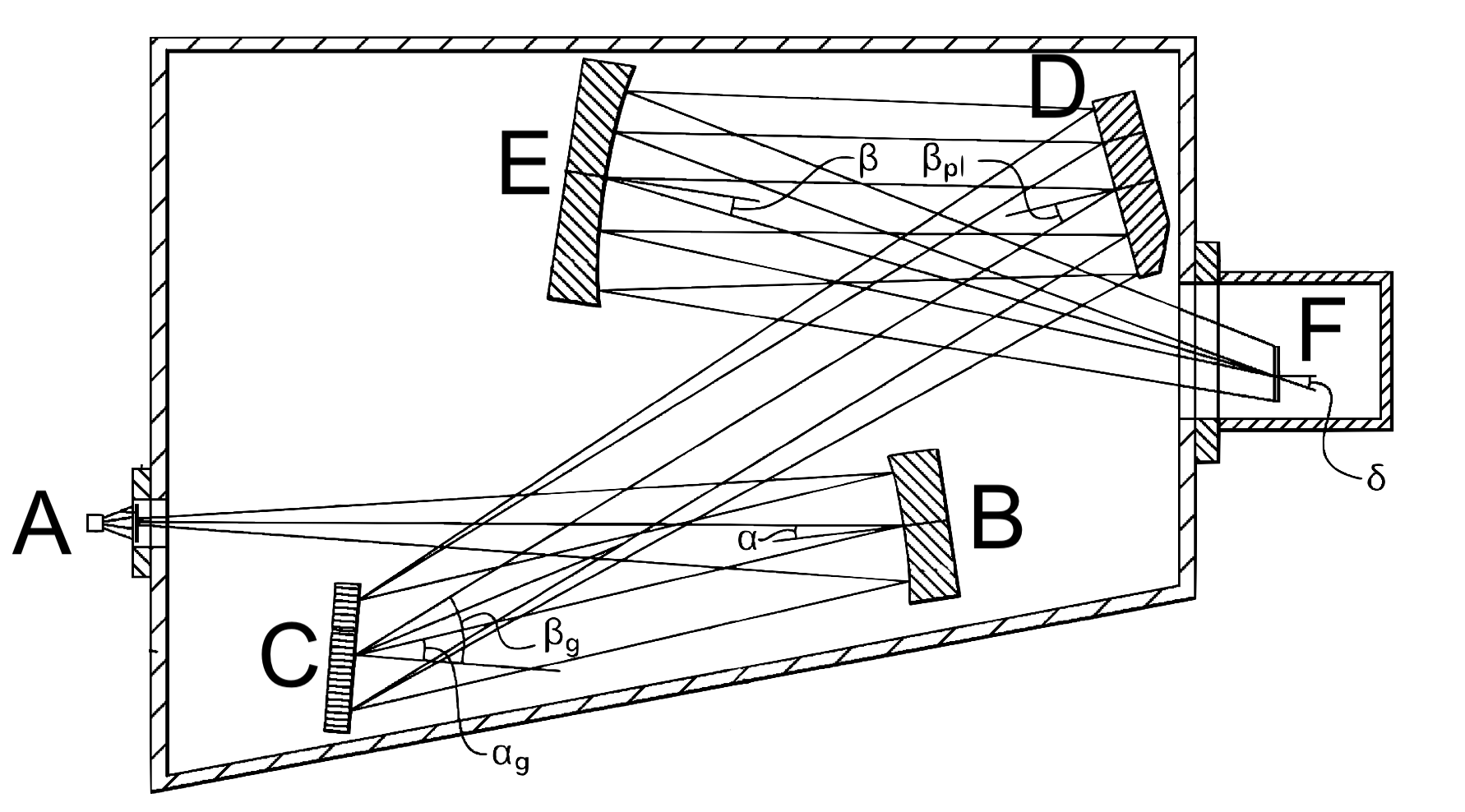}}
\caption{\sloppy{\nohyphens{A simplified schematic of the internal configuration of the {\em{Princeton Instruments}} IsoPlane{\textsuperscript{\tiny\textregistered}}~320 spectrograph housing. `A' represents the linearised fibre array that is aligned with the spectrograph slit, `B' is a concave toroidal-shaped collimating mirror with an off-axis angle, $\alpha$, `C' is the chosen diffraction grating (either 4320~lines/mm, 3600~lines/mm, or 2400~lines/mm) with angles of incidence and refraction, $\alpha_{g}$ and $\beta_{g}$, respectively, `D' is the aspheric aberration correction mirror with an angle of incidence, $\beta_{pl}$, `E' is the concave focusing mirror with an angle of incidence, $\beta$, and `F' is the spectral imaging detector placed at a maximum angle, $\delta$, to the focal plane. The background image is adapted from {\em{Princeton Instruments}} hardware specification documents. }}}
\label{fig:CT_spectrograph_schematic}
\end{figure}

\section{Instrumentation}
\label{sec:instrumentation}
In 2018, a relatively small budget ($\sim$\$120k~USD) was secured to design, build, transport and test an IFU, which would act as a prototype instrument for the upcoming Indian NLST facility. The primary science objectives of the NLST revolve around studying the nature and subsequent consequences of solar magnetism \citep{2010IAUS..264..499H, 2010AN....331..628H}. As outlined by \citet{2008JApA...29..345S}, the NLST aims to investigate the role of magnetohydrodynamic (MHD) waves in the supply of energy flux across different layers of the solar atmosphere, including the determination of their periods across a wide range of spatial and temporal scales. This scientific objective is supported by others, including the study of dynamically evolving structures that rely on high-cadence observations, examination of large-scale features (including pores, sunspots, and entire active regions) and their role in triggering rapid plasma motions, such as those associated with flares, filament eruptions, coronal mass ejections, etc.

Recent work by \citet{2016ExA....42..271R} examined the typical evolutionary timescales of dynamic features in the solar atmosphere and stipulated that a temporal cadence of $\sim 2-3$~s is required to study such processes with the NLST. While investigating potential instrumentation for the NLST, \citet{2008JApA...29..345S} stated that a spectral resolution of a few~m{\AA} is required to adequately resolve the velocity signatures ($\sim100$~m/s) associated with MHD waves in the solar photosphere. However, for chromospheric MHD wave signatures, which motivated the design of the {\sc{francis}} instrument and often have wave amplitudes exceeding $10$~km/s \citep{JessLRSP2023}, the spectral resolution required for chromospheric lines (e.g., Ca~{\sc{ii}}~H/K) can often be relaxed to several tens of~m{\AA} \citep{https://doi.org/10.1002/2016JA022871}. Furthermore, the observation of dynamic MHD wave phenomena, such as the development of umbral flashes in sunspot atmospheres, requires both high time resolution (consistent with $\sim 2-3$~s as stated above) and simultaneous sampling of the entire sunspot umbra, hence making an IFU the logical choice of instrument. Finally, \citet{2010ASSP...19..156S} and \citet{2016ExA....42..271R} provide a list of crucial spectral lines that are required to be observed as part of NLST operations to address the scientific objectives, including the Ca~{\sc{ii}}~K, H$\beta$, and H$\alpha$ absorption lines. As a result, the {\sc{francis}} instrument prototype has the overarching goals of being able to obtain complete spectral mapping of a two-dimensional field of view with a temporal cadence of $<3$~s, a spectral resolution no more than a few tens of~m{\AA}, with sensitivity to key chromospheric spectral lines with wavelengths as low as $\approx390$~nm, including Ca~{\sc{ii}}~K, H$\beta$, and H$\alpha$.

For simplicity and to minimise the number of additional components needed (e.g., microlenses and additional mirrors), a fibre-fed IFU was chosen over its image slicer and microlens array counterparts. With a budget of $\sim$\$120k~USD, it was necessary to obtain all components required for the production of the final instrument, including:
\begin{itemize}
\item Fibre optics (including the array assembly and mapping);
\item Spectrograph housing (including adjustable entrance slits);
\item Diffraction gratings (including movable mounts);
\item Digital camera for acquiring observed spectra;
\item Polarisation optics;
\item High-pass and bandpass filters (including filter wheels);
\item Calibration lamps spanning near-UV and optical wavelengths;
\item Acquisition PC and data storage infrastructure;
\item Control software (spectrograph, camera, polarisation optics, etc.);
\item Water cooling and circulation hardware;
\item Transportation cases; and
\item Shipping and travel costs associated with the commissioning time.
\end{itemize}

Below, we breakdown all of the components selected for the final instrument prototype, before discussing first-light commissioning observations in Section~{\ref{sec:commissioning}}.

\subsection{Spectrograph}
\label{sec:spectrograph}
As a result of the limited budget, it was decided to centre the build on the well-established Czerny-Turner spectrograph housing produced by {\em{Princeton Instruments}}\footnote{{\em{Princeton Instruments}} are part of the Teledyne Imaging Group --- \href{www.teledyneimaging.com}{www.teledyneimaging.com}}. The {\em{Princeton Instruments}} IsoPlane{\textsuperscript{\tiny\textregistered}}~320 spectrograph housing was chosen due to its rugged cast steel construction and previous successful scientific applications in laboratory and optical astrophysics \citep[e.g.,][]{2016RScI...87kD616W, 2021RScI...92f3002K, 2022AcSpe.18806341C}. 

In this configuration, light is incident on the slit plane as a divergent beam, which propagates to a concave toroidal-shaped collimating mirror with an off-axis angle, $\alpha$ (see Figure~{\ref{fig:CT_spectrograph_schematic}}). The light is subsequently reflected as a collimated beam and directed towards the chosen diffraction grating. For the prototype IFU presented here, 3 unique holographic reflection gratings were selected, comprising of 4320~lines/mm, 3600~lines/mm, and 2400~lines/mm variants. Holographic gratings, unlike ruled diffraction gratings, are produced by exposing the underlying photosensitive material to the interference pattern produced by two interfering laser beams. The interference pattern creates a finely structured (periodic) pattern on the surface, which is then chemically treated to expose a sinusoidal surface pattern capable of precise control over light dispersion \citep{1969OptCo...1....5L, 2004PASP..116..403B}. Holographic gratings were selected here due to the optical techniques used in their manufacture not introducing the spacing errors and/or surface irregularities often found in traditional ruled gratings. This helps to not only provide consistent linear dispersion properties, but also suppresses stray light and ghosting artefacts when high-density gratings are required \citep{2013SPIE.8870E..0HS}. The 4320~lines/mm, 3600~lines/mm, and 2400~lines/mm gratings employed in the {\sc{francis}} instrument were manufactured by {\em{Richardson Gratings}}\footnote{{\em{Richardson Gratings}} are part of the Newport Corporation --- \href{www.newport.com}{www.newport.com}}, with the area of each grating equal to $102\times102$~mm$^{2}$. 

Following reflection off the concave toroidal-shaped mirror (labelled `B' in Figure~{\ref{fig:CT_spectrograph_schematic}}), the collimated beam strikes the grating with an angle of incidence, $\alpha_{g}$, and is diffracted as a dispersed beam with an angle of refraction, $\beta_{g}$. The diffracted beam from the chosen holographic reflection grating is directed towards an aspheric aberration correction mirror (which is rotationally symmetric) with an angle of incidence, $\beta_{pl}$, before being deflected towards an aspheric concave focusing mirror with an angle of incidence, $\beta$. Following reflection from the final focusing mirror, the convergent beam forms images of the dispersed input light on to a detector placed in the focal plane. Here, the detector may be placed with a maximum angle, $\delta$, to the imaging plane, in order to capture spectral images that are close to being free from axial and field aberrations (i.e., is approximately anastigmatic). A simplified schematic of the {\em{Princeton Instruments}} IsoPlane{\textsuperscript{\tiny\textregistered}}~320 spectrograph housing, including internal optics and the diffraction grating, is shown in Figure~{\ref{fig:CT_spectrograph_schematic}}.

\subsection{Fibre bundle}
\label{sec:fibrebundle}
To feed the spectrograph, an optical fibre bundle was designed to capture an approximately square two-dimensional field-of-view, then linearise the fibres into a one-dimensional array that can be placed against the entrance slit of the spectrograph. Initially, when the project began, the available bench space at the NLST was unknown due to the ongoing design phase. Thus, a $1.5$~m long optical fibre bundle was chosen to ensure it had the capability to function with different configurations at the future telescope site. Shorter fibre bundles would help minimise the risk of stress-related fibre breakages \citep{ELABDI2008222}, but this risk was deemed necessary as part of the prototype trialling of the instrument. However, to help protect the fibre bundle from excess stress being exerted on the fibres, the final bundle was encased using shrink-fit PVC monocoil tubing.

\begin{figure} 
\centerline{
\includegraphics[width=0.49\textwidth, trim= 110mm 120mm 55.5mm 120mm, clip=]{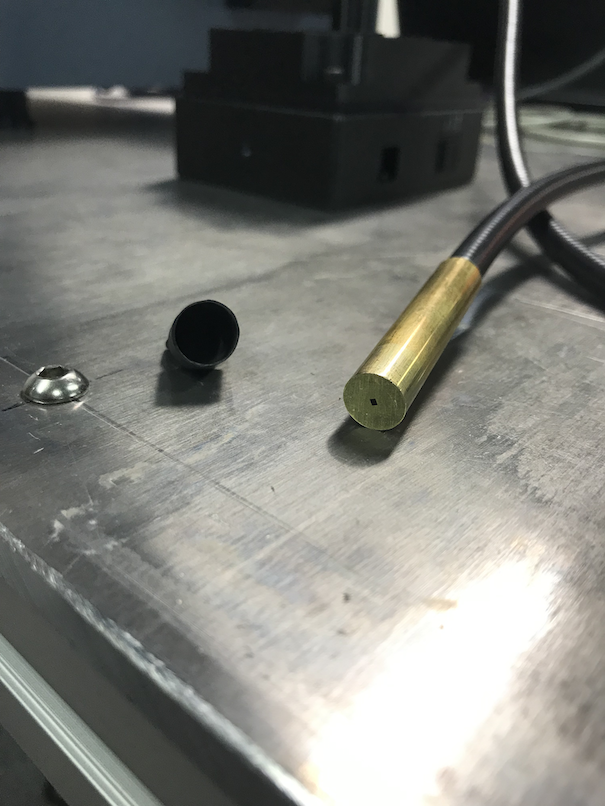}
\includegraphics[width=0.49\textwidth, trim= 133mm 65mm 72mm 75mm, clip=]{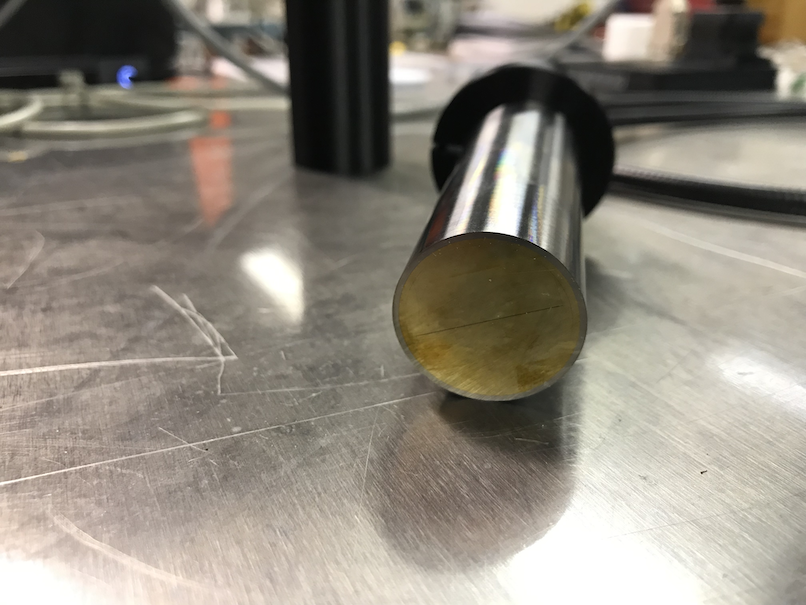}
}
\caption{\sloppy{\nohyphens{{\em{Left:}} A view of the two-dimensional fibre end, where the 400 optical fibres are arranged in a closely packed $20\times20$ hexagonal configuration, which is seen as the dark rectangular region towards the centre of the brass ferrule. The surface area covered by the fibre array is $1.082$~mm$^{2}$, with light transmission provided by the cores of the optical fibres with an effective area of $0.503$~mm$^{2}$, hence providing a filling factor of approximately $46.5$\%. The brass end of the ferrule is polished to a flatness that is $<1~\mu$m, enabling a back reflection to be captured by an additional camera to help with fibre alignment. {\em{Right:}} The linearised end of the fibre bundle, where all 400 fibres are arranged in a one-dimensional configuration that is $22.0$~mm long, which is seen in the image as the narrow dark band spanning the centre of the brass ferrule. The linearisation of the fibre array enables the propagated light to pass through the entrance slit of the spectrograph and, with the mapping of the fibres known, two-dimensional spectral images to be reformed following calibration.  }}}
\label{fig:fibre}
\end{figure}

As the primary goal was to obtain spectra in the near-UV and blue portion of the electromagnetic spectrum, we were able to utilise thinner cladding thicknesses around the optical fibre cores \citep{Goto:15}, yet still retain total internal reflection along the fibres for maximum ray (critical) angles $<10^{\circ}$. Circular cross-section fibres were provided by {\em{FiberTech Optica}}\footnote{{\em{FiberTech Optica}} --- \href{https://fibertech-optica.com}{https://fibertech-optica.com}}, comprising of a $40~\mu$m diameter silica core, a $4~\mu$m thick silica cladding, and a $3.5~\mu$m thick polyimide coating, providing a total diameter of $55~\mu$m for each fibre and a numerical aperture, $NA=0.22\pm0.02$. At the two-dimensional array end of the fibre bundle, a $20\times20$ closely packed hexagonal configuration was chosen to maximise the filling factor of the fibres to light, providing a total of 400 individual fibres to be placed in the focal plane of the incident beam. The final dimensions of the two-dimensional array is $1.127\times0.960$~mm$^{2}$, providing an occupied area of $1.082$~mm$^{2}$. Each fibre core has a surface area of $1.257\times10^{-3}$~mm$^{2}$, hence 400 fibres cover a total area of $0.503$~mm$^{2}$, providing an overall filling factor of $46.5$\%. This is a comparable filling factor to the DL-NIRSP IFU commissioned on DKIST \citep{2022SoPh..297..137J}.

\begin{table}[!t]
\caption{Overview of prototype {\sc{francis}} fibre characteristics employed during the August~2022 commissioning phase.}
\label{tab:francisoverview}
\begin{tabular}{| l | c |}    
\hline
Fibre parameter & Value \\
\hline
Number of fibres & 400 \\
Bundle configuration & $20\times20$ \\
Fibre cross-sectional shape & Circular \\
Fibre material & 100\% multi-mode silica \\
Fibre core diameter & 40.0~$\mu$m \\
Fibre cladding thickness & 4.0~$\mu$m \\
Fibre coating thickness & 3.5~$\mu$m \\
Total fibre diameter & 55.0~$\mu$m \\
Fibre packing configuration & Hexagonal \\
{\bf{2D fibre end}} & \\
\hspace{5mm}Dimensions of bundle group & $1.127\times0.960$~mm$^{2}$ \\
\hspace{5mm}Bundle group termination & Stainless steel ferrule with brass insert \\ 
\hspace{5mm}Dimensions of bundle group ferrule & 10~mm diameter / 50~mm length \\
{\bf{Linear fibre end}} & \\
\hspace{5mm}Dimensions of linearised fibre group & $22{\,}000\times55$~$\mu$m$^{2}$ \\
\hspace{5mm}Linearised fibre group termination & Stainless steel ferrule \\
\hspace{5mm}Dimensions of linearised group ferrule & 28~mm diameter / 100~mm length \\
Numerical aperture ($NA$) & $0.22\pm0.02$ \\
Overall length of fibre array & 1.5~m \\
Fibre array sheathing & Black PVC monocoil tubing \\
\hline
\end{tabular} \\
\end{table}

The two-dimensional fibre array is housed at the centre of a circular brass ferrule with a diameter of $10$~mm (see the left panel of Figure~{\ref{fig:fibre}}). The brass end is polished to a flatness that is $<1~\mu$m, which enables the ferrule to provide a back-reflection image that can be captured by another camera to assist with the co-alignment between the observed spectra and the targeted field-of-view. For this purpose, a ROSA \citep{2010SoPh..261..363J} CCD camera was employed to capture the back-reflection image, which allows use of the master ROSA synchronisation trigger to ensure simultaneity between contextual/back-reflection imaging and the acquired spectra\footnote{Note --- the back-reflection setup is affectionately known as the {\em{Detecting the Orientation of the bUndle Group camera}} ({\sc{DOUGcam}}), in honour of the tireless assistance provided by D.~Gilliam during the initial commissioning run of {\sc{francis}}.}. Over the $1.5$~m length of the optical fibre, the $20\times20$ configuration of the fibre bundle is linearised into a one-dimensional array prior to being mounted at the entrance slit of the spectrograph (see the right panel of Figure~{\ref{fig:fibre}}). Here, all 400 fibres arranged side-by-side provide an array length equal to $22.0$~mm, which required a larger brass ferrule to accommodate this increased dimension over the original two-dimensional layout that utilised a $10$~mm diameter brass ferrule. 

As discussed by \citet{lpor.202000553}, multi-mode fibres are sensitive to various types of fluctuations, such as thermal, acoustic, or mechanical vibrations, which can induce modal interference as a result of changes in the phase of the guided modes. Here, the permitted number of transversal electromagnetic modes within an idealised circular cross-section fibre is proportional to the square of the fibre-core radius \citep{doi:10.1080/713820379, 1980OptL....5..270H}. Experimental characterisation of optical fibres by \citet{2001PASP..113..851B} revealed that modal noise is reduced when the energy is spread out over more permitted modes. Furthermore, \citet{2001PASP..113..851B} found that lower wavelengths provide higher numbers of available modes within the same fibre. Hence, the combination of large-diameter fibre cores and the blueward wavelength regime intrinsic to the {\sc{francis}} instrument is ideal for minimising transmitted modal noise within the fibres. To further minimise modal interference, the fibre bundle is securely clamped to the optical bench and remains fixed throughout the duration of any solar observing campaign, which helps to mitigate time-varying modal interference patterns arising from telescope vibrations and/or accidental movements of the fibre bundle. Furthermore, to protect the brass ferrule from damage when being mechanically clamped to the entrance slit of the spectrograph, it was compression fit inside a larger ($28$~mm diameter) stainless steel tube, which is visible around the outside of the brass insert in the right panel of Figure~{\ref{fig:fibre}}. 

Finally, the ferrule housing the linearised array is positioned in-line with the entrance slit, which is operated with a slit width of $50~\mu$m to ensure all light from the individual fibres is allowed to propagate into the spectrograph. Leading on from the discussions above concerning modal interference patterns with multi-mode fibres, it must be highlighted that inherent fibre properties (e.g., imperfections at the core-cladding interface) can produce time-varying speckle patterns at the fibre exit that are modulated by any unconstrained mechanical stresses and/or temperature shifts \citep{Epworth:79}. As described by \citet{1981OptL....6..324G}, and experimentally demonstrated by \citet{2011MNRAS.417..689L}, the fraction of the illuminated fibre end face that is ultimately transmitted to the imaging detector plays an important role in ensuring the response of the instrument remains stable with time. It is important to note that a reduction in the transmitted fraction of light coming from the fibre does not necessarily only stem from the use of a narrow slit at the entrance to the spectrograph (i.e., a slit width less than the diameter of the fibre core), but will also result from overfilling the diffraction grating in an attempt to increase the spectral resolution. For the initial commissioning of the {\sc{francis}} instrument, we chose to implement a slit width ($50~\mu$m) that was larger than the diameter of the fibre cores ($40~\mu$m) to ensure modal interference noise was minimised in the captured spectra. This approach, in addition to the scrambling processes \citep[see, e.g.,][]{2008SPIE.7018E..4WA} inherent to fibres, also help to minimise uncertainties from different spatial and angular information arising from variable seeing conditions, since the light that is coupled into the fibres largely results in a similar response from the spectrograph and thus helps to stabilise the signal \citep{2011MNRAS.417..689L}. An overview of the optical fibre bundle employed is provided in Table~{\ref{tab:francisoverview}}. 

\begin{figure}[!t]
\centerline{
\includegraphics[width=0.9\textwidth, trim= 0mm 0mm 0mm 0mm, clip=]{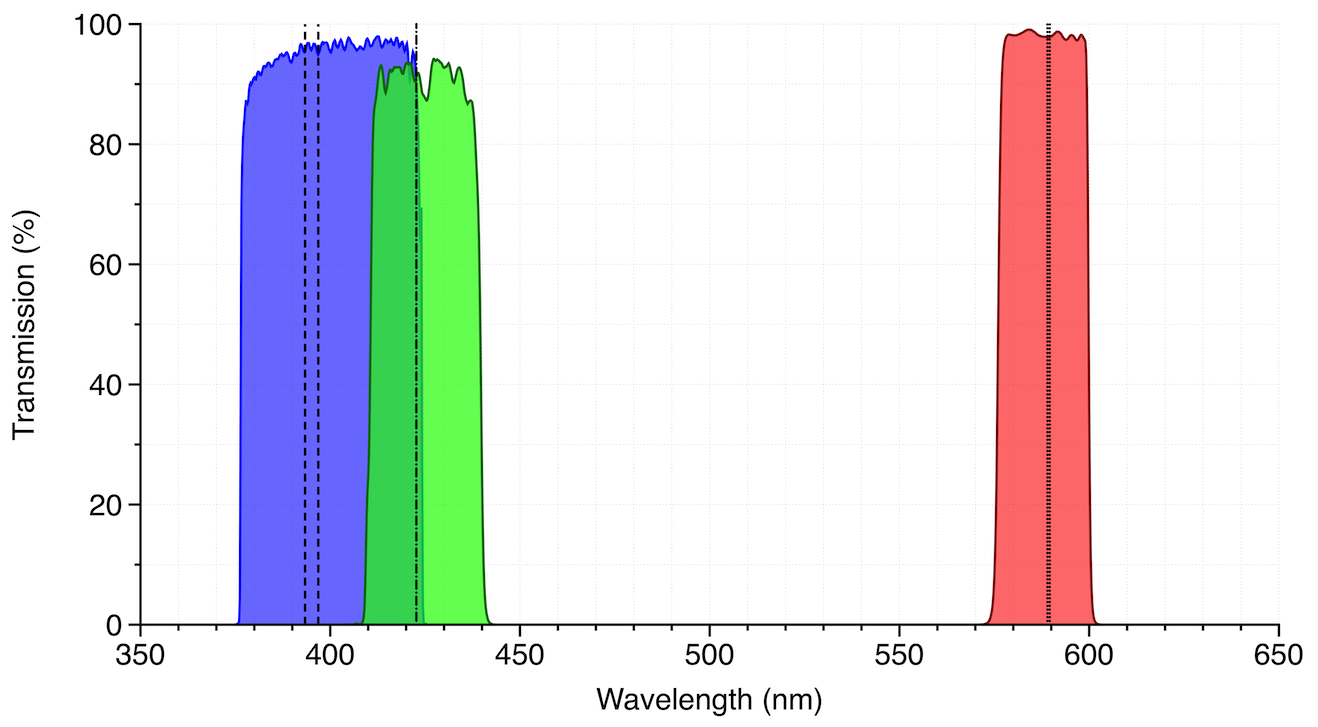}
}
\caption{\sloppy{\nohyphens{Transmission profiles of the three different bandpass filters selected to prevent conflicting spectral orders from contaminating the final spectral images. The shaded blue, green, and red bandpass filters span $376-424$~nm, $411-439$~nm, and $574-600$~nm, respectively. Each filter allows for the isolation of prioritised spectral lines, including the Ca~{\sc{ii}}~H \& K lines ($396.847$ and $393.366$~nm, respectively, highlighted using vertical dashed black lines), the Ca~{\sc{i}} line ($422.673$~nm, indicated with the vertical dot-dash line), and the Na~{\sc{i}}~D$_{1}$ \& D$_{2}$ line pair ($589.592$~nm and $588.995$~nm, respectively, highlighted using the vertical dotted black lines). Each bandpass filter has transmissions $>90$\% at the wavelengths of interest. }}}
\label{fig:Semrock}
\end{figure}

\subsection{Wavelength order filters}
\label{sec:wavelengthorderfilters}
The transmission profile of the DST spans the near ultraviolet ($\sim350$~nm) through to the far infrared. Light incident on a diffraction grating produces multiple orders, whereby monochromatic light at a certain wavelength appears at more than one angle of diffraction. As a result, the superposition of different orders of diffracted light on the detector can lead to ambiguous spectral data. Hence, bandpass filter sets were purchased to help eliminate conflicting spectral orders and allow for unambiguous capture of spectral signatures from solar sources. 

Specific spectral lines were prioritised for the purposes of obtaining suitable bandpass filters. In total, $3$ lines were chosen: (1) Ca~{\sc{ii}}~H/K at $\sim395$~nm, (2) Ca~{\sc{i}} at $\sim423$~nm, and (3) Na~{\sc{i}}~D$_{1}$/D$_{2}$ at $\sim589$~nm, with the final filters manufactured by {\em{Semrock}}\footnote{{\em{Semrock}} are part of the IDEX Health \& Science corporation --- \href{https://www.idex-hs.com/semrock}{https://www.idex-hs.com/semrock}}. Each filter was mounted in a circular housing with an aperture diameter of $50$~mm to accommodate a wide range of future optical beam sizes. At the wavelengths corresponding to the prioritised spectral lines, each bandpass filter has a transmission exceeding $90$\%, which helps to maximise light reaching the spectrograph (see Figure~{\ref{fig:Semrock}}). The selected filter is placed upstream of the two-dimensional fibre head, thus pre-selecting the wavelengths of interest before being guided into the spectrograph by the fibre bundle. 

\subsection{Camera and data storage}
\label{sec:camera}
\begin{figure}[!t]
\centerline{
\includegraphics[width=0.8\textwidth, trim= 0mm 0mm 0mm 0mm, clip=]{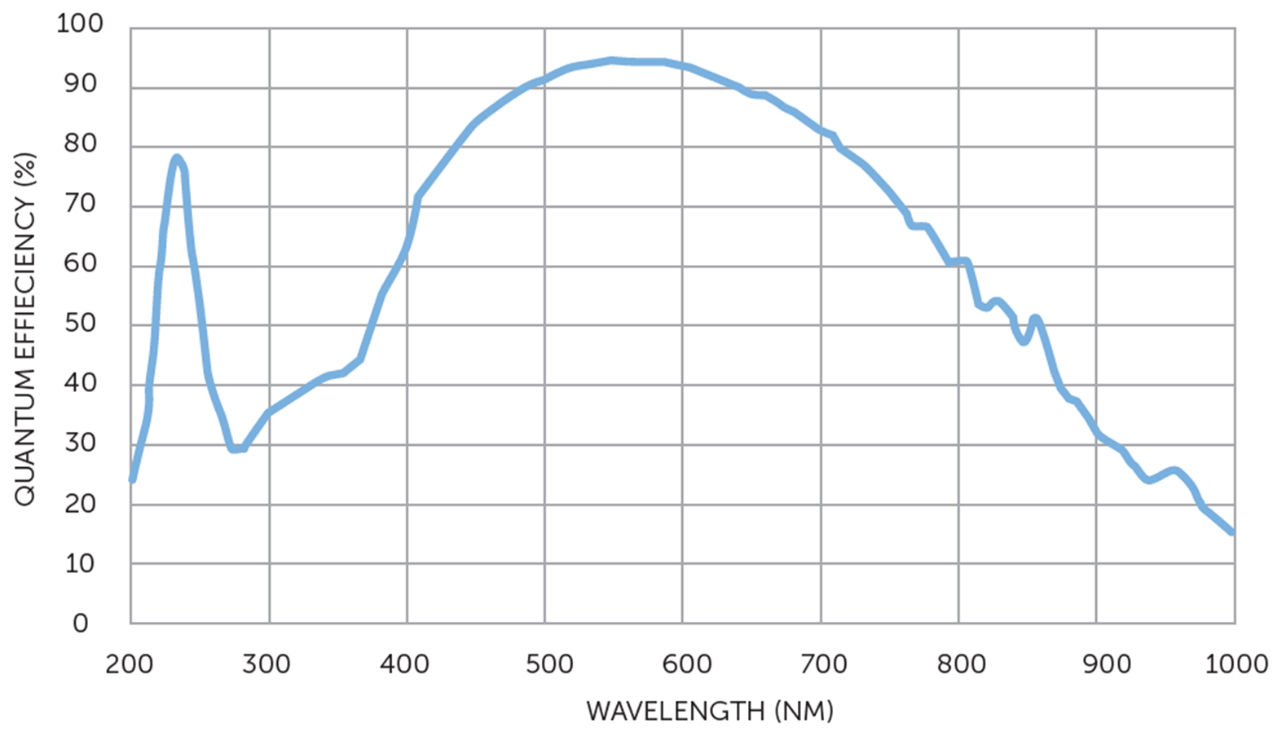}
}
\caption{\sloppy{\nohyphens{The quantum efficiencies (QEs) of the {\em{Princeton Instruments}} Kuro~2048B scientific CMOS camera chosen to image the fibre-resolved spectra. {\sc{francis}} operates within the wavelength range of $390-700$~nm, hence this camera provides QEs~$>60$~\% across the useable wavelength domain. This image is kindly reproduced from the Teledyne Imaging Group resource centre (\href{https://www.photometrics.com/learn/imaging-topics/quantum-efficiency}{https://www.photometrics.com/learn/imaging-topics/quantum-efficiency}).  }}}
\label{fig:QE}
\end{figure}

To image the resulting spectra, a {\em{Princeton Instruments}} Kuro~2048B scientific CMOS camera was selected. This camera has a back-illuminated $2048\times2048$~pixel$^{2}$ imaging array and utilises $11\times11~\mu$m$^{2}$ pixels to form an imaging area that is $22.53\times22.53$~mm$^{2}$. A large imaging area was necessary to accommodate all 400 optical fibres once re-formatted into a linearised array. As highlighted in Table~{\ref{tab:francisoverview}}, each fibre has an outer diameter of $55~\mu$m, resulting in a final length of $22.0$~mm once linearised prior to entry into the spectrograph. Due to the spectrograph employing a $1:1$ magnification, each imaged fibre also has a total diameter of $55~\mu$m, resulting in its information being spread across $\approx5$ detector pixels, thus creating an oversampled final spectral image. However, importantly, the imaging chip length of $22.53$~mm is able to accommodate all 400 optical fibres within a single frame. The Kuro~2048B camera has a well depth of $80{\,}000$~e$^{-}$ and provides $16$-bit readout at a maximum rate of $23$~frames per second. A readout noise of $1.5$~e$^{-}$ (rms) is maintained using liquid-assisted cooling, providing a dark current of approximately $0.7$~e$^{-}$/pixel/s at $-25\,^{\circ}$C. For this setup, a {\em{Koolance}}\footnote{{\em{Koolance}} --- \href{https://koolance.com}{https://koolance.com}} EX2-1055 liquid cooling system was employed, which has a cooling capacity of $900$~W (or $3071$~BTU/hr) with a $25\,^{\circ}$C liquid-ambient delta temperature at a circulation rate of $4.5$~litres/min. A peak quantum efficiency (QE) of $>95$~\% is found at approximately $560$~nm, with QEs~$>60$~\% across the entire wavelength range sampled by {\sc{francis}} ($390-700$~nm; see Figure~{\ref{fig:QE}} for more details).

The camera utilises a rolling shutter and communicates to the storage PC using a USB~3.0 interface. Basic setup (e.g., exposure times, windowing and binning functions, etc.) is controlled via {\em{Princeton Instruments}} LightField{\textsuperscript{\tiny\textregistered}} software that is installed on the main storage PC. Here, the PC employed is a simple Microsoft Windows-based laptop that is connected via a USB~3.1 (10~GB/s) interface to an external hard disk enclosure running two 4~TB solid-state disks in a RAID-0 configuration. Data is saved directly from LightField{\textsuperscript{\tiny\textregistered}} on to the 8~TB disk, enabling several hours of unbroken observations to be accumulated. Additionally, the ROSA synchronisation box \citep[see][for more information]{2010SoPh..261..363J} can be connected to the Kuro~2048B camera by way of a TTL voltage signal, allowing the camera sequence to be triggered by the master ROSA machine for microsecond precision spectra/imaging capabilities.

\subsection{Polarisation optics}
\label{sec:polarisationoptics}
To enable polarimetric measurements of the incident light, a number of additional polarisation optics can be placed upstream of the fibre bundle. As discussed by \citet{okamoto2006optical}, it is not typical for the polarisation state of light to be maintained while propagating through fibre optics. Aspects that are often difficult to alleviate completely, including mechanical vibrations, temperature variations, and slight bending of the fibre bundle can readily manipulate the final polarisation state of the output light \citep{2022SoPh..297..137J}. As a result, it was decided that any polarisation optics introduced should be placed upstream of the fibres, hence pre-selecting the polarisation state to be transmitted and imaged, thus minimising potential impacts caused by the fibres themselves. To accomplish this, 2 liquid crystal variable retarders (LCVRs) were purchased from {\em{Meadowlark Optics}}\footnote{{\em{Meadowlark Optics}} --- \href{https://www.meadowlark.com/}{https://www.meadowlark.com/}}, each with a UV coating to allow operation within the wavelength range of $390-600$~nm and a large clear aperture of $40.6$~mm to facilitate future optical setups that may employ large beam sizes. The LCVRs are also temperature controlled via a digital interface, and offer typical response times of $\approx5$~ms to switch from one-half to zero waves (i.e., low to high voltage) and $\approx20$~ms to switch from zero to one-half wave (i.e., high to low voltage).

\begin{figure}[!t]
\centerline{
\includegraphics[width=1.0\textwidth, trim= 0mm 0mm 0mm 0mm, clip=]{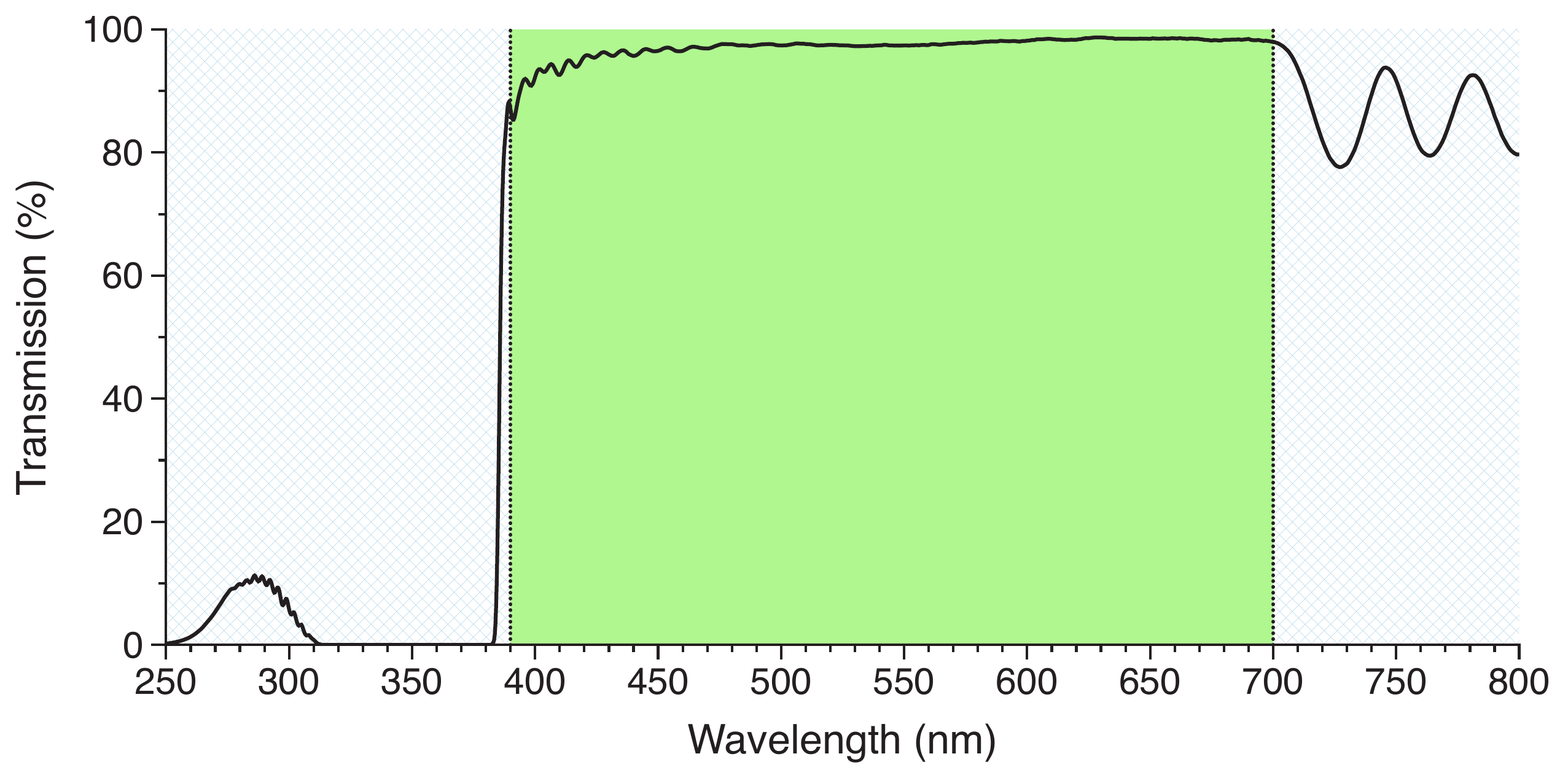}
}
\caption{\sloppy{\nohyphens{The transmission curve of the custom high-pass filter manufactured to protect the polarisation optics from wavelengths $\le380$~nm, which can damage the organic material found in the LCVRs. The typical spectral range of {\sc{francis}} is highlighted using the shaded green region that is bounded by the vertical black dotted lines, while the blue hashed regions denote spectral windows that are outside of the initial instrument specifications. The bump of enhanced transmission between $260-310$~nm is not a concern since it falls below the typical atmopsheric cut-off wavelength found at ground-based telescope sites. The average transmission of this high-pass filter is $97.1$\% across the spectral range of $390-700$~nm. }}}
\label{fig:HPF}
\end{figure}

\begin{figure}[!t]
\centerline{
\includegraphics[width=1.0\textwidth, trim= 0mm 0mm 0mm 0mm, clip=]{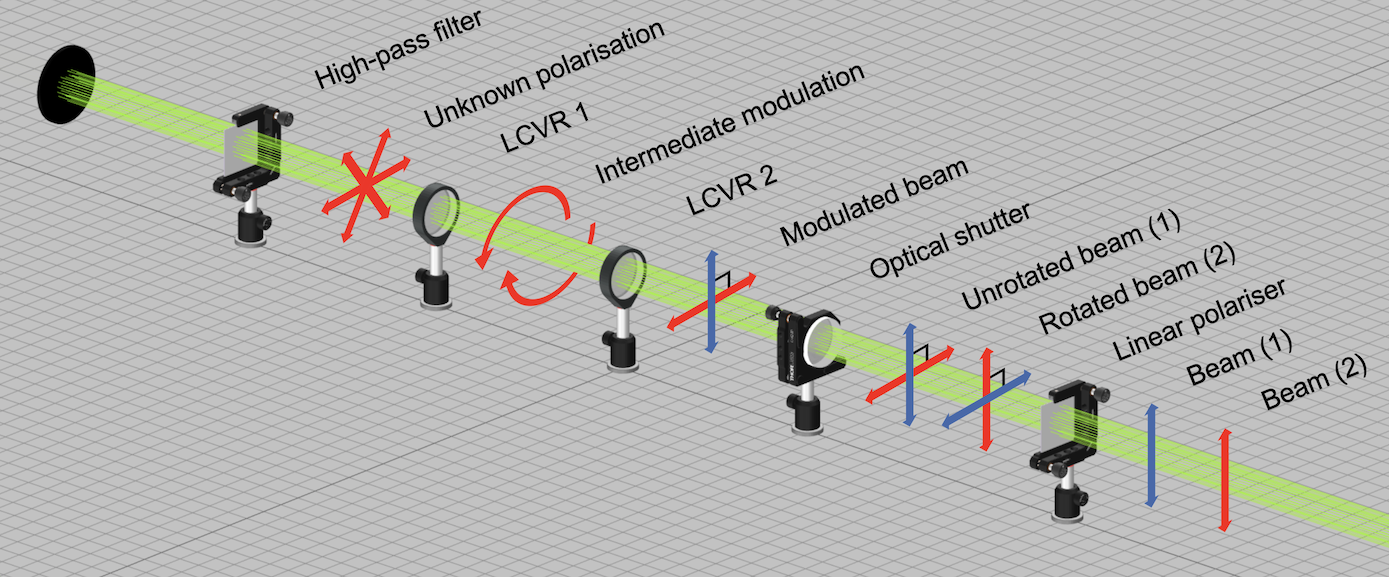}
}
\caption{\sloppy{\nohyphens{A schematic of the components making up the polarisation optics for the {\sc{francis}} spectrograph. Incident light from the telescope is denoted by the black circular shape in the upper-left corner. The light rays, highlighted using solid green tracers, then pass through a high-pass filter to remove wavelengths $<380$~nm that are damaging to the organic material within the LCVRs. The unpolarised light then passes through the 2 LCVRs to produce a modulated beam, where the polarisation state is still unknown, but modified by the LCVRs in a pre-determined way. Two possible components of light output by the LCVRs are shown for visualisation purposes using blue and red arrows, each of which are polarised perpendicular to one another. A twisted nematic liquid crystal cell (i.e., an optical shutter) is then used to rotate the beam by either $0^{\circ}$ or $90^{\circ}$, producing the unrotated beam (1) and the rotated beam (2), respectively, which are indicated in the figure (note: a conventional dual-beam setup employs a polarising beam splitter or a Wollaston prism to negate the need for an optical shutter; see the main text for more information). Finally, the beam can then be analysed by passing through the linear polariser to produce a measurable intensity signal corresponding to the modulated polarisation state produced by the combination of the LCVRs, optical shutter, and linear polariser. Here, beam (1) and beam (2) can be selected independently from one another by using the optical shutter to rotate the beam by $0^{\circ}$ and $90^{\circ}$, respectively. The modulated light then forms an image on the two-dimensional fibre array, for propagation into the spectrograph for subsequent Stokes imaging. Optical component images provided courtesy of {\em{3DOptix}} (\href{https://3doptix.com}{https://3doptix.com}). }}}
\label{fig:polarisationoptics}
\end{figure}

However, care must be taken with the LCVRs as their organic properties means they can become degraded if exposed to wavelengths $<380$~nm. To protect the LCVRs, a custom high-pass filter was sourced from {\em{Laser 2000}}\footnote{{\em{Laser 2000}} --- \href{https://photonics.laser2000.co.uk}{https://photonics.laser2000.co.uk}} that offers $\approx0$\% transmission for wavelengths $<250$~nm, or an equivalent optical density (OD; where the OD is the negative of the base-10 logarithm of the transmission that varies between 0 and 1) of $\mathrm{OD}>6$ (see Figure~{\ref{fig:HPF}}). There is a slight increase in transmission between $260-310$~nm that peaks at $\approx10$\% ($\mathrm{OD}\ge0.977$), although this is beneath the atmospheric cut-off wavelength of $\sim310$~nm found at ground-based telescope facilities \citep{1998JGR...10331541E, 2019APJAS..55..165L}. Transmission remains $\approx0$\% ($\mathrm{OD}>6$) up to $\approx382$~nm, before rising steeply to $>83$\% ($\mathrm{OD}\le0.080$) at $\approx388$~nm. Transmission remains $>87$\% ($\mathrm{OD}\le0.060$) across the operational spectral range of $390-700$~nm as shown in Figure~{\ref{fig:HPF}}, with a mean transmission of $97.1$\% ($\mathrm{OD}=0.012$).

Typically, once the light has passed through the 2 LCVRs, a dual-beam configuration will separate the polarisation states of the modulated beam, using either a polarising beam splitter or a Wollaston prism, and image these states independently \citep[e.g.,][]{2008BASI...36...99N, 2022SoPh..297...22D, 2022SoPh..297..137J}. However, implementing a dual-beam setup here would be beyond the scope of the current prototype, especially if additional imaging detectors and optics were required. Even if the single {\em{Princeton Instruments}} Kuro~2048B camera was employed to capture both modulated states in the most simplistic configuration, this would halve the available spectral range, which is something we wanted to avoid. As a result, for the {\sc{francis}} prototype, we decided to implement a combination of both a twisted nematic liquid crystal cell (i.e., an optical shutter, also obtained from {\em{Meadowlark Optics}}) and a linear polariser to sequentially alternate each modulated polarisation state through the spectrograph. In this configuration, the optical shutter is able to rotate the output beam from the LCVRs by either $0^{\circ}$ or $90^{\circ}$ through the application of a high-frequency voltage source. With $0$\,V applied, the optical beam is rotated by $90^{\circ}$, while the application of a $8$\,V/$2$\,kHz AC square-wave signal causes the optical beam to pass straight through the optical shutter without rotation. Timescales are $0.4$~ms and $5.0$~ms for $0^{\circ}\rightarrow90^{\circ}$ and $90^{\circ}\rightarrow0^{\circ}$ beam rotations, respectively. Like the LCVRs, the optical shutter is temperature controlled with a clear aperture of $40.6$~mm to accommodate future telescope facilities operating with large beam sizes and has a UV coating to enable use with wavelengths as low as $390$~nm. Once the modulated beam has been rotated (or not), an ultrabroadband ($300-2700$~nm operating range) linear polariser is used to analyse the output beam by creating a measurable intensity signal corresponding to the modulated polarisation state produced by the combination of the LCVRs, optical shutter, and linear polariser, which is then focused on the two-dimensional ($20\times20$) fibre array before being propagated into the spectrograph for subsequent spectral imaging. A schematic of the polarisation optics are shown in Figure~{\ref{fig:polarisationoptics}} for easier visualisation. 

The polarisation optics include three components (2~LCVRs and 1~optical shutter) that require computer control to ensure the beam modulation is synchronised with the spectrograph (if adjusting the chosen grating and/or central wavelength of the spectral range) and acquisition camera (to make sure spectral images are not obtained during the modulation transition phases). Thankfully, the digital control interfaces provided by {\em{Meadowlark Optics}} (for the 2~LCVRs and the optical shutter) and {\em{Princeton Instruments}} (for the Kuro~2048B sCMOS camera) come complete with the necessary software development kit (SDK) to enable the instrument hardware to be paired with and controlled by the LabVIEW{\textsuperscript{\tiny\textregistered}} software provided by {\em{National Instruments}}. LabVIEW{\textsuperscript{\tiny\textregistered}} is a graphical programming environment that enables the automation of LCVR/optical shutter configurations, diffraction grating rotations, and image acquisitions to ensure that the liquid crystals have settled in their selected states before exposing the camera to the modulated incident light. While it is possible to configure the liquid crystals, diffraction gratings, and camera settings manually, it is significantly more robust and streamlined to pair all pieces of hardware (using their respective SDKs) under the same software umbrella. This type of solar instrumentation automation using the LabVIEW{\textsuperscript{\tiny\textregistered}} environment has been successfully demonstrated in a number of previous projects \citep[e.g.,][]{Kontani2010, 2011A&A...534A.105B, 2017MNRAS.465.1601B, 10.1093/pasj/psaa006}.

\section{Instrument commissioning and science verification}
\label{sec:commissioning}
The {\sc{francis}} instrument was designed and constructed, with final laboratory tests being conducted on-time towards the start of 2022. However, at this stage, construction of the Indian NLST had not yet commenced. As a result, it was decided that initial science verification should be performed on another facility, notably the Dunn Solar Telescope (DST), in the Sacramento Peak mountains, New Mexico, USA during the summer of 2022. Due to busy scheduling at the DST, we were only able to be granted 10~days of telescope time during the latter stages of August 2022. Thankfully, following a few days of intermittent sunshine and poor seeing conditions while the instrument was being setup and focused on the optical bench, the weather improved and we were presented with a few continuous hours of cloudless skies and excellent atmospheric seeing. For the first stage of instrument commissioning presented here, the polarisation optics were not utilised, which was a consequence of the limited (10-day) observing time available to the team to setup, test, refine, benchmark, then pack away the complete instrument. However, as discussed in Section~{\ref{sec:currentstatus}} below, addition of the polarisation optics is a priority for the next observing campaign with {\sc{francis}}.

\subsection{Data overview}
\label{sec:dataoverview}
The data sequence presented here as part of the science verification process was obtained between 18:03 -- 18:39~UT on 2022 August 29. A sunspot group within active region NOAA~13089 was observed under excellent seeing conditions, which was positioned at heliocentric co-ordinates ($-86''$, $-491''$), or S23.9E5.7 in the conventional heliographic co-ordinate system. High-order adaptive optics \citep{2004SPIE.5490...34R} were utilised to further improve the image quality. Due to the finite number of optical fibres available, it was decided that the $20\times20$ two-dimensional array bundle should cover an approximate $30\times30$~arcsec$^{2}$ field-of-view. This configuration would place $\approx1{\,}.{\!\!}''5$ across the diameter of each fibre, providing a ideal compromise between high spatial sampling and overall field-of-view coverage. Of course, placing $\approx1{\,}.{\!\!}''5$ across the entire $55\,\mu$m fibre diameter means that only the transmitting core will sample $\approx1{\,}.{\!\!}''1$ across its $40\,\mu$m diameter. Nevertheless, this sampling interval would ensure the complete field-of-view sampled by {\sc{francis}} was approximately $30\times30$~arcsec$^{2}$. To accomplish this, the focal plane of the DST was required to have a spatial sampling of $\approx29{\,}.{\!\!}''3$/mm, which was facilitated through use of a focusing lens with a $200$~mm focal length (acting on an $\approx22$~mm pupil), resulting in an $f$/9 beam placed on the {\sc{francis}} ferrule. Importantly, this produced a ray angle $<2^{\circ}$, which is significantly lower than the maximum permissible for the fibre bundle ($10^{\circ}$). 

For the {\sc{francis}} data presented here, the $2400$~lines/mm holographic diffraction grating was implemented to sample the Na~{\sc{i}}~D$_{1}$ \& D$_{2}$ absorption line pair located at $589.592$~nm and $588.995$~nm, respectively. A $25$~ms exposure time was used for the Na~{\sc{i}}~D$_{1}$/D$_{2}$ observations to acquire Stokes~$I$ spectra of the sunspot region, which provided a spectral imaging rate of $23$~s$^{-1}$. Hence, $50\,000$ spatially-resolved spectral images were obtained across the $36$~min duration of the observing sequence. Immediately after the Na~{\sc{i}}~D$_{1}$/D$_{2}$ acquisitions were complete, the $3600$~lines/mm diffraction grating was employed with a $100$~ms exposure time to provide Stokes~$I$ sunspot spectra, centred on the Ca~{\sc{ii}}~H/K absorption line pair at $396.847$~nm and $393.366$~nm, respectively, with an imaging rate of $9$~s$^{-1}$, hence obtaining $200$~spectral images in just over $20$~s of observing time. 

In addition to the {\sc{francis}} instrument, the Rapid Oscillations in the Solar Atmosphere \citep[ROSA;][]{2010SoPh..261..363J} and Hydrogen-Alpha Rapid Dynamics camera \citep[HARDcam;][]{2012ApJ...757..160J} imaging systems were utilised to provide simultaneous contextual imaging of the field-of-view captured by the spectrograph. ROSA images consisted of blue continuum (5.2{\,}nm bandpass filter centred at 417.0{\,}nm), G-band (0.9{\,}nm bandpass filter centred at 430.5{\,}nm), and Ca~{\sc{ii}}~K (0.1{\,}nm bandpass filter centred at 393.3{\,}nm) filtergrams taken with platescales of $0{\,}.{\!\!}''077$ per pixel, providing field-of-view sizes equal to $77''\times77''$. The G-band and 417.0{\,}nm continuum channels were acquired at frame rates of 30.3~s$^{-1}$, while the Ca~{\sc{ii}}~K images were obtained at frame rates of 4.3~s$^{-1}$. HARDcam images provided H$\alpha$ (0.025{\,}nm filter centred on the line core at 656.281{\,}nm) filtergrams with a platescale of $0{\,}.{\!\!}''084$ per pixel, providing a field-of-view size equal to $172''\times172''$, which was obtained at a frame rate of 35~s$^{-1}$. Following the methods described by \citet{2008A&A...488..375W}, speckle reconstruction algorithms were implemented to improve the final imaging data products. For the present science verification of {\sc{francis}} data products, only sample G-band and Ca~{\sc{ii}}~K images have been employed for co-alignment purposes, with the remainder of the contemporaneous imaging data left for future follow-up science projects. 

\begin{figure}[!t]
\centerline{
\includegraphics[width=0.495\textwidth, trim= 0mm 0mm 0mm 0mm, clip=]{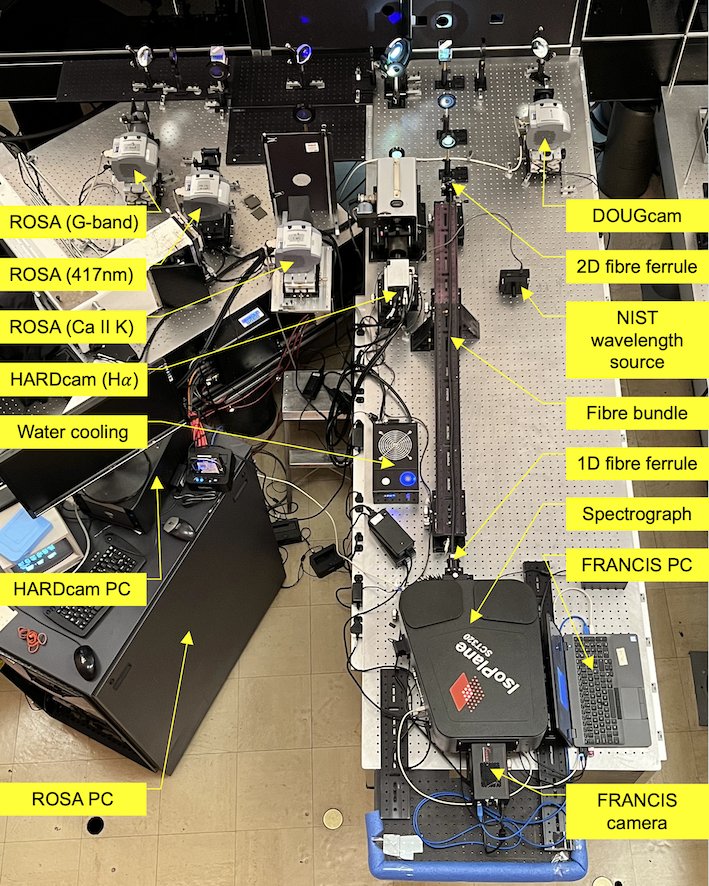}
\includegraphics[width=0.495\textwidth, trim= 0mm 0mm 0mm 43mm, clip=]{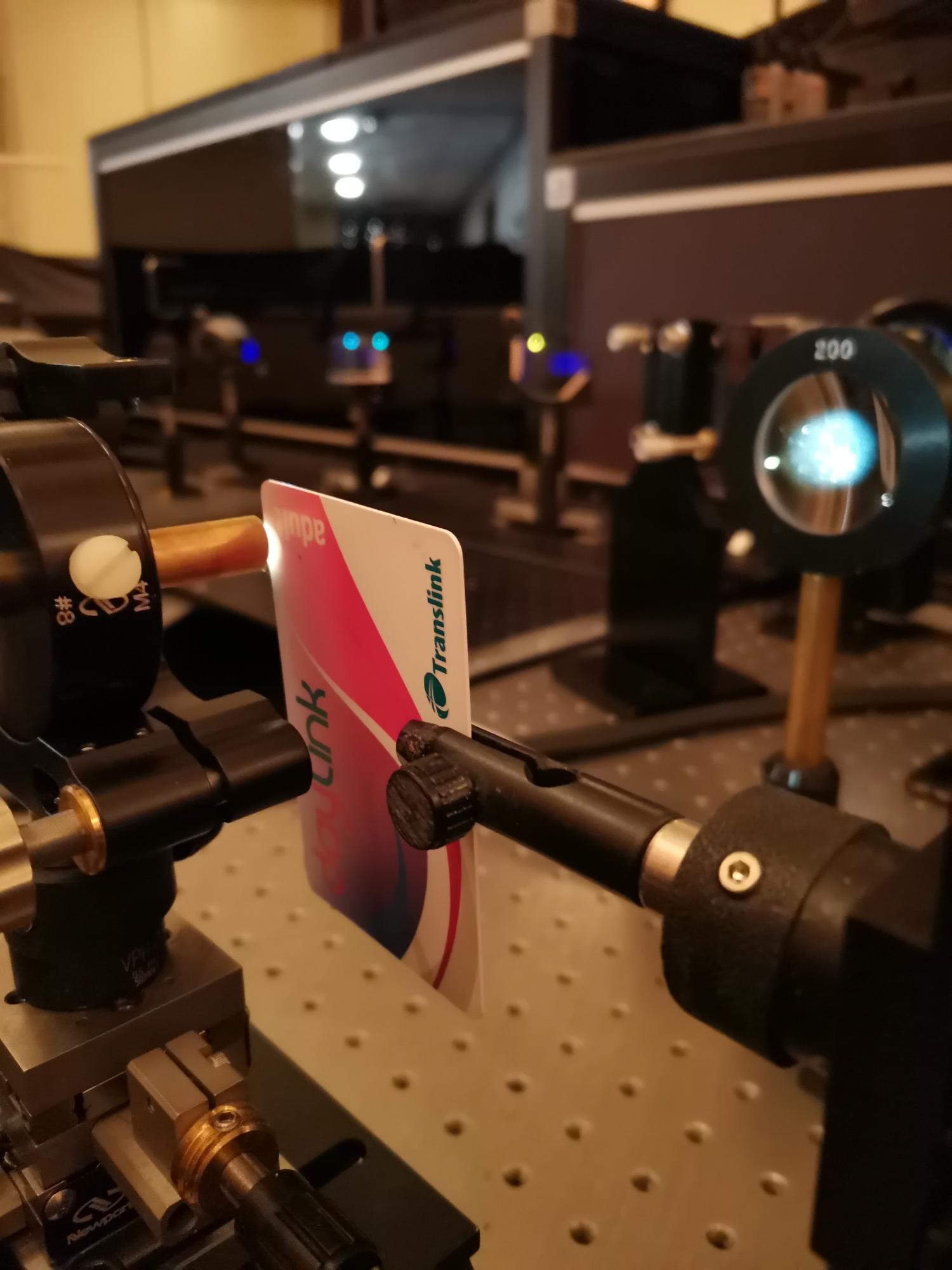}
}
\caption{\sloppy{\nohyphens{The optical layout employed for the commissioning of the {\sc{francis}} instrument during the summer of 2022 at the DST (left panel). Highlighted, using yellow boxes, are the core components of the observational setup, including the placement of contextual ROSA/HARDcam/{\sc{DOUGcam}} imagers, the control and data storage PCs, the fibre bundle (including the orientation of two-dimensional and linearised array ends), and the spectrograph and associated imaging hardware. The right panel shows the method employed to map the two-dimensional fibre bundle to its linearised array prior to connection to the spectrograph. Here, an optical blocker, in the form of a rigid bus pass, is attached to a bi-directional platform controlled by micrometers, which provides the ability to translate the optical blocker in front of the fibre array one row/column at a time. It is then possible to observe what corresponding rows disappear from the spectral image captured by the {\sc{francis}} sCMOS camera and relate these to the individual rows/columns occulted by the optical blocker, hence creating a process to re-map the linearised fibre spectra back into a two-dimensional field of view. Note, this process only needs to be completed once due to the use of the same fibre/camera orientation for all future observing campaigns. }}}
\label{fig:fibrealignment}
\end{figure}

The Helioseismic and Magnetic Imager \citep[HMI;][]{2012SoPh..275..229S} present on the Solar Dynamics Observatory \citep[SDO;][]{2012SoPh..275....3P} provided a contextual continuum image that was employed to co-align the images obtained from the DST with the full-disk HMI observations. The SDO/HMI continuum images were processed following the standard {\texttt{hmi\_prep}} routine within {\sc{sswidl}}, following which $80\times80$~arcsec$^{2}$ subfields were extracted from the processed data with a central pointing close to that of the ground-based image sequences. Using the SDO/HMI continuum context image to define absolute solar coordinates, the ROSA G-band observations were subjected to the cross-correlation techniques described by \citet{2013ApJ...779..168J, 2016NatPh..12..179J, 2017ApJ...842...59J, 2020NatAs...4..220J} to provide sub-pixel co-alignment accuracy between the imaging sequences, ultimately providing an accuracy down to approximately one tenth of an SDO/HMI pixel, corresponding to $\approx0{\,}.{\!\!}''05$ ($\approx36$~km). 

Finally, the polished brass fibre ferrule, which houses the two-dimensional ($20\times20$) fibre array, produced a back-reflection image that was captured by {\sc{DOUGcam}} (see Section~{\ref{sec:fibrebundle}} for more information). Here, {\sc{DOUGcam}} employed the same imaging optics as the rest of the ROSA channels, providing a consistent imaging platescale that enabled a straightforward way to map the two-dimensional fibre array directly on to physical features on the solar surface. A top-down perspective of the observational setup and optical layout is shown in the left panel of Figure~{\ref{fig:fibrealignment}}.

\begin{figure}[!t]
\centerline{
\includegraphics[width=0.8\textwidth, trim= 0mm 0mm 0mm 0mm, clip=]{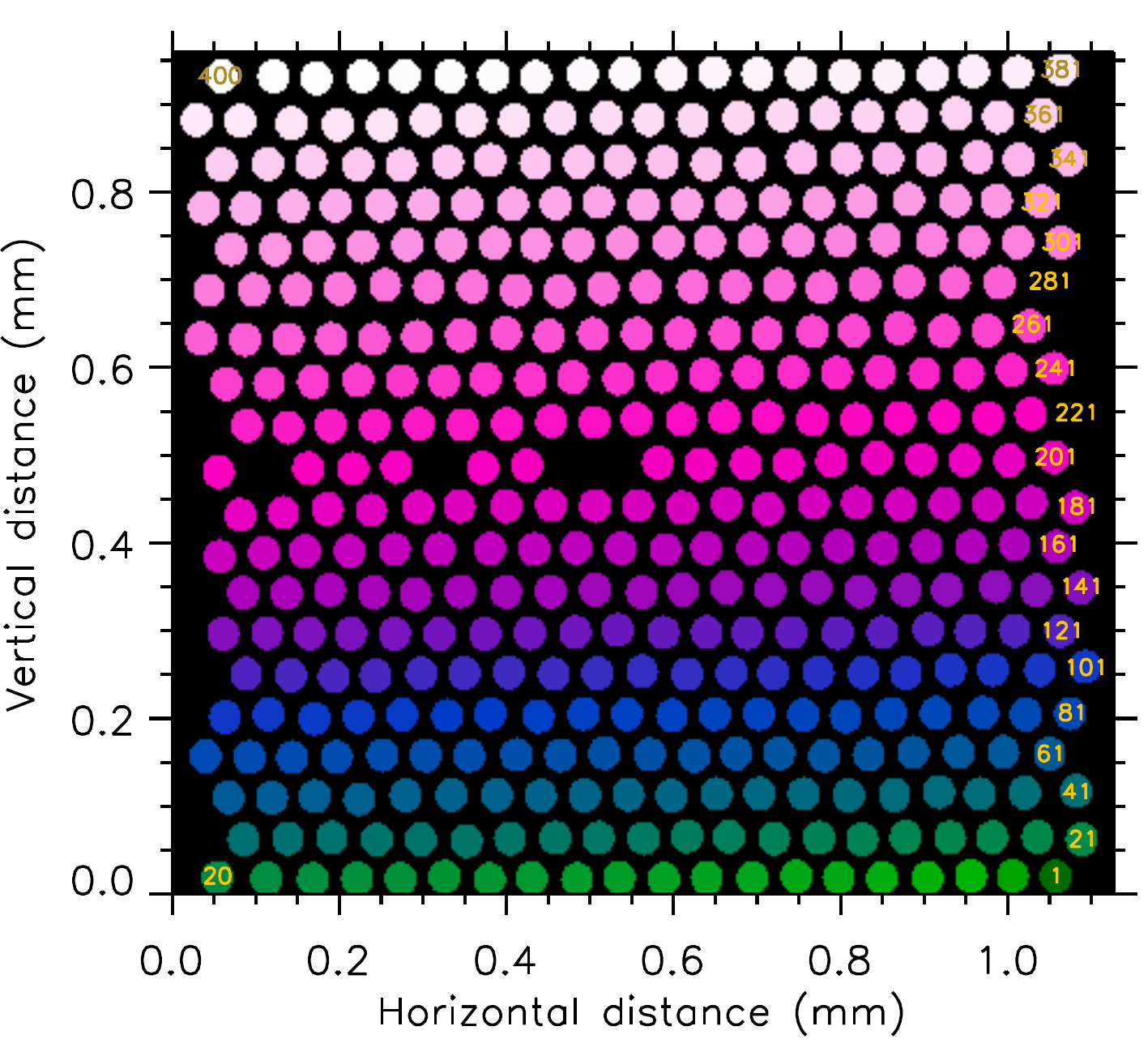}
}
\caption{\sloppy{\nohyphens{The completed mapping process to translate the linearised array back into the two-dimensional fibre group. Here, the green$\rightarrow$blue$\rightarrow$magenta$\rightarrow$white colour scale represents each sequential fibre core along the linearised array, which is displayed as a function of the spatial dimensions (in mm) along the $20\times20$ fibre head. For added visual clarity, the numbering system used is also highlighted at the beginning of each row, as well as at the ends of the first and last rows. It can be seen that fibre numbers $211$, $212$, $215$, $219$, $221$, and $281$ are absent, which indicates these are damaged fibres and unable to propagate light efficiently to the end of the bundle.   }}}
\label{fig:fibrenumbers}
\end{figure}

\begin{figure} 
\centerline{
\includegraphics[width=0.54\textwidth, trim= 0mm 23mm 0mm 0mm, clip=]{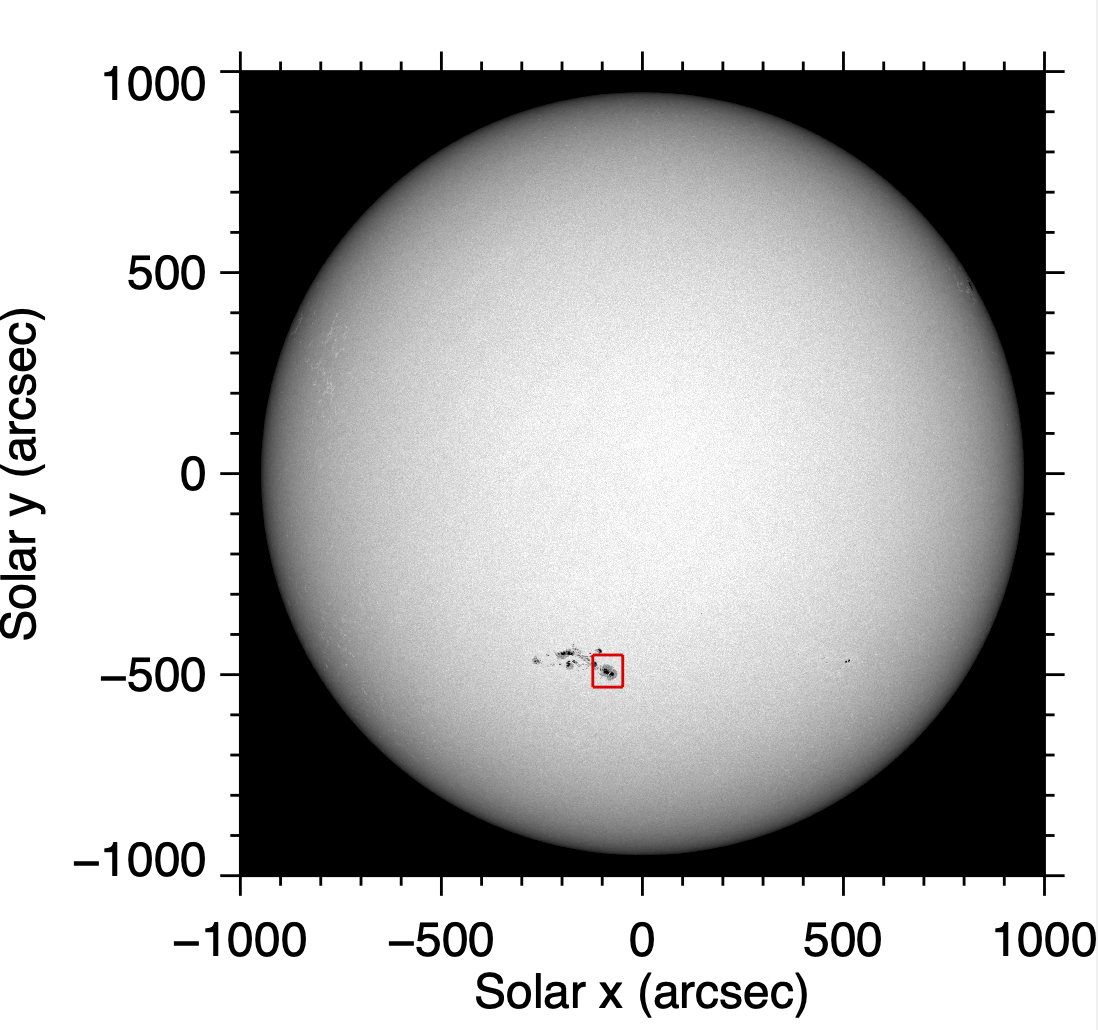}
\includegraphics[width=0.490\textwidth, trim= 17mm 24mm 0mm 0mm, clip=]{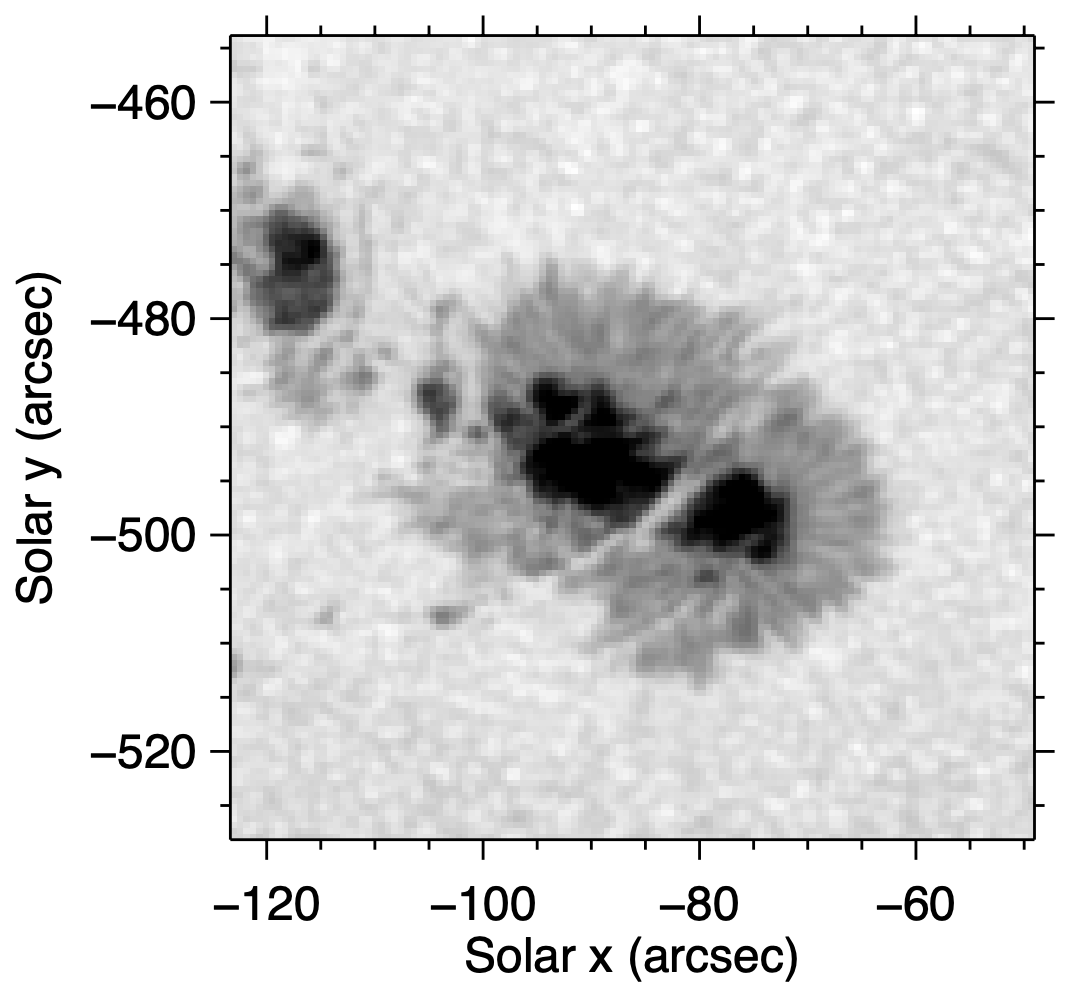}
}
\centerline{
\includegraphics[width=0.52\textwidth, trim= 0mm 20mm 0mm 0mm, clip=]{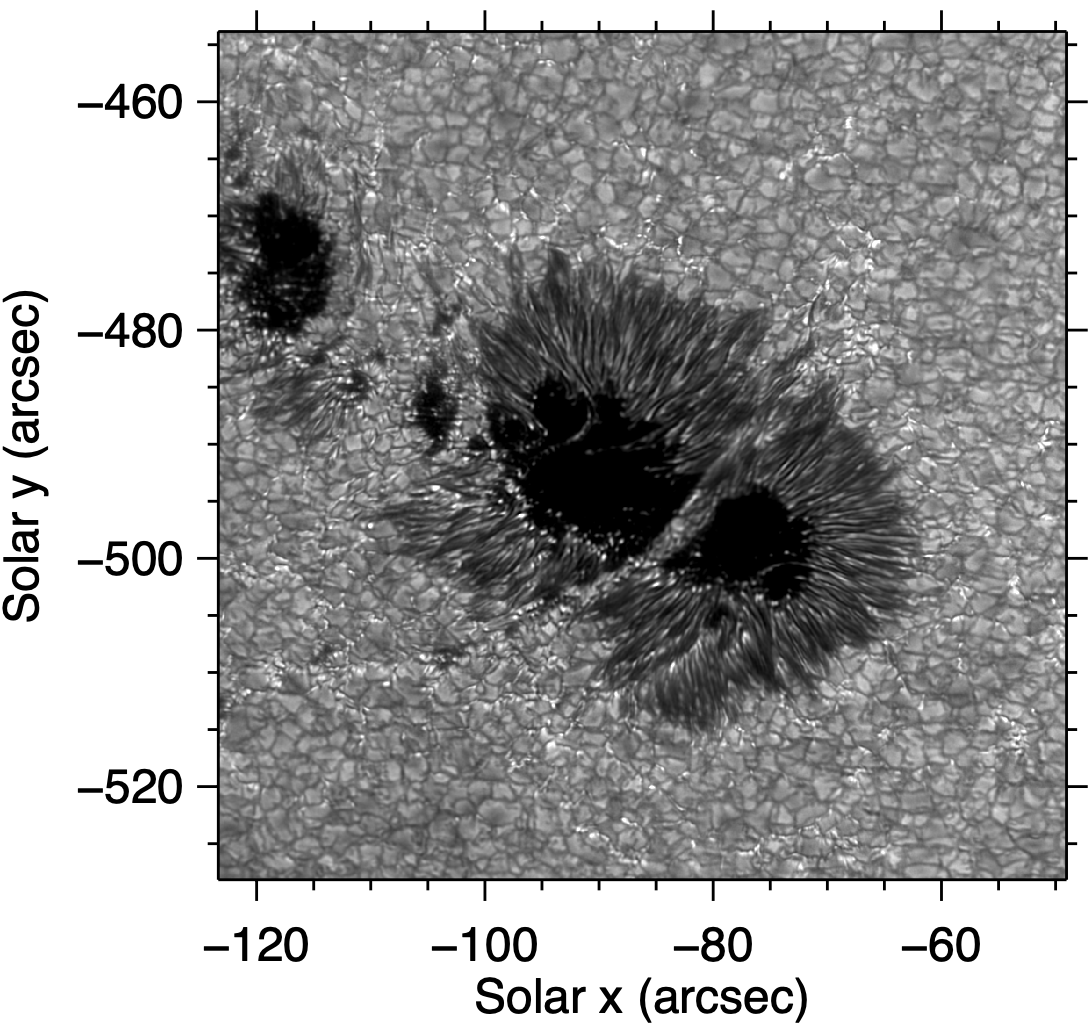}
\includegraphics[width=0.493\textwidth, trim= 12mm 20mm 0mm 0mm, clip=]{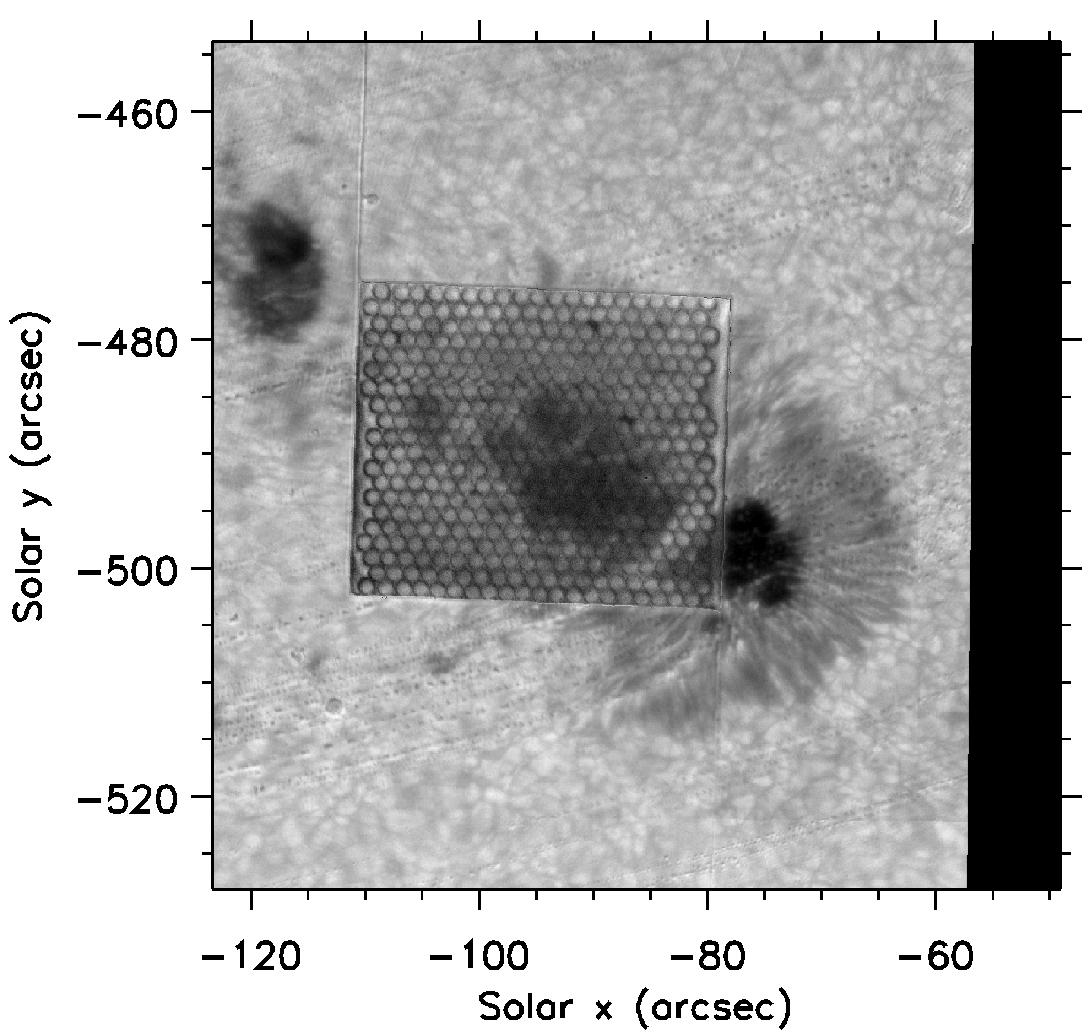}
}
\centerline{
\includegraphics[width=0.52\textwidth, trim= 0mm 0mm 0mm 0mm, clip=]{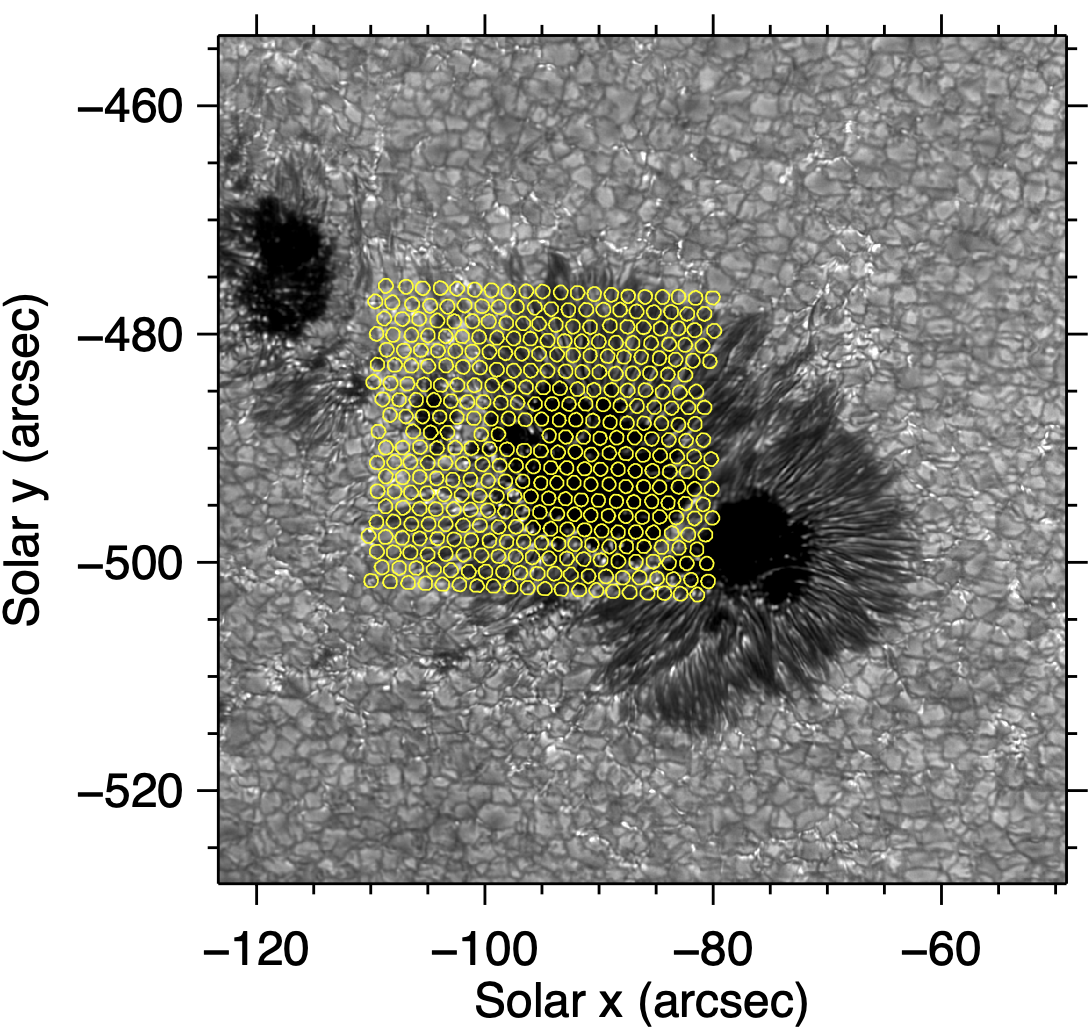}
\includegraphics[width=0.493\textwidth, trim= 12mm 0mm 0mm 0mm, clip]{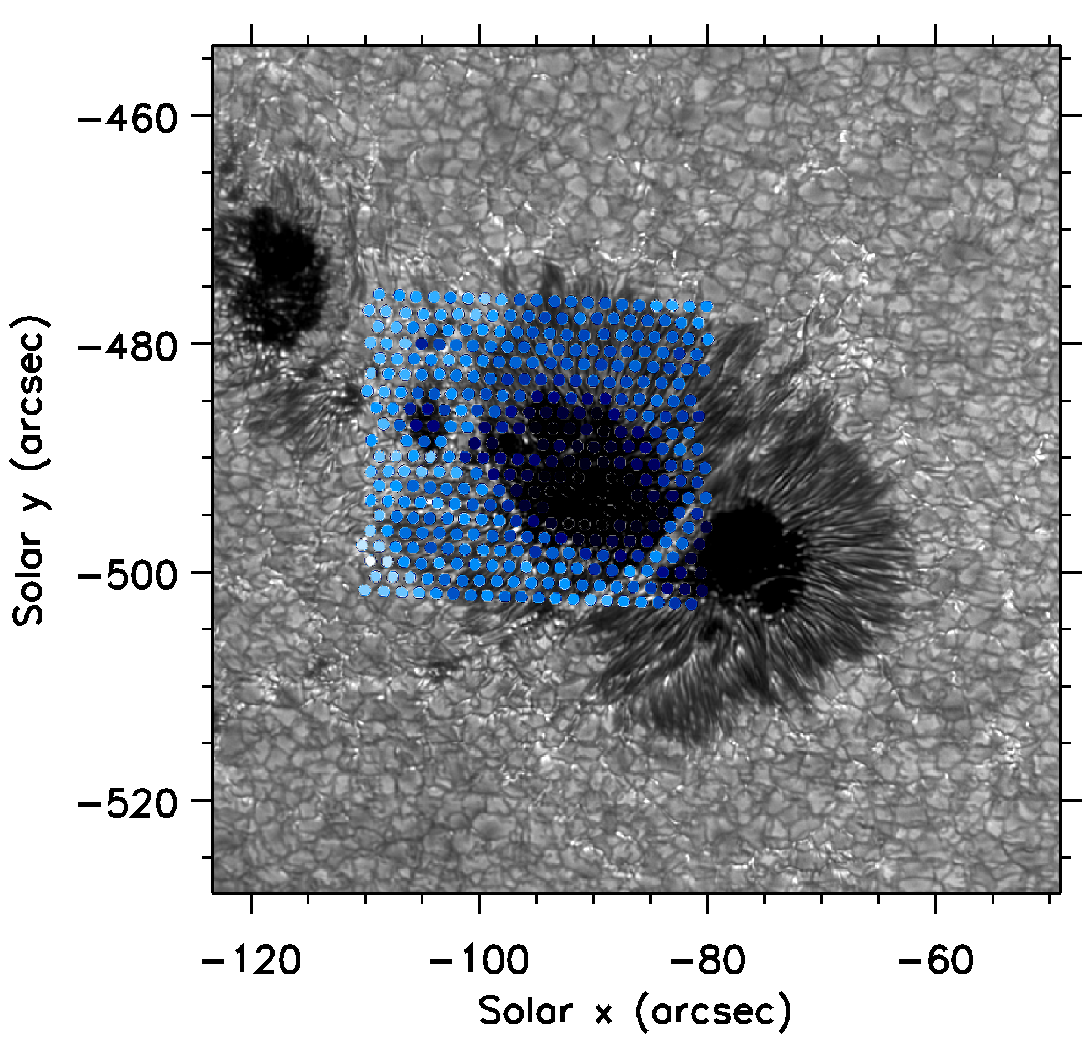}
}
\caption{\sloppy{\nohyphens{A full disk continuum intensity image acquired by SDO/HMI at 17:12~UT (upper left). The red box denotes the field-of-view captured by the ROSA instrument at the DST. The upper-right panel displays a sub-field of the full SDO/HMI image that is now identically sized to that of the ROSA time series, where the middle-left panel displays a sample ROSA G-band image acquired simultaneously at 17:12~UT. The middle-right panel depicts a simultaneous {\sc{DOUGcam}} image, which was created by imaging the light reflected off the brass fibre ferrule, enabling the locations of each individual fibre to be mapped to its corresponding location on the solar surface. Note, the intensities corresponding to back reflections from the fibre bundle have been rescaled for visual clarity, since the majority of light is transmitted down the fibres and not reflected to the {\sc{DOUGcam}} imager. The lower-left panel overplots the positioning of the individual {\sc{francis}} fibres using solid yellow contours on top of the underlying ROSA G-band image, while the lower-right panel interlaces a re-formatted two-dimensional continuum image derived from {\sc{francis}} spectra (using a black$\rightarrow$blue$\rightarrow$white colour scale) on top of the black-and-white ROSA G-band context image. }}}
\label{fig:FOV}
\end{figure}

\subsection{Fibre mapping}
\label{sec:fibremapping}
In order to establish the mapping between the head of the fibre bundle and the linearised array that is placed at the entrance to the spectrograph, a method is required to illuminate individual fibres and track the corresponding brightening at the opposite end of the fibre group. Typically, this process can be accomplished in a laboratory environment by, e.g., computationally rearranging the array based on intensity mapping between observed/simulated data \citep{2019arXiv190500391L} or back-illuminating the fibre bundle using precisely focused LED light sources \citep{2015AJ....149...77D}. In the case of {\sc{francis}}, this process was accomplished during the commissioning phase of the instrument. Here, an optical blocker\footnote{Initially cardboard was chosen to act as the optical blocker, but the edges of the cardboard were not sufficiently sharp to abruptly block light to the individual fibres. Instead, a bus pass from the Northern Ireland company, {\em{Translink}}, was chosen due to its sharp plastic edges, which helped to mask individual rows/columns with high degrees of precision (see the right panel of Figure~{\ref{fig:fibrealignment}}).} was attached to a bi-directional platform that could be translated in the horizontal and vertical directions using precise micrometers, before being placed in front of the two-dimensional end of the fibre bundle (see the right panel of Figure~{\ref{fig:fibrealignment}}). Next, the horizontal and vertical micrometers were adjusted to progressively block out portions of the light reaching the fibre array, with data saved after each platform shift to investigate which corresponding rows of the imaged spectra subsequently disappeared. By translating the optical blocker first horizontally, and then vertically across the face of the two-dimensional fibre array, it was possible to map not only the ordering of the fibres onto the linearised array, but also to investigate whether any of the $400$ fibres were damaged.

Following this process, it was found that the uppermost (i.e., fibre number `1') and lowermost (i.e., fibre number `400') rows of the spectral image acquired by the {\em{Princeton Instruments}} Kuro~2048B sCMOS camera corresponded to the lower-right and upper-left fibres, respectively, once re-mapped back into its two-dimensional array. The complete re-mapping lookup image is shown in Figure~{\ref{fig:fibrenumbers}}, where the beginning of each row is highlighted by its corresponding fibre number. During this process it was noticed that $6$ fibres appeared to be damaged, with almost zero transmission along the length of the fibre. Once the re-mapping was complete, these fibres were found to correspond to fibre numbers $211$, $212$, $215$, $219$, $221$ and $281$, as can be seen in Figure~{\ref{fig:fibrenumbers}}. Three of the six fibres (numbered $219$, $221$ and $281$) are close to the edges of the two-dimensional fibre array and are, therefore, on the periphery of the selected field-of-view. However, it is unfortunate that the three other damaged fibres (numbered $211$, $212$ and $215$) are relatively central within the two-dimensional array and, as a result, care must be taken to ensure these damaged fibres do not compromise the overall science objectives of the experiment when selecting the central solar pointing of the fibre head. As discussed in Section~{\ref{sec:fibrebundle}}, the $6$ damaged fibres may have arisen from the relatively long fibre lengths chosen for this prototype instrument. However, future fibre bundles will likely be able to utilise shorter lengths ($\sim0.5$~m instead of the current $\approx1.5$~m), which will help to make the overall fibre array more robust.

With the linear/two-dimensional fibre re-mapping complete, it is possible to use the back-reflection image from the fibre ferrule, which was imaged by {\sc{DOUGcam}}, to select regions of interest from the observed field-of-view and extract their associated spectra for subsequent analyses. Figure~{\ref{fig:FOV}} provides contextual information for the data products utilised in the present science verification study, including the fields-of-view captured by SDO/HMI, ROSA, and {\sc{DOUGcam}}, in addition to the spatial orientation of the two-dimensional {\sc{francis}} fibre head.

\subsection{Wavelength calibration}
\label{sec:wavelengthcalibration}
Many options exist to calibrate the wavelength axis of an acquired spectrum, including the comparison with a standard reference such as the solar atlas compiled by \citet{1984sfat.book.....K} using the Fourier Transform Spectrometer \citep[FTS;][]{1978fsoo.conf...33B, 1979MmArc.106...33B} at the McMath-Pierce facility at Kitt Peak Observatory, USA. However, this process can become more difficult if the observations acquired are from regions away from solar disk centre and/or include spectral lines that are more dynamic in their evolution. As such, to calibrate the wavelength axis of spectra recorded by the {\sc{francis}} instrument, we utilised the {\em{Princeton Instruments}} IntelliCal{\textsuperscript{\tiny\textregistered}} calibration source, which includes both a mercury (Hg) and neon-argon (Ne-Ar) lamp that have been strictly calibrated by National Institute of Standards and Technology (NIST) processes.

\begin{figure}[!t]
\begin{minipage}{\textwidth}
  \centering
  \raisebox{-0.5\height}{\includegraphics[height=6cm]{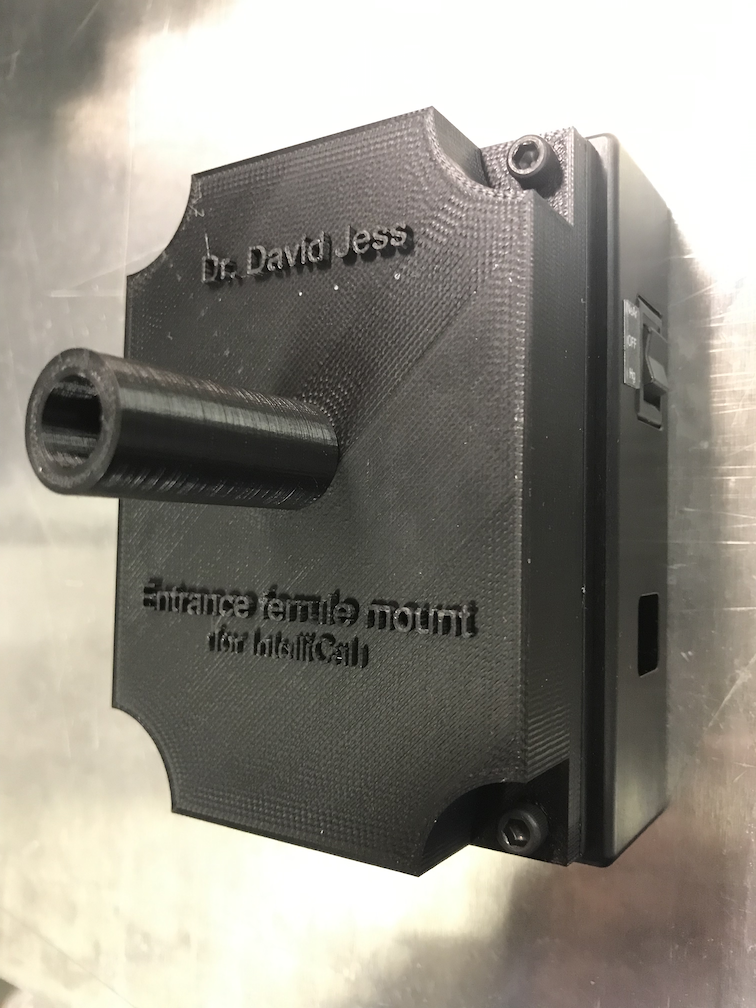}}
  \raisebox{-0.5\height}{\includegraphics[height=12cm]{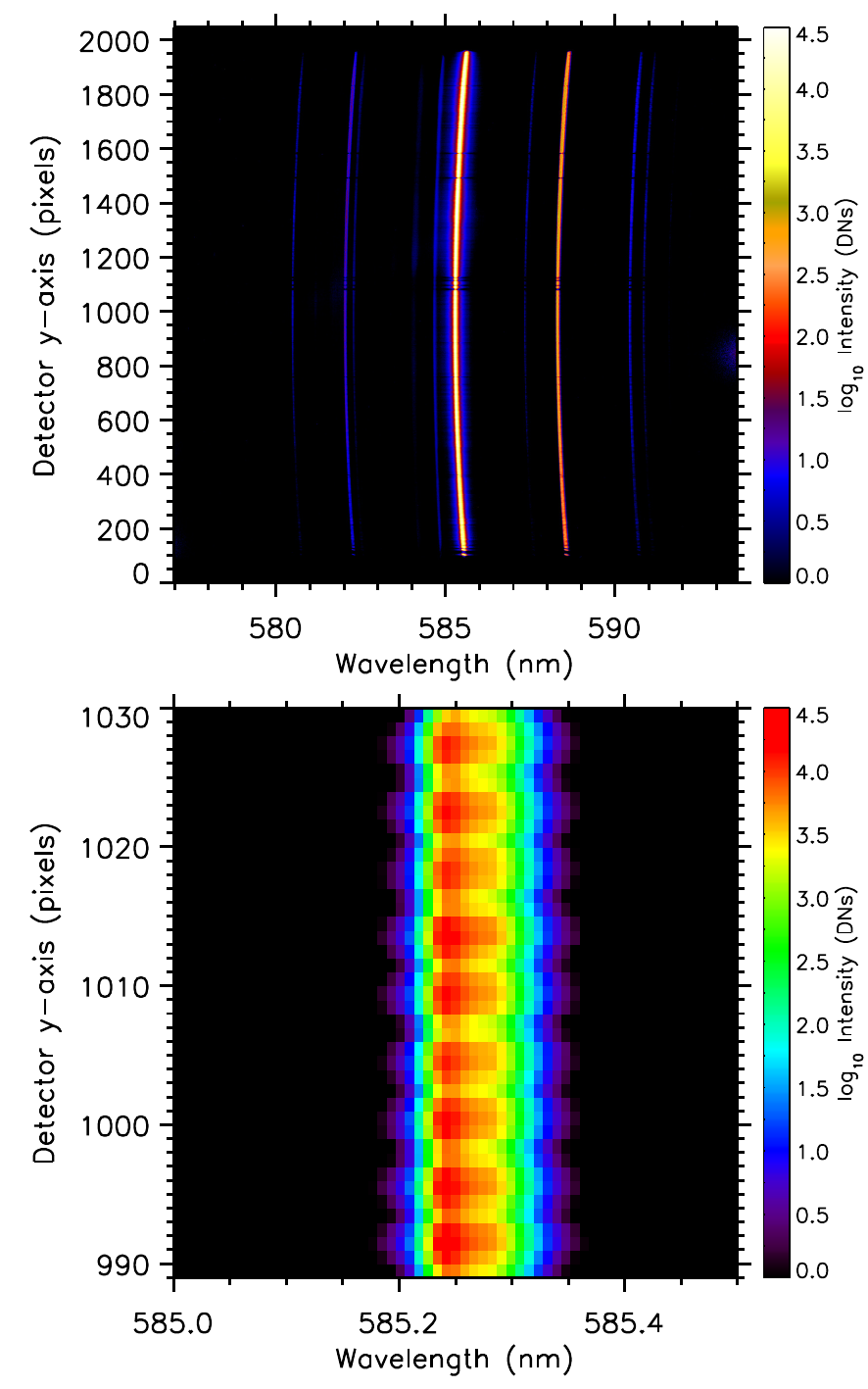}}
\end{minipage}
\caption{\sloppy{\nohyphens{The left panel shows the custom 3D printed mount that enables the two-dimensional fibre head to be exposed to light from the IntelliCal{\textsuperscript{\tiny\textregistered}} Hg/Ne-Ar calibrated reference lamps. The fibre head slots into the cavity seen towards the left-hand side of the image, while the Hg/Ne-Ar lamp bolts directly to the back of the mount, which minimises contamination from unknown sources during the wavelength calibration process. The upper-right panel displays a complete spectral image of Ne-Ar emission lines obtained by the Kuro~2048B sCMOS camera using an exposure time of 10~s and displayed on a log-scale for added visual clarity, with the most prominent emission lines being visible at $585.249$~nm and $588.190$~nm. The colour bar provides an indication of the detector counts accumulated during the calibration process. Note that the observed bending of the spectral lines is a result of the short focal length, off-axis optics used within the spectrograph housing and the use of plane diffraction gratings, which is discussed in more detail in Section~{\ref{sec:wavelengthcalibration}}. The lower-right panel reveals a zoom-in to the spectral image to show the visibility of individual fibres along the vertical axis of the camera. }}}
\label{fig:intellical}
\end{figure}

The IntelliCal{\textsuperscript{\tiny\textregistered}} system works in conjunction with the LightField{\textsuperscript{\tiny\textregistered}} software to minimise the need for user input. However, first it was necessary to subject the two-dimensional fibre head to the output light from the calibration lamp, without external contamination from stray light or unknown sources. To achieve this, a custom mount was 3D printed that allowed the brass ferrule to be inserted into a chamber that was illuminated by the Hg/Ne-Ar lamps. The left panel of Figure~{\ref{fig:intellical}} displays the 3D printed mount, which enables the calibration lamps to be bolted securely to the housing, while offering an insertion cavity for the fibre ferrule. The custom mount was secured to a multi-axis platform fixed to the optical bench, which allowed the IntelliCal{\textsuperscript{\tiny\textregistered}} system to be easily placed into the optical setup without the need to move any existing components. This meant that wavelength calibration could be accomplished/checked multiple times each day using the automated approaches available within the LightField{\textsuperscript{\tiny\textregistered}} software.

Specifically, the IntelliCal{\textsuperscript{\tiny\textregistered}} system employs an automated calibration routine in which a non-linear least-squares refinement algorithm is derived that minimises spectral intensity residuals with respect to a theoretical model of the Czerny-Turner spectrograph\footnote{More details of the process can be viewed in the dedicated IntelliCal{\textsuperscript{\tiny\textregistered}} technical note --- \href{https://www.princetoninstruments.com/wp-content/uploads/2020/04/TechNote\_FullyAutoWavlengthCalibrationDataAccuracy.pdf}{https://www.princetoninstruments.com/wp-content/uploads/2020/04/TechNote\_FullyAutoWavlengthCalibrationDataAccuracy.pdf}}. The system simulates the entire observed spectrum so that the number of observables is always equal to the number of horizontal pixels in the detector array, here being equal to $2048$ in the case of the {\em{Princeton Instruments}} Kuro~2048B sCMOS camera. Following the application of the IntelliCal{\textsuperscript{\tiny\textregistered}} approach, a typical rms error of the derived wavelength axis is found to be on the order of $0.002$~nm. However, for completeness, spectral images of the Hg/Ne-Ar lamps were also obtained using 10~s exposures to manually repeat the wavelength calibration process once the data had been downloaded. The right panel of Figure~{\ref{fig:intellical}} displays a sample spectrum obtained using the Ne-Ar lamp, alongside a zoom-in to reveal the individual fibres that are sampled along the vertical axis of the image. 

To accurately calibrate the wavelength array using the IntelliCal{\textsuperscript{\tiny\textregistered}} Hg/Ne-Ar reference lamps, it is important that there is sufficient overlap between the selected wavelength range used for science observations and the window employed for capturing the calibrated emission lines. As the spectral lines of interest for data verification were the Na~{\sc{i}}~D$_{1}$ \& D$_{2}$ absorption line pair located at $589.592$~nm and $588.995$~nm, respectively, the strong Ne-Ar emission lines at $585.249$~nm, $588.190$~nm and $594.483$~nm were selected to manually calibrate the wavelength domain since they overlap completely with the wavelengths chosen for scientific purposes. Slight curvature of the spectral lines, as seen in the upper-right panel of Figure~{\ref{fig:intellical}}, which is caused by the effects of off-axis optics within the spectrograph housing and the use of plane diffraction gratings \citep{doi:10.1366/000370203322554527}, is a common phenomenon in compact spectrographs with short focal lengths \citep{2022SCPMA..6589603Q}. As the spectral curvature is symmetric about the mid-point along the $y$-axis of the slit image, we created an initial calibration spectrum by integrating across the central $5$ fibres, spanning the detector pixels $1012 \le y{\mathrm{-pixels}} \le 1034$. This spectrum was then plotted as a function of the wavelength array provided via the IntelliCal{\textsuperscript{\tiny\textregistered}} process, with the dominant emission lines at $585.249$~nm, $588.190$~nm and $594.483$~nm fitted with a Gaussian and their central wavelengths compared to the expected values from the Ne-Ar source. The fitted Gaussian centroids were within $\pm0.001$~nm of the expected emission peak wavelengths, which is consistent with the $0.002$~nm rms wavelength uncertainty provided by the IntelliCal{\textsuperscript{\tiny\textregistered}} software that sampled more (faint) features within the selected spectral window. Thus, for the $2400$~lines/mm diffraction grating chosen, with a central science wavelength of $589.000$~nm, a wavelength sampling of $\Delta\lambda=0.008$~nm and an overall spectral range spanning $580.758 \le \lambda \le 597.250$~nm was provided. 

To calibrate the wavelength domain corresponding to the Ca~{\sc{ii}}~H/K spectra obtained using the $3600$~lines/mm grating, an identical approach was followed, only now using the strong Hg emission lines at $404.656$~nm and $407.784$~nm produced by the IntelliCal{\textsuperscript{\tiny\textregistered}} calibration lamp. While these wavelengths do not completely overlap with the wavelength region selected for Ca~{\sc{ii}}~H/K science observations, they are the closest (in wavelength) strong emission features present within the spectral signatures of the lamp. As with the Ne-Ar emission features described above, the fitted Gaussian centroids of the $404.656$~nm and $407.784$~nm emission lines were within $\pm0.001$~nm of the expected peak wavelengths provided by the IntelliCal{\textsuperscript{\tiny\textregistered}} system. As such, for the $3600$~lines/mm grating, a wavelength sampling of $\Delta\lambda=0.005$~nm was established, providing an overall spectral range spanning $389.547 \le \lambda \le 400.459$~nm for the scientific Ca~{\sc{ii}}~H/K observations.

\subsection{Resolving power}
\label{sec:resolvingpower}
The resolving power, $R$, of a spectrograph is a defining characteristic that indicates how closely (in wavelength space) two features can be, yet still be resolved. The classical definition for the resolving power is,
\begin{equation}
    R = \frac{\lambda}{\Delta\lambda} \ ,
\label{eqn:resolvingpower}
\end{equation}
where $\lambda$ is the wavelength of the features of interest and $\Delta\lambda$ is the minimum wavelength separation these features need to have to still be distinguishable. As a result, the selection of what defines $\Delta\lambda$ is rather arbitrary. Traditionally, the Rayleigh criterion \citep{doi:10.1080/14786447908639684} has been employed, although there are ongoing debates regarding exactly how close Airy disks need to be before they become impossible to distinguish apart, which is fuelled by the fundamental works of \citet{1916ApJ....44...76S}, \citet{PhysRev.29.478}, and \citet{doi:10.1080/14786443708561813}. 

In recent years, it has become common practice to adopt the measured full-width at half-maximum (FWHM) as the definition of $\Delta\lambda$ in Equation~{\ref{eqn:resolvingpower}} \citep{2013PASA...30...48R}. Hence, to get an accurate measurement of the FWHM of a spectral line, it is desirable to utilise a feature that is isolated, unblended, and unbroadened by mechanisms other than instrumental effects. As such, the Hg/Ne-Ar calibration lamp offers an ideal spectrum from which to measure the FWHMs across a variety of different wavelengths and diffraction gratings. For the $2400$~lines/mm grating, the strong Ne emission line at $585.249$~nm was chosen as this was also within the wavelength window selected for the main solar observations. For the $3600$~lines/mm and $4320$~lines/mm gratings, the strong Hg emission lines at $404.656$~nm and $365.016$~nm were selected, respectively, where the former emission feature is close to the wavelength window selected for solar observations in the Ca~{\sc{ii}}~H/K region.

\begin{figure}[!t]
\centerline{
\includegraphics[width=\textwidth, trim= 0mm 0mm 0mm 0mm, clip=]{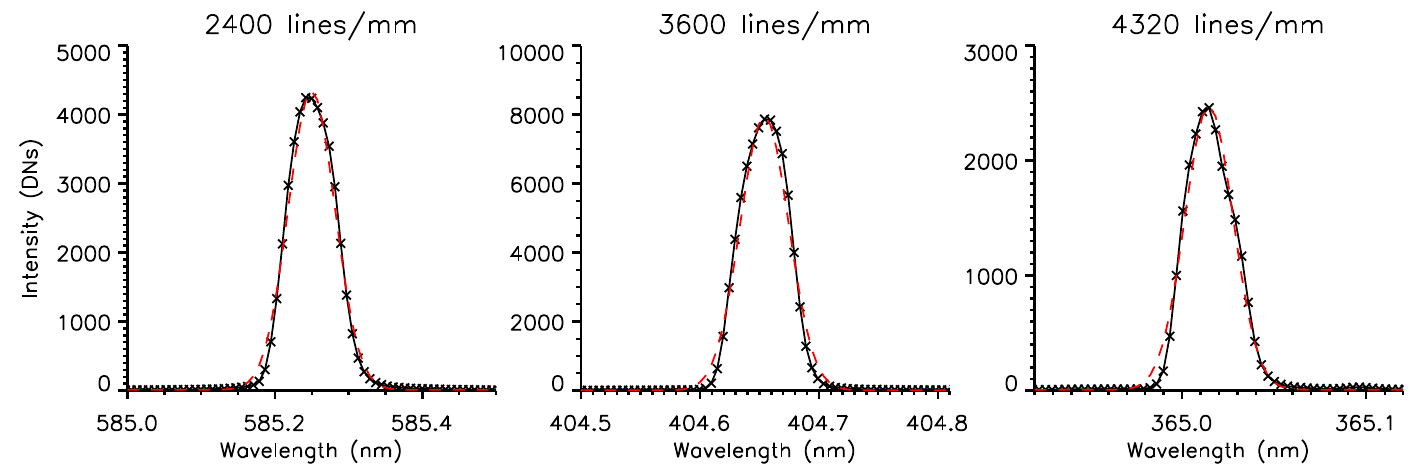}
}
\caption{\sloppy{\nohyphens{The solid black lines are example spectral profiles captured by the same fibre core for the $585.249$~nm (left panel), $404.656$~nm (middle panel), and $365.016$~nm (right panel) calibration lamp emission lines using the $2400$~lines/mm, $3600$~lines/mm, and $4320$~lines/mm diffraction gratings, respectively. The dashed red lines denote the fitted Gaussian for each emission profile, which determines the associated FWHM for that feature. }}}
\label{fig:resolvingpower}
\end{figure}

For each selected emission line, spectra from all $394$ operational fibres were fitted with a Gaussian to extract both the fitted line centroids (i.e., a measure of the absolute wavelength calibration and the quality of the spectral curvature correction -- see Section~{\ref{sec:datareduction}} for more information on the spectral curvature correction steps taken) and their corresponding FWHMs. For the $2400$~lines/mm grating, the chosen Ne emission line at $585.249$~nm was found to have a measured line centre of $585.249\pm0.001$~nm and a FWHM of $0.037\pm0.006$~nm, providing a spectral resolving power, $R_{2400}=15\,817\pm2742$. Similarly, for the $3600$~lines/mm grating, the chosen Hg emission line at $404.656$~nm was found to have a measured line centre of $404.656\pm0.001$~nm and a FWHM of $0.024\pm0.002$~nm, providing a spectral resolving power, $R_{3600}=16\,860\pm1058$. Finally, for the $4320$~lines/mm grating, the chosen Hg emission line at $365.016$~nm was found to have a measured line centre of $365.016\pm0.001$~nm and a FWHM of $0.018\pm0.001$~nm, providing a spectral resolving power, $R_{4320}=20\,278\pm1020$. Figure~{\ref{fig:resolvingpower}} shows sample emission spectra captured by a single fibre using the $2400$~lines/mm (left panel), $3600$~lines/mm (middle panel), and $4320$~lines/mm (right panel) diffraction gratings, where the fitted Gaussians are also shown using a dashed red line in each panel. The numerical measurements are also outlined in Table~{\ref{tab:resolvingpower}}.

The resolving power can also be estimated theoretically by computing the FWHM of the instrumental profile following \citet{Mohammadi2010},
\begin{equation}
\label{eqn:theoryresolvingpower}
    \mathrm{FWHM} = \frac{d\lambda}{dl} \times \mathrm{slit~width} \ ,
\end{equation}
where $d\lambda/dl$ is the linear dispersion of the spectrograph (nm/$\mu$m) under the assumption that the entrance slit width is larger than the pixel size of the imaging detector. For the first-light observations with {\sc{francis}} we employed a $50~\mu$m slit width, which is almost a factor of $5$ larger than the pixel size on the imaging camera ($11~\mu$m), hence allowing this simplified estimate to be used. Hence, for the $2400$~lines/mm grating, which provides a linear dispersion of $d\lambda/dl = 0.00072$~nm/$\mu$m, a slit width of $50~\mu$m equates to a theoretical FWHM of $0.036$~nm and a resolving power of $R_{2400}=16\,306$. Similar calculations for the $3600$~lines/mm and $4320$~lines/mm gratings, which have associated linear dispersions of $0.00045$~nm/$\mu$m and $0.00032$~nm/$\mu$m, provide theoretical FWHMs of $0.023$~nm and $0.016$~nm, alongside theoretical resolving powers of $R_{3600}=17\,921$ and $R_{4320}=25\,090$, respectively (see Table~{\ref{tab:resolvingpower}}). 

From inspection of Figure~{\ref{fig:resolvingpower}}, alongside the information provided in Table~{\ref{tab:resolvingpower}}, it is clear that the instrument is almost achieving its theoretical resolving power across the three different diffraction gratings. Specifically, for the $2400$~lines/mm, $3600$~lines/mm, and $4320$~lines/mm gratings, the instrument reached $97.0$\,\%, $94.1$\,\%, and $89.6$\,\%, respectively, of its theoretical maximum resolving power values. The reduction in relative resolving power from the lowest ($2400$~lines/mm) to highest ($4320$~lines/mm) resolution diffraction gratings is likely the consequence of a slight optical coma in the imaged spectra \citep[seen as increased broadening in the red wing of the observed spectral line;][]{Tu2021} and/or a residual astigmatism caused by imperfect focusing of both the tangential (wavelength resolution optimised) and sagittal (fibre resolution optimised) focal planes provided by the toroidal mirror within the spectrograph optics (see Figure~{\ref{fig:CT_spectrograph_schematic}}). In the case of Czerny-Turner spectrograph configurations, coma is introduced from chief rays reflecting from mirrors rotated about its sagittal axis, which results in broadening along the tangential (wavelength) directions of the image. On the other hand, an astigmatism is commonly observed as the asymmetric broadening of the image along the sagittal plane (i.e., along the fibre direction of the image) when a detector is positioned for the tightest sagittal focus to provide maximum wavelength resolution \citep{Foreman:68, Lee:10}. Typically, the toroidal mirror used in Czerny-Turner configurations helps to correct for astigmatism by enabling both the tangential and sagittal foci to cross at the centre of the focal plane. However, while this works very well for the wavelength selected during the design phase, it is natural for slight astigmatisms to manifest as user-selected wavelength windows move away from the original design conditions.

\begin{table}[!t]
\caption{Measurements of the line centre and FWHM for the strongest spectral lines that are part of the Hg/Ne-Ar calibration lamps using the $2400$~lines/mm, $3600$~lines/mm, and $4320$~lines/mm diffraction gratings. The measuring and theoretical resolving powers, $R$, are also displayed. All of the numerical values provided in this table are for a slit width of $50~\mu$m, which was employed during the initial commissioning phase of the {\sc{francis}} instrument. }
\label{tab:resolvingpower}
\begin{tabular}{| l | c | c | c | c | c | c |}    
\hline
{\tiny{Dispersion}} & {\tiny{Emission}} &  & {\tiny{Measured}} & {\tiny{Measured}} & {\tiny{Theoretical}} & {\tiny{Theoretical}} \\
{\tiny{grating}} & {\tiny{line}} & {\tiny{Measured line}} & {\tiny{FWHM}} & {\tiny{resolving}} & {\tiny{FWHM}} & {\tiny{resolving}} \\
{\tiny{(lines/mm)}} & {\tiny{used (nm)}} & {\tiny{centre (nm)}} & {\tiny{(nm)}} & {\tiny{power ($R$)}} & {\tiny{(nm)}} & {\tiny{power ($R$)}} \\
\hline
 &  & {\tiny{$585.249$}} & {\tiny{$0.037$}} & {\tiny{$15\,817$}} &  &  \\
{\tiny{$2400$}} & {\tiny{$585.249$}} & {\tiny{$\pm$}} & {\tiny{$\pm$}} & {\tiny{$\pm$}} & {\tiny{$0.036$}} & {\tiny{$16\,306$}} \\
 &  & {\tiny{$0.001$}} & {\tiny{$0.006$}} & {\tiny{$2742$}} &  &  \\
\hline
 &  & {\tiny{$404.656$}} & {\tiny{$0.024$}} & {\tiny{$16\,860$}} &  &  \\
{\tiny{$3600$}} & {\tiny{$404.656$}} & {\tiny{$\pm$}} & {\tiny{$\pm$}} & {\tiny{$\pm$}} & {\tiny{$0.023$}} & {\tiny{$17\,921$}} \\
 &  & {\tiny{$0.001$}} & {\tiny{$0.002$}} & {\tiny{$1058$}} &  &  \\
\hline
 &  & {\tiny{$365.016$}} & {\tiny{$0.018$}} & {\tiny{$20\,278$}} &  &  \\
{\tiny{$4320$}} & {\tiny{$365.016$}} & {\tiny{$\pm$}} & {\tiny{$\pm$}} & {\tiny{$\pm$}} & {\tiny{$0.016$}} & {\tiny{$22\,631$}} \\
 &  & {\tiny{$0.001$}} & {\tiny{$0.001$}} & {\tiny{$1020$}} &  &  \\
\hline
\end{tabular}
\end{table}

Recently, \citet{2022arXiv221116635P} demonstrated a formal method to help remove observed astigmatisms, whereby a spectral image can be modelled as an array of localised basis elements with coefficients, helping to render it as a very large, sparse matrix, which can be solved using linear least squares optimisation. This approach has recently been applied to the Spectral Imaging of the Coronal Environment \citep[SPICE;][]{2020A&A...642A..14S} instrument onboard Solar Orbiter \citep{2013SoPh..285...25M}, with great success demonstrated in removing similar artefacts observed in SPICE spectral images. Importantly, this technique provides an alternative approach to traditional deconvolution methods \citep[e.g.,][]{2013ApJ...765..144P}, which may be important if the point spread function of the instrument varies across the image plane (something that convolution approaches cannot account for). As such, future scientific applications of the {\sc{francis}} instrument will utilise the methods put forward by \citet{2022arXiv221116635P} to help minimise the effects of instrument astigmatism seen in Figure~{\ref{fig:resolvingpower}}. 

Importantly, future observing campaigns with {\sc{francis}} will be able to utilise a narrower slit for science observations. Initially, the $50~\mu$m slit width was chosen so that all light propagating down the $40~\mu$m (core diameter) fibres would pass through into the spectrograph, hence minimising modal interference noise. However, this resulted in an exposure times of $25$~ms and $100$~ms for the Na~{\sc{i}}~D$_{1}$/D$_{2}$ and Ca~{\sc{ii}}~H/K observations, which provided associated spectral frame rates of $23$~s$^{-1}$ and $9$~s$^{-1}$, respectively. Importantly, in the case of the Na~{\sc{i}}~D$_{1}$/D$_{2}$ observations running at a maximal frame rate of $23$~s$^{-1}$, an exposure time of $\approx40$~ms would still enable an identical data rate to be achieved, although would likely result in saturation of the detector due to the increased exposure time. As a result, it is possible to combine slightly longer exposure times with a reduced slit width to obtain better spectral resolutions without negatively impacting the overall cadence of the science observations. For example, reducing the slit width by just over half (i.e., $50~\mu$m $\rightarrow 20~\mu$m) would provide theoretical FWHMs of $0.014$~nm, $0.009$~nm, and $0.006$~nm for the $2400$~lines/mm, $3600$~lines/mm, and $4320$~lines/mm diffraction gratings, ultimately producing resolving powers of $R_{2400}=40\,765$, $R_{3600}=44\,800$, and $R_{4320}=56\,580$, respectively. It is possible to reduce the slit width even more to help further improve the spectral resolution. However, as noted above, Equation~{\ref{eqn:theoryresolvingpower}} is only valid for slit widths that are larger than the pixel size on the detector. As a result, calculating the theoretical resolving power for even narrower slit widths requires more sophisticated approximations, such as those described by \citet{2014JOSAA..31.2002C}.

Of course, reducing the slit width to values smaller than the diameter of the fibre cores will result in a fraction of the illuminated fibre end face not transmitting fully to the imaging detector. As discussed in Section~{\ref{sec:fibrebundle}}, this may increase the measured modal interference noise if the speckle pattern at the fibre exit evolves on time scales similar to (or longer than) the camera exposure times. This would have the unwanted consequence of varying the point spread function, and hence response of the instrument, which would make spectral flat-fielding difficult and differential photometry (necessary for polarimetric measurements) extremely challenging. Many high-resolution instruments in the nighttime astrophysical community have worked on promising solutions to these difficulties, including the use of high-frequency fibre agitators to `mix' the modal interference patterns on much shorter timescales and help maximise the resulting signal-to-noise ratio \citep[e.g.,][]{2018ApJ...853..181P}. As a result, care must be taken when re-configuring the slit width of the spectrograph to ensure instrument sensitivity (particularly to spectropolarimetric measurements) is not compromised. 

For comparison, resolving powers in the range of $R \sim 40\,000 - 55\,000$ are similar to, if not higher than some current dual Fabry-P{\'{e}}rot interferometers, such as the CHROMIS \citep{2017psio.confE..85S} and CRISP \citep{2008ApJ...689L..69S} instruments that have resolving powers on the order of $R \sim 20\,000$ and $R \sim 60\,000$, respectively \citep{2017ApJ...851L...6R, 2021A&A...653A..68L}, but of course without the excellent spatial resolution that comes from imaging-based instrument setups. Contrarily, some Fabry-P{\'{e}}rot imaging instruments are optimised for very high spectral resolutions, with the Interferometric Bidimensional Spectrometer \citep[IBIS;][]{2006SoPh..236..415C} having a design resolution (measured from the FWHM of the instrumental profile) on the order of $R \sim 200\,000$ \citep{2008A&A...481..897R}. Of course, while a Fabry-P{\'{e}}rot instrument is in operation, it is necessary to scan the selected spectral line with the minimum number of wavelength points to maintain a fast line scan, while also preserving as much of the spectral sensitivity as possible \citep{2023A&A...673A..35D}. Resolving powers on the order of $R \sim 40\,000 - 55\,000$ are consistent with modern filtergraphs, such as the space-based HMI instrument \citep{2012SoPh..275..229S}, which are capable of high-precision Doppler studies of the lower solar atmosphere. On the other hand, existing slit-based spectrographs can obtain very high resolving powers, such as $R\sim210\,000$ for the FIRS instrument, $R\sim190\,000$ for the GRIS IFU, $R\sim180\,000$ for the new Visible Spectro-Polarimeter \citep[ViSP;][]{deWijn2022} on DKIST, and $R\sim170\,000$ for the Spectro-Polarimeter for Infrared and Optical Regions \citep[SPINOR;][]{2006SoPh..235...55S}. These instruments obtain increased resolving powers by using traditional low- or medium-density ruled diffraction gratings (typically a few tens to a few hundreds of lines/mm), but observing the resulting dispersed light in higher spectral orders. In order to maintain efficiency when observing in higher diffraction orders, these instruments typically employ steeper blaze angles (e.g., $63^{\circ}$ in the case of the GRIS instrument) to ensure optimal dispersion of light. For the diffraction gratings employed for use with the {\sc{francis}} instrument, which are holographic in nature, the efficiency at a particular wavelength will be determined by the depth of the grooves formed by holographic methods (i.e., the modulation). As such, future upgrades to the {\sc{francis}} instrument may include the use of traditional ruled diffraction gratings operating at high blaze angles to improve the efficiency of higher-order wavelength dispersion captured by the detectors, hence helping to improve the spectral resolution. Importantly, as highlighted by \citet{https://doi.org/10.1002/2016JA022871}, it is still possible to undertake high-precision chromospheric polarimetry studies with resolving powers on the order of $R\sim20\,000$ (particularly true for Stokes~$V$ measurements of circularly polarised light), hence supporting the future capabilities of the {\sc{francis}} instrument. 

\subsection{Data reduction}
\label{sec:datareduction}
As described in Section~{\ref{sec:dataoverview}}, $50\,000$ fibre-resolved spectral images centred on the Na~{\sc{i}}~D$_{1}$ \& D$_{2}$ absorption line pair located at $589.592$~nm and $588.995$~nm, respectively, were obtained, in addition to a short $\approx20$~s burst ($200$~spectral images) of Ca~{\sc{ii}}~H/K spectra at $396.847$~nm and $393.366$~nm, respectively. To calibrate these data into science-ready products, a reduction pipeline was created to streamline the processing steps required. Each step of the data reduction pipeline was applied independently to each of the Na~{\sc{i}}~D$_{1}$/D$_{2}$ and Ca~{\sc{ii}}~H/K data products to ensure consistency and repeatability of their respective calibrations. 

\begin{figure} 
\centerline{
\includegraphics[width=\textwidth, trim= 0mm 0mm 0mm 3mm, clip=]{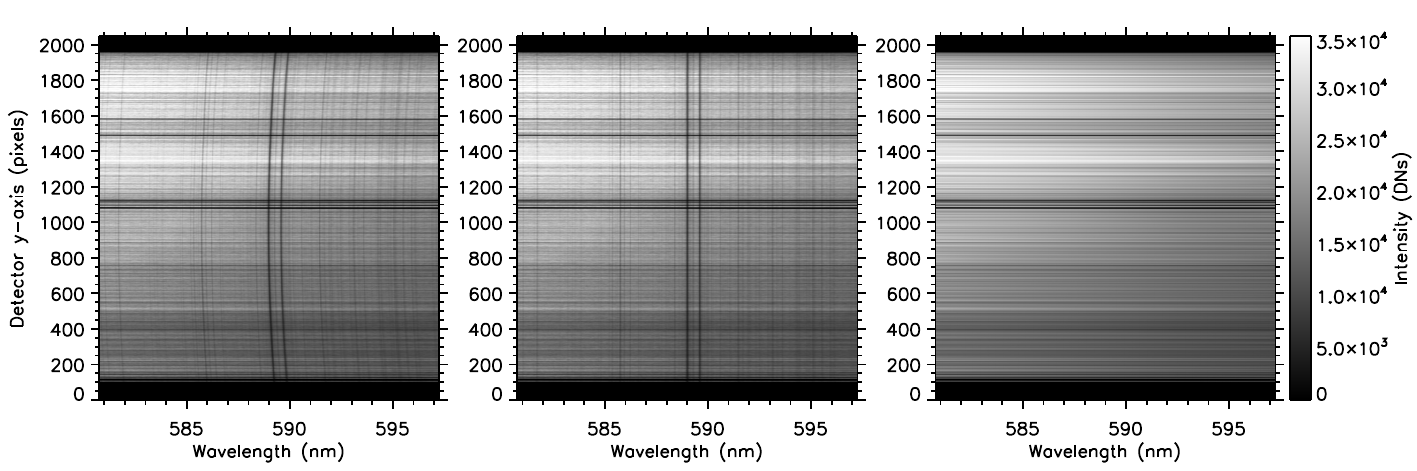}
}
\caption{\sloppy{\nohyphens{The three stages of deriving a flat-field frame for the {\sc{francis}} instrument, shown here for the case of the Na~{\sc{i}}~D$_{1}$/D$_{2}$ spectral window. The left panel shows an average spectral image obtained by summing $1000$ individual exposures, each with an exposure time of $25$~ms to match that of the science observations, where the telescope was de-focused and performed a random slew across a quiescent region of the solar surface with a $\mu$-angle consistent with the science field-of-view. The middle panel displays the same average spectral image, only now with the spectral curvature corrected through use of the quadratic function derived from data obtained with the Ne-Ar calibration lamp. Finally, the right panel depicts the final flat-field frame, where each of the spectral lines have been removed. The colour bar depicts the image intensities in detector numbers, where the {\em{Princeton Instruments}} Kuro~2048B camera saturates at $\approx64\,000$ DNs (i.e., 16-bit).  }}}
\label{fig:flatfields}
\end{figure}

First, an average dark frame was produced that was obtained with an identical exposure time to that of the science observations ($25$~ms for Na~{\sc{i}}~D$_{1}$/D$_{2}$; $100$~ms for Ca~{\sc{ii}}~H/K) by averaging over $1000$ frames captured under dark conditions (i.e., the telescope beam blocked from the fibre head). Next, an average flat field was created by de-focusing the telescope and slewing randomly across the solar surface, albeit slightly offset from the active region under investigation to ensure the solar $\mu$-angle, and hence background light level, was preserved. A total of $1000$ exposures, each with an identical exposure time ($25$~ms for Na~{\sc{i}}~D$_{1}$/D$_{2}$; $100$~ms for Ca~{\sc{ii}}~H/K) to that of the science observations, were obtained. These observations were subsequently averaged to create the flat-field frame, where the corresponding Na~{\sc{i}}~D$_{1}$/D$_{2}$ image can be visualised in the left panel of Figure~{\ref{fig:flatfields}}. 

Typically, for imaging observations, it is possible to simply subtract the average dark frame and divide by the average flat field to produce an image that is devoid of thermal and chip sensitivity degradations. However, this is not directly possible for spectral observations since the average flat field will still contain spectral signatures in the form of solar absorption lines. Furthermore, as discussed in Section~{\ref{sec:wavelengthcalibration}} and shown in the upper-right panel of Figure~{\ref{fig:intellical}}, the acquired spectral images have an associated spectral curvature due to the spectrograph optics utilised. Hence, it is necessary to compensate for the slit image curvature along the $y$-axis before attempting to remove the presence of absorption lines from the initial flat field. The offset between the observed and absolute wavelengths (caused by the spectral curvature) can be modelled by a quadratic function \citep{2022SCPMA..6589603Q},
\begin{equation}
\label{eqn:slitcurvature}
\lambda^{\prime} = C\lambda(y-y_{0})^{2} + \lambda \ ,
\end{equation}
where $\lambda^{\prime}$ and $\lambda$ are the observed and absolute wavelengths at specific $y$ coordinates along the slit, $y_{0}$ is the $y$-pixel position corresponding to the apex of the curvature (i.e., the point of symmetry for the spectral curvature), and $C$ is a constant. Equation~{\ref{eqn:slitcurvature}} clearly demonstrates that stronger spectral curvature is experienced at longer wavelengths, hence will be more apparent towards the upper end of the spectral range covered by {\sc{francis}} (i.e., $\sim700$~nm).

To calculate the curvature correction function, the spectra from each horizontal row of the Ne-Ar (for Na~{\sc{i}}~D$_{1}$/D$_{2}$ science observations) and Hg (for Ca~{\sc{ii}}~H/K science observations) slit images were extracted, with the strong emission peaks fitted using a Gaussian profile. The wavelength offsets, as a function of $y$-pixel position, enabled the quadratic correction function (see Equation~{\ref{eqn:slitcurvature}}) to be derived, allowing the wavelength-dependent spectral curvature to be corrected across the entire spectral imaging domain. Applying the curvature correction function to the average flat-field frame produces the intermediate image shown in the middle panel of Figure~{\ref{fig:flatfields}} for the Na~{\sc{i}}~D$_{1}$/D$_{2}$ observations. Comparing the left and middle panels of Figure~{\ref{fig:flatfields}}, it is clear that the spectral curvature has been corrected but that the spectral lines are still present, which need to be removed to form the final flat-field image. We must note that Equation~{\ref{eqn:slitcurvature}} corrects for the spectral curvature, but does not compensate for any residual spectral skew that may be present. However, during initial setup of the {\sc{francis}} instrument, small adjustments to the orientation of the grating mount were performed to ensure the wavelength dispersion was along the horizontal rows of the imaging detector, thus minimising any misalignment between the entrance slit and the direction of the diffraction grating \citep[see, e.g.,][for more discussions on spectral skew corrections]{1991sopo.work....3L, 1992SPIE.1746...22E, 2011ApSpe..65...85E}.

To remove spectral lines from the final flat-field image, standard processes are adopted \citep[see, e.g.,][]{2010MmSAI..81..763J, 2013RAA....13.1240W}, whereby first a spectrum is extracted for each $y$-pixel along the detector. Next, a number of continuum locations are identified that are away from obvious absorption profiles. As the flat-field image has already been corrected for spectral curvature, the wavelengths of the continuum locations will remain constant across all $y$-pixels of the detector. A second-order polynomial is then fitted to the continuum data points, with the line of best fit tracking the spectral sensitivity across the given wavelength range, in this case $580.758 \le \lambda \le 597.250$~nm for the Na~{\sc{i}}~D$_{1}$/D$_{2}$ observations and $389.547 \le \lambda \le 400.459$~nm for the Ca~{\sc{ii}}~H/K data products. The fitted continuum for the Na~{\sc{i}}~D$_{1}$/D$_{2}$ observations is shown in the right panel of Figure~{\ref{fig:flatfields}}. To finalise the flat-field image, the fitted continua (e.g., the right panel of Figure~{\ref{fig:flatfields}} in the case of Na~{\sc{i}}~D$_{1}$/D$_{2}$ observations) are divided by their median value to produce a flat-field image correction map that can be applied to subsequent science observations to correct for instrument spectral sensitivity variations.

It must be noted that the panels shown in Figure~{\ref{fig:flatfields}} reveal a number of rows that have relatively low intensities. This does not necessarily mean that the instrument has poor light sensitivity associated with these $y$-pixels. Instead, these rows of decreased intensity will correspond to the detector pixels where no fibre core is imaged and, as a result, minimal light is propagated through the entrance slit of the spectrograph and dispersed onto these rows of the imaging detector. Some of the light present along these darker rows will naturally arise from internal scattering within the optics or from around the periphery of the fibre cladding, hence it is not expected for them to have completely zero counts. Regardless, these rows are excluded when the fibre-core spectra are extracted for further scientific study and analyses. However, they are retained here to simplify and streamline the flat-fielding processes applied to scientific observations.

Once the science spectral images have been calibrated, the next step is to extract spectra corresponding to each of the $394$ working fibres. As discussed in Section~{\ref{sec:fibremapping}}, fibres $211$, $212$, $215$, $219$, $221$, and $281$ transmit almost zero light down the length of the fibre bundle, hence it is not possible to extract spectra for these specific fibre cores (see Figure~{\ref{fig:fibrenumbers}} for their precise location on the two-dimensional fibre head). To pinpoint the $y$-pixel on the spectral image corresponding to each fibre core, the intensities of the brightest emission line ($585.249$~nm) from the Ne-Ar calibration lamp (see Figure~{\ref{fig:intellical}}) were extracted as a function of the detector $y$-pixel. Plotting this one-dimensional lightcurve as a function of the $2048$ pixels along the $y$-axis of the camera allowed $394$ intensity peaks to be clearly identified, which highlighted the $y$-pixel location of each fibre core. For completeness, the same process was performed on the brightest emission line ($404.656$~nm) from the Hg calibration lamp. Due to the fibre ferrule remaining fixed in position at the entrance slit of the spectrograph, the established $y$-pixel locations of the fibre cores remained identical, irrespective of what wavelength region and/or calibration lamp was employed. 

\begin{figure} 
\centerline{
\includegraphics[width=\textwidth, trim= 0mm 0mm 0mm 3mm, clip=]{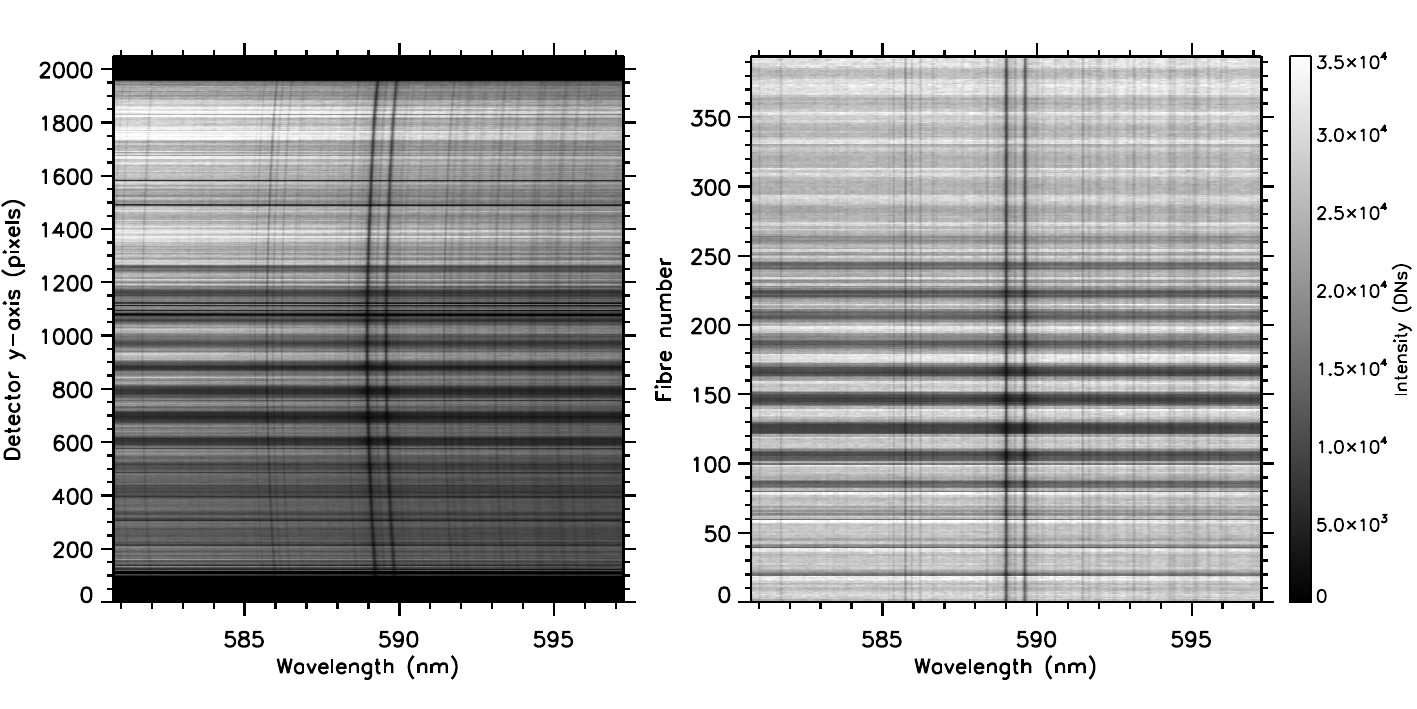}
}
\centerline{
\includegraphics[width=\textwidth, trim= 0mm 0mm 0mm 7mm, clip=]{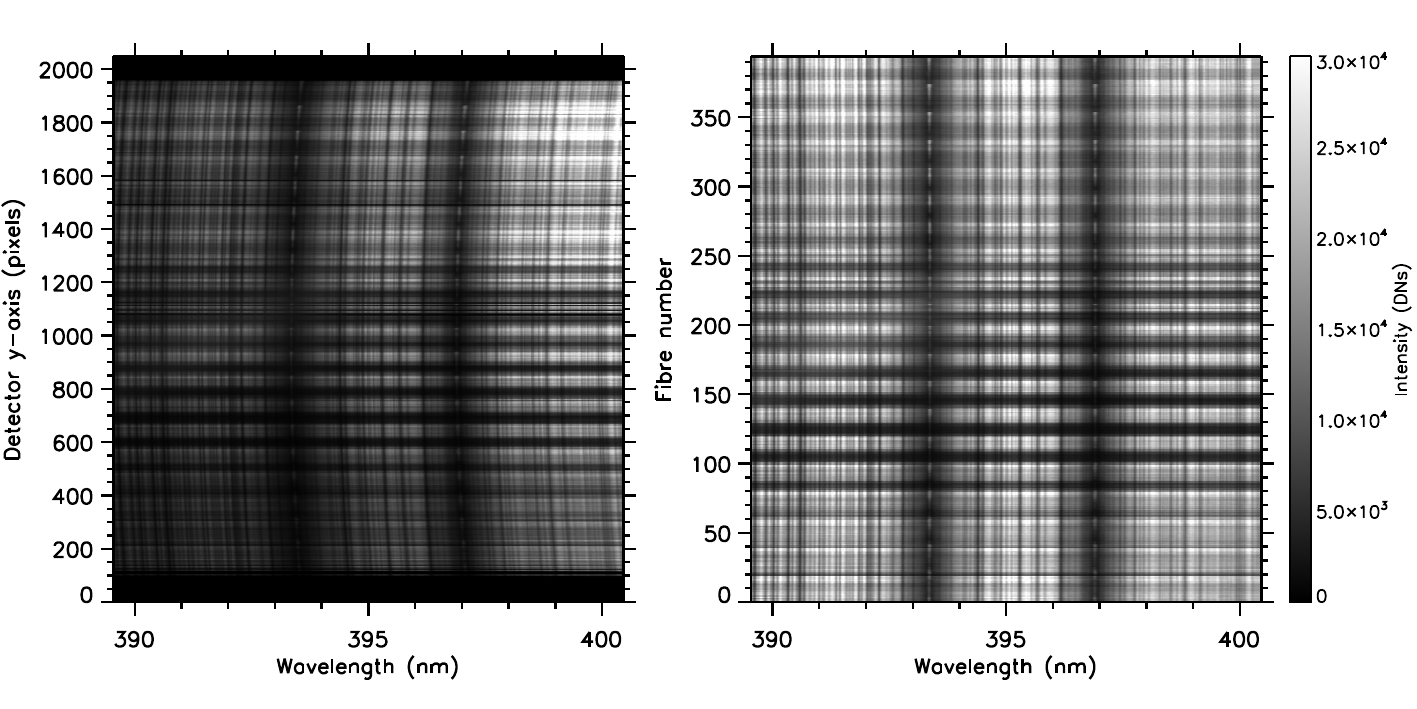}
}
\caption{\sloppy{\nohyphens{Raw spectral images of the sunspot that was part of NOAA~13089, captured on 2022 August~29 in the Na~{\sc{i}}~D$_{1}$/D$_{2}$ (upper left; $2400$~lines/mm grating) and Ca~{\sc{ii}}~H/K (lower left; $3600$~lines/mm grating) wavelength windows. The right panels depict the same spectral images, only after the processes of dark-frame subtraction, spectral curvature correction, flat-fielding, and fibre extraction have been applied. The colour bar depicts the image intensities in detector numbers, where the {\em{Princeton Instruments}} Kuro~2048B camera saturates at $\approx64\,000$ DNs (i.e., 16-bit). The dark horizontal bands seen in the spectral images correspond to fibre cores that sampled the sunspot umbra, hence have continuum intensities that are $\sim40$\% of their non-sunspot counterparts.   }}}
\label{fig:dataaverages}
\end{figure}

It must be noted that each fibre core is $40~\mu$m in diameter (see Table~{\ref{tab:francisoverview}}), while the {\em{Princeton Instruments}} Kuro~2048B camera employed has $11\times11~\mu$m$^{2}$ pixels. Hence, each fibre core will span over $3$ complete pixels, which is consistent with the zoom-in image shown in the lower-right panel of Figure~{\ref{fig:intellical}}. As a result, selecting a single $y$-pixel from the detector to extract a spectrum from a chosen fibre core will underestimate the total flux propagated by the fibre. Thus, for each of the $394$ identified intensity peaks, 
the immediately neighbouring pixels (i.e., $\pm1$~pixel) were used to extract the associated science spectra from each fibre core. To do this, all $3$ $y$-pixels associated with each fibre core location were averaged to produce a final two-dimensional array that is $2048\times394$ pixels$^{2}$ in size for each processed science spectral image (i.e., $\lambda~\times$ fibre number). These smaller spectral arrays (i.e., $2048\times394$ pixels$^{2}$ instead of the raw $2048\times2048$ pixels$^{2}$ sizes) are useful to minimise the amount of disk space required to hold the processed files, especially considering the Na~{\sc{i}}~D$_{1}$/D$_{2}$ science observations presented here were acquired with a frame rate of $23$~s$^{-1}$. 

Figure~{\ref{fig:dataaverages}} displays `before' and `after' spectral science images of the sunspot that was part of NOAA~$13089$ on 2022 August~29. The left panels of Figure~{\ref{fig:dataaverages}} are the raw spectral images obtained in the Na~{\sc{i}}~D$_{1}$/D$_{2}$ (upper left) and Ca~{\sc{ii}}~H/K (lower left) wavelength windows, while the right panels display the same images following the application of dark-field subtraction, spectral curvature correction, flat-fielding, and fibre extraction. Dark horizontal bands are visible in the processed spectral images shown in the right panels of Figure~{\ref{fig:dataaverages}}. These correspond to fibre cores that sampled the sunspot umbra, where the associated continuum intensities are $\sim40$\% of their counterparts that observed non-umbral locations.

With the linear/two-dimensional fibre mapping known (see Figure~{\ref{fig:fibrenumbers}}) and the science observations processed onto a common wavelength axis, it is possible to re-create two-dimensional images at precise wavelengths chosen by the research team. The lower-right panel of Figure~{\ref{fig:FOV}} overplots the ROSA G-band context image with a two-dimensional {\sc{francis}} image obtained at a continuum wavelength just blueward of the Na~{\sc{i}}~D$_{2}$ spectral line. From Figure~{\ref{fig:FOV}}, it is clear that the spatial re-mapping of {\sc{francis}} fibres has worked incredibly well, allowing precise spectral measurements to be studied alongside the imaging observations, both of which were obtained with high temporal cadences ($\ge23$~s$^{-1}$). 

\subsection{Data verification}
\label{sec:dataverification}
As discussed in Section~{\ref{sec:datareduction}}, the {\sc{francis}} data acquired during science observations is reduced into a science-ready format via means of dark current subtraction, correcting for spectral curvature, flat-fielding, and transforming the dispersive $x$-axis array of the imaging detector onto a calibrated wavelength grid. Once these steps have been performed, it is possible to extract spectra corresponding to each individual fibre core and examine the temporal, spatial and/or spectral evolution of the solar features of interest.

\begin{figure} 
\centerline{
\includegraphics[width=0.8\textwidth, trim= 0mm 0mm 0mm 0mm, clip=]{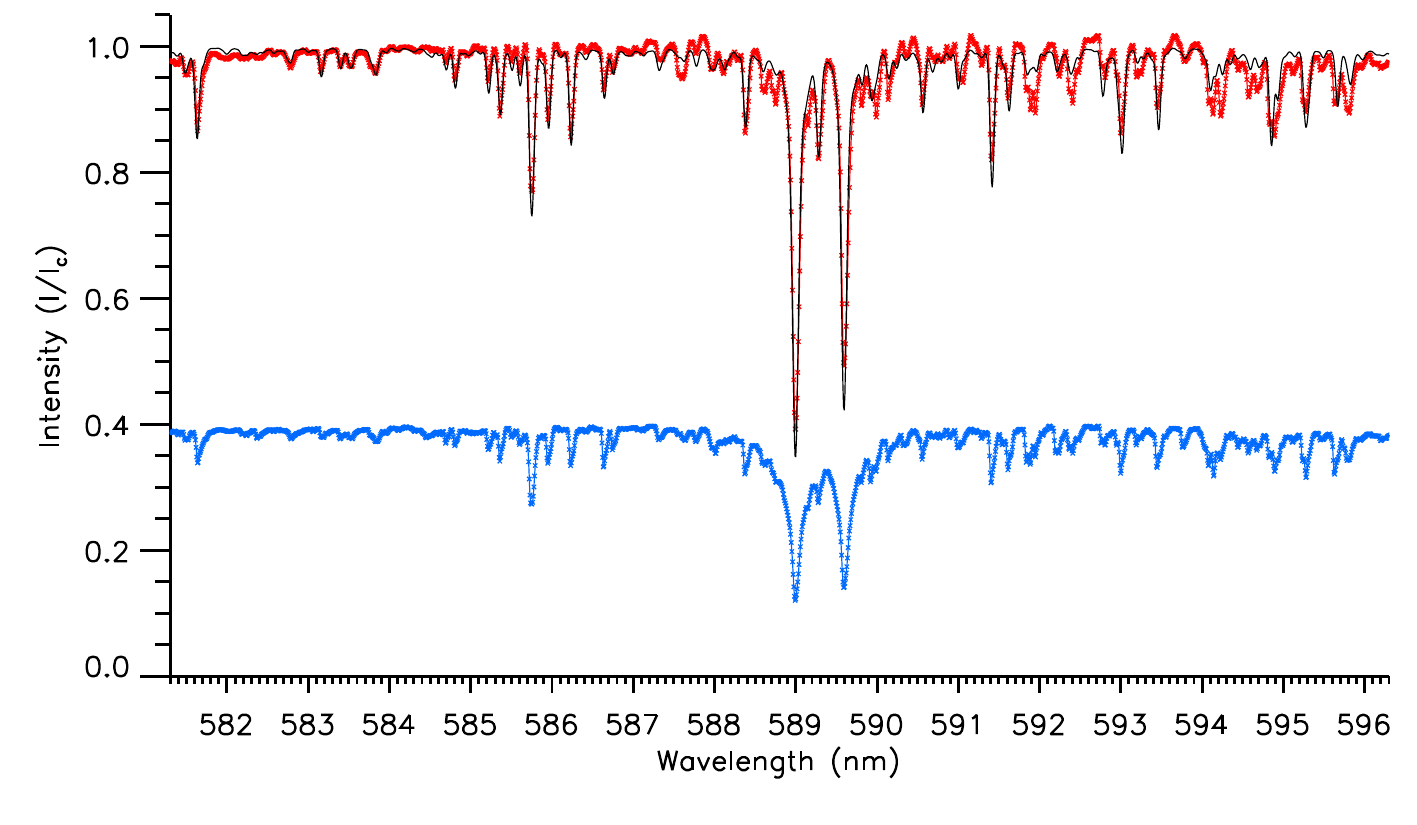}
}
\centerline{
\includegraphics[width=0.8\textwidth, trim= 0mm 0mm 0mm 0mm, clip=]{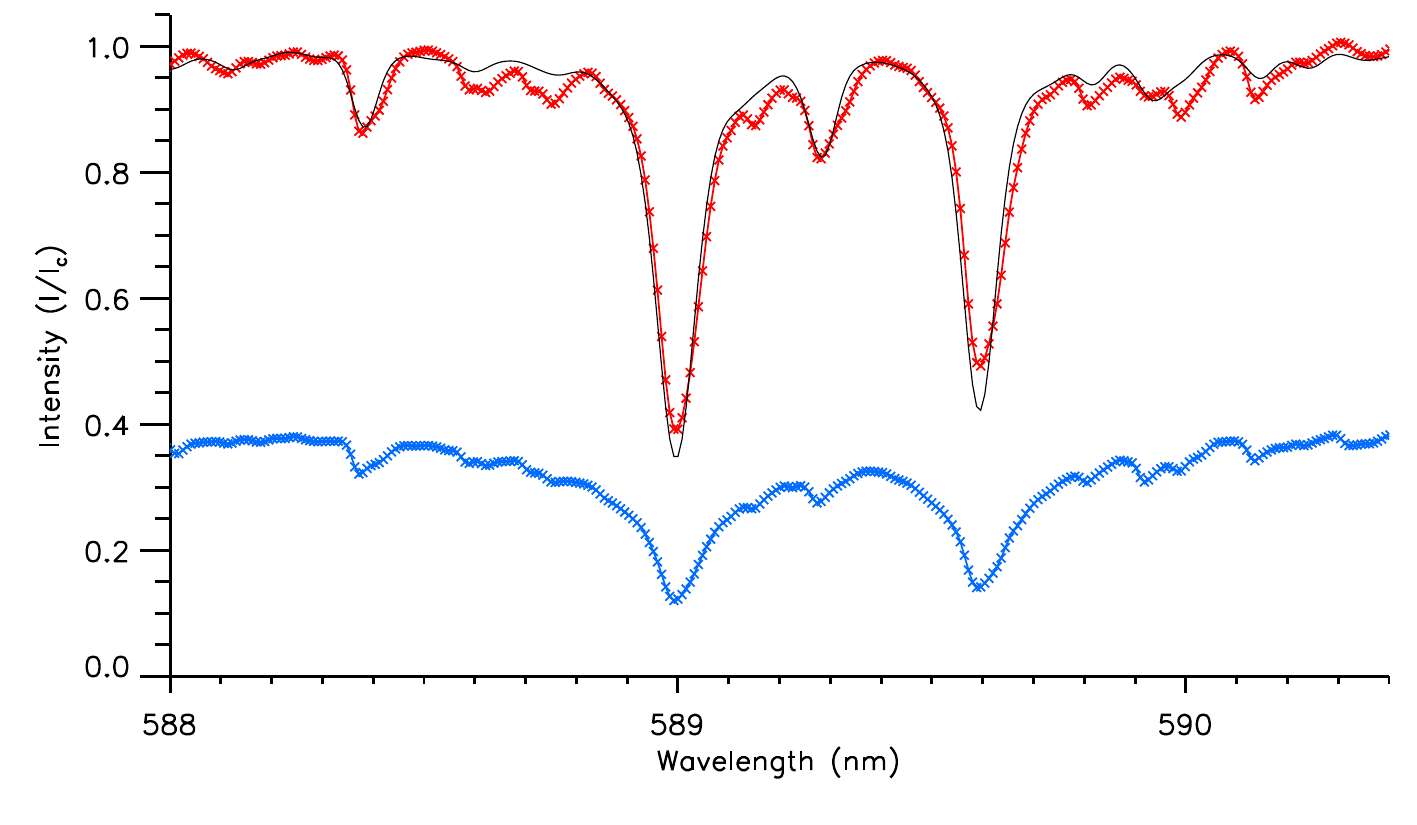}
}
\caption{\sloppy{\nohyphens{Sample spectra captured using the $2400$~lines/mm grating and centred on the Na~{\sc{i}}~D$_{1}$ \& D$_{2}$ absorption line pair, spanning the complete wavelength range of $581.3-596.3$~nm (upper panel). Solid red and blue lines correspond to sample quiet Sun and umbral locations, respectively, while the solid black line displays the quiet Sun FTS spectrum for comparison. The intensity of each spectrum is normalised to the quiet Sun continuum intensity, $I_{\textrm{c}}$, for easier comparison. The lower panel displays a zoom-in to the Na~{\sc{i}}~D$_{1}$ \& D$_{2}$ line pair for better visual clarity. In both panels, individual spectral measurements are indicated using the coloured crosses, with a wavelength sampling, $\Delta\lambda=0.008$~nm\,pixel$^{-1}$, employed across the detector.  }}}
\label{fig:spectra_Na}
\end{figure}

\begin{figure} 
\centerline{
\includegraphics[height=4.5cm, trim= 0mm 0mm 3mm 0mm, clip=]{FOV_ROSA.png}
\includegraphics[height=4.5cm, trim= 36mm 0mm 3mm 0mm, clip=]{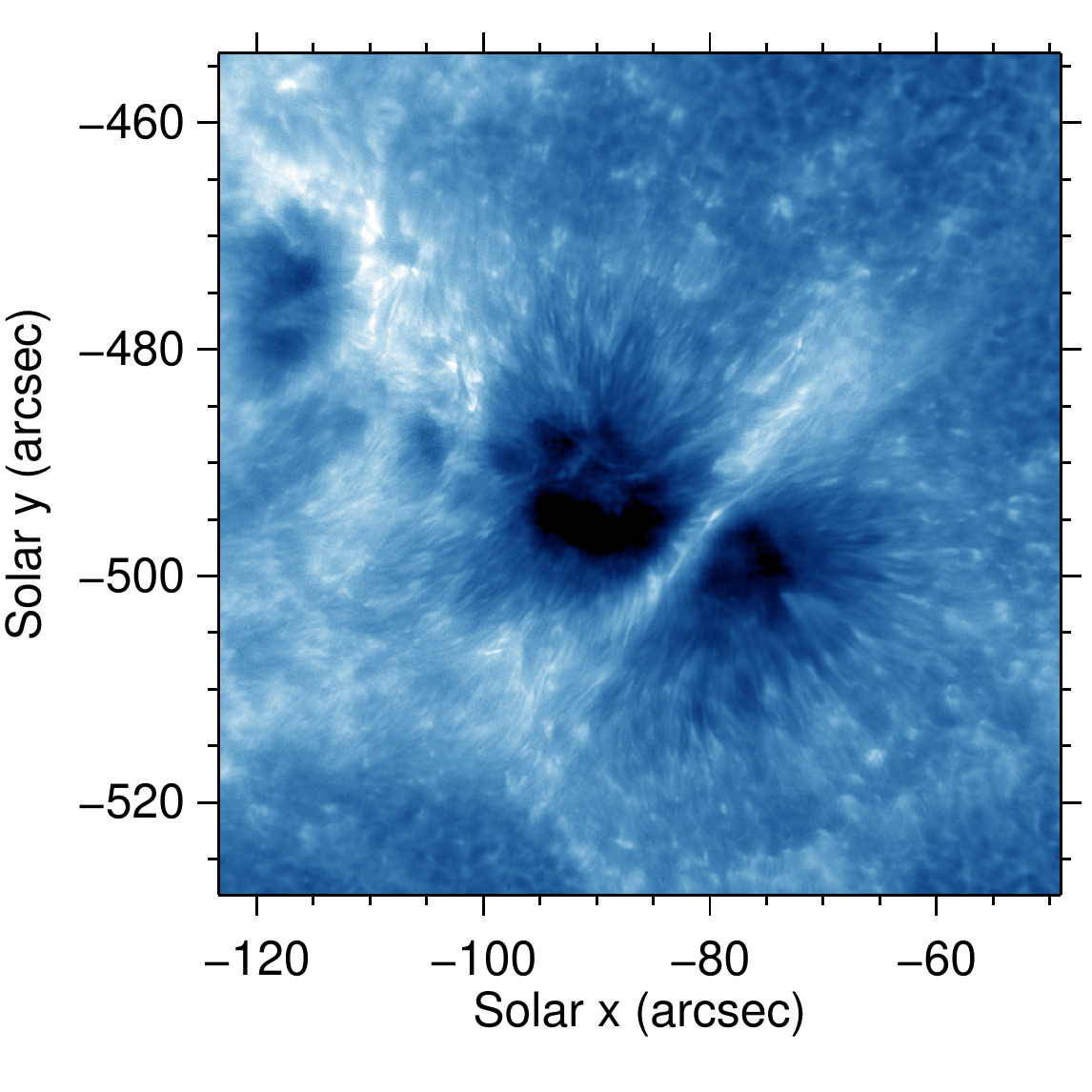}
\includegraphics[height=4.5cm, trim= 36mm 0mm 3mm 0mm, clip=]{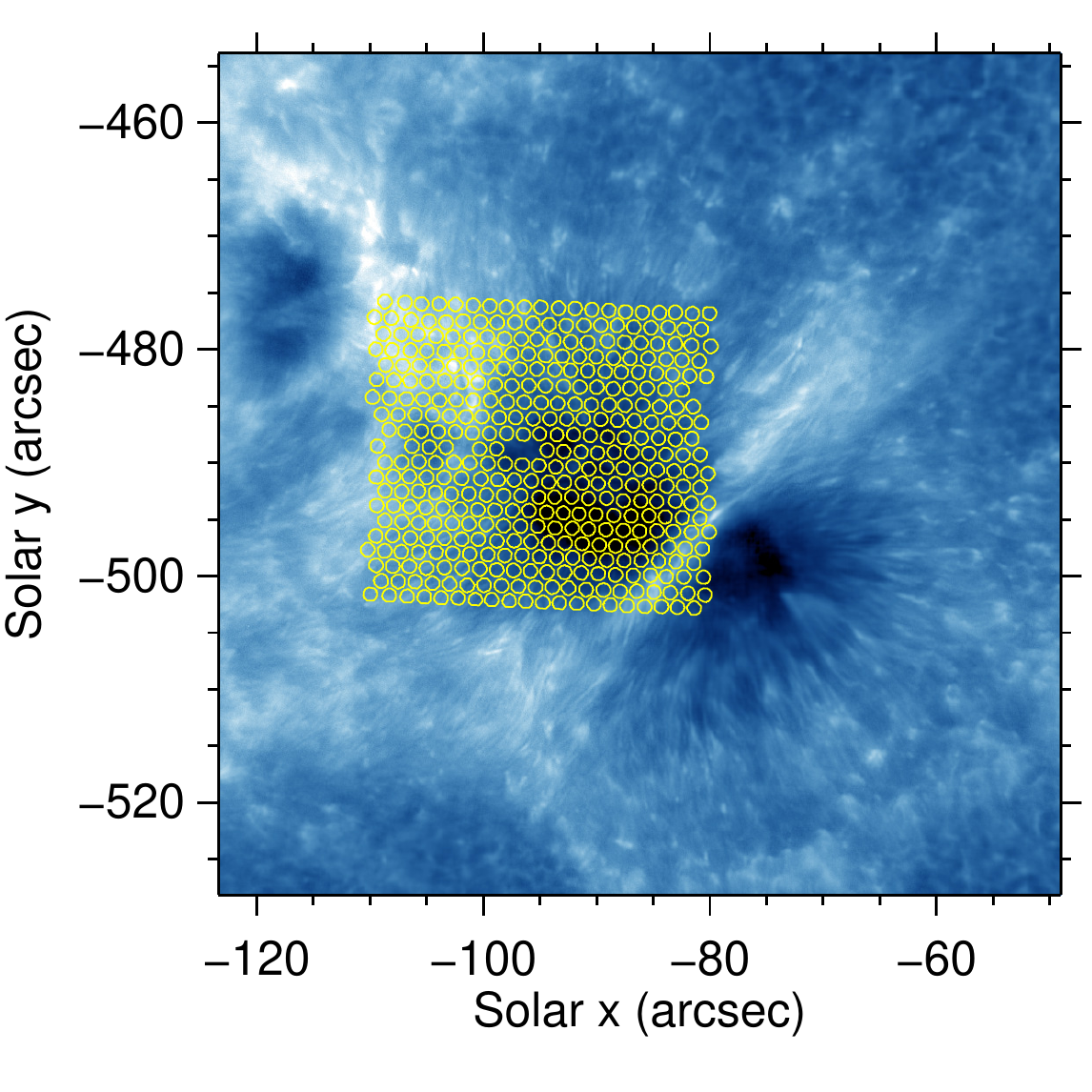}
}
\centerline{
\includegraphics[width=1.0\textwidth, trim= 0mm 0mm 0mm 0mm, clip=]{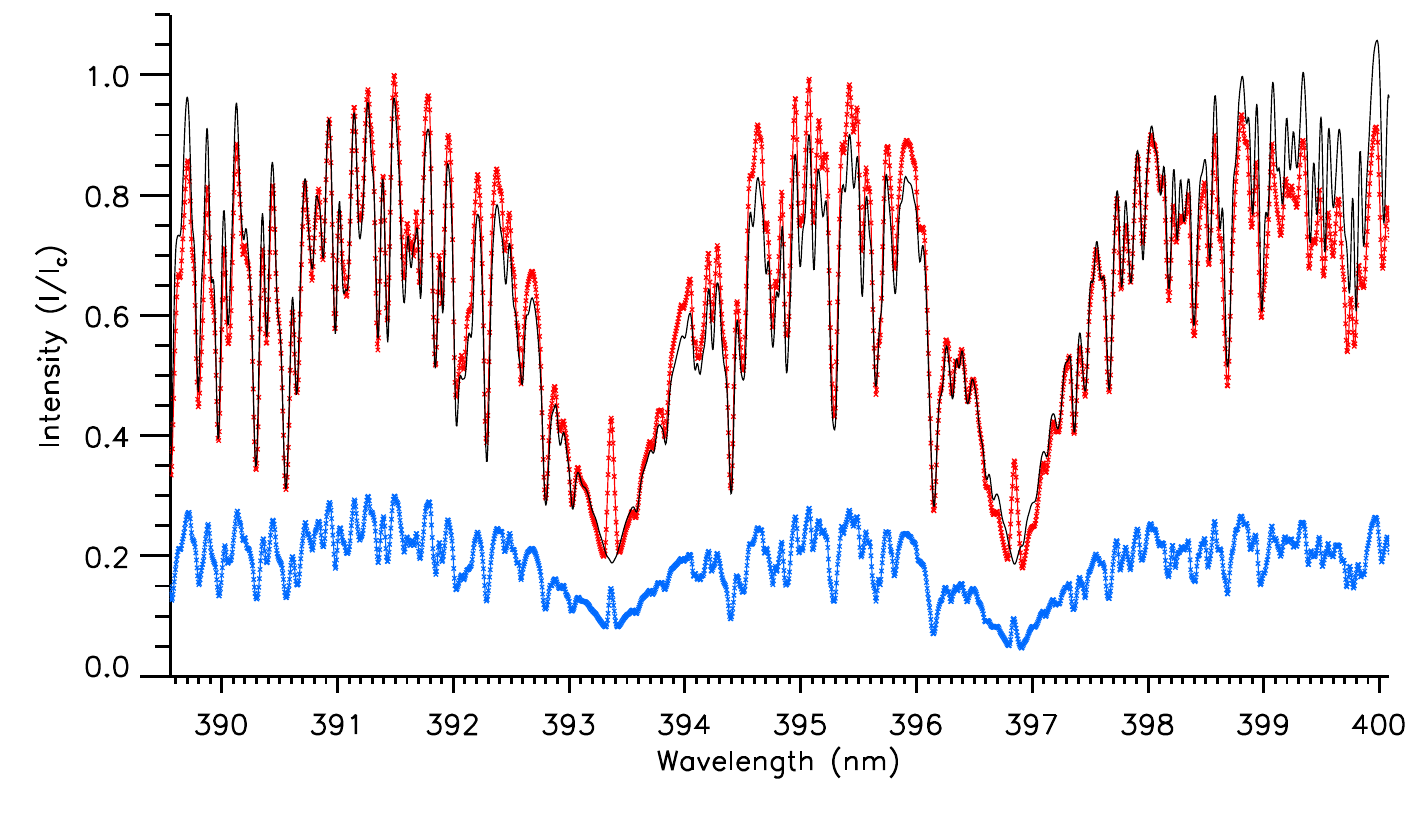}
}
\caption{\sloppy{\nohyphens{The upper panels display the observational field-of-view associated with the Ca~{\sc{ii}}~H/K spectra, where the left and middle panels display sample G-band and Ca~{\sc{ii}}~K images, respectively, obtained by the ROSA instrument that operated simultaneously with {\sc{francis}}. The upper-right panel also shows the ROSA Ca~{\sc{ii}}~K field, only overplotted with the positioning of the {\sc{francis}} fibre bundle using solid yellow contours. Sample spectra captured using the $3600$~lines/mm grating and centred on the Ca~{\sc{ii}}~H/K line pair, spanning the complete wavelength range of $389.55-400.08$~nm, is shown in the lower panel. Solid red and blue lines correspond to sample quiet (i.e., non-umbral) Sun and umbral locations, respectively, while the solid black line displays the disk-averaged solar flux atlas spectrum for comparison. The intensity of each spectrum is normalised to the quiet Sun continuum intensity, $I_{\textrm{c}}$, for easier comparison. Individual spectral measurements are indicated using the coloured crosses, with a wavelength sampling, $\Delta\lambda=0.005$~nm\,pixel$^{-1}$, employed across the detector.  }}}
\label{fig:spectra_Ca}
\end{figure}

Figure~{\ref{fig:spectra_Na}} displays sample spectra acquired during the observations of the sunspot present in NOAA~13089 on 2022 August 29. The upper panel of Figure~{\ref{fig:spectra_Na}} displays the spectral window corresponding to the $2400$~lines/mm grating centred on the Na~{\sc{i}}~D$_{1}$ \& D$_{2}$ absorption line pair, with a wavelength sampling of $\Delta\lambda=0.008$~nm\,pixel$^{-1}$ employed across the detector. The lower panel of Figure~{\ref{fig:spectra_Na}} provides a zoom-in to the central wavelength region encompassing the Na~{\sc{i}}~D$_{1}$ \& D$_{2}$ absorption lines. In each panel, the solid red line corresponds to spectra extracted from non-umbral locations, while the solid blue line reveals spectra obtained from the sunspot umbral core. It can be seen that the umbral continuum has intensities that are $\sim40$\% of its non-umbral counterpart, which is consistent with previous observations \citep[e.g.,][to name but a few examples]{Gosain2011, 2012A&A...541A..60R, 2018NatPh..14..480G, 2022ApJ...938..143G, 2022ApJ...936...37L} and validates the success of the flat-fielding processing outlined in Section~{\ref{sec:datareduction}}. Furthermore, in the lower panel of Figure~{\ref{fig:spectra_Na}}, increased line broadening can be seen in the spectrum extracted from an umbral location, which is consistent with Zeeman effects previously observed \citep{1897ApJ.....5..332Z, 10.1119/1.4975109, Zhang2020}. For comparison, the solid black lines used in Figure~{\ref{fig:spectra_Na}} display the FTS atlas spectrum \citep{1984sfat.book.....K, 2005MSAIS...8..189K} that has been degraded to match the instrumental resolving power ($R_{2400} \approx 15\,817$) of {\sc{francis}}. Very close agreement between the quiescent {\sc{francis}} spectrum and that provided from the FTS atlas reiterates the precise wavelength calibration performed during the data reduction phase. 

In a similar manner to the upper panel of Figure~{\ref{fig:spectra_Na}}, the lower panel of Figure~{\ref{fig:spectra_Ca}} displays non-umbral and umbral spectra using solid red and blue lines, respectively, for the wavelength window obtained using the $3600$~lines/mm diffraction grating centred on the Ca~{\sc{ii}}~H/K absorption line pair. Here, a wavelength sampling of ${\Delta\lambda=0.005}$~nm\,pixel$^{-1}$ is employed across the detector, providing spectral coverage spanning approximately $390 - 400$~nm. The FTS solar spectrum is again degraded to match the resolving power ($R_{3600} \approx 16\,860$) of the {\sc{francis}} observations and overplotted using a solid black line. The FTS atlas is a disk-integrated flux spectrum and, as a result, does not display any self-reversals at the core of the Ca~{\sc{ii}}~H/K lines. On the other hand, self-reversals are clearly visible in Figure~{\ref{fig:spectra_Ca}}, particularly for the spectrum extracted from the umbral location, where intense shock fronts (and their associated strong self-reversals) are commonly expected \citep[e.g.,][to name but a few examples employing Ca~{\sc{ii}} spectra]{1969SoPh....7..351B, 2000Sci...288.1396S, 2013A&A...556A.115D, 2017ApJ...845..102H, 2018ApJ...860...28H, 2020ApJ...892...49H, 2021RSPTA.37900171M}. Even in the non-umbral location (red line in Figure~{\ref{fig:spectra_Ca}}), the extracted spectrum demonstrates self-reversals at the cores of the Ca~{\sc{ii}}~H \& K lines, indicating that these locations are still not entirely quiescent, which is a common feature regularly observed in the immediate vicinity of solar active regions \citep{2018A&A...611A..62B, 2018A&A...612A..28L} and often linked to a measure of magnetic activity \citep[see, e.g., the $S$-index;][]{1968ApJ...153..221W}. As beam-splitters (instead of dichroics) were utilised for the {\sc{francis}} commissioning observations, it was possible to simultaneously capture high-cadence Ca~{\sc{ii}}~K imaging with ROSA while the {\sc{francis}} Ca~{\sc{ii}}~H/K spectra were being obtained. The upper panels of Figure~{\ref{fig:spectra_Ca}} display the two-dimensional fibre array coverage across the sunspot structure, which provides a significant number of fibres sampling uniquely umbral, penumbral, and quiescent (non-umbral) locations. 

Overall, Figures~{\ref{fig:FOV}}, {\ref{fig:spectra_Na}} \& {\ref{fig:spectra_Ca}} display the capabilities of the {\sc{francis}} instrument in its current form. We have demonstrated its suitability to perform in-depth studies of the lower solar atmosphere by combining a spatially-resolved, two-dimensional field of view with excellent spectral coverage and rapid temporal sampling. In its present form, the {\sc{francis}} instrument significantly exceeds the scientific requirements for temporal cadence of $\sim3$~s that were defined for the NLST by \citet{2016ExA....42..271R}. Furthermore, we have demonstrated excellent data fidelity in key spectral lines defined in the NLST science objectives \citep[e.g.,][]{2010ASSP...19..156S, 2016ExA....42..271R}, namely for the Ca~{\sc{ii}}~H/K and Na~{\sc{i}}~D$_{1}$/D$_{2}$ absorption features. Finally, employing the $4320$~lines/mm grating, a spectral resolving power of $R_{4320} \approx 20\,278$ was achieved using a slit width of $50~\mu$m, which results in a spectral line FWHM of $\approx0.018$~nm. While this is not quite at the scientific requirements level of a several tens of~m{\AA} for chromospheric MHD wave diagnostics, we have shown how reducing the slit width down to narrower values will significantly improve the subsequent FWHM values to $\approx0.006$~nm, which is consistent with the near-UV spectrograph onboard IRIS \citep{2014SoPh..289.2733D}. As a result, the performance of {\sc{francis}}, demonstrated during the commissioning phase, is encouraging as a baseline for future instrument upgrades, which will be discussed in more detail in Section~{\ref{sec:currentstatus}} below.  

\subsection{Current status and future upgrades}
\label{sec:currentstatus}
At present, the {\sc{francis}} instrument is available as a common-user facility at the Dunn Solar Telescope (DST), which is operated by the Sunspot Solar Observatory Consortium (SSOC) that is led by New Mexico State University (NMSU). As the Indian National Large Solar Telescope (NLST) is still in the planning phase, it is not clear when {\sc{francis}} will be transferred to the NLST for integration testing. Thankfully, due to the rugged cast steel construction of the spectrograph and hardened flight cases, it will be possible to move {\sc{francis}} to any other available telescope facility, even if temporarily, to undertake scientific operations.

\begin{figure} 
\centerline{
\includegraphics[width=0.48\textwidth, trim= 160mm 0mm 180mm 0mm, clip=]{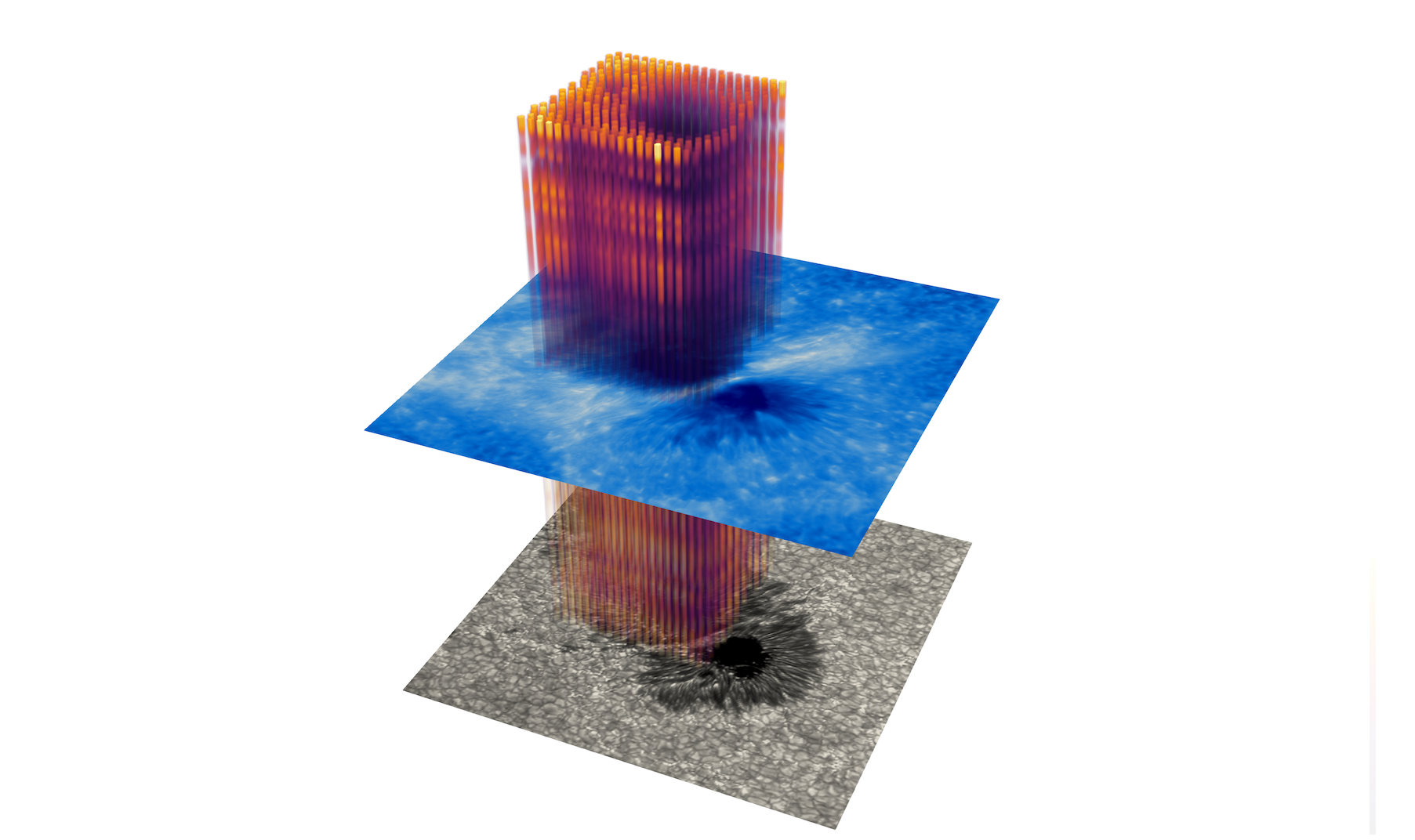}
\includegraphics[width=0.48\textwidth, trim= 160mm 0mm 180mm 0mm, clip=]{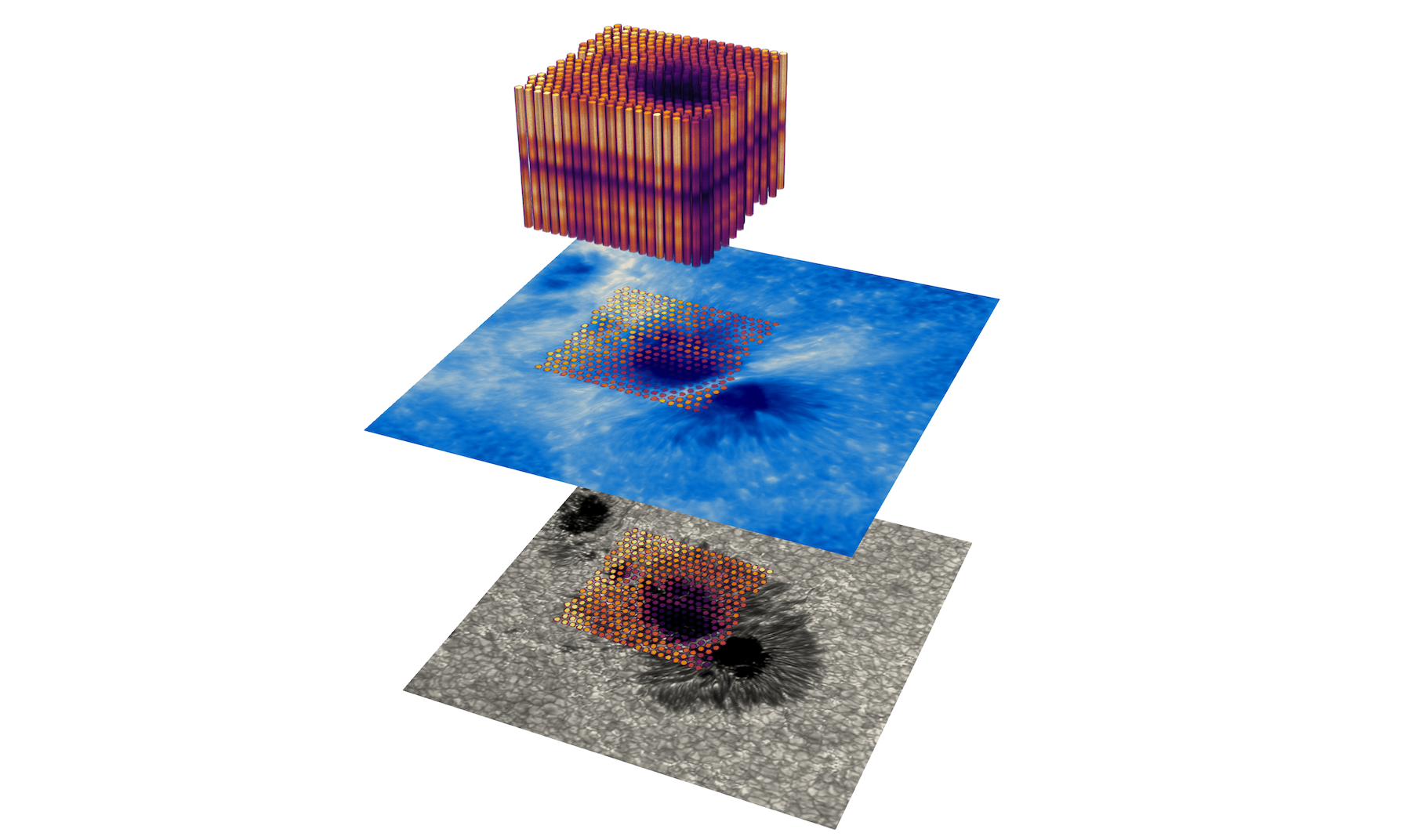}
}
\caption{\sloppy{\nohyphens{Three-dimensional visualisations of {\sc{francis}} data, alongside complementary high-cadence images obtained by the ROSA instrument. Here, ROSA G-band and Ca~{\sc{ii}}~K line-core images are shown using black/white and blue colour tables, respectively. In the left panel, the spatially-resolved {\sc{francis}} spectra, extracted from a window centred on the Ca~{\sc{ii}}~K spectral line at $393.366$~nm, is displayed with the wavelength axis orientated along the vertical direction, where the Ca~{\sc{ii}}~K line core is placed in the same plane as the corresponding Ca~{\sc{ii}}~K ROSA image. The right panel depicts spatially-resolved intensities extracted from the {\sc{francis}} spectra at photospheric (Ca~{\sc{ii}}~K continuum) and chromospheric (Ca~{\sc{ii}}~K line core) heights, which are placed on top of the ROSA G-band and Ca~{\sc{ii}}~K line-core images, respectively. Hovering above the ROSA Ca~{\sc{ii}}~K image is a sample spectral cube ($x, y, \lambda$) acquired instantaneously by the {\sc{francis}} instrument, where the spectral morphology of the underlying sunspot is clearly visible through the examination of spatially-resolved absorption lines that are seen along the vertical (wavelength) domain of the depicted sample cube. These types of novel visualisation tools will greatly assist with the identification, extraction, and analyses of dynamic solar features manifesting in {\sc{francis}} data products.  }}}
\label{fig:futurework}
\end{figure}

As discussed in Section~{\ref{sec:dataverification}}, we have demonstrated the excellent data fidelity provided by the {\sc{francis}} instrument and, hence, its suitability for exciting research projects involving observations of the lower solar atmosphere. In particular, the extensive wavelength range provides many isolated spectral lines, formed over a wide range of heights in the solar atmosphere, which enables a multitude of wave propagation studies to be undertaken through the analyses of oscillatory phase lags bridging different spectral features \citep[e.g., continuing the works of][to name but a few examples]{1981A&A...102..147K, 1983A&A...123..263U, 1991A&A...241..212M, 2006ApJ...640.1153C, 2009ApJ...692.1211C, 2012ApJ...746..183J, 2017ApJS..229....7G, 2017ApJS..229...10J, 2017ApJS..229....9J, 10.1093/mnras/sty1861, 2022ApJ...938..143G, 2022A&A...665L...2G}. Furthermore, the rapid cadences associated with {\sc{francis}} data products will enable the development of highly dynamic shock phenomena within the chromosphere to be examined with unprecedented resolution. With this in mind, shocks developing in Ca~{\sc{ii}} bright grains \citep[e.g.,][]{1992ApJ...397L..59C, 1997ApJ...481..500C, 2022A&A...668A.153M}, sunspot umbral observations \citep[e.g.,][]{2003A&A...403..277R, 2013A&A...556A.115D, 2018NatPh..14..480G, 2018ApJ...860...28H, 2018A&A...619A..63J, 2019ApJ...882..161A, 2020ApJ...892...49H}, and other small-scale atmospheric dynamics \citep[e.g.,][]{2021RSPTA.37900185E, Wang_2021, HILLIER20231962} can be examined with high temporal resolution to better identify the mechanisms responsible for non-linear wave behaviour. Finally, the simultaneous spatial sampling intrinsic to IFU designs means that the spectral signatures of reconnection phenomena (e.g., through the manifestation of solar flares) can be probed more readily since the field-of-view of the {\sc{francis}} instrument can be overlaid on top of the magnetic neutral line to maximise the capture of associated flare kernels, alongside the effects of chromospheric evaporation \citep[e.g.,][]{1998A&AS..129..553X, Hao2017, 10.1093/pasj/psy105, 2021ApJ...915...16M, 10.1093/mnras/stac2570}. Importantly, the ease of visualisation of {\sc{francis}} data will enable scientific discoveries to be identified, analysed, and interpreted in a timely manner. Such novel visualisation capabilities are shown in Figure~{\ref{fig:futurework}}, where we demonstrate how complementary high-cadence imaging and IFU spectroscopy (and future spectropolarimetry) are able to work in harmony.

In the short term, the {\sc{francis}} instrument will continue to be developed and refined by introducing the (single-beam) polarisation optics (see Section~{\ref{sec:polarisationoptics}}), developing a robust correction for residual wavelength-dependent astigmatisms (see Section~{\ref{sec:resolvingpower}}), and examining the effect on exposure times and spectral sensitivity through use of narrower entrance slits. At present, we are also laboratory testing the use of digital micromirror devices \citep[DMDs;][]{Dudley2003, Lee2008}, which are semi-conductor-based optical micro-electrical-mechanical systems consisting of hundreds of thousands of highly reflective tilted micromirrors. In particular, we are investigating the suitability of DMDs to enable high-speed (upwards of $1000$~Hz) switching between orthogonal polarisation states, which may provide better polarimetric precision over its optical shutter counterpart, while still operating within a single-beam configuration to preserve the large wavelength coverage. Similar methods to rapidly modulate the polarisation states have been implemented by the Zurich Imaging Polarimeters \citep[ZIMPOL;][]{10.1117/12.200596} and the Fast SpectroPolarimeter \citep[FSP;][]{2016A&A...590A..89I} instruments, which are able to utilise fast charge transfer in the silicon chip architecture and ferroelectric liquid crystals, respectively, to push polarimetric modulation rates in excess of several hundred measured states each second. Importantly, DMDs consume relatively little electrical power and reliability tests have shown very long lifetimes, with greater than $100{\,}000$ operating hours and more than 1~trillion mirror cycles before failure \citep{Douglass1998}. This makes DMDs a very attractive technology for both ground-based and space-borne hardware due to their reliability, small size, and application longevity. 

In the longer term, we will examine the cost effectiveness of implementing a ruled diffraction grating with steep blaze angles (as an alternative to the holographic gratings presently employed) to obtain spectra in higher spectral orders to further improve the resolving power of the {\sc{francis}} instrument. The optical design of the {\sc{francis}} instrument is also ideally configured to test the feasibility of using volume holographic gratings \citep[VHGs;][]{10.1117/1.1775230} as the dispersive elements for high-resolution solar spectropolarimetry. Holographic gratings (such as those currently employed) and VHGs are two types of gratings created using laser interference patterns. Holographic gratings are surface relief structures, while VHGs are formed within the volume of the component itself by introducing a periodic modulation of the refractive indices embedded throughout the photosensitive material. When compared to standard holographic diffraction gratings, the volumetric period modulation of the refractive indices within VHGs enables them to diffract light over a wide range of wavelengths with even higher efficiencies \citep{Barden_2000}. Additionally, the {\sc{francis}} design may allow future evaluations of profile classifications in near-real time based on a diffractive neural network inscribed within the VHG itself \citep{1990Natur.343..325P}. To further improve the spectral resolving power, we will investigate a revised optical design to increase the $f$-number of the optics, which will naturally boost the spectral resolution achievable, although it is noted that this would be a major revision and may be more suited to a second generation iteration of the instrument.

\section{Conclusion}
\label{sec:conclusion}
In recent years, the solar physics community has expressed a strong need for the acquisition of spectra with high spatial, spectral, and temporal resolutions, including the capture of polarimetric information for use with modern inversion routines. Furthermore, to combat potential analysis difficulties associated with the non-simultaneity of traditional slit-based rastering spectrographs and Fabry-P{\'{e}}rot interferometers, there has been a monumental push towards the creation of integral field units (IFUs) to enable contemporaneous high-resolution spectra to be obtained from a well-resolved, two-dimensional field of view. To date, the majority of solar IFU designs have focused on the mid-optical to infrared regions of the electromagnetic spectrum, hence creating the motivation for an IFU dedicated to the near-UV.

Here, we describe the scientific goals, construction, commissioning, and initial science verification of a new prototype IFU designed to operate within the spectral range spanning approximately $390-700$~nm. The Fibre-Resolved opticAl and Near-ultraviolet Czerny-Turner Imaging Spectropolarimeter ({\sc{francis}}) is a fibre-fed IFU that employs a compact design, holographic diffraction gratings and a high-performance, large-format scientific CMOS camera to obtain spectra from a re-mappable 400-fibre ($20\times20$) imaging array. Commissioning of the {\sc{francis}} instrument took place in August 2022 at the Dunn Solar Telescope in the Sacramento Peak mountains, NM, USA, with the initial science verification focused on the intensity and wavelength calibrations of Stokes~$I$ spectral images linked to spectral windows incorporating the Ca~{\sc{ii}}~H/K and Na~{\sc{i}}~D$_{1}$/D$_{2}$ Fraunhofer lines. 

Following the successful commissioning, imminent upgrades to the {\sc{francis}} IFU include the addition of polarisation optics to allow the high-speed acquisition of complementary Stokes~$Q/U/V$ spectra, which will enable modern multi-line spectropolarimetric inversion routines to be applied to the data. As technology progressively becomes miniaturised, more reliable and, ultimately, cheaper, longer-term upgrades will focus on the implementation of high-speed digital micromirror devices (DMDs) and higher-order ruled diffraction gratings to further increase the spectral sensitivity (and speed of acquisition) of the measured polarimetry. 

Ultimately, the future of IFUs for solar physics research is very promising. The latest (and future) generations of solar telescopes, including the ground-based DKIST, EST, and Indian NLST, as well as upcoming space-borne missions \citep[e.g., the Brazilian Galileo Space Solar Telescope, GSST;][]{2019AGUFMSH13C3447A}, all have IFUs at the forefront of their proposed scientific hardware. Hence, the {\sc{francis}} instrument helps to pave this path by demonstrating its suitability for near-UV observations with simultaneous high spectral, spatial and temporal resolutions. Therefore, we hope that the {\sc{francis}} instrument will inspire and motivate next generation instrument developments as the community looks forward to the scientific discoveries that await us over the next decade.

%

%

%
\begin{acks}
D.B.J. and S.D.T.G. acknowledge support from the UK Space Agency for a National Space Technology Programme (NSTP) Technology for Space Science award (SSc~009). 
D.B.J. also wishes to thank the UK Science and Technology Facilities Council (STFC) for the award of an Ernest Rutherford Fellowship (ST/K004220/1) where this project was initially conceived. 
D.B.J. and S.D.T.G. are also grateful to STFC for additional funding via the grant awards ST/T00021X/1 and ST/X000923/1. 
D.B.J. and W.B. also wish to thank The Leverhulme Trust for grant RPG-2019-371. 
The authors would like to offer gratitude to Prof. Mihalis Mathioudakis and Prof. Francis Keenan, who offered expert guidance throughout the project, including both on scientific and financial issues, with the overall success of the project largely attributed to their assistance.
D.B.J. would also like to thank Dr Hans-Peter Doerr and Dr Michiel van Noort for discussions linked to fibre interference patterns, which helped direct {\sc{francis}} instrument development.  
The authors wish to acknowledge scientific discussions with the Waves in the Lower Solar Atmosphere (WaLSA; \href{https://www.WaLSA.team}{https://www.WaLSA.team}) team, which has been supported by the Research Council of Norway (project no. 262622), The Royal Society \citep[award no. Hooke18b/SCTM;][]{2021RSPTA.37900169J}, and the International Space Science Institute (ISSI Team~502). 
Finally, D.B.J. would like to thank the anonymous referee for their invaluable suggestions and comments that improved the overall quality of the manuscript. 
\end{acks}

\begin{authorcontribution}
D.B.J. defined the instrument goals, designed the optical configuration, assembled and calibrated the instrument in laboratory conditions, assisted with the commissioning observations, reduced and verified the data, and wrote the manuscript. 
S.D.T.G. assisted with laboratory calibrations, performed the commissioning observations, helped reduce the scientific data, and wrote sections of the manuscript. 
W.B. took part in laboratory calibrations, assisted with the acquisition of commissioning observations, developed the fibre remapping algorithms, and prepared figures for the manuscript.
J.L. worked extensively in the laboratory to align the instrument optics and calibrate the initial spectra. 
S.J. and P.H.K. provided scientific vision, assisted with text development, and examined the final spectra for inclusion within the manuscript.
L.E.A.V., A.D.L. and F.L.G. assisted with spectrograph calibration knowledge, particularly in relation to the use of holographic diffraction gratings, and enabled additional laboratory testing to be performed within Brazil's Instituto Nacional de Investigaci{\'{o}}n Espacial del Brasil (INPE) facility.
D.J.C. assisted with advertising the instrument capabilities within the United States and examined the calibrated spectra for any irregularities. 
D.G. was the (invaluable) Chief Observer during the commissioning observations of {\sc{francis}} during August~2022 and pioneered the optical configuration of {\sc{DOUGcam}}, which has since become a permanent and complementary part of the {\sc{francis}} hardware. 
D.B. provided initial motivation for the project and enabled D.B.J. to become part of the Indian NLST project. 
All authors provided contributions towards the final manuscript in the form of figures, text, and proofreading. 
\end{authorcontribution}
\begin{fundinginformation}
UK Space Agency (UKSA): SSc~009.
UK Science and Technology Facilities Council (STFC): ST/K004220/1, ST/T00021X/1, ST/X000923/1.
The Leverhulme Trust: RPG-2019-371.
The Royal Society: Hooke18b/SCTM.
\end{fundinginformation}
\begin{dataavailability}
Due to the data volume (approaching 5~TB), the {\sc{francis}} commissioning data is available, in various stages of reduction format, directly from the lead author (David Jess; \href{mailto:d.jess@qub.ac.uk}{d.jess@qub.ac.uk}). Similarly, complementary ROSA/HARDcam imaging observations can also be provided to interested parties upon request via the same email address (data volume exceeding 10~TB). 
\end{dataavailability}
%
%

%
%
\bibliographystyle{spr-mp-sola}
\bibliography{FRANCIS}  

\begin{thebibliography}{188}
\ifx\bisbn     \undefined \def\bisbn  #1{ISBN #1}\fi
\ifx\binits    \undefined \def\binits#1{#1}\fi
\ifx\bauthor   \undefined \def\bauthor#1{#1}\fi
\ifx\batitle   \undefined \def\batitle#1{#1}\fi
\ifx\bjtitle   \undefined \def\bjtitle#1{\textit{#1}}\fi
\ifx\bvolume   \undefined \def\bvolume#1{\textbf{#1}}\fi
\ifx\byear     \undefined \def\byear#1{#1}\fi
\ifx\bissue    \undefined \def\bissue#1{#1}\fi
\ifx\bfpage    \undefined \def\bfpage#1{#1}\fi
\ifx\blpage    \undefined \def\blpage #1{#1}\fi
\ifx\burl      \undefined \def\burl#1{#1}\fi
\ifx\href      \undefined \def\href#1#2{#2}\fi
\ifx\betal     \undefined \def\betal{et al.}\fi
\ifx\bctitle   \undefined \def\bctitle#1{#1}\fi
\ifx\beditor   \undefined \def\beditor#1{#1}\fi
\ifx\bbtitle   \undefined \def\bbtitle#1{\textit{#1}}\fi
\ifx\bedition  \undefined \def\bedition#1{#1}\fi
\ifx\bseriesno \undefined \def\bseriesno#1{\textbf{#1}}\fi
\ifx\blocation \undefined \def\blocation#1{#1}\fi
\ifx\bsertitle \undefined \def\bsertitle#1{\textit{#1}}\fi
\ifx\bsnm      \undefined \def\bsnm#1{#1}\fi
\ifx\bsuffix   \undefined \def\bsuffix#1{#1}\fi
\ifx\bparticle \undefined \def\bparticle#1{#1}\fi
\ifx\barticle  \undefined \def\barticle#1{}\fi
\ifx\binstitute  \undefined \def\binstitute#1{#1}\fi
\ifx\bpublisher  \undefined \def\bpublisher#1{#1}\fi
\ifx\doiurl    \undefined \def\doiurl#1{\href{#1}{DOI}}\fi
\makeatletter
\def\safeHref#1#2#3{\in@{http}{#2}\ifin@\href{#2}{#3}\else\href{#1#2}{#3}\fi}
\makeatother
\ifx\adsurl    \undefined
  \def\adsurl#1{\safeHref{https://ui.adsabs.harvard.edu/abs/}{#1}{ADS}}\fi
\ifx\arxivurl  \undefined
  \def\arxivurl#1{\safeHref{http://arxiv.org/abs/}{#1}{arXiv}}\fi
\ifx\botherref \undefined \def\botherref#1{}\fi
\ifx\url       \undefined \def\url#1{#1}\fi
\ifx\bchapter  \undefined \def\bchapter#1{}\fi
\ifx\bbook     \undefined \def\bbook#1{}\fi
\ifx\bcomment  \undefined \def\bcomment#1{#1}\fi
\ifx\oauthor   \undefined \def\oauthor#1{#1}\fi
\ifx\citeauthoryear \undefined\def \citeauthoryear#1{#1}\fi
\def\endbibitem {}
\ifx\bconflocation  \undefined \def\bconflocation#1{#1} \fi

\bibitem[\protect\citeauthoryear{{Abbasvand}
  et~al.}{2020}]{2020A&A...642A..52A}
\begin{barticle}
\bauthor{\bsnm{{Abbasvand}}, \binits{V.}},
\bauthor{\bsnm{{Sobotka}}, \binits{M.}},
\bauthor{\bsnm{{{\v{S}}vanda}}, \binits{M.}},
\bauthor{\bsnm{{Heinzel}}, \binits{P.}},
\bauthor{\bsnm{{Garc{\'\i}a-Rivas}}, \binits{M.}},
\bauthor{\bsnm{{Denker}}, \binits{C.}},
\bauthor{\bsnm{{Balthasar}}, \binits{H.}},
\bauthor{\bsnm{{Verma}}, \binits{M.}},
\bauthor{\bsnm{{Kontogiannis}}, \binits{I.}},
\bauthor{\bsnm{{Koza}}, \binits{J.}},
\bauthor{\bsnm{{Korda}}, \binits{D.}},
\bauthor{\bsnm{{Kuckein}}, \binits{C.}}:
\byear{2020},
\batitle{{Observational study of chromospheric heating by acoustic waves}}.
\bjtitle{\aap}
\bvolume{642},
\bfpage{A52}.
\doiurl{https://doi.org/10.1051/0004-6361/202038559}.
\adsurl{2020A&A...642A..52A}.
\end{barticle}
\endbibitem

\bibitem[\protect\citeauthoryear{{Abell} et~al.}{2009}]{2009arXiv0912.0201L}
\begin{botherref}
\oauthor{\bsnm{{Abell}}, \binits{P.A.}},
\oauthor{\bsnm{{Allison}}, \binits{J.}},
\oauthor{\bsnm{{Anderson}}, \binits{S.F.}},
\oauthor{\bsnm{{Andrew}}, \binits{J.R.}},
\oauthor{\bsnm{{Angel}}, \binits{J.R.P.}},
\oauthor{\bsnm{{Armus}}, \binits{L.}},
\oauthor{\bsnm{{Arnett}}, \binits{D.}},
\oauthor{\bsnm{{Asztalos}}, \binits{S.J.}},
\oauthor{\bsnm{{Axelrod}}, \binits{T.S.}},
\oauthor{\bsnm{{Bailey}}, \binits{S.}},
\oauthor{\bsnm{{Ballantyne}}, \binits{D.R.}},
\oauthor{\bsnm{{Bankert}}, \binits{J.R.}},
\oauthor{\bsnm{{Barkhouse}}, \binits{W.A.}},
\oauthor{\bsnm{{Barr}}, \binits{J.D.}},
\oauthor{\bsnm{{Barrientos}}, \binits{L.F.}},
\oauthor{\bsnm{{Barth}}, \binits{A.J.}},
\oauthor{\bsnm{{Bartlett}}, \binits{J.G.}},
\oauthor{\bsnm{{Becker}}, \binits{A.C.}},
\oauthor{\bsnm{{Becla}}, \binits{J.}},
\oauthor{\bsnm{{Beers}}, \binits{T.C.}},
\oauthor{\bsnm{{Bernstein}}, \binits{J.P.}},
\oauthor{\bsnm{{Biswas}}, \binits{R.}},
\oauthor{\bsnm{{Blanton}}, \binits{M.R.}},
\oauthor{\bsnm{{Bloom}}, \binits{J.S.}},
\oauthor{\bsnm{{Bochanski}}, \binits{J.J.}},
\oauthor{\bsnm{{Boeshaar}}, \binits{P.}},
\oauthor{\bsnm{{Borne}}, \binits{K.D.}},
\oauthor{\bsnm{{Bradac}}, \binits{M.}},
\oauthor{\bsnm{{Brandt}}, \binits{W.N.}},
\oauthor{\bsnm{{Bridge}}, \binits{C.R.}},
\oauthor{\bsnm{{Brown}}, \binits{M.E.}},
\oauthor{\bsnm{{Brunner}}, \binits{R.J.}},
\oauthor{\bsnm{{Bullock}}, \binits{J.S.}},
\oauthor{\bsnm{{Burgasser}}, \binits{A.J.}},
\oauthor{\bsnm{{Burge}}, \binits{J.H.}},
\oauthor{\bsnm{{Burke}}, \binits{D.L.}},
\oauthor{\bsnm{{Cargile}}, \binits{P.A.}},
\oauthor{\bsnm{{Chandrasekharan}}, \binits{S.}},
\oauthor{\bsnm{{Chartas}}, \binits{G.}},
\oauthor{\bsnm{{Chesley}}, \binits{S.R.}},
\oauthor{\bsnm{{Chu}}, \binits{Y.-H.}},
\oauthor{\bsnm{{Cinabro}}, \binits{D.}},
\oauthor{\bsnm{{Claire}}, \binits{M.W.}},
\oauthor{\bsnm{{Claver}}, \binits{C.F.}},
\oauthor{\bsnm{{Clowe}}, \binits{D.}},
\oauthor{\bsnm{{Connolly}}, \binits{A.J.}},
\oauthor{\bsnm{{Cook}}, \binits{K.H.}},
\oauthor{\bsnm{{Cooke}}, \binits{J.}},
\oauthor{\bsnm{{Cooray}}, \binits{A.}},
\oauthor{\bsnm{{Covey}}, \binits{K.R.}},
\oauthor{\bsnm{{Culliton}}, \binits{C.S.}},
\oauthor{\bsnm{{de Jong}}, \binits{R.}},
\oauthor{\bsnm{{de Vries}}, \binits{W.H.}},
\oauthor{\bsnm{{Debattista}}, \binits{V.P.}},
\oauthor{\bsnm{{Delgado}}, \binits{F.}},
\oauthor{\bsnm{{Dell'Antonio}}, \binits{I.P.}},
\oauthor{\bsnm{{Dhital}}, \binits{S.}},
\oauthor{\bsnm{{Di Stefano}}, \binits{R.}},
\oauthor{\bsnm{{Dickinson}}, \binits{M.}},
\oauthor{\bsnm{{Dilday}}, \binits{B.}},
\oauthor{\bsnm{{Djorgovski}}, \binits{S.G.}},
\oauthor{\bsnm{{Dobler}}, \binits{G.}},
\oauthor{\bsnm{{Donalek}}, \binits{C.}},
\oauthor{\bsnm{{Dubois-Felsmann}}, \binits{G.}},
\oauthor{\bsnm{{Durech}}, \binits{J.}},
\oauthor{\bsnm{{Eliasdottir}}, \binits{A.}},
\oauthor{\bsnm{{Eracleous}}, \binits{M.}},
\oauthor{\bsnm{{Eyer}}, \binits{L.}},
\oauthor{\bsnm{{Falco}}, \binits{E.E.}},
\oauthor{\bsnm{{Fan}}, \binits{X.}},
\oauthor{\bsnm{{Fassnacht}}, \binits{C.D.}},
\oauthor{\bsnm{{Ferguson}}, \binits{H.C.}},
\oauthor{\bsnm{{Fernandez}}, \binits{Y.R.}},
\oauthor{\bsnm{{Fields}}, \binits{B.D.}},
\oauthor{\bsnm{{Finkbeiner}}, \binits{D.}},
\oauthor{\bsnm{{Figueroa}}, \binits{E.E.}},
\oauthor{\bsnm{{Fox}}, \binits{D.B.}},
\oauthor{\bsnm{{Francke}}, \binits{H.}},
\oauthor{\bsnm{{Frank}}, \binits{J.S.}},
\oauthor{\bsnm{{Frieman}}, \binits{J.}},
\oauthor{\bsnm{{Fromenteau}}, \binits{S.}},
\oauthor{\bsnm{{Furqan}}, \binits{M.}},
\oauthor{\bsnm{{Galaz}}, \binits{G.}},
\oauthor{\bsnm{{Gal-Yam}}, \binits{A.}},
\oauthor{\bsnm{{Garnavich}}, \binits{P.}},
\oauthor{\bsnm{{Gawiser}}, \binits{E.}},
\oauthor{\bsnm{{Geary}}, \binits{J.}},
\oauthor{\bsnm{{Gee}}, \binits{P.}},
\oauthor{\bsnm{{Gibson}}, \binits{R.R.}},
\oauthor{\bsnm{{Gilmore}}, \binits{K.}},
\oauthor{\bsnm{{Grace}}, \binits{E.A.}},
\oauthor{\bsnm{{Green}}, \binits{R.F.}},
\oauthor{\bsnm{{Gressler}}, \binits{W.J.}},
\oauthor{\bsnm{{Grillmair}}, \binits{C.J.}},
\oauthor{\bsnm{{Habib}}, \binits{S.}},
\oauthor{\bsnm{{Haggerty}}, \binits{J.S.}},
\oauthor{\bsnm{{Hamuy}}, \binits{M.}},
\oauthor{\bsnm{{Harris}}, \binits{A.W.}},
\oauthor{\bsnm{{Hawley}}, \binits{S.L.}},
\oauthor{\bsnm{{Heavens}}, \binits{A.F.}},
\oauthor{\bsnm{{Hebb}}, \binits{L.}},
\oauthor{\bsnm{{Henry}}, \binits{T.J.}},
\oauthor{\bsnm{{Hileman}}, \binits{E.}},
\oauthor{\bsnm{{Hilton}}, \binits{E.J.}},
\oauthor{\bsnm{{Hoadley}}, \binits{K.}},
\oauthor{\bsnm{{Holberg}}, \binits{J.B.}},
\oauthor{\bsnm{{Holman}}, \binits{M.J.}},
\oauthor{\bsnm{{Howell}}, \binits{S.B.}},
\oauthor{\bsnm{{Infante}}, \binits{L.}},
\oauthor{\bsnm{{Ivezic}}, \binits{Z.}},
\oauthor{\bsnm{{Jacoby}}, \binits{S.H.}},
\oauthor{\bsnm{{Jain}}, \binits{B.}},
\oauthor{\bsnm{{R}}},
\oauthor{\bsnm{{Jedicke}}},
\oauthor{\bsnm{{Jee}}, \binits{M.J.}},
\oauthor{\bsnm{{Garrett Jernigan}}, \binits{J.}},
\oauthor{\bsnm{{Jha}}, \binits{S.W.}},
\oauthor{\bsnm{{Johnston}}, \binits{K.V.}},
\oauthor{\bsnm{{Jones}}, \binits{R.L.}},
\oauthor{\bsnm{{Juric}}, \binits{M.}},
\oauthor{\bsnm{{Kaasalainen}}, \binits{M.}},
\oauthor{\bsnm{{Styliani}}},
\oauthor{\bsnm{{Kafka}}},
\oauthor{\bsnm{{Kahn}}, \binits{S.M.}},
\oauthor{\bsnm{{Kaib}}, \binits{N.A.}},
\oauthor{\bsnm{{Kalirai}}, \binits{J.}},
\oauthor{\bsnm{{Kantor}}, \binits{J.}},
\oauthor{\bsnm{{Kasliwal}}, \binits{M.M.}},
\oauthor{\bsnm{{Keeton}}, \binits{C.R.}},
\oauthor{\bsnm{{Kessler}}, \binits{R.}},
\oauthor{\bsnm{{Knezevic}}, \binits{Z.}},
\oauthor{\bsnm{{Kowalski}}, \binits{A.}},
\oauthor{\bsnm{{Krabbendam}}, \binits{V.L.}},
\oauthor{\bsnm{{Krughoff}}, \binits{K.S.}},
\oauthor{\bsnm{{Kulkarni}}, \binits{S.}},
\oauthor{\bsnm{{Kuhlman}}, \binits{S.}},
\oauthor{\bsnm{{Lacy}}, \binits{M.}},
\oauthor{\bsnm{{Lepine}}, \binits{S.}},
\oauthor{\bsnm{{Liang}}, \binits{M.}},
\oauthor{\bsnm{{Lien}}, \binits{A.}},
\oauthor{\bsnm{{Lira}}, \binits{P.}},
\oauthor{\bsnm{{Long}}, \binits{K.S.}},
\oauthor{\bsnm{{Lorenz}}, \binits{S.}},
\oauthor{\bsnm{{Lotz}}, \binits{J.M.}},
\oauthor{\bsnm{{Lupton}}, \binits{R.H.}},
\oauthor{\bsnm{{Lutz}}, \binits{J.}},
\oauthor{\bsnm{{Macri}}, \binits{L.M.}},
\oauthor{\bsnm{{Mahabal}}, \binits{A.A.}},
\oauthor{\bsnm{{Mandelbaum}}, \binits{R.}},
\oauthor{\bsnm{{Marshall}}, \binits{P.}},
\oauthor{\bsnm{{May}}, \binits{M.}},
\oauthor{\bsnm{{McGehee}}, \binits{P.M.}},
\oauthor{\bsnm{{Meadows}}, \binits{B.T.}},
\oauthor{\bsnm{{Meert}}, \binits{A.}},
\oauthor{\bsnm{{Milani}}, \binits{A.}},
\oauthor{\bsnm{{Miller}}, \binits{C.J.}},
\oauthor{\bsnm{{Miller}}, \binits{M.}},
\oauthor{\bsnm{{Mills}}, \binits{D.}},
\oauthor{\bsnm{{Minniti}}, \binits{D.}},
\oauthor{\bsnm{{Monet}}, \binits{D.}},
\oauthor{\bsnm{{Mukadam}}, \binits{A.S.}},
\oauthor{\bsnm{{Nakar}}, \binits{E.}},
\oauthor{\bsnm{{Neill}}, \binits{D.R.}},
\oauthor{\bsnm{{Newman}}, \binits{J.A.}},
\oauthor{\bsnm{{Nikolaev}}, \binits{S.}},
\oauthor{\bsnm{{Nordby}}, \binits{M.}},
\oauthor{\bsnm{{O'Connor}}, \binits{P.}},
\oauthor{\bsnm{{Oguri}}, \binits{M.}},
\oauthor{\bsnm{{Oliver}}, \binits{J.}},
\oauthor{\bsnm{{Olivier}}, \binits{S.S.}},
\oauthor{\bsnm{{Olsen}}, \binits{J.K.}},
\oauthor{\bsnm{{Olsen}}, \binits{K.}},
\oauthor{\bsnm{{Olszewski}}, \binits{E.W.}},
\oauthor{\bsnm{{Oluseyi}}, \binits{H.}},
\oauthor{\bsnm{{Padilla}}, \binits{N.D.}},
\oauthor{\bsnm{{Parker}}, \binits{A.}},
\oauthor{\bsnm{{Pepper}}, \binits{J.}},
\oauthor{\bsnm{{Peterson}}, \binits{J.R.}},
\oauthor{\bsnm{{Petry}}, \binits{C.}},
\oauthor{\bsnm{{Pinto}}, \binits{P.A.}},
\oauthor{\bsnm{{Pizagno}}, \binits{J.L.}},
\oauthor{\bsnm{{Popescu}}, \binits{B.}},
\oauthor{\bsnm{{Prsa}}, \binits{A.}},
\oauthor{\bsnm{{Radcka}}, \binits{V.}},
\oauthor{\bsnm{{Raddick}}, \binits{M.J.}},
\oauthor{\bsnm{{Rasmussen}}, \binits{A.}},
\oauthor{\bsnm{{Rau}}, \binits{A.}},
\oauthor{\bsnm{{Rho}}, \binits{J.}},
\oauthor{\bsnm{{Rhoads}}, \binits{J.E.}},
\oauthor{\bsnm{{Richards}}, \binits{G.T.}},
\oauthor{\bsnm{{Ridgway}}, \binits{S.T.}},
\oauthor{\bsnm{{Robertson}}, \binits{B.E.}},
\oauthor{\bsnm{{Roskar}}, \binits{R.}},
\oauthor{\bsnm{{Saha}}, \binits{A.}},
\oauthor{\bsnm{{Sarajedini}}, \binits{A.}},
\oauthor{\bsnm{{Scannapieco}}, \binits{E.}},
\oauthor{\bsnm{{Schalk}}, \binits{T.}},
\oauthor{\bsnm{{Schindler}}, \binits{R.}},
\oauthor{\bsnm{{Schmidt}}, \binits{S.}},
\oauthor{\bsnm{{Schmidt}}, \binits{S.}},
\oauthor{\bsnm{{Schneider}}, \binits{D.P.}},
\oauthor{\bsnm{{Schumacher}}, \binits{G.}},
\oauthor{\bsnm{{Scranton}}, \binits{R.}},
\oauthor{\bsnm{{Sebag}}, \binits{J.}},
\oauthor{\bsnm{{Seppala}}, \binits{L.G.}},
\oauthor{\bsnm{{Shemmer}}, \binits{O.}},
\oauthor{\bsnm{{Simon}}, \binits{J.D.}},
\oauthor{\bsnm{{Sivertz}}, \binits{M.}},
\oauthor{\bsnm{{Smith}}, \binits{H.A.}},
\oauthor{\bsnm{{Allyn Smith}}, \binits{J.}},
\oauthor{\bsnm{{Smith}}, \binits{N.}},
\oauthor{\bsnm{{Spitz}}, \binits{A.H.}},
\oauthor{\bsnm{{Stanford}}, \binits{A.}},
\oauthor{\bsnm{{Stassun}}, \binits{K.G.}},
\oauthor{\bsnm{{Strader}}, \binits{J.}},
\oauthor{\bsnm{{Strauss}}, \binits{M.A.}},
\oauthor{\bsnm{{Stubbs}}, \binits{C.W.}},
\oauthor{\bsnm{{Sweeney}}, \binits{D.W.}},
\oauthor{\bsnm{{Szalay}}, \binits{A.}},
\oauthor{\bsnm{{Szkody}}, \binits{P.}},
\oauthor{\bsnm{{Takada}}, \binits{M.}},
\oauthor{\bsnm{{Thorman}}, \binits{P.}},
\oauthor{\bsnm{{Trilling}}, \binits{D.E.}},
\oauthor{\bsnm{{Trimble}}, \binits{V.}},
\oauthor{\bsnm{{Tyson}}, \binits{A.}},
\oauthor{\bsnm{{Van Berg}}, \binits{R.}},
\oauthor{\bsnm{{Vanden Berk}}, \binits{D.}},
\oauthor{\bsnm{{VanderPlas}}, \binits{J.}},
\oauthor{\bsnm{{Verde}}, \binits{L.}},
\oauthor{\bsnm{{Vrsnak}}, \binits{B.}},
\oauthor{\bsnm{{Walkowicz}}, \binits{L.M.}},
\oauthor{\bsnm{{Wandelt}}, \binits{B.D.}},
\oauthor{\bsnm{{Wang}}, \binits{S.}},
\oauthor{\bsnm{{Wang}}, \binits{Y.}},
\oauthor{\bsnm{{Warner}}, \binits{M.}},
\oauthor{\bsnm{{Wechsler}}, \binits{R.H.}},
\oauthor{\bsnm{{West}}, \binits{A.A.}},
\oauthor{\bsnm{{Wiecha}}, \binits{O.}},
\oauthor{\bsnm{{Williams}}, \binits{B.F.}},
\oauthor{\bsnm{{Willman}}, \binits{B.}},
\oauthor{\bsnm{{Wittman}}, \binits{D.}},
\oauthor{\bsnm{{Wolff}}, \binits{S.C.}},
\oauthor{\bsnm{{Wood-Vasey}}, \binits{W.M.}},
\oauthor{\bsnm{{Wozniak}}, \binits{P.}},
\oauthor{\bsnm{{Young}}, \binits{P.}},
\oauthor{\bsnm{{Zentner}}, \binits{A.}},
\oauthor{\bsnm{{Zhan}}, \binits{H.}}:
2009,
{LSST Science Book, Version 2.0}.
\textit{arXiv e-prints},
arXiv:0912.0201.
\doiurl{https://doi.org/10.48550/arXiv.0912.0201}.
\adsurl{2009arXiv0912.0201L}.
\end{botherref}
\endbibitem

\bibitem[\protect\citeauthoryear{{Allington-Smith}}{2006}]{2006NewAR..50..244A}
\begin{barticle}
\bauthor{\bsnm{{Allington-Smith}}, \binits{J.}}:
\byear{2006},
\batitle{{Basic principles of integral field spectroscopy}}.
\bjtitle{\nar}
\bvolume{50},
\bfpage{244}.
\doiurl{https://doi.org/10.1016/j.newar.2006.02.024}.
\adsurl{2006NewAR..50..244A}.
\end{barticle}
\endbibitem

\bibitem[\protect\citeauthoryear{Anan et~al.}{2018}]{10.1093/pasj/psy105}
\begin{barticle}
\bauthor{\bsnm{Anan}, \binits{T.}},
\bauthor{\bsnm{Yoneya}, \binits{T.}},
\bauthor{\bsnm{Ichimoto}, \binits{K.}},
\bauthor{\bsnm{UeNo}, \binits{S.}},
\bauthor{\bsnm{Shiota}, \binits{D.}},
\bauthor{\bsnm{Nozawa}, \binits{S.}},
\bauthor{\bsnm{Takasao}, \binits{S.}},
\bauthor{\bsnm{Kawate}, \binits{T.}}:
\byear{2018},
\batitle{{Measurement of vector magnetic field in a flare kernel with a
  spectropolarimetric observation in He~{\sc{i}}~10830{\AA}}}.
\bjtitle{Publications of the Astronomical Society of Japan}
\bvolume{70},
\bfpage{101}.
\doiurl{https://doi.org/10.1093/pasj/psy105}.
\burl{https://doi.org/10.1093/pasj/psy105}.
\end{barticle}
\endbibitem

\bibitem[\protect\citeauthoryear{{Anan} et~al.}{2019}]{2019ApJ...882..161A}
\begin{barticle}
\bauthor{\bsnm{{Anan}}, \binits{T.}},
\bauthor{\bsnm{{Schad}}, \binits{T.A.}},
\bauthor{\bsnm{{Jaeggli}}, \binits{S.A.}},
\bauthor{\bsnm{{Tarr}}, \binits{L.A.}}:
\byear{2019},
\batitle{{Shock Heating Energy of Umbral Flashes Measured with Integral Field
  Unit Spectroscopy}}.
\bjtitle{\apj}
\bvolume{882},
\bfpage{161}.
\doiurl{https://doi.org/10.3847/1538-4357/ab357f}.
\adsurl{2019ApJ...882..161A}.
\end{barticle}
\endbibitem

\bibitem[\protect\citeauthoryear{{Avila} and
  {Singh}}{2008}]{2008SPIE.7018E..4WA}
\begin{bchapter}
\bauthor{\bsnm{{Avila}}, \binits{G.}},
\bauthor{\bsnm{{Singh}}, \binits{P.}}:
\byear{2008},
\bctitle{{Optical fiber scrambling and light pipes for high accuracy radial
  velocities measurements}}.
In: \beditor{\bsnm{{Atad-Ettedgui}}, \binits{E.}},
\beditor{\bsnm{{Lemke}}, \binits{D.}} (eds.)
\bbtitle{Advanced Optical and Mechanical Technologies in Telescopes and
  Instrumentation},
\bsertitle{Society of Photo-Optical Instrumentation Engineers (SPIE) Conference
  Series}
\bseriesno{7018},
\bfpage{70184W}.
\doiurl{https://doi.org/10.1117/12.789509}.
\adsurl{2008SPIE.7018E..4WA}.
\end{bchapter}
\endbibitem

\bibitem[\protect\citeauthoryear{Bail{\'e}n, Orozco~Su{\'a}rez, and del
  Toro~Iniesta}{2023}]{Bailen2023}
\begin{barticle}
\bauthor{\bsnm{Bail{\'e}n}, \binits{F.J.}},
\bauthor{\bsnm{Orozco~Su{\'a}rez}, \binits{D.}},
\bauthor{\bparticle{del} \bsnm{Toro~Iniesta}, \binits{J.C.}}:
\byear{2023},
\batitle{Fabry-P{\'e}rot etalons in solar astronomy. A review}.
\bjtitle{Astrophysics and Space Science}
\bvolume{368},
\bfpage{55}.
\end{barticle}
\endbibitem

\bibitem[\protect\citeauthoryear{{Bailey}, {Cotton}, and
  {Kedziora-Chudczer}}{2017}]{2017MNRAS.465.1601B}
\begin{barticle}
\bauthor{\bsnm{{Bailey}}, \binits{J.}},
\bauthor{\bsnm{{Cotton}}, \binits{D.V.}},
\bauthor{\bsnm{{Kedziora-Chudczer}}, \binits{L.}}:
\byear{2017},
\batitle{{A high-precision polarimeter for small telescopes}}.
\bjtitle{\mnras}
\bvolume{465},
\bfpage{1601}.
\doiurl{https://doi.org/10.1093/mnras/stw2886}.
\adsurl{2017MNRAS.465.1601B}.
\end{barticle}
\endbibitem

\bibitem[\protect\citeauthoryear{{Baldry}, {Bland-Hawthorn}, and
  {Robertson}}{2004}]{2004PASP..116..403B}
\begin{barticle}
\bauthor{\bsnm{{Baldry}}, \binits{I.K.}},
\bauthor{\bsnm{{Bland-Hawthorn}}, \binits{J.}},
\bauthor{\bsnm{{Robertson}}, \binits{J.G.}}:
\byear{2004},
\batitle{{Volume Phase Holographic Gratings: Polarization Properties and
  Diffraction Efficiency}}.
\bjtitle{\pasp}
\bvolume{116},
\bfpage{403}.
\doiurl{https://doi.org/10.1086/383622}.
\adsurl{2004PASP..116..403B}.
\end{barticle}
\endbibitem

\bibitem[\protect\citeauthoryear{Barden et~al.}{2000}]{Barden_2000}
\begin{barticle}
\bauthor{\bsnm{Barden}, \binits{S.C.}},
\bauthor{\bsnm{Arns}, \binits{J.A.}},
\bauthor{\bsnm{Colburn}, \binits{W.S.}},
\bauthor{\bsnm{Williams}, \binits{J.B.}}:
\byear{2000},
\batitle{Volume‐Phase Holographic Gratings and the Efficiency of Three Simple
  Volume‐Phase Holographic Gratings}.
\bjtitle{Publications of the Astronomical Society of the Pacific}
\bvolume{112},
\bfpage{809}.
\doiurl{https://doi.org/10.1086/316576}.
\burl{https://dx.doi.org/10.1086/316576}.
\end{barticle}
\endbibitem

\bibitem[\protect\citeauthoryear{{Barthol} et~al.}{2011}]{2011SoPh..268....1B}
\begin{barticle}
\bauthor{\bsnm{{Barthol}}, \binits{P.}},
\bauthor{\bsnm{{Gandorfer}}, \binits{A.}},
\bauthor{\bsnm{{Solanki}}, \binits{S.K.}},
\bauthor{\bsnm{{Sch{\"u}ssler}}, \binits{M.}},
\bauthor{\bsnm{{Chares}}, \binits{B.}},
\bauthor{\bsnm{{Curdt}}, \binits{W.}},
\bauthor{\bsnm{{Deutsch}}, \binits{W.}},
\bauthor{\bsnm{{Feller}}, \binits{A.}},
\bauthor{\bsnm{{Germerott}}, \binits{D.}},
\bauthor{\bsnm{{Grauf}}, \binits{B.}},
\bauthor{\bsnm{{Heerlein}}, \binits{K.}},
\bauthor{\bsnm{{Hirzberger}}, \binits{J.}},
\bauthor{\bsnm{{Kolleck}}, \binits{M.}},
\bauthor{\bsnm{{Meller}}, \binits{R.}},
\bauthor{\bsnm{{M{\"u}ller}}, \binits{R.}},
\bauthor{\bsnm{{Riethm{\"u}ller}}, \binits{T.L.}},
\bauthor{\bsnm{{Tomasch}}, \binits{G.}},
\bauthor{\bsnm{{Kn{\"o}lker}}, \binits{M.}},
\bauthor{\bsnm{{Lites}}, \binits{B.W.}},
\bauthor{\bsnm{{Card}}, \binits{G.}},
\bauthor{\bsnm{{Elmore}}, \binits{D.}},
\bauthor{\bsnm{{Fox}}, \binits{J.}},
\bauthor{\bsnm{{Lecinski}}, \binits{A.}},
\bauthor{\bsnm{{Nelson}}, \binits{P.}},
\bauthor{\bsnm{{Summers}}, \binits{R.}},
\bauthor{\bsnm{{Watt}}, \binits{A.}},
\bauthor{\bsnm{{Mart{\'\i}nez Pillet}}, \binits{V.}},
\bauthor{\bsnm{{Bonet}}, \binits{J.A.}},
\bauthor{\bsnm{{Schmidt}}, \binits{W.}},
\bauthor{\bsnm{{Berkefeld}}, \binits{T.}},
\bauthor{\bsnm{{Title}}, \binits{A.M.}},
\bauthor{\bsnm{{Domingo}}, \binits{V.}},
\bauthor{\bsnm{{Gasent Blesa}}, \binits{J.L.}},
\bauthor{\bsnm{{Del Toro Iniesta}}, \binits{J.C.}},
\bauthor{\bsnm{{L{\'o}pez Jim{\'e}nez}}, \binits{A.}},
\bauthor{\bsnm{{{\'A}lvarez-Herrero}}, \binits{A.}},
\bauthor{\bsnm{{Sabau-Graziati}}, \binits{L.}},
\bauthor{\bsnm{{Widani}}, \binits{C.}},
\bauthor{\bsnm{{Haberler}}, \binits{P.}},
\bauthor{\bsnm{{H{\"a}rtel}}, \binits{K.}},
\bauthor{\bsnm{{Kampf}}, \binits{D.}},
\bauthor{\bsnm{{Levin}}, \binits{T.}},
\bauthor{\bsnm{{P{\'e}rez Grand e}}, \binits{I.}},
\bauthor{\bsnm{{Sanz-Andr{\'e}s}}, \binits{A.}},
\bauthor{\bsnm{{Schmidt}}, \binits{E.}}:
\byear{2011},
\batitle{{The {\sc Sunrise} Mission}}.
\bjtitle{\solphys}
\bvolume{268},
\bfpage{1}.
\doiurl{https://doi.org/10.1007/s11207-010-9662-9}.
\adsurl{2011SoPh..268....1B}.
\end{barticle}
\endbibitem

\bibitem[\protect\citeauthoryear{{Baudrand} and
  {Walker}}{2001}]{2001PASP..113..851B}
\begin{barticle}
\bauthor{\bsnm{{Baudrand}}, \binits{J.}},
\bauthor{\bsnm{{Walker}}, \binits{G.A.H.}}:
\byear{2001},
\batitle{{Modal Noise in High-Resolution, Fiber-fed Spectra: A Study and Simple
  Cure}}.
\bjtitle{\pasp}
\bvolume{113},
\bfpage{851}.
\doiurl{https://doi.org/10.1086/322143}.
\adsurl{2001PASP..113..851B}.
\end{barticle}
\endbibitem

\bibitem[\protect\citeauthoryear{{Beckers} and
  {Tallant}}{1969}]{1969SoPh....7..351B}
\begin{barticle}
\bauthor{\bsnm{{Beckers}}, \binits{J.M.}},
\bauthor{\bsnm{{Tallant}}, \binits{P.E.}}:
\byear{1969},
\batitle{{Chromospheric Inhomogeneities in Sunspot Umbrae}}.
\bjtitle{\solphys}
\bvolume{7},
\bfpage{351}.
\doiurl{https://doi.org/10.1007/BF00146140}.
\adsurl{1969SoPh....7..351B}.
\end{barticle}
\endbibitem

\bibitem[\protect\citeauthoryear{{Berkefeld}
  et~al.}{2011}]{2011SoPh..268..103B}
\begin{barticle}
\bauthor{\bsnm{{Berkefeld}}, \binits{T.}},
\bauthor{\bsnm{{Schmidt}}, \binits{W.}},
\bauthor{\bsnm{{Soltau}}, \binits{D.}},
\bauthor{\bsnm{{Bell}}, \binits{A.}},
\bauthor{\bsnm{{Doerr}}, \binits{H.P.}},
\bauthor{\bsnm{{Feger}}, \binits{B.}},
\bauthor{\bsnm{{Friedlein}}, \binits{R.}},
\bauthor{\bsnm{{Gerber}}, \binits{K.}},
\bauthor{\bsnm{{Heidecke}}, \binits{F.}},
\bauthor{\bsnm{{Kentischer}}, \binits{T.}},
\bauthor{\bsnm{{v. D. L{\"u}he}}, \binits{O.}},
\bauthor{\bsnm{{Sigwarth}}, \binits{M.}},
\bauthor{\bsnm{{W{\"a}lde}}, \binits{E.}},
\bauthor{\bsnm{{Barthol}}, \binits{P.}},
\bauthor{\bsnm{{Deutsch}}, \binits{W.}},
\bauthor{\bsnm{{Gandorfer}}, \binits{A.}},
\bauthor{\bsnm{{Germerott}}, \binits{D.}},
\bauthor{\bsnm{{Grauf}}, \binits{B.}},
\bauthor{\bsnm{{Meller}}, \binits{R.}},
\bauthor{\bsnm{{{\'A}lvarez-Herrero}}, \binits{A.}},
\bauthor{\bsnm{{Kn{\"o}lker}}, \binits{M.}},
\bauthor{\bsnm{{Mart{\'\i}nez Pillet}}, \binits{V.}},
\bauthor{\bsnm{{Solanki}}, \binits{S.K.}},
\bauthor{\bsnm{{Title}}, \binits{A.M.}}:
\byear{2011},
\batitle{{The Wave-Front Correction System for the {\sc Sunrise} Balloon-Borne
  Solar Observatory}}.
\bjtitle{\solphys}
\bvolume{268},
\bfpage{103}.
\doiurl{https://doi.org/10.1007/s11207-010-9676-3}.
\adsurl{2011SoPh..268..103B}.
\end{barticle}
\endbibitem

\bibitem[\protect\citeauthoryear{{Bethge} et~al.}{2011}]{2011A&A...534A.105B}
\begin{barticle}
\bauthor{\bsnm{{Bethge}}, \binits{C.}},
\bauthor{\bsnm{{Peter}}, \binits{H.}},
\bauthor{\bsnm{{Kentischer}}, \binits{T.J.}},
\bauthor{\bsnm{{Halbgewachs}}, \binits{C.}},
\bauthor{\bsnm{{Elmore}}, \binits{D.F.}},
\bauthor{\bsnm{{Beck}}, \binits{C.}}:
\byear{2011},
\batitle{{The Chromospheric Telescope}}.
\bjtitle{\aap}
\bvolume{534},
\bfpage{A105}.
\doiurl{https://doi.org/10.1051/0004-6361/201117456}.
\adsurl{2011A&A...534A.105B}.
\end{barticle}
\endbibitem

\bibitem[\protect\citeauthoryear{{Bj{\o}rgen}
  et~al.}{2018}]{2018A&A...611A..62B}
\begin{barticle}
\bauthor{\bsnm{{Bj{\o}rgen}}, \binits{J.P.}},
\bauthor{\bsnm{{Sukhorukov}}, \binits{A.V.}},
\bauthor{\bsnm{{Leenaarts}}, \binits{J.}},
\bauthor{\bsnm{{Carlsson}}, \binits{M.}},
\bauthor{\bsnm{{de la Cruz Rodr{\'\i}guez}}, \binits{J.}},
\bauthor{\bsnm{{Scharmer}}, \binits{G.B.}},
\bauthor{\bsnm{{Hansteen}}, \binits{V.H.}}:
\byear{2018},
\batitle{{Three-dimensional modeling of the Ca II H and K lines in the solar
  atmosphere}}.
\bjtitle{\aap}
\bvolume{611},
\bfpage{A62}.
\doiurl{https://doi.org/10.1051/0004-6361/201731926}.
\adsurl{2018A&A...611A..62B}.
\end{barticle}
\endbibitem

\bibitem[\protect\citeauthoryear{{Brault}}{1978}]{1978fsoo.conf...33B}
\begin{bchapter}
\bauthor{\bsnm{{Brault}}, \binits{J.W.}}:
\byear{1978},
\bctitle{{Solar Fourier Transform Spectroscopy}}.
In: \beditor{\bsnm{{Godoli}}, \binits{G.}} (ed.)
\bbtitle{Future solar optical observations needs and constraints}
\bseriesno{106},
\bfpage{33}.
\adsurl{1978fsoo.conf...33B}.
\end{bchapter}
\endbibitem

\bibitem[\protect\citeauthoryear{{Brault}}{1979}]{1979MmArc.106...33B}
\begin{barticle}
\bauthor{\bsnm{{Brault}}, \binits{J.W.}}:
\byear{1979},
\batitle{{Solar Fourier transform spectroscopy.}}
\bjtitle{Osservazioni e memorie dell'Osservatorio astrofisico di Arcetri}
\bvolume{106},
\bfpage{33}.
\adsurl{1979MmArc.106...33B}.
\end{barticle}
\endbibitem

\bibitem[\protect\citeauthoryear{{Buxton} and
  {Oxon}}{1937}]{doi:10.1080/14786443708561813}
\begin{barticle}
\bauthor{\bsnm{{Buxton}}, \binits{A.}},
\bauthor{\bsnm{{Oxon}}, \binits{M.A.}}:
\byear{1937},
\batitle{XLI. Note on optical resolution}.
\bjtitle{The London, Edinburgh, and Dublin Philosophical Magazine and Journal
  of Science}
\bvolume{23},
\bfpage{440}.
\doiurl{https://doi.org/10.1080/14786443708561813}.
\burl{https://doi.org/10.1080/14786443708561813}.
\end{barticle}
\endbibitem

\bibitem[\protect\citeauthoryear{{Calcines}, {L{\'o}pez}, and
  {Collados}}{2013}]{2013JAI.....250009C}
\begin{barticle}
\bauthor{\bsnm{{Calcines}}, \binits{A.}},
\bauthor{\bsnm{{L{\'o}pez}}, \binits{R.L.}},
\bauthor{\bsnm{{Collados}}, \binits{M.}}:
\byear{2013},
\batitle{{MuSICa: the Multi-Slit Image Slicer for the est Spectrograph}}.
\bjtitle{Journal of Astronomical Instrumentation}
\bvolume{2},
\bfpage{1350009}.
\doiurl{https://doi.org/10.1142/S2251171713500098}.
\adsurl{2013JAI.....250009C}.
\end{barticle}
\endbibitem

\bibitem[\protect\citeauthoryear{{Capozzi} et~al.}{2020}]{2020A&A...641A..63C}
\begin{barticle}
\bauthor{\bsnm{{Capozzi}}, \binits{E.}},
\bauthor{\bsnm{{Ballester}}, \binits{E.A.}},
\bauthor{\bsnm{{Belluzzi}}, \binits{L.}},
\bauthor{\bsnm{{Bianda}}, \binits{M.}},
\bauthor{\bsnm{{Dhara}}, \binits{S.K.}},
\bauthor{\bsnm{{Ramelli}}, \binits{R.}}:
\byear{2020},
\batitle{{Observational indications of magneto-optical effects in the
  scattering polarization wings of the Ca I 4227 {\r{A}} line}}.
\bjtitle{\aap}
\bvolume{641},
\bfpage{A63}.
\doiurl{https://doi.org/10.1051/0004-6361/202038455}.
\adsurl{2020A&A...641A..63C}.
\end{barticle}
\endbibitem

\bibitem[\protect\citeauthoryear{{Carlsson} and
  {Stein}}{1992}]{1992ApJ...397L..59C}
\begin{barticle}
\bauthor{\bsnm{{Carlsson}}, \binits{M.}},
\bauthor{\bsnm{{Stein}}, \binits{R.F.}}:
\byear{1992},
\batitle{{Non-LTE Radiating Acoustic Shocks and CA II K2V Bright Points}}.
\bjtitle{\apjl}
\bvolume{397},
\bfpage{L59}.
\doiurl{https://doi.org/10.1086/186544}.
\adsurl{1992ApJ...397L..59C}.
\end{barticle}
\endbibitem

\bibitem[\protect\citeauthoryear{{Carlsson} and
  {Stein}}{1997}]{1997ApJ...481..500C}
\begin{barticle}
\bauthor{\bsnm{{Carlsson}}, \binits{M.}},
\bauthor{\bsnm{{Stein}}, \binits{R.F.}}:
\byear{1997},
\batitle{{Formation of Solar Calcium H and K Bright Grains}}.
\bjtitle{\apj}
\bvolume{481},
\bfpage{500}.
\doiurl{https://doi.org/10.1086/304043}.
\adsurl{1997ApJ...481..500C}.
\end{barticle}
\endbibitem

\bibitem[\protect\citeauthoryear{{Casini} and {de
  Wijn}}{2014}]{2014JOSAA..31.2002C}
\begin{barticle}
\bauthor{\bsnm{{Casini}}, \binits{R.}},
\bauthor{\bsnm{{de Wijn}}, \binits{A.G.}}:
\byear{2014},
\batitle{{On the instrument profile of slit spectrographs}}.
\bjtitle{Journal of the Optical Society of America A}
\bvolume{31},
\bfpage{2002}.
\doiurl{https://doi.org/10.1364/JOSAA.31.002002}.
\adsurl{2014JOSAA..31.2002C}.
\end{barticle}
\endbibitem

\bibitem[\protect\citeauthoryear{{Cavallini}}{2006}]{2006SoPh..236..415C}
\begin{barticle}
\bauthor{\bsnm{{Cavallini}}, \binits{F.}}:
\byear{2006},
\batitle{{IBIS: A New Post-Focus Instrument for Solar Imaging Spectroscopy}}.
\bjtitle{\solphys}
\bvolume{236},
\bfpage{415}.
\doiurl{https://doi.org/10.1007/s11207-006-0103-8}.
\adsurl{2006SoPh..236..415C}.
\end{barticle}
\endbibitem

\bibitem[\protect\citeauthoryear{{Centeno}, {Collados}, and {Trujillo
  Bueno}}{2006}]{2006ApJ...640.1153C}
\begin{barticle}
\bauthor{\bsnm{{Centeno}}, \binits{R.}},
\bauthor{\bsnm{{Collados}}, \binits{M.}},
\bauthor{\bsnm{{Trujillo Bueno}}, \binits{J.}}:
\byear{2006},
\batitle{{Spectropolarimetric Investigation of the Propagation of
  Magnetoacoustic Waves and Shock Formation in Sunspot Atmospheres}}.
\bjtitle{\apj}
\bvolume{640},
\bfpage{1153}.
\doiurl{https://doi.org/10.1086/500185}.
\adsurl{2006ApJ...640.1153C}.
\end{barticle}
\endbibitem

\bibitem[\protect\citeauthoryear{{Centeno}, {Collados}, and {Trujillo
  Bueno}}{2009}]{2009ApJ...692.1211C}
\begin{barticle}
\bauthor{\bsnm{{Centeno}}, \binits{R.}},
\bauthor{\bsnm{{Collados}}, \binits{M.}},
\bauthor{\bsnm{{Trujillo Bueno}}, \binits{J.}}:
\byear{2009},
\batitle{{Wave Propagation and Shock Formation in Different Magnetic
  Structures}}.
\bjtitle{\apj}
\bvolume{692},
\bfpage{1211}.
\doiurl{https://doi.org/10.1088/0004-637X/692/2/1211}.
\adsurl{2009ApJ...692.1211C}.
\end{barticle}
\endbibitem

\bibitem[\protect\citeauthoryear{{Chae} et~al.}{2017}]{2017ApJ...836...18C}
\begin{barticle}
\bauthor{\bsnm{{Chae}}, \binits{J.}},
\bauthor{\bsnm{{Lee}}, \binits{J.}},
\bauthor{\bsnm{{Cho}}, \binits{K.}},
\bauthor{\bsnm{{Song}}, \binits{D.}},
\bauthor{\bsnm{{Cho}}, \binits{K.}},
\bauthor{\bsnm{{Yurchyshyn}}, \binits{V.}}:
\byear{2017},
\batitle{{Photospheric Origin of Three-minute Oscillations in a Sunspot}}.
\bjtitle{\apj}
\bvolume{836},
\bfpage{18}.
\doiurl{https://doi.org/10.3847/1538-4357/836/1/18}.
\adsurl{2017ApJ...836...18C}.
\end{barticle}
\endbibitem

\bibitem[\protect\citeauthoryear{{Collados} et~al.}{2012}]{2012AN....333..872C}
\begin{barticle}
\bauthor{\bsnm{{Collados}}, \binits{M.}},
\bauthor{\bsnm{{L{\'o}pez}}, \binits{R.}},
\bauthor{\bsnm{{P{\'a}ez}}, \binits{E.}},
\bauthor{\bsnm{{Hern{\'a}ndez}}, \binits{E.}},
\bauthor{\bsnm{{Reyes}}, \binits{M.}},
\bauthor{\bsnm{{Calcines}}, \binits{A.}},
\bauthor{\bsnm{{Ballesteros}}, \binits{E.}},
\bauthor{\bsnm{{D{\'\i}az}}, \binits{J.J.}},
\bauthor{\bsnm{{Denker}}, \binits{C.}},
\bauthor{\bsnm{{Lagg}}, \binits{A.}},
\bauthor{\bsnm{{Schlichenmaier}}, \binits{R.}},
\bauthor{\bsnm{{Schmidt}}, \binits{W.}},
\bauthor{\bsnm{{Solanki}}, \binits{S.K.}},
\bauthor{\bsnm{{Strassmeier}}, \binits{K.G.}},
\bauthor{\bsnm{{von der L{\"u}he}}, \binits{O.}},
\bauthor{\bsnm{{Volkmer}}, \binits{R.}}:
\byear{2012},
\batitle{{GRIS: The GREGOR Infrared Spectrograph}}.
\bjtitle{Astronomische Nachrichten}
\bvolume{333},
\bfpage{872}.
\doiurl{https://doi.org/10.1002/asna.201211738}.
\adsurl{2012AN....333..872C}.
\end{barticle}
\endbibitem

\bibitem[\protect\citeauthoryear{{Cousin} et~al.}{2022}]{2022AcSpe.18806341C}
\begin{barticle}
\bauthor{\bsnm{{Cousin}}, \binits{A.}},
\bauthor{\bsnm{{Sautter}}, \binits{V.}},
\bauthor{\bsnm{{Fabre}}, \binits{C.}},
\bauthor{\bsnm{{Dromart}}, \binits{G.}},
\bauthor{\bsnm{{Montagnac}}, \binits{G.}},
\bauthor{\bsnm{{Drouet}}, \binits{C.}},
\bauthor{\bsnm{{Meslin}}, \binits{P.Y.}},
\bauthor{\bsnm{{Gasnault}}, \binits{O.}},
\bauthor{\bsnm{{Beyssac}}, \binits{O.}},
\bauthor{\bsnm{{Bernard}}, \binits{S.}},
\bauthor{\bsnm{{Cloutis}}, \binits{E.}},
\bauthor{\bsnm{{Forni}}, \binits{O.}},
\bauthor{\bsnm{{Beck}}, \binits{P.}},
\bauthor{\bsnm{{Fouchet}}, \binits{T.}},
\bauthor{\bsnm{{Johnson}}, \binits{J.R.}},
\bauthor{\bsnm{{Lasue}}, \binits{J.}},
\bauthor{\bsnm{{Ollila}}, \binits{A.M.}},
\bauthor{\bsnm{{De Parseval}}, \binits{P.}},
\bauthor{\bsnm{{Gouy}}, \binits{S.}},
\bauthor{\bsnm{{Caron}}, \binits{B.}},
\bauthor{\bsnm{{Madariaga}}, \binits{J.M.}},
\bauthor{\bsnm{{Arana}}, \binits{G.}},
\bauthor{\bsnm{{Madsen}}, \binits{M.B.}},
\bauthor{\bsnm{{Laserna}}, \binits{J.}},
\bauthor{\bsnm{{Moros}}, \binits{J.}},
\bauthor{\bsnm{{Manrique}}, \binits{J.A.}},
\bauthor{\bsnm{{Lopez-Reyes}}, \binits{G.}},
\bauthor{\bsnm{{Rull}}, \binits{F.}},
\bauthor{\bsnm{{Maurice}}, \binits{S.}},
\bauthor{\bsnm{{Wiens}}, \binits{R.C.}}:
\byear{2022},
\batitle{{SuperCam calibration targets on board the perseverance rover:
  Fabrication and quantitative characterization}}.
\bjtitle{Spectrochimica Acta}
\bvolume{188},
\bfpage{106341}.
\doiurl{https://doi.org/10.1016/j.sab.2021.106341}.
\adsurl{2022AcSpe.18806341C}.
\end{barticle}
\endbibitem

\bibitem[\protect\citeauthoryear{Daino, Marchis, and
  Piazzolla}{1980}]{doi:10.1080/713820379}
\begin{barticle}
\bauthor{\bsnm{Daino}, \binits{B.}},
\bauthor{\bsnm{Marchis}, \binits{G.D.}},
\bauthor{\bsnm{Piazzolla}, \binits{S.}}:
\byear{1980},
\batitle{Speckle and Modal Noise in Optical Fibres Theory and Experiment}.
\bjtitle{Optica Acta: International Journal of Optics}
\bvolume{27},
\bfpage{1151}.
\doiurl{https://doi.org/10.1080/713820379}.
\end{barticle}
\endbibitem

\bibitem[\protect\citeauthoryear{{de la Cruz Rodr{\'\i}guez}
  et~al.}{2013}]{2013A&A...556A.115D}
\begin{barticle}
\bauthor{\bsnm{{de la Cruz Rodr{\'\i}guez}}, \binits{J.}},
\bauthor{\bsnm{{Rouppe van der Voort}}, \binits{L.}},
\bauthor{\bsnm{{Socas-Navarro}}, \binits{H.}},
\bauthor{\bsnm{{van Noort}}, \binits{M.}}:
\byear{2013},
\batitle{{Physical properties of a sunspot chromosphere with umbral flashes}}.
\bjtitle{\aap}
\bvolume{556},
\bfpage{A115}.
\doiurl{https://doi.org/10.1051/0004-6361/201321629}.
\adsurl{2013A&A...556A.115D}.
\end{barticle}
\endbibitem

\bibitem[\protect\citeauthoryear{{De Pontieu}
  et~al.}{2014}]{2014SoPh..289.2733D}
\begin{barticle}
\bauthor{\bsnm{{De Pontieu}}, \binits{B.}},
\bauthor{\bsnm{{Title}}, \binits{A.M.}},
\bauthor{\bsnm{{Lemen}}, \binits{J.R.}},
\bauthor{\bsnm{{Kushner}}, \binits{G.D.}},
\bauthor{\bsnm{{Akin}}, \binits{D.J.}},
\bauthor{\bsnm{{Allard}}, \binits{B.}},
\bauthor{\bsnm{{Berger}}, \binits{T.}},
\bauthor{\bsnm{{Boerner}}, \binits{P.}},
\bauthor{\bsnm{{Cheung}}, \binits{M.}},
\bauthor{\bsnm{{Chou}}, \binits{C.}},
\bauthor{\bsnm{{Drake}}, \binits{J.F.}},
\bauthor{\bsnm{{Duncan}}, \binits{D.W.}},
\bauthor{\bsnm{{Freeland}}, \binits{S.}},
\bauthor{\bsnm{{Heyman}}, \binits{G.F.}},
\bauthor{\bsnm{{Hoffman}}, \binits{C.}},
\bauthor{\bsnm{{Hurlburt}}, \binits{N.E.}},
\bauthor{\bsnm{{Lindgren}}, \binits{R.W.}},
\bauthor{\bsnm{{Mathur}}, \binits{D.}},
\bauthor{\bsnm{{Rehse}}, \binits{R.}},
\bauthor{\bsnm{{Sabolish}}, \binits{D.}},
\bauthor{\bsnm{{Seguin}}, \binits{R.}},
\bauthor{\bsnm{{Schrijver}}, \binits{C.J.}},
\bauthor{\bsnm{{Tarbell}}, \binits{T.D.}},
\bauthor{\bsnm{{W{\"u}lser}}, \binits{J.-P.}},
\bauthor{\bsnm{{Wolfson}}, \binits{C.J.}},
\bauthor{\bsnm{{Yanari}}, \binits{C.}},
\bauthor{\bsnm{{Mudge}}, \binits{J.}},
\bauthor{\bsnm{{Nguyen-Phuc}}, \binits{N.}},
\bauthor{\bsnm{{Timmons}}, \binits{R.}},
\bauthor{\bsnm{{van Bezooijen}}, \binits{R.}},
\bauthor{\bsnm{{Weingrod}}, \binits{I.}},
\bauthor{\bsnm{{Brookner}}, \binits{R.}},
\bauthor{\bsnm{{Butcher}}, \binits{G.}},
\bauthor{\bsnm{{Dougherty}}, \binits{B.}},
\bauthor{\bsnm{{Eder}}, \binits{J.}},
\bauthor{\bsnm{{Knagenhjelm}}, \binits{V.}},
\bauthor{\bsnm{{Larsen}}, \binits{S.}},
\bauthor{\bsnm{{Mansir}}, \binits{D.}},
\bauthor{\bsnm{{Phan}}, \binits{L.}},
\bauthor{\bsnm{{Boyle}}, \binits{P.}},
\bauthor{\bsnm{{Cheimets}}, \binits{P.N.}},
\bauthor{\bsnm{{DeLuca}}, \binits{E.E.}},
\bauthor{\bsnm{{Golub}}, \binits{L.}},
\bauthor{\bsnm{{Gates}}, \binits{R.}},
\bauthor{\bsnm{{Hertz}}, \binits{E.}},
\bauthor{\bsnm{{McKillop}}, \binits{S.}},
\bauthor{\bsnm{{Park}}, \binits{S.}},
\bauthor{\bsnm{{Perry}}, \binits{T.}},
\bauthor{\bsnm{{Podgorski}}, \binits{W.A.}},
\bauthor{\bsnm{{Reeves}}, \binits{K.}},
\bauthor{\bsnm{{Saar}}, \binits{S.}},
\bauthor{\bsnm{{Testa}}, \binits{P.}},
\bauthor{\bsnm{{Tian}}, \binits{H.}},
\bauthor{\bsnm{{Weber}}, \binits{M.}},
\bauthor{\bsnm{{Dunn}}, \binits{C.}},
\bauthor{\bsnm{{Eccles}}, \binits{S.}},
\bauthor{\bsnm{{Jaeggli}}, \binits{S.A.}},
\bauthor{\bsnm{{Kankelborg}}, \binits{C.C.}},
\bauthor{\bsnm{{Mashburn}}, \binits{K.}},
\bauthor{\bsnm{{Pust}}, \binits{N.}},
\bauthor{\bsnm{{Springer}}, \binits{L.}},
\bauthor{\bsnm{{Carvalho}}, \binits{R.}},
\bauthor{\bsnm{{Kleint}}, \binits{L.}},
\bauthor{\bsnm{{Marmie}}, \binits{J.}},
\bauthor{\bsnm{{Mazmanian}}, \binits{E.}},
\bauthor{\bsnm{{Pereira}}, \binits{T.M.D.}},
\bauthor{\bsnm{{Sawyer}}, \binits{S.}},
\bauthor{\bsnm{{Strong}}, \binits{J.}},
\bauthor{\bsnm{{Worden}}, \binits{S.P.}},
\bauthor{\bsnm{{Carlsson}}, \binits{M.}},
\bauthor{\bsnm{{Hansteen}}, \binits{V.H.}},
\bauthor{\bsnm{{Leenaarts}}, \binits{J.}},
\bauthor{\bsnm{{Wiesmann}}, \binits{M.}},
\bauthor{\bsnm{{Aloise}}, \binits{J.}},
\bauthor{\bsnm{{Chu}}, \binits{K.-C.}},
\bauthor{\bsnm{{Bush}}, \binits{R.I.}},
\bauthor{\bsnm{{Scherrer}}, \binits{P.H.}},
\bauthor{\bsnm{{Brekke}}, \binits{P.}},
\bauthor{\bsnm{{Martinez-Sykora}}, \binits{J.}},
\bauthor{\bsnm{{Lites}}, \binits{B.W.}},
\bauthor{\bsnm{{McIntosh}}, \binits{S.W.}},
\bauthor{\bsnm{{Uitenbroek}}, \binits{H.}},
\bauthor{\bsnm{{Okamoto}}, \binits{T.J.}},
\bauthor{\bsnm{{Gummin}}, \binits{M.A.}},
\bauthor{\bsnm{{Auker}}, \binits{G.}},
\bauthor{\bsnm{{Jerram}}, \binits{P.}},
\bauthor{\bsnm{{Pool}}, \binits{P.}},
\bauthor{\bsnm{{Waltham}}, \binits{N.}}:
\byear{2014},
\batitle{{The Interface Region Imaging Spectrograph (IRIS)}}.
\bjtitle{\solphys}
\bvolume{289},
\bfpage{2733}.
\doiurl{https://doi.org/10.1007/s11207-014-0485-y}.
\adsurl{2014SoPh..289.2733D}.
\end{barticle}
\endbibitem

\bibitem[\protect\citeauthoryear{{de Wijn} et~al.}{2012}]{2012SPIE.8446E..6XD}
\begin{bchapter}
\bauthor{\bsnm{{de Wijn}}, \binits{A.G.}},
\bauthor{\bsnm{{Casini}}, \binits{R.}},
\bauthor{\bsnm{{Nelson}}, \binits{P.G.}},
\bauthor{\bsnm{{Huang}}, \binits{P.}}:
\byear{2012},
\bctitle{{Preliminary design of the visible spectro-polarimeter for the
  Advanced Technology Solar Telescope}}.
In: \beditor{\bsnm{{McLean}}, \binits{I.S.}},
\beditor{\bsnm{{Ramsay}}, \binits{S.K.}},
\beditor{\bsnm{{Takami}}, \binits{H.}} (eds.)
\bbtitle{Ground-based and Airborne Instrumentation for Astronomy IV},
\bsertitle{Society of Photo-Optical Instrumentation Engineers (SPIE) Conference
  Series}
\bseriesno{8446},
\bfpage{84466X}.
\doiurl{https://doi.org/10.1117/12.926497}.
\adsurl{2012SPIE.8446E..6XD}.
\end{bchapter}
\endbibitem

\bibitem[\protect\citeauthoryear{{de Wijn} et~al.}{2022}]{2022SoPh..297...22D}
\begin{barticle}
\bauthor{\bsnm{{de Wijn}}, \binits{A.G.}},
\bauthor{\bsnm{{Casini}}, \binits{R.}},
\bauthor{\bsnm{{Carlile}}, \binits{A.}},
\bauthor{\bsnm{{Lecinski}}, \binits{A.R.}},
\bauthor{\bsnm{{Sewell}}, \binits{S.}},
\bauthor{\bsnm{{Zmarzly}}, \binits{P.}},
\bauthor{\bsnm{{Eigenbrot}}, \binits{A.D.}},
\bauthor{\bsnm{{Beck}}, \binits{C.}},
\bauthor{\bsnm{{W{\"o}ger}}, \binits{F.}},
\bauthor{\bsnm{{Kn{\"o}lker}}, \binits{M.}}:
\byear{2022},
\batitle{{The Visible Spectro-Polarimeter of the Daniel K. Inouye Solar
  Telescope}}.
\bjtitle{\solphys}
\bvolume{297},
\bfpage{22}.
\doiurl{https://doi.org/10.1007/s11207-022-01954-1}.
\adsurl{2022SoPh..297...22D}.
\end{barticle}
\endbibitem

\bibitem[\protect\citeauthoryear{de~Wijn et~al.}{2022}]{deWijn2022}
\begin{barticle}
\bauthor{\bparticle{de} \bsnm{Wijn}, \binits{A.G.}},
\bauthor{\bsnm{Casini}, \binits{R.}},
\bauthor{\bsnm{Carlile}, \binits{A.}},
\bauthor{\bsnm{Lecinski}, \binits{A.R.}},
\bauthor{\bsnm{Sewell}, \binits{S.}},
\bauthor{\bsnm{Zmarzly}, \binits{P.}},
\bauthor{\bsnm{Eigenbrot}, \binits{A.D.}},
\bauthor{\bsnm{Beck}, \binits{C.}},
\bauthor{\bsnm{W{\"o}ger}, \binits{F.}},
\bauthor{\bsnm{Kn{\"o}lker}, \binits{M.}}:
\byear{2022},
\batitle{The Visible Spectro-Polarimeter of the Daniel K. Inouye Solar
  Telescope}.
\bjtitle{Solar Physics}
\bvolume{297},
\bfpage{22}.
\bisbn{1573-093X}.
\doiurl{https://doi.org/10.1007/s11207-022-01954-1}.
\burl{https://doi.org/10.1007/s11207-022-01954-1}.
\end{barticle}
\endbibitem

\bibitem[\protect\citeauthoryear{{Dhananjay}}{2014}]{2014MNRAS.437.2092D}
\begin{barticle}
\bauthor{\bsnm{{Dhananjay}}, \binits{K.}}:
\byear{2014},
\batitle{{Site evaluation study for the Indian National Large Solar Telescope
  using microthermal measurements}}.
\bjtitle{\mnras}
\bvolume{437},
\bfpage{2092}.
\doiurl{https://doi.org/10.1093/mnras/stt1985}.
\adsurl{2014MNRAS.437.2092D}.
\end{barticle}
\endbibitem

\bibitem[\protect\citeauthoryear{{D{\'\i}az Baso}
  et~al.}{2023}]{2023A&A...673A..35D}
\begin{barticle}
\bauthor{\bsnm{{D{\'\i}az Baso}}, \binits{C.J.}},
\bauthor{\bsnm{{Rouppe van der Voort}}, \binits{L.}},
\bauthor{\bsnm{{de la Cruz Rodr{\'\i}guez}}, \binits{J.}},
\bauthor{\bsnm{{Leenaarts}}, \binits{J.}}:
\byear{2023},
\batitle{{Designing wavelength sampling for Fabry-P{\'e}rot observations.
  Information-based spectral sampling}}.
\bjtitle{\aap}
\bvolume{673},
\bfpage{A35}.
\doiurl{https://doi.org/10.1051/0004-6361/202346230}.
\adsurl{2023A&A...673A..35D}.
\end{barticle}
\endbibitem

\bibitem[\protect\citeauthoryear{{Dominguez-Tagle}
  et~al.}{2022}]{2022JAI....1150014D}
\begin{barticle}
\bauthor{\bsnm{{Dominguez-Tagle}}, \binits{C.}},
\bauthor{\bsnm{{Collados}}, \binits{M.}},
\bauthor{\bsnm{{Lopez}}, \binits{R.}},
\bauthor{\bsnm{{Cedillo}}, \binits{J.J.V.}},
\bauthor{\bsnm{{Esteves}}, \binits{M.A.}},
\bauthor{\bsnm{{Grassin}}, \binits{O.}},
\bauthor{\bsnm{{Vega}}, \binits{N.}},
\bauthor{\bsnm{{Mato}}, \binits{A.}},
\bauthor{\bsnm{{Quintero}}, \binits{J.}},
\bauthor{\bsnm{{Rodriguez}}, \binits{H.}},
\bauthor{\bsnm{{Regalado}}, \binits{S.}},
\bauthor{\bsnm{{Gonzalez}}, \binits{F.}}:
\byear{2022},
\batitle{{First Light of the Integral Field Unit of GRIS on the GREGOR Solar
  Telescope}}.
\bjtitle{Journal of Astronomical Instrumentation}
\bvolume{11},
\bfpage{2250014}.
\doiurl{https://doi.org/10.1142/S2251171722500143}.
\adsurl{2022JAI....1150014D}.
\end{barticle}
\endbibitem

\bibitem[\protect\citeauthoryear{Douglass}{1998}]{Douglass1998}
\begin{bchapter}
\bauthor{\bsnm{Douglass}, \binits{M.R.}}:
\byear{1998},
\bctitle{Lifetime estimates and unique failure mechanisms of the Digital
  Micromirror Device (DMD)}.
In: \bbtitle{1998 IEEE International Reliability Physics Symposium Proceedings.
  36th Annual (Cat. No.98CH36173)},
\bfpage{9}.
\doiurl{https://doi.org/10.1109/RELPHY.1998.670436}.
\end{bchapter}
\endbibitem

\bibitem[\protect\citeauthoryear{{Drory} et~al.}{2015}]{2015AJ....149...77D}
\begin{barticle}
\bauthor{\bsnm{{Drory}}, \binits{N.}},
\bauthor{\bsnm{{MacDonald}}, \binits{N.}},
\bauthor{\bsnm{{Bershady}}, \binits{M.A.}},
\bauthor{\bsnm{{Bundy}}, \binits{K.}},
\bauthor{\bsnm{{Gunn}}, \binits{J.}},
\bauthor{\bsnm{{Law}}, \binits{D.R.}},
\bauthor{\bsnm{{Smith}}, \binits{M.}},
\bauthor{\bsnm{{Stoll}}, \binits{R.}},
\bauthor{\bsnm{{Tremonti}}, \binits{C.A.}},
\bauthor{\bsnm{{Wake}}, \binits{D.A.}},
\bauthor{\bsnm{{Yan}}, \binits{R.}},
\bauthor{\bsnm{{Weijmans}}, \binits{A.M.}},
\bauthor{\bsnm{{Byler}}, \binits{N.}},
\bauthor{\bsnm{{Cherinka}}, \binits{B.}},
\bauthor{\bsnm{{Cope}}, \binits{F.}},
\bauthor{\bsnm{{Eigenbrot}}, \binits{A.}},
\bauthor{\bsnm{{Harding}}, \binits{P.}},
\bauthor{\bsnm{{Holder}}, \binits{D.}},
\bauthor{\bsnm{{Huehnerhoff}}, \binits{J.}},
\bauthor{\bsnm{{Jaehnig}}, \binits{K.}},
\bauthor{\bsnm{{Jansen}}, \binits{T.C.}},
\bauthor{\bsnm{{Klaene}}, \binits{M.}},
\bauthor{\bsnm{{Paat}}, \binits{A.M.}},
\bauthor{\bsnm{{Percival}}, \binits{J.}},
\bauthor{\bsnm{{Sayres}}, \binits{C.}}:
\byear{2015},
\batitle{{The MaNGA Integral Field Unit Fiber Feed System for the Sloan 2.5 m
  Telescope}}.
\bjtitle{\aj}
\bvolume{149},
\bfpage{77}.
\doiurl{https://doi.org/10.1088/0004-6256/149/2/77}.
\adsurl{2015AJ....149...77D}.
\end{barticle}
\endbibitem

\bibitem[\protect\citeauthoryear{Dudley, Duncan, and
  Slaughter}{2003}]{Dudley2003}
\begin{bchapter}
\bauthor{\bsnm{Dudley}, \binits{D.}},
\bauthor{\bsnm{Duncan}, \binits{W.M.}},
\bauthor{\bsnm{Slaughter}, \binits{J.}}:
\byear{2003},
\bctitle{{Emerging digital micromirror device (DMD) applications}}.
In: \beditor{\bsnm{Urey}, \binits{H.}} (ed.)
\bbtitle{MOEMS Display and Imaging Systems}
\bseriesno{4985},
\bpublisher{SPIE},
\bfpage{14 }.
\bcomment{International Society for Optics and Photonics}.
\doiurl{https://doi.org/10.1117/12.480761}.
\burl{https://doi.org/10.1117/12.480761}.
\end{bchapter}
\endbibitem

\bibitem[\protect\citeauthoryear{{Dunn}}{1969}]{Dunn1969}
\begin{barticle}
\bauthor{\bsnm{{Dunn}}, \binits{R.B.}}:
\byear{1969},
\batitle{{Sacramento Peak's New Solar Telescope}}.
\bjtitle{\skytel}
\bvolume{38},
\bfpage{368}.
\adsurl{1969S&T....38..368D}.
\end{barticle}
\endbibitem

\bibitem[\protect\citeauthoryear{{Eklund} et~al.}{2021}]{2021RSPTA.37900185E}
\begin{barticle}
\bauthor{\bsnm{{Eklund}}, \binits{H.}},
\bauthor{\bsnm{{Wedemeyer}}, \binits{S.}},
\bauthor{\bsnm{{Snow}}, \binits{B.}},
\bauthor{\bsnm{{Jess}}, \binits{D.B.}},
\bauthor{\bsnm{{Jafarzadeh}}, \binits{S.}},
\bauthor{\bsnm{{Grant}}, \binits{S.D.T.}},
\bauthor{\bsnm{{Carlsson}}, \binits{M.}},
\bauthor{\bsnm{{Szydlarski}}, \binits{M.}}:
\byear{2021},
\batitle{{Characterization of shock wave signatures at millimetre wavelengths
  from Bifrost simulations}}.
\bjtitle{Philosophical Transactions of the Royal Society of London Series A}
\bvolume{379},
\bfpage{20200185}.
\doiurl{https://doi.org/10.1098/rsta.2020.0185}.
\adsurl{2021RSPTA.37900185E}.
\end{barticle}
\endbibitem

\bibitem[\protect\citeauthoryear{{El Abdi} et~al.}{2008}]{ELABDI2008222}
\begin{barticle}
\bauthor{\bsnm{{El Abdi}}, \binits{R.}},
\bauthor{\bsnm{Rujinsky}, \binits{A.}},
\bauthor{\bsnm{Borda}, \binits{C.}},
\bauthor{\bsnm{Severin}, \binits{I.}},
\bauthor{\bsnm{Poulain}, \binits{M.}}:
\byear{2008},
\batitle{New method for strength improvement of silica optical fibers}.
\bjtitle{Optics and Lasers in Engineering}
\bvolume{46},
\bfpage{222}.
\doiurl{https://doi.org/10.1016/j.optlaseng.2007.10.005}.
\burl{https://www.sciencedirect.com/science/article/pii/S0143816607001856}.
\end{barticle}
\endbibitem

\bibitem[\protect\citeauthoryear{{Elmore} et~al.}{1992}]{1992SPIE.1746...22E}
\begin{bchapter}
\bauthor{\bsnm{{Elmore}}, \binits{D.F.}},
\bauthor{\bsnm{{Lites}}, \binits{B.W.}},
\bauthor{\bsnm{{Tomczyk}}, \binits{S.}},
\bauthor{\bsnm{{Skumanich}}, \binits{A.P.}},
\bauthor{\bsnm{{Dunn}}, \binits{R.B.}},
\bauthor{\bsnm{{Schuenke}}, \binits{J.A.}},
\bauthor{\bsnm{{Streander}}, \binits{K.V.}},
\bauthor{\bsnm{{Leach}}, \binits{T.W.}},
\bauthor{\bsnm{{Chambellan}}, \binits{C.W.}},
\bauthor{\bsnm{{Hull}}, \binits{H.K.}}:
\byear{1992},
\bctitle{{The Advanced Stokes Polarimeter - A new instrument for solar magnetic
  field research}}.
In: \beditor{\bsnm{{Goldstein}}, \binits{D.H.}},
\beditor{\bsnm{{Chipman}}, \binits{R.A.}} (eds.)
\bbtitle{Polarization Analysis and Measurement},
\bsertitle{Society of Photo-Optical Instrumentation Engineers (SPIE) Conference
  Series}
\bseriesno{1746},
\bfpage{22}.
\doiurl{https://doi.org/10.1117/12.138795}.
\adsurl{1992SPIE.1746...22E}.
\end{bchapter}
\endbibitem

\bibitem[\protect\citeauthoryear{Epworth}{1979}]{Epworth:79}
\begin{bchapter}
\bauthor{\bsnm{Epworth}, \binits{R.E.}}:
\byear{1979},
\bctitle{ThD1 Phenomenon of modal noise in fiber systems}.
In: \bbtitle{Optical Fiber Communication},
\bpublisher{Optica Publishing Group},
\bfpage{ThD1}.
\doiurl{https://doi.org/10.1364/OFC.1979.ThD1}.
\burl{https://opg.optica.org/abstract.cfm?URI=OFC-1979-ThD1}.
\end{bchapter}
\endbibitem

\bibitem[\protect\citeauthoryear{{Erlick} et~al.}{1998}]{1998JGR...10331541E}
\begin{barticle}
\bauthor{\bsnm{{Erlick}}, \binits{C.}},
\bauthor{\bsnm{{Frederick}}, \binits{J.E.}},
\bauthor{\bsnm{{Saxena}}, \binits{V.K.}},
\bauthor{\bsnm{{Wenny}}, \binits{B.N.}}:
\byear{1998},
\batitle{{Atmospheric transmission in the ultraviolet and visible: Aerosols in
  cloudy atmospheres}}.
\bjtitle{\jgr}
\bvolume{103},
\bfpage{31,541}.
\doiurl{https://doi.org/10.1029/1998JD200053}.
\adsurl{1998JGR...10331541E}.
\end{barticle}
\endbibitem

\bibitem[\protect\citeauthoryear{{Esmonde-White}, {Esmonde-White}, and
  {Morris}}{2011}]{2011ApSpe..65...85E}
\begin{barticle}
\bauthor{\bsnm{{Esmonde-White}}, \binits{F.W.L.}},
\bauthor{\bsnm{{Esmonde-White}}, \binits{K.A.}},
\bauthor{\bsnm{{Morris}}, \binits{M.D.}}:
\byear{2011},
\batitle{{Minor Distortions with Major Consequences: Correcting Distortions in
  Imaging Spectrographs}}.
\bjtitle{Applied Spectroscopy}
\bvolume{65},
\bfpage{85}.
\doiurl{https://doi.org/10.1366/10-06040}.
\adsurl{2011ApSpe..65...85E}.
\end{barticle}
\endbibitem

\bibitem[\protect\citeauthoryear{{Felipe}, {Socas-Navarro}, and
  {Przybylski}}{2018}]{2018A&A...614A..73F}
\begin{barticle}
\bauthor{\bsnm{{Felipe}}, \binits{T.}},
\bauthor{\bsnm{{Socas-Navarro}}, \binits{H.}},
\bauthor{\bsnm{{Przybylski}}, \binits{D.}}:
\byear{2018},
\batitle{{Inversions of synthetic umbral flashes: Effects of scanning time on
  the inferred atmospheres}}.
\bjtitle{\aap}
\bvolume{614},
\bfpage{A73}.
\doiurl{https://doi.org/10.1051/0004-6361/201732169}.
\adsurl{2018A&A...614A..73F}.
\end{barticle}
\endbibitem

\bibitem[\protect\citeauthoryear{Foreman}{1968}]{Foreman:68}
\begin{barticle}
\bauthor{\bsnm{Foreman}, \binits{W.T.}}:
\byear{1968},
\batitle{Lens Correction of Astigmatism in a Czerny-Turner Spectrograph}.
\bjtitle{Appl. Opt.}
\bvolume{7},
\bfpage{1053}.
\doiurl{https://doi.org/10.1364/AO.7.001053}.
\burl{https://opg.optica.org/ao/abstract.cfm?URI=ao-7-6-1053}.
\end{barticle}
\endbibitem

\bibitem[\protect\citeauthoryear{{Fox} et~al.}{2016}]{2016SSRv..204....7F}
\begin{barticle}
\bauthor{\bsnm{{Fox}}, \binits{N.J.}},
\bauthor{\bsnm{{Velli}}, \binits{M.C.}},
\bauthor{\bsnm{{Bale}}, \binits{S.D.}},
\bauthor{\bsnm{{Decker}}, \binits{R.}},
\bauthor{\bsnm{{Driesman}}, \binits{A.}},
\bauthor{\bsnm{{Howard}}, \binits{R.A.}},
\bauthor{\bsnm{{Kasper}}, \binits{J.C.}},
\bauthor{\bsnm{{Kinnison}}, \binits{J.}},
\bauthor{\bsnm{{Kusterer}}, \binits{M.}},
\bauthor{\bsnm{{Lario}}, \binits{D.}},
\bauthor{\bsnm{{Lockwood}}, \binits{M.K.}},
\bauthor{\bsnm{{McComas}}, \binits{D.J.}},
\bauthor{\bsnm{{Raouafi}}, \binits{N.E.}},
\bauthor{\bsnm{{Szabo}}, \binits{A.}}:
\byear{2016},
\batitle{{The Solar Probe Plus Mission: Humanity's First Visit to Our Star}}.
\bjtitle{\ssr}
\bvolume{204},
\bfpage{7}.
\doiurl{https://doi.org/10.1007/s11214-015-0211-6}.
\adsurl{2016SSRv..204....7F}.
\end{barticle}
\endbibitem

\bibitem[\protect\citeauthoryear{{Gafeira} et~al.}{2017}]{2017ApJS..229....7G}
\begin{barticle}
\bauthor{\bsnm{{Gafeira}}, \binits{R.}},
\bauthor{\bsnm{{Jafarzadeh}}, \binits{S.}},
\bauthor{\bsnm{{Solanki}}, \binits{S.K.}},
\bauthor{\bsnm{{Lagg}}, \binits{A.}},
\bauthor{\bsnm{{van Noort}}, \binits{M.}},
\bauthor{\bsnm{{Barthol}}, \binits{P.}},
\bauthor{\bsnm{{Blanco Rodr{\'\i}guez}}, \binits{J.}},
\bauthor{\bsnm{{del Toro Iniesta}}, \binits{J.C.}},
\bauthor{\bsnm{{Gandorfer}}, \binits{A.}},
\bauthor{\bsnm{{Gizon}}, \binits{L.}},
\bauthor{\bsnm{{Hirzberger}}, \binits{J.}},
\bauthor{\bsnm{{Kn{\"o}lker}}, \binits{M.}},
\bauthor{\bsnm{{Orozco Su{\'a}rez}}, \binits{D.}},
\bauthor{\bsnm{{Riethm{\"u}ller}}, \binits{T.L.}},
\bauthor{\bsnm{{Schmidt}}, \binits{W.}}:
\byear{2017},
\batitle{{Oscillations on Width and Intensity of Slender Ca II H Fibrils from
  Sunrise/SuFI}}.
\bjtitle{\apjs}
\bvolume{229},
\bfpage{7}.
\doiurl{https://doi.org/10.3847/1538-4365/229/1/7}.
\adsurl{2017ApJS..229....7G}.
\end{barticle}
\endbibitem

\bibitem[\protect\citeauthoryear{{Goode} et~al.}{2010}]{2010AN....331..620G}
\begin{barticle}
\bauthor{\bsnm{{Goode}}, \binits{P.R.}},
\bauthor{\bsnm{{Coulter}}, \binits{R.}},
\bauthor{\bsnm{{Gorceix}}, \binits{N.}},
\bauthor{\bsnm{{Yurchyshyn}}, \binits{V.}},
\bauthor{\bsnm{{Cao}}, \binits{W.}}:
\byear{2010},
\batitle{{The NST: First results and some lessons for ATST and EST}}.
\bjtitle{Astronomische Nachrichten}
\bvolume{331},
\bfpage{620}.
\doiurl{https://doi.org/10.1002/asna.201011387}.
\adsurl{2010AN....331..620G}.
\end{barticle}
\endbibitem

\bibitem[\protect\citeauthoryear{{Goodman} and
  {Rawson}}{1981}]{1981OptL....6..324G}
\begin{barticle}
\bauthor{\bsnm{{Goodman}}, \binits{J.W.}},
\bauthor{\bsnm{{Rawson}}, \binits{E.G.}}:
\byear{1981},
\batitle{{Statistics of modal noise in fibers: a case of constrained speckle}}.
\bjtitle{Optics Letters}
\bvolume{6},
\bfpage{324}.
\doiurl{https://doi.org/10.1364/OL.6.000324}.
\adsurl{1981OptL....6..324G}.
\end{barticle}
\endbibitem

\bibitem[\protect\citeauthoryear{Gosain, Mathew, and
  Venkatakrishnan}{2011}]{Gosain2011}
\begin{barticle}
\bauthor{\bsnm{Gosain}, \binits{S.}},
\bauthor{\bsnm{Mathew}, \binits{S.K.}},
\bauthor{\bsnm{Venkatakrishnan}, \binits{P.}}:
\byear{2011},
\batitle{Acoustic Power Absorption and its Relation to Vector Magnetic Field of
  a Sunspot}.
\bjtitle{Solar Physics}
\bvolume{268},
\bfpage{335}.
\bisbn{1573-093X}.
\doiurl{https://doi.org/10.1007/s11207-010-9625-1}.
\burl{https://doi.org/10.1007/s11207-010-9625-1}.
\end{barticle}
\endbibitem

\bibitem[\protect\citeauthoryear{Goto et~al.}{2015}]{Goto:15}
\begin{barticle}
\bauthor{\bsnm{Goto}, \binits{Y.}},
\bauthor{\bsnm{Nakajima}, \binits{K.}},
\bauthor{\bsnm{Matsui}, \binits{T.}},
\bauthor{\bsnm{Kurashima}, \binits{T.}},
\bauthor{\bsnm{Yamamoto}, \binits{F.}}:
\byear{2015},
\batitle{Influence of Cladding Thickness on Transmission Loss and its
  Relationship With Multicore Fiber Structure}.
\bjtitle{J. Lightwave Technol.}
\bvolume{33},
\bfpage{4942}.
\burl{https://opg.optica.org/jlt/abstract.cfm?URI=jlt-33-23-4942}.
\end{barticle}
\endbibitem

\bibitem[\protect\citeauthoryear{{Grant} et~al.}{2015}]{2015ApJ...806..132G}
\begin{barticle}
\bauthor{\bsnm{{Grant}}, \binits{S.D.T.}},
\bauthor{\bsnm{{Jess}}, \binits{D.B.}},
\bauthor{\bsnm{{Moreels}}, \binits{M.G.}},
\bauthor{\bsnm{{Morton}}, \binits{R.J.}},
\bauthor{\bsnm{{Christian}}, \binits{D.J.}},
\bauthor{\bsnm{{Giagkiozis}}, \binits{I.}},
\bauthor{\bsnm{{Verth}}, \binits{G.}},
\bauthor{\bsnm{{Fedun}}, \binits{V.}},
\bauthor{\bsnm{{Keys}}, \binits{P.H.}},
\bauthor{\bsnm{{Van Doorsselaere}}, \binits{T.}},
\bauthor{\bsnm{{Erd{\'e}lyi}}, \binits{R.}}:
\byear{2015},
\batitle{{Wave Damping Observed in Upwardly Propagating Sausage-mode
  Oscillations Contained within a Magnetic Pore}}.
\bjtitle{\apj}
\bvolume{806},
\bfpage{132}.
\doiurl{https://doi.org/10.1088/0004-637X/806/1/132}.
\adsurl{2015ApJ...806..132G}.
\end{barticle}
\endbibitem

\bibitem[\protect\citeauthoryear{{Grant} et~al.}{2018}]{2018NatPh..14..480G}
\begin{barticle}
\bauthor{\bsnm{{Grant}}, \binits{S.D.T.}},
\bauthor{\bsnm{{Jess}}, \binits{D.B.}},
\bauthor{\bsnm{{Zaqarashvili}}, \binits{T.V.}},
\bauthor{\bsnm{{Beck}}, \binits{C.}},
\bauthor{\bsnm{{Socas-Navarro}}, \binits{H.}},
\bauthor{\bsnm{{Aschwanden}}, \binits{M.J.}},
\bauthor{\bsnm{{Keys}}, \binits{P.H.}},
\bauthor{\bsnm{{Christian}}, \binits{D.J.}},
\bauthor{\bsnm{{Houston}}, \binits{S.J.}},
\bauthor{\bsnm{{Hewitt}}, \binits{R.L.}}:
\byear{2018},
\batitle{{Alfv{\'e}n wave dissipation in the solar chromosphere}}.
\bjtitle{Nature Physics}
\bvolume{14},
\bfpage{480}.
\doiurl{https://doi.org/10.1038/s41567-018-0058-3}.
\adsurl{2018NatPh..14..480G}.
\end{barticle}
\endbibitem

\bibitem[\protect\citeauthoryear{{Grant} et~al.}{2022}]{2022ApJ...938..143G}
\begin{barticle}
\bauthor{\bsnm{{Grant}}, \binits{S.D.T.}},
\bauthor{\bsnm{{Jess}}, \binits{D.B.}},
\bauthor{\bsnm{{Stangalini}}, \binits{M.}},
\bauthor{\bsnm{{Jafarzadeh}}, \binits{S.}},
\bauthor{\bsnm{{Fedun}}, \binits{V.}},
\bauthor{\bsnm{{Verth}}, \binits{G.}},
\bauthor{\bsnm{{Keys}}, \binits{P.H.}},
\bauthor{\bsnm{{Rajaguru}}, \binits{S.P.}},
\bauthor{\bsnm{{Uitenbroek}}, \binits{H.}},
\bauthor{\bsnm{{MacBride}}, \binits{C.D.}},
\bauthor{\bsnm{{Bate}}, \binits{W.}},
\bauthor{\bsnm{{Gilchrist-Millar}}, \binits{C.A.}}:
\byear{2022},
\batitle{{The Propagation of Coherent Waves Across Multiple Solar Magnetic
  Pores}}.
\bjtitle{\apj}
\bvolume{938},
\bfpage{143}.
\doiurl{https://doi.org/10.3847/1538-4357/ac91ca}.
\adsurl{2022ApJ...938..143G}.
\end{barticle}
\endbibitem

\bibitem[\protect\citeauthoryear{{Guevara G{\'o}mez}
  et~al.}{2022}]{2022A&A...665L...2G}
\begin{barticle}
\bauthor{\bsnm{{Guevara G{\'o}mez}}, \binits{J.C.}},
\bauthor{\bsnm{{Jafarzadeh}}, \binits{S.}},
\bauthor{\bsnm{{Wedemeyer}}, \binits{S.}},
\bauthor{\bsnm{{Szydlarski}}, \binits{M.}}:
\byear{2022},
\batitle{{Propagation of transverse waves in the solar chromosphere probed at
  different heights with ALMA sub-bands}}.
\bjtitle{\aap}
\bvolume{665},
\bfpage{L2}.
\doiurl{https://doi.org/10.1051/0004-6361/202244387}.
\adsurl{2022A&A...665L...2G}.
\end{barticle}
\endbibitem

\bibitem[\protect\citeauthoryear{{Hagen} and
  {Kudenov}}{2013}]{2013OptEn..52i0901H}
\begin{barticle}
\bauthor{\bsnm{{Hagen}}, \binits{N.}},
\bauthor{\bsnm{{Kudenov}}, \binits{M.W.}}:
\byear{2013},
\batitle{{Review of snapshot spectral imaging technologies}}.
\bjtitle{Optical Engineering}
\bvolume{52},
\bfpage{090901}.
\doiurl{https://doi.org/10.1117/1.OE.52.9.090901}.
\adsurl{2013OptEn..52i0901H}.
\end{barticle}
\endbibitem

\bibitem[\protect\citeauthoryear{Hao et~al.}{2017}]{Hao2017}
\begin{barticle}
\bauthor{\bsnm{Hao}, \binits{Q.}},
\bauthor{\bsnm{Yang}, \binits{K.}},
\bauthor{\bsnm{Cheng}, \binits{X.}},
\bauthor{\bsnm{Guo}, \binits{Y.}},
\bauthor{\bsnm{Fang}, \binits{C.}},
\bauthor{\bsnm{Ding}, \binits{M.D.}},
\bauthor{\bsnm{Chen}, \binits{P.F.}},
\bauthor{\bsnm{Li}, \binits{Z.}}:
\byear{2017},
\batitle{A circular white-light flare with impulsive and gradual white-light
  kernels}.
\bjtitle{Nature Communications}
\bvolume{8},
\bfpage{2202}.
\bisbn{2041-1723}.
\doiurl{https://doi.org/10.1038/s41467-017-02343-0}.
\burl{https://doi.org/10.1038/s41467-017-02343-0}.
\end{barticle}
\endbibitem

\bibitem[\protect\citeauthoryear{{Hasan}}{2010}]{2010IAUS..264..499H}
\begin{bchapter}
\bauthor{\bsnm{{Hasan}}, \binits{S.S.}}:
\byear{2010},
\bctitle{{The Indian National Large Solar Telescope (NLST)}}.
In: \beditor{\bsnm{{Kosovichev}}, \binits{A.G.}},
\beditor{\bsnm{{Andrei}}, \binits{A.H.}},
\beditor{\bsnm{{Rozelot}}, \binits{J.-P.}} (eds.)
\bbtitle{Solar and Stellar Variability: Impact on Earth and Planets}
\bseriesno{264},
\bfpage{499}.
\doiurl{https://doi.org/10.1017/S1743921309993206}.
\adsurl{2010IAUS..264..499H}.
\end{bchapter}
\endbibitem

\bibitem[\protect\citeauthoryear{{Hasan} et~al.}{2010}]{2010AN....331..628H}
\begin{barticle}
\bauthor{\bsnm{{Hasan}}, \binits{S.S.}},
\bauthor{\bsnm{{Soltau}}, \binits{D.}},
\bauthor{\bsnm{{K{\"a}rcher}}, \binits{H.}},
\bauthor{\bsnm{{S{\"u}{\ss}}}, \binits{M.}},
\bauthor{\bsnm{{Berkefeld}}, \binits{T.}}:
\byear{2010},
\batitle{{NLST: India's National Large Solar Telescope}}.
\bjtitle{Astronomische Nachrichten}
\bvolume{331},
\bfpage{628}.
\doiurl{https://doi.org/10.1002/asna.201011389}.
\adsurl{2010AN....331..628H}.
\end{barticle}
\endbibitem

\bibitem[\protect\citeauthoryear{{Henriques}
  et~al.}{2017}]{2017ApJ...845..102H}
\begin{barticle}
\bauthor{\bsnm{{Henriques}}, \binits{V.M.J.}},
\bauthor{\bsnm{{Mathioudakis}}, \binits{M.}},
\bauthor{\bsnm{{Socas-Navarro}}, \binits{H.}},
\bauthor{\bsnm{{de la Cruz Rodr{\'\i}guez}}, \binits{J.}}:
\byear{2017},
\batitle{{A Hot Downflowing Model Atmosphere for Umbral Flashes and the
  Physical Properties of Their Dark Fibrils}}.
\bjtitle{\apj}
\bvolume{845},
\bfpage{102}.
\doiurl{https://doi.org/10.3847/1538-4357/aa7ca4}.
\adsurl{2017ApJ...845..102H}.
\end{barticle}
\endbibitem

\bibitem[\protect\citeauthoryear{{Hill}, {Tremblay}, and
  {Kawasaki}}{1980}]{1980OptL....5..270H}
\begin{barticle}
\bauthor{\bsnm{{Hill}}, \binits{K.O.}},
\bauthor{\bsnm{{Tremblay}}, \binits{Y.}},
\bauthor{\bsnm{{Kawasaki}}, \binits{B.S.}}:
\byear{1980},
\batitle{{Modal noise in multimode fiber links: theory and experiment}}.
\bjtitle{Optics Letters}
\bvolume{5},
\bfpage{270}.
\doiurl{https://doi.org/10.1364/OL.5.000270}.
\adsurl{1980OptL....5..270H}.
\end{barticle}
\endbibitem

\bibitem[\protect\citeauthoryear{Hillier and Snow}{2023}]{HILLIER20231962}
\begin{barticle}
\bauthor{\bsnm{Hillier}, \binits{A.}},
\bauthor{\bsnm{Snow}, \binits{B.}}:
\byear{2023},
\batitle{Shocks and instabilities in the partially ionised solar atmosphere}.
\bjtitle{Advances in Space Research}
\bvolume{71},
\bfpage{1962}.
\bcomment{Recent progress in the physics of the Sun and heliosphere}.
\doiurl{https://doi.org/10.1016/j.asr.2022.08.079}.
\burl{https://www.sciencedirect.com/science/article/pii/S0273117722008158}.
\end{barticle}
\endbibitem

\bibitem[\protect\citeauthoryear{{Houston} et~al.}{2018}]{2018ApJ...860...28H}
\begin{barticle}
\bauthor{\bsnm{{Houston}}, \binits{S.J.}},
\bauthor{\bsnm{{Jess}}, \binits{D.B.}},
\bauthor{\bsnm{{Asensio Ramos}}, \binits{A.}},
\bauthor{\bsnm{{Grant}}, \binits{S.D.T.}},
\bauthor{\bsnm{{Beck}}, \binits{C.}},
\bauthor{\bsnm{{Norton}}, \binits{A.A.}},
\bauthor{\bsnm{{Krishna Prasad}}, \binits{S.}}:
\byear{2018},
\batitle{{The Magnetic Response of the Solar Atmosphere to Umbral Flashes}}.
\bjtitle{\apj}
\bvolume{860},
\bfpage{28}.
\doiurl{https://doi.org/10.3847/1538-4357/aab366}.
\adsurl{2018ApJ...860...28H}.
\end{barticle}
\endbibitem

\bibitem[\protect\citeauthoryear{{Houston} et~al.}{2020}]{2020ApJ...892...49H}
\begin{barticle}
\bauthor{\bsnm{{Houston}}, \binits{S.J.}},
\bauthor{\bsnm{{Jess}}, \binits{D.B.}},
\bauthor{\bsnm{{Keppens}}, \binits{R.}},
\bauthor{\bsnm{{Stangalini}}, \binits{M.}},
\bauthor{\bsnm{{Keys}}, \binits{P.H.}},
\bauthor{\bsnm{{Grant}}, \binits{S.D.T.}},
\bauthor{\bsnm{{Jafarzadeh}}, \binits{S.}},
\bauthor{\bsnm{{McFetridge}}, \binits{L.M.}},
\bauthor{\bsnm{{Murabito}}, \binits{M.}},
\bauthor{\bsnm{{Ermolli}}, \binits{I.}},
\bauthor{\bsnm{{Giorgi}}, \binits{F.}}:
\byear{2020},
\batitle{{Magnetohydrodynamic Nonlinearities in Sunspot Atmospheres:
  Chromospheric Detections of Intermediate Shocks}}.
\bjtitle{\apj}
\bvolume{892},
\bfpage{49}.
\doiurl{https://doi.org/10.3847/1538-4357/ab7a90}.
\adsurl{2020ApJ...892...49H}.
\end{barticle}
\endbibitem

\bibitem[\protect\citeauthoryear{Houston}{1927}]{PhysRev.29.478}
\begin{barticle}
\bauthor{\bsnm{Houston}, \binits{W.V.}}:
\byear{1927},
\batitle{A Compound Interferometer for Fine Structure Work}.
\bjtitle{Phys. Rev.}
\bvolume{29},
\bfpage{478}.
\doiurl{https://doi.org/10.1103/PhysRev.29.478}.
\burl{https://link.aps.org/doi/10.1103/PhysRev.29.478}.
\end{barticle}
\endbibitem

\bibitem[\protect\citeauthoryear{{Iglesias} and
  {Feller}}{2019}]{2019OptEn..58h2417I}
\begin{barticle}
\bauthor{\bsnm{{Iglesias}}, \binits{F.A.}},
\bauthor{\bsnm{{Feller}}, \binits{A.}}:
\byear{2019},
\batitle{{Instrumentation for solar spectropolarimetry: state of the art and
  prospects}}.
\bjtitle{Optical Engineering}
\bvolume{58},
\bfpage{082417}.
\doiurl{https://doi.org/10.1117/1.OE.58.8.082417}.
\adsurl{2019OptEn..58h2417I}.
\end{barticle}
\endbibitem

\bibitem[\protect\citeauthoryear{{Iglesias} et~al.}{2016}]{2016A&A...590A..89I}
\begin{barticle}
\bauthor{\bsnm{{Iglesias}}, \binits{F.A.}},
\bauthor{\bsnm{{Feller}}, \binits{A.}},
\bauthor{\bsnm{{Nagaraju}}, \binits{K.}},
\bauthor{\bsnm{{Solanki}}, \binits{S.K.}}:
\byear{2016},
\batitle{{High-resolution, high-sensitivity, ground-based solar
  spectropolarimetry with a new fast imaging polarimeter. I. Prototype
  characterization}}.
\bjtitle{\aap}
\bvolume{590},
\bfpage{A89}.
\doiurl{https://doi.org/10.1051/0004-6361/201628376}.
\adsurl{2016A&A...590A..89I}.
\end{barticle}
\endbibitem

\bibitem[\protect\citeauthoryear{{Jaeggli} et~al.}{2010}]{2010MmSAI..81..763J}
\begin{barticle}
\bauthor{\bsnm{{Jaeggli}}, \binits{S.A.}},
\bauthor{\bsnm{{Lin}}, \binits{H.}},
\bauthor{\bsnm{{Mickey}}, \binits{D.L.}},
\bauthor{\bsnm{{Kuhn}}, \binits{J.R.}},
\bauthor{\bsnm{{Hegwer}}, \binits{S.L.}},
\bauthor{\bsnm{{Rimmele}}, \binits{T.R.}},
\bauthor{\bsnm{{Penn}}, \binits{M.J.}}:
\byear{2010},
\batitle{{FIRS: a new instrument for photospheric and chromospheric studies at
  the DST.}}
\bjtitle{\memsai}
\bvolume{81},
\bfpage{763}.
\adsurl{2010MmSAI..81..763J}.
\end{barticle}
\endbibitem

\bibitem[\protect\citeauthoryear{{Jaeggli} et~al.}{2022}]{2022SoPh..297..137J}
\begin{barticle}
\bauthor{\bsnm{{Jaeggli}}, \binits{S.A.}},
\bauthor{\bsnm{{Lin}}, \binits{H.}},
\bauthor{\bsnm{{Onaka}}, \binits{P.}},
\bauthor{\bsnm{{Yamada}}, \binits{H.}},
\bauthor{\bsnm{{Anan}}, \binits{T.}},
\bauthor{\bsnm{{Bonnet}}, \binits{M.}},
\bauthor{\bsnm{{Ching}}, \binits{G.}},
\bauthor{\bsnm{{Huang}}, \binits{X.-P.}},
\bauthor{\bsnm{{Kramar}}, \binits{M.}},
\bauthor{\bsnm{{McGregor}}, \binits{H.}},
\bauthor{\bsnm{{Nitta}}, \binits{G.}},
\bauthor{\bsnm{{Rae}}, \binits{C.}},
\bauthor{\bsnm{{Robertson}}, \binits{L.}},
\bauthor{\bsnm{{Schad}}, \binits{T.A.}},
\bauthor{\bsnm{{Toyama}}, \binits{P.}},
\bauthor{\bsnm{{Young}}, \binits{J.}},
\bauthor{\bsnm{{Berst}}, \binits{C.}},
\bauthor{\bsnm{{Harrington}}, \binits{D.M.}},
\bauthor{\bsnm{{Liang}}, \binits{M.}},
\bauthor{\bsnm{{Puentes}}, \binits{M.}},
\bauthor{\bsnm{{Sekulic}}, \binits{P.}},
\bauthor{\bsnm{{Smith}}, \binits{B.}},
\bauthor{\bsnm{{Sueoka}}, \binits{S.R.}}:
\byear{2022},
\batitle{{The Diffraction-Limited Near-Infrared Spectropolarimeter (DL-NIRSP)
  of the Daniel K. Inouye Solar Telescope (DKIST)}}.
\bjtitle{\solphys}
\bvolume{297},
\bfpage{137}.
\doiurl{https://doi.org/10.1007/s11207-022-02062-w}.
\adsurl{2022SoPh..297..137J}.
\end{barticle}
\endbibitem

\bibitem[\protect\citeauthoryear{{Jafarzadeh}
  et~al.}{2013}]{2013A&A...549A.116J}
\begin{barticle}
\bauthor{\bsnm{{Jafarzadeh}}, \binits{S.}},
\bauthor{\bsnm{{Solanki}}, \binits{S.K.}},
\bauthor{\bsnm{{Feller}}, \binits{A.}},
\bauthor{\bsnm{{Lagg}}, \binits{A.}},
\bauthor{\bsnm{{Pietarila}}, \binits{A.}},
\bauthor{\bsnm{{Danilovic}}, \binits{S.}},
\bauthor{\bsnm{{Riethm{\"u}ller}}, \binits{T.L.}},
\bauthor{\bsnm{{Mart{\'\i}nez Pillet}}, \binits{V.}}:
\byear{2013},
\batitle{{Structure and dynamics of isolated internetwork Ca II H bright points
  observed by SUNRISE}}.
\bjtitle{\aap}
\bvolume{549},
\bfpage{A116}.
\doiurl{https://doi.org/10.1051/0004-6361/201220089}.
\adsurl{2013A&A...549A.116J}.
\end{barticle}
\endbibitem

\bibitem[\protect\citeauthoryear{{Jafarzadeh}
  et~al.}{2017a}]{2017ApJS..229...10J}
\begin{barticle}
\bauthor{\bsnm{{Jafarzadeh}}, \binits{S.}},
\bauthor{\bsnm{{Solanki}}, \binits{S.K.}},
\bauthor{\bsnm{{Stangalini}}, \binits{M.}},
\bauthor{\bsnm{{Steiner}}, \binits{O.}},
\bauthor{\bsnm{{Cameron}}, \binits{R.H.}},
\bauthor{\bsnm{{Danilovic}}, \binits{S.}}:
\byear{2017}a,
\batitle{{High-frequency Oscillations in Small Magnetic Elements Observed with
  Sunrise/SuFI}}.
\bjtitle{\apjs}
\bvolume{229},
\bfpage{10}.
\doiurl{https://doi.org/10.3847/1538-4365/229/1/10}.
\adsurl{2017ApJS..229...10J}.
\end{barticle}
\endbibitem

\bibitem[\protect\citeauthoryear{{Jafarzadeh}
  et~al.}{2017b}]{2017ApJS..229...11J}
\begin{barticle}
\bauthor{\bsnm{{Jafarzadeh}}, \binits{S.}},
\bauthor{\bsnm{{Rutten}}, \binits{R.J.}},
\bauthor{\bsnm{{Solanki}}, \binits{S.K.}},
\bauthor{\bsnm{{Wiegelmann}}, \binits{T.}},
\bauthor{\bsnm{{Riethm{\"u}ller}}, \binits{T.L.}},
\bauthor{\bsnm{{van Noort}}, \binits{M.}},
\bauthor{\bsnm{{Szydlarski}}, \binits{M.}},
\bauthor{\bsnm{{Blanco Rodr{\'\i}guez}}, \binits{J.}},
\bauthor{\bsnm{{Barthol}}, \binits{P.}},
\bauthor{\bsnm{{del Toro Iniesta}}, \binits{J.C.}},
\bauthor{\bsnm{{Gandorfer}}, \binits{A.}},
\bauthor{\bsnm{{Gizon}}, \binits{L.}},
\bauthor{\bsnm{{Hirzberger}}, \binits{J.}},
\bauthor{\bsnm{{Kn{\"o}lker}}, \binits{M.}},
\bauthor{\bsnm{{Mart{\'\i}nez Pillet}}, \binits{V.}},
\bauthor{\bsnm{{Orozco Su{\'a}rez}}, \binits{D.}},
\bauthor{\bsnm{{Schmidt}}, \binits{W.}}:
\byear{2017}b,
\batitle{{Slender Ca II H Fibrils Mapping Magnetic Fields in the Low Solar
  Chromosphere}}.
\bjtitle{\apjs}
\bvolume{229},
\bfpage{11}.
\doiurl{https://doi.org/10.3847/1538-4365/229/1/11}.
\adsurl{2017ApJS..229...11J}.
\end{barticle}
\endbibitem

\bibitem[\protect\citeauthoryear{{Jafarzadeh}
  et~al.}{2017c}]{2017ApJS..229....9J}
\begin{barticle}
\bauthor{\bsnm{{Jafarzadeh}}, \binits{S.}},
\bauthor{\bsnm{{Solanki}}, \binits{S.K.}},
\bauthor{\bsnm{{Gafeira}}, \binits{R.}},
\bauthor{\bsnm{{van Noort}}, \binits{M.}},
\bauthor{\bsnm{{Barthol}}, \binits{P.}},
\bauthor{\bsnm{{Blanco Rodr{\'\i}guez}}, \binits{J.}},
\bauthor{\bsnm{{del Toro Iniesta}}, \binits{J.C.}},
\bauthor{\bsnm{{Gandorfer}}, \binits{A.}},
\bauthor{\bsnm{{Gizon}}, \binits{L.}},
\bauthor{\bsnm{{Hirzberger}}, \binits{J.}},
\bauthor{\bsnm{{Kn{\"o}lker}}, \binits{M.}},
\bauthor{\bsnm{{Orozco Su{\'a}rez}}, \binits{D.}},
\bauthor{\bsnm{{Riethm{\"u}ller}}, \binits{T.L.}},
\bauthor{\bsnm{{Schmidt}}, \binits{W.}}:
\byear{2017}c,
\batitle{{Transverse Oscillations in Slender Ca II H Fibrils Observed with
  Sunrise/SuFI}}.
\bjtitle{\apjs}
\bvolume{229},
\bfpage{9}.
\doiurl{https://doi.org/10.3847/1538-4365/229/1/9}.
\adsurl{2017ApJS..229....9J}.
\end{barticle}
\endbibitem

\bibitem[\protect\citeauthoryear{{Jakobsen} et~al.}{2022}]{2022A&A...661A..80J}
\begin{barticle}
\bauthor{\bsnm{{Jakobsen}}, \binits{P.}},
\bauthor{\bsnm{{Ferruit}}, \binits{P.}},
\bauthor{\bsnm{{Alves de Oliveira}}, \binits{C.}},
\bauthor{\bsnm{{Arribas}}, \binits{S.}},
\bauthor{\bsnm{{Bagnasco}}, \binits{G.}},
\bauthor{\bsnm{{Barho}}, \binits{R.}},
\bauthor{\bsnm{{Beck}}, \binits{T.L.}},
\bauthor{\bsnm{{Birkmann}}, \binits{S.}},
\bauthor{\bsnm{{B{\"o}ker}}, \binits{T.}},
\bauthor{\bsnm{{Bunker}}, \binits{A.J.}},
\bauthor{\bsnm{{Charlot}}, \binits{S.}},
\bauthor{\bsnm{{de Jong}}, \binits{P.}},
\bauthor{\bsnm{{de Marchi}}, \binits{G.}},
\bauthor{\bsnm{{Ehrenwinkler}}, \binits{R.}},
\bauthor{\bsnm{{Falcolini}}, \binits{M.}},
\bauthor{\bsnm{{Fels}}, \binits{R.}},
\bauthor{\bsnm{{Franx}}, \binits{M.}},
\bauthor{\bsnm{{Franz}}, \binits{D.}},
\bauthor{\bsnm{{Funke}}, \binits{M.}},
\bauthor{\bsnm{{Giardino}}, \binits{G.}},
\bauthor{\bsnm{{Gnata}}, \binits{X.}},
\bauthor{\bsnm{{Holota}}, \binits{W.}},
\bauthor{\bsnm{{Honnen}}, \binits{K.}},
\bauthor{\bsnm{{Jensen}}, \binits{P.L.}},
\bauthor{\bsnm{{Jentsch}}, \binits{M.}},
\bauthor{\bsnm{{Johnson}}, \binits{T.}},
\bauthor{\bsnm{{Jollet}}, \binits{D.}},
\bauthor{\bsnm{{Karl}}, \binits{H.}},
\bauthor{\bsnm{{Kling}}, \binits{G.}},
\bauthor{\bsnm{{K{\"o}hler}}, \binits{J.}},
\bauthor{\bsnm{{Kolm}}, \binits{M.-G.}},
\bauthor{\bsnm{{Kumari}}, \binits{N.}},
\bauthor{\bsnm{{Lander}}, \binits{M.E.}},
\bauthor{\bsnm{{Lemke}}, \binits{R.}},
\bauthor{\bsnm{{L{\'o}pez-Caniego}}, \binits{M.}},
\bauthor{\bsnm{{L{\"u}tzgendorf}}, \binits{N.}},
\bauthor{\bsnm{{Maiolino}}, \binits{R.}},
\bauthor{\bsnm{{Manjavacas}}, \binits{E.}},
\bauthor{\bsnm{{Marston}}, \binits{A.}},
\bauthor{\bsnm{{Maschmann}}, \binits{M.}},
\bauthor{\bsnm{{Maurer}}, \binits{R.}},
\bauthor{\bsnm{{Messerschmidt}}, \binits{B.}},
\bauthor{\bsnm{{Moseley}}, \binits{S.H.}},
\bauthor{\bsnm{{Mosner}}, \binits{P.}},
\bauthor{\bsnm{{Mott}}, \binits{D.B.}},
\bauthor{\bsnm{{Muzerolle}}, \binits{J.}},
\bauthor{\bsnm{{Pirzkal}}, \binits{N.}},
\bauthor{\bsnm{{Pittet}}, \binits{J.-F.}},
\bauthor{\bsnm{{Plitzke}}, \binits{A.}},
\bauthor{\bsnm{{Posselt}}, \binits{W.}},
\bauthor{\bsnm{{Rapp}}, \binits{B.}},
\bauthor{\bsnm{{Rauscher}}, \binits{B.J.}},
\bauthor{\bsnm{{Rawle}}, \binits{T.}},
\bauthor{\bsnm{{Rix}}, \binits{H.-W.}},
\bauthor{\bsnm{{R{\"o}del}}, \binits{A.}},
\bauthor{\bsnm{{Rumler}}, \binits{P.}},
\bauthor{\bsnm{{Sabbi}}, \binits{E.}},
\bauthor{\bsnm{{Salvignol}}, \binits{J.-C.}},
\bauthor{\bsnm{{Schmid}}, \binits{T.}},
\bauthor{\bsnm{{Sirianni}}, \binits{M.}},
\bauthor{\bsnm{{Smith}}, \binits{C.}},
\bauthor{\bsnm{{Strada}}, \binits{P.}},
\bauthor{\bsnm{{te Plate}}, \binits{M.}},
\bauthor{\bsnm{{Valenti}}, \binits{J.}},
\bauthor{\bsnm{{Wettemann}}, \binits{T.}},
\bauthor{\bsnm{{Wiehe}}, \binits{T.}},
\bauthor{\bsnm{{Wiesmayer}}, \binits{M.}},
\bauthor{\bsnm{{Willott}}, \binits{C.J.}},
\bauthor{\bsnm{{Wright}}, \binits{R.}},
\bauthor{\bsnm{{Zeidler}}, \binits{P.}},
\bauthor{\bsnm{{Zincke}}, \binits{C.}}:
\byear{2022},
\batitle{{The Near-Infrared Spectrograph (NIRSpec) on the James Webb Space
  Telescope. I. Overview of the instrument and its capabilities}}.
\bjtitle{\aap}
\bvolume{661},
\bfpage{A80}.
\doiurl{https://doi.org/10.1051/0004-6361/202142663}.
\adsurl{2022A&A...661A..80J}.
\end{barticle}
\endbibitem

\bibitem[\protect\citeauthoryear{{Jess} et~al.}{2010a}]{2010ApJ...719L.134J}
\begin{barticle}
\bauthor{\bsnm{{Jess}}, \binits{D.B.}},
\bauthor{\bsnm{{Mathioudakis}}, \binits{M.}},
\bauthor{\bsnm{{Christian}}, \binits{D.J.}},
\bauthor{\bsnm{{Crockett}}, \binits{P.J.}},
\bauthor{\bsnm{{Keenan}}, \binits{F.P.}}:
\byear{2010}a,
\batitle{{A Study of Magnetic Bright Points in the Na I D$_{1}$ Line}}.
\bjtitle{\apjl}
\bvolume{719},
\bfpage{L134}.
\doiurl{https://doi.org/10.1088/2041-8205/719/2/L134}.
\adsurl{2010ApJ...719L.134J}.
\end{barticle}
\endbibitem

\bibitem[\protect\citeauthoryear{{Jess} et~al.}{2010b}]{2010SoPh..261..363J}
\begin{barticle}
\bauthor{\bsnm{{Jess}}, \binits{D.B.}},
\bauthor{\bsnm{{Mathioudakis}}, \binits{M.}},
\bauthor{\bsnm{{Christian}}, \binits{D.J.}},
\bauthor{\bsnm{{Keenan}}, \binits{F.P.}},
\bauthor{\bsnm{{Ryans}}, \binits{R.S.I.}},
\bauthor{\bsnm{{Crockett}}, \binits{P.J.}}:
\byear{2010}b,
\batitle{{ROSA: A High-cadence, Synchronized Multi-camera Solar Imaging
  System}}.
\bjtitle{\solphys}
\bvolume{261},
\bfpage{363}.
\doiurl{https://doi.org/10.1007/s11207-009-9500-0}.
\adsurl{2010SoPh..261..363J}.
\end{barticle}
\endbibitem

\bibitem[\protect\citeauthoryear{{Jess} et~al.}{2012a}]{2012ApJ...746..183J}
\begin{barticle}
\bauthor{\bsnm{{Jess}}, \binits{D.B.}},
\bauthor{\bsnm{{Shelyag}}, \binits{S.}},
\bauthor{\bsnm{{Mathioudakis}}, \binits{M.}},
\bauthor{\bsnm{{Keys}}, \binits{P.H.}},
\bauthor{\bsnm{{Christian}}, \binits{D.J.}},
\bauthor{\bsnm{{Keenan}}, \binits{F.P.}}:
\byear{2012}a,
\batitle{{Propagating Wave Phenomena Detected in Observations and Simulations
  of the Lower Solar Atmosphere}}.
\bjtitle{\apj}
\bvolume{746},
\bfpage{183}.
\doiurl{https://doi.org/10.1088/0004-637X/746/2/183}.
\adsurl{2012ApJ...746..183J}.
\end{barticle}
\endbibitem

\bibitem[\protect\citeauthoryear{{Jess} et~al.}{2012b}]{2012ApJ...757..160J}
\begin{barticle}
\bauthor{\bsnm{{Jess}}, \binits{D.B.}},
\bauthor{\bsnm{{De Moortel}}, \binits{I.}},
\bauthor{\bsnm{{Mathioudakis}}, \binits{M.}},
\bauthor{\bsnm{{Christian}}, \binits{D.J.}},
\bauthor{\bsnm{{Reardon}}, \binits{K.P.}},
\bauthor{\bsnm{{Keys}}, \binits{P.H.}},
\bauthor{\bsnm{{Keenan}}, \binits{F.P.}}:
\byear{2012}b,
\batitle{{The Source of 3 Minute Magnetoacoustic Oscillations in Coronal
  Fans}}.
\bjtitle{\apj}
\bvolume{757},
\bfpage{160}.
\doiurl{https://doi.org/10.1088/0004-637X/757/2/160}.
\adsurl{2012ApJ...757..160J}.
\end{barticle}
\endbibitem

\bibitem[\protect\citeauthoryear{{Jess} et~al.}{2013}]{2013ApJ...779..168J}
\begin{barticle}
\bauthor{\bsnm{{Jess}}, \binits{D.B.}},
\bauthor{\bsnm{{Reznikova}}, \binits{V.E.}},
\bauthor{\bsnm{{Van Doorsselaere}}, \binits{T.}},
\bauthor{\bsnm{{Keys}}, \binits{P.H.}},
\bauthor{\bsnm{{Mackay}}, \binits{D.H.}}:
\byear{2013},
\batitle{{The Influence of the Magnetic Field on Running Penumbral Waves in the
  Solar Chromosphere}}.
\bjtitle{\apj}
\bvolume{779},
\bfpage{168}.
\doiurl{https://doi.org/10.1088/0004-637X/779/2/168}.
\adsurl{2013ApJ...779..168J}.
\end{barticle}
\endbibitem

\bibitem[\protect\citeauthoryear{{Jess} et~al.}{2016}]{2016NatPh..12..179J}
\begin{barticle}
\bauthor{\bsnm{{Jess}}, \binits{D.B.}},
\bauthor{\bsnm{{Reznikova}}, \binits{V.E.}},
\bauthor{\bsnm{{Ryans}}, \binits{R.S.I.}},
\bauthor{\bsnm{{Christian}}, \binits{D.J.}},
\bauthor{\bsnm{{Keys}}, \binits{P.H.}},
\bauthor{\bsnm{{Mathioudakis}}, \binits{M.}},
\bauthor{\bsnm{{Mackay}}, \binits{D.H.}},
\bauthor{\bsnm{{Krishna Prasad}}, \binits{S.}},
\bauthor{\bsnm{{Banerjee}}, \binits{D.}},
\bauthor{\bsnm{{Grant}}, \binits{S.D.T.}},
\bauthor{\bsnm{{Yau}}, \binits{S.}},
\bauthor{\bsnm{{Diamond}}, \binits{C.}}:
\byear{2016},
\batitle{{Solar coronal magnetic fields derived using seismology techniques
  applied to omnipresent sunspot waves}}.
\bjtitle{Nature Physics}
\bvolume{12},
\bfpage{179}.
\doiurl{https://doi.org/10.1038/nphys3544}.
\adsurl{2016NatPh..12..179J}.
\end{barticle}
\endbibitem

\bibitem[\protect\citeauthoryear{{Jess} et~al.}{2017}]{2017ApJ...842...59J}
\begin{barticle}
\bauthor{\bsnm{{Jess}}, \binits{D.B.}},
\bauthor{\bsnm{{Van Doorsselaere}}, \binits{T.}},
\bauthor{\bsnm{{Verth}}, \binits{G.}},
\bauthor{\bsnm{{Fedun}}, \binits{V.}},
\bauthor{\bsnm{{Krishna Prasad}}, \binits{S.}},
\bauthor{\bsnm{{Erd{\'e}lyi}}, \binits{R.}},
\bauthor{\bsnm{{Keys}}, \binits{P.H.}},
\bauthor{\bsnm{{Grant}}, \binits{S.D.T.}},
\bauthor{\bsnm{{Uitenbroek}}, \binits{H.}},
\bauthor{\bsnm{{Christian}}, \binits{D.J.}}:
\byear{2017},
\batitle{{An Inside Look at Sunspot Oscillations with Higher Azimuthal
  Wavenumbers}}.
\bjtitle{\apj}
\bvolume{842},
\bfpage{59}.
\doiurl{https://doi.org/10.3847/1538-4357/aa73d6}.
\adsurl{2017ApJ...842...59J}.
\end{barticle}
\endbibitem

\bibitem[\protect\citeauthoryear{{Jess} et~al.}{2020}]{2020NatAs...4..220J}
\begin{barticle}
\bauthor{\bsnm{{Jess}}, \binits{D.B.}},
\bauthor{\bsnm{{Snow}}, \binits{B.}},
\bauthor{\bsnm{{Houston}}, \binits{S.J.}},
\bauthor{\bsnm{{Botha}}, \binits{G.J.J.}},
\bauthor{\bsnm{{Fleck}}, \binits{B.}},
\bauthor{\bsnm{{Krishna Prasad}}, \binits{S.}},
\bauthor{\bsnm{{Asensio Ramos}}, \binits{A.}},
\bauthor{\bsnm{{Morton}}, \binits{R.J.}},
\bauthor{\bsnm{{Keys}}, \binits{P.H.}},
\bauthor{\bsnm{{Jafarzadeh}}, \binits{S.}},
\bauthor{\bsnm{{Stangalini}}, \binits{M.}},
\bauthor{\bsnm{{Grant}}, \binits{S.D.T.}},
\bauthor{\bsnm{{Christian}}, \binits{D.J.}}:
\byear{2020},
\batitle{{A chromospheric resonance cavity in a sunspot mapped with
  seismology}}.
\bjtitle{Nature Astronomy}
\bvolume{4},
\bfpage{220}.
\doiurl{https://doi.org/10.1038/s41550-019-0945-2}.
\adsurl{2020NatAs...4..220J}.
\end{barticle}
\endbibitem

\bibitem[\protect\citeauthoryear{{Jess} et~al.}{2021}]{2021RSPTA.37900169J}
\begin{barticle}
\bauthor{\bsnm{{Jess}}, \binits{D.B.}},
\bauthor{\bsnm{{Keys}}, \binits{P.H.}},
\bauthor{\bsnm{{Stangalini}}, \binits{M.}},
\bauthor{\bsnm{{Jafarzadeh}}, \binits{S.}}:
\byear{2021},
\batitle{{High-resolution wave dynamics in the lower solar atmosphere}}.
\bjtitle{Philosophical Transactions of the Royal Society of London Series A}
\bvolume{379},
\bfpage{20200169}.
\doiurl{https://doi.org/10.1098/rsta.2020.0169}.
\adsurl{2021RSPTA.37900169J}.
\end{barticle}
\endbibitem

\bibitem[\protect\citeauthoryear{Jess et~al.}{2023}]{JessLRSP2023}
\begin{barticle}
\bauthor{\bsnm{Jess}, \binits{D.B.}},
\bauthor{\bsnm{Jafarzadeh}, \binits{S.}},
\bauthor{\bsnm{Keys}, \binits{P.H.}},
\bauthor{\bsnm{Stangalini}, \binits{M.}},
\bauthor{\bsnm{Verth}, \binits{G.}},
\bauthor{\bsnm{Grant}, \binits{S.D.T.}}:
\byear{2023},
\batitle{Waves in the lower solar atmosphere: the dawn of next-generation solar
  telescopes}.
\bjtitle{Living Reviews in Solar Physics}
\bvolume{20},
\bfpage{1}.
\bisbn{1614-4961}.
\doiurl{https://doi.org/10.1007/s41116-022-00035-6}.
\burl{https://doi.org/10.1007/s41116-022-00035-6}.
\end{barticle}
\endbibitem

\bibitem[\protect\citeauthoryear{{Joshi} and {de la Cruz
  Rodr{\'\i}guez}}{2018}]{2018A&A...619A..63J}
\begin{barticle}
\bauthor{\bsnm{{Joshi}}, \binits{J.}},
\bauthor{\bsnm{{de la Cruz Rodr{\'\i}guez}}, \binits{J.}}:
\byear{2018},
\batitle{{Magnetic field variations associated with umbral flashes and
  penumbral waves}}.
\bjtitle{\aap}
\bvolume{619},
\bfpage{A63}.
\doiurl{https://doi.org/10.1051/0004-6361/201832955}.
\adsurl{2018A&A...619A..63J}.
\end{barticle}
\endbibitem

\bibitem[\protect\citeauthoryear{{Kafle} et~al.}{2021}]{2021RScI...92f3002K}
\begin{barticle}
\bauthor{\bsnm{{Kafle}}, \binits{N.}},
\bauthor{\bsnm{{Elliott}}, \binits{D.}},
\bauthor{\bsnm{{Garren}}, \binits{E.W.}},
\bauthor{\bsnm{{He}}, \binits{Z.}},
\bauthor{\bsnm{{Gebhart}}, \binits{T.E.}},
\bauthor{\bsnm{{Zhang}}, \binits{Z.}},
\bauthor{\bsnm{{Biewer}}, \binits{T.M.}}:
\byear{2021},
\batitle{{Design and implementation of a portable diagnostic system for Thomson
  scattering and optical emission spectroscopy measurements}}.
\bjtitle{Review of Scientific Instruments}
\bvolume{92},
\bfpage{063002}.
\doiurl{https://doi.org/10.1063/5.0043818}.
\adsurl{2021RScI...92f3002K}.
\end{barticle}
\endbibitem

\bibitem[\protect\citeauthoryear{Kayshap et~al.}{2018}]{10.1093/mnras/sty1861}
\begin{barticle}
\bauthor{\bsnm{Kayshap}, \binits{P.}},
\bauthor{\bsnm{Murawski}, \binits{K.}},
\bauthor{\bsnm{Srivastava}, \binits{A.K.}},
\bauthor{\bsnm{Musielak}, \binits{Z.E.}},
\bauthor{\bsnm{Dwivedi}, \binits{B.N.}}:
\byear{2018},
\batitle{{Vertical propagation of acoustic waves in the solar internetworkas
  observed by IRIS}}.
\bjtitle{Monthly Notices of the Royal Astronomical Society}
\bvolume{479},
\bfpage{5512}.
\doiurl{https://doi.org/10.1093/mnras/sty1861}.
\burl{https://doi.org/10.1093/mnras/sty1861}.
\end{barticle}
\endbibitem

\bibitem[\protect\citeauthoryear{{Kiselman} et~al.}{2011}]{2011A&A...535A..14K}
\begin{barticle}
\bauthor{\bsnm{{Kiselman}}, \binits{D.}},
\bauthor{\bsnm{{Pereira}}, \binits{T.M.D.}},
\bauthor{\bsnm{{Gustafsson}}, \binits{B.}},
\bauthor{\bsnm{{Asplund}}, \binits{M.}},
\bauthor{\bsnm{{Mel{\'e}ndez}}, \binits{J.}},
\bauthor{\bsnm{{Langhans}}, \binits{K.}}:
\byear{2011},
\batitle{{Is the solar spectrum latitude-dependent?. An investigation with
  SST/TRIPPEL}}.
\bjtitle{\aap}
\bvolume{535},
\bfpage{A14}.
\doiurl{https://doi.org/10.1051/0004-6361/201117553}.
\adsurl{2011A&A...535A..14K}.
\end{barticle}
\endbibitem

\bibitem[\protect\citeauthoryear{{Kneer}, {Mattig}, and
  {Uexkuell}}{1981}]{1981A&A...102..147K}
\begin{barticle}
\bauthor{\bsnm{{Kneer}}, \binits{F.}},
\bauthor{\bsnm{{Mattig}}, \binits{W.}},
\bauthor{\bsnm{{Uexkuell}}, \binits{M.V.}}:
\byear{1981},
\batitle{{The chromosphere above sunspot umbrae. III - Spatial and temporal
  variations of chromospheric lines}}.
\bjtitle{\aap}
\bvolume{102},
\bfpage{147}.
\adsurl{1981A&A...102..147K}.
\end{barticle}
\endbibitem

\bibitem[\protect\citeauthoryear{Kotani et~al.}{2010}]{Kontani2010}
\begin{barticle}
\bauthor{\bsnm{Kotani}, \binits{T.}},
\bauthor{\bsnm{K{\"u}veler}, \binits{G.}},
\bauthor{\bsnm{Dao}, \binits{V.D.}},
\bauthor{\bsnm{Zuber}, \binits{A.}},
\bauthor{\bsnm{Ramelli}, \binits{R.}}:
\byear{2010},
\batitle{Robotic and Non-Robotic Control of Astrophysical Instruments}.
\bjtitle{Advances in Astronomy}
\bvolume{2010},
\bfpage{620424}.
\bisbn{1687-7969}.
\doiurl{https://doi.org/10.1155/2010/620424}.
\burl{https://doi.org/10.1155/2010/620424}.
\end{barticle}
\endbibitem

\bibitem[\protect\citeauthoryear{{Kowalski} et~al.}{2022}]{2022ApJ...928..190K}
\begin{barticle}
\bauthor{\bsnm{{Kowalski}}, \binits{A.F.}},
\bauthor{\bsnm{{Allred}}, \binits{J.C.}},
\bauthor{\bsnm{{Carlsson}}, \binits{M.}},
\bauthor{\bsnm{{Kerr}}, \binits{G.S.}},
\bauthor{\bsnm{{Tremblay}}, \binits{P.-E.}},
\bauthor{\bsnm{{Namekata}}, \binits{K.}},
\bauthor{\bsnm{{Kuridze}}, \binits{D.}},
\bauthor{\bsnm{{Uitenbroek}}, \binits{H.}}:
\byear{2022},
\batitle{{The Atmospheric Response to High Nonthermal Electron-beam Fluxes in
  Solar Flares. II. Hydrogen-broadening Predictions for Solar Flare
  Observations with the Daniel K. Inouye Solar Telescope}}.
\bjtitle{\apj}
\bvolume{928},
\bfpage{190}.
\doiurl{https://doi.org/10.3847/1538-4357/ac5174}.
\adsurl{2022ApJ...928..190K}.
\end{barticle}
\endbibitem

\bibitem[\protect\citeauthoryear{{Kuridze} et~al.}{2016}]{2016ApJ...832..147K}
\begin{barticle}
\bauthor{\bsnm{{Kuridze}}, \binits{D.}},
\bauthor{\bsnm{{Mathioudakis}}, \binits{M.}},
\bauthor{\bsnm{{Christian}}, \binits{D.J.}},
\bauthor{\bsnm{{Kowalski}}, \binits{A.F.}},
\bauthor{\bsnm{{Jess}}, \binits{D.B.}},
\bauthor{\bsnm{{Grant}}, \binits{S.D.T.}},
\bauthor{\bsnm{{Kawate}}, \binits{T.}},
\bauthor{\bsnm{{Sim{\~o}es}}, \binits{P.J.A.}},
\bauthor{\bsnm{{Allred}}, \binits{J.C.}},
\bauthor{\bsnm{{Keenan}}, \binits{F.P.}}:
\byear{2016},
\batitle{{Observations and Simulations of the Na I D$_{1}$ Line Profiles in an
  M-class Solar Flare}}.
\bjtitle{\apj}
\bvolume{832},
\bfpage{147}.
\doiurl{https://doi.org/10.3847/0004-637X/832/2/147}.
\adsurl{2016ApJ...832..147K}.
\end{barticle}
\endbibitem

\bibitem[\protect\citeauthoryear{{Kuridze} et~al.}{2022}]{2022ApJ...937...56K}
\begin{barticle}
\bauthor{\bsnm{{Kuridze}}, \binits{D.}},
\bauthor{\bsnm{{Heinzel}}, \binits{P.}},
\bauthor{\bsnm{{Koza}}, \binits{J.}},
\bauthor{\bsnm{{Oliver}}, \binits{R.}}:
\byear{2022},
\batitle{{Dark Off-limb Gap: Manifestation of a Temperature Minimum and the
  Dynamic Nature of the Chromosphere}}.
\bjtitle{\apj}
\bvolume{937},
\bfpage{56}.
\doiurl{https://doi.org/10.3847/1538-4357/ac8d8e}.
\adsurl{2022ApJ...937...56K}.
\end{barticle}
\endbibitem

\bibitem[\protect\citeauthoryear{{Kurucz}}{2005}]{2005MSAIS...8..189K}
\begin{barticle}
\bauthor{\bsnm{{Kurucz}}, \binits{R.L.}}:
\byear{2005},
\batitle{{New atlases for solar flux, irradiance, central intensity, and limb
  intensity}}.
\bjtitle{Memorie della Societa Astronomica Italiana Supplementi}
\bvolume{8},
\bfpage{189}.
\adsurl{2005MSAIS...8..189K}.
\end{barticle}
\endbibitem

\bibitem[\protect\citeauthoryear{{Kurucz} et~al.}{1984}]{1984sfat.book.....K}
\begin{bbook}
\bauthor{\bsnm{{Kurucz}}, \binits{R.L.}},
\bauthor{\bsnm{{Furenlid}}, \binits{I.}},
\bauthor{\bsnm{{Brault}}, \binits{J.}},
\bauthor{\bsnm{{Testerman}}, \binits{L.}}:
\byear{1984},
\bbtitle{{Solar flux atlas from 296 to 1300 nm}}.
\adsurl{1984sfat.book.....K}.
\end{bbook}
\endbibitem

\bibitem[\protect\citeauthoryear{{Labeyrie} and
  {Flamand}}{1969}]{1969OptCo...1....5L}
\begin{barticle}
\bauthor{\bsnm{{Labeyrie}}, \binits{A.}},
\bauthor{\bsnm{{Flamand}}, \binits{J.}}:
\byear{1969},
\batitle{{Spectrographic performance of holographically made diffraction
  gratings}}.
\bjtitle{Optics Communications}
\bvolume{1},
\bfpage{5}.
\doiurl{https://doi.org/10.1016/0030-4018(69)90068-6}.
\adsurl{1969OptCo...1....5L}.
\end{barticle}
\endbibitem

\bibitem[\protect\citeauthoryear{{Landi
  Degl'Innocenti}}{1984}]{1984SoPh...91....1L}
\begin{barticle}
\bauthor{\bsnm{{Landi Degl'Innocenti}}, \binits{E.}}:
\byear{1984},
\batitle{{Polarization in Spectral Lines - Part Three - Resonance Polarization
  in the Non-Magnetic Collisionless Regime}}.
\bjtitle{\solphys}
\bvolume{91},
\bfpage{1}.
\doiurl{https://doi.org/10.1007/BF00213606}.
\adsurl{1984SoPh...91....1L}.
\end{barticle}
\endbibitem

\bibitem[\protect\citeauthoryear{Lee}{2008}]{Lee2008}
\begin{botherref}
\oauthor{\bsnm{Lee}, \binits{B.}}:
2008,
Introduction to$\pm$12 degree orthogonal digital micromirror devices (dmds).
\textit{Texas Instruments},
2018.
\end{botherref}
\endbibitem

\bibitem[\protect\citeauthoryear{{Lee} et~al.}{2019}]{2019APJAS..55..165L}
\begin{barticle}
\bauthor{\bsnm{{Lee}}, \binits{H.}},
\bauthor{\bsnm{{Kim}}, \binits{W.}},
\bauthor{\bsnm{{Lee}}, \binits{Y.G.}},
\bauthor{\bsnm{{Kim}}, \binits{J.}},
\bauthor{\bsnm{{Cho}}, \binits{H.K.}}:
\byear{2019},
\batitle{{Atmospheric Transmission of Ultraviolet and Total Solar Radiation by
  Clouds, Aerosols, and Ozone in Seoul, Korea: a Comparison of Semi-Empirical
  Model Predictions with Observations}}.
\bjtitle{Asia-Pacific Journal of Atmospheric Sciences}
\bvolume{55},
\bfpage{165}.
\doiurl{https://doi.org/10.1007/s13143-018-0082-3}.
\adsurl{2019APJAS..55..165L}.
\end{barticle}
\endbibitem

\bibitem[\protect\citeauthoryear{Lee, Thompson, and Rolland}{2010}]{Lee:10}
\begin{barticle}
\bauthor{\bsnm{Lee}, \binits{K.-S.}},
\bauthor{\bsnm{Thompson}, \binits{K.P.}},
\bauthor{\bsnm{Rolland}, \binits{J.P.}}:
\byear{2010},
\batitle{Broadband astigmatism-corrected Czerny--Turner spectrometer}.
\bjtitle{Opt. Express}
\bvolume{18},
\bfpage{23378}.
\doiurl{https://doi.org/10.1364/OE.18.023378}.
\burl{https://opg.optica.org/oe/abstract.cfm?URI=oe-18-22-23378}.
\end{barticle}
\endbibitem

\bibitem[\protect\citeauthoryear{{Leenaarts}
  et~al.}{2018}]{2018A&A...612A..28L}
\begin{barticle}
\bauthor{\bsnm{{Leenaarts}}, \binits{J.}},
\bauthor{\bsnm{{de la Cruz Rodr{\'\i}guez}}, \binits{J.}},
\bauthor{\bsnm{{Danilovic}}, \binits{S.}},
\bauthor{\bsnm{{Scharmer}}, \binits{G.}},
\bauthor{\bsnm{{Carlsson}}, \binits{M.}}:
\byear{2018},
\batitle{{Chromospheric heating during flux emergence in the solar
  atmosphere}}.
\bjtitle{\aap}
\bvolume{612},
\bfpage{A28}.
\doiurl{https://doi.org/10.1051/0004-6361/201732027}.
\adsurl{2018A&A...612A..28L}.
\end{barticle}
\endbibitem

\bibitem[\protect\citeauthoryear{{Lemke} et~al.}{2011}]{2011MNRAS.417..689L}
\begin{barticle}
\bauthor{\bsnm{{Lemke}}, \binits{U.}},
\bauthor{\bsnm{{Corbett}}, \binits{J.}},
\bauthor{\bsnm{{Allington-Smith}}, \binits{J.}},
\bauthor{\bsnm{{Murray}}, \binits{G.}}:
\byear{2011},
\batitle{{Modal noise prediction in fibre spectroscopy - I. Visibility and the
  coherent model}}.
\bjtitle{\mnras}
\bvolume{417},
\bfpage{689}.
\doiurl{https://doi.org/10.1111/j.1365-2966.2011.19312.x}.
\adsurl{2011MNRAS.417..689L}.
\end{barticle}
\endbibitem

\bibitem[\protect\citeauthoryear{{Li} et~al.}{2019}]{2019arXiv190500391L}
\begin{botherref}
\oauthor{\bsnm{{Li}}, \binits{Q.}},
\oauthor{\bsnm{{Lin}}, \binits{J.}},
\oauthor{\bsnm{{Clancy}}, \binits{N.T.}},
\oauthor{\bsnm{{Elson}}, \binits{D.S.}}:
2019,
{Estimation of Tissue Oxygen Saturation from RGB images and Sparse
  Hyperspectral Signals based on Conditional Generative Adversarial Network}.
\textit{arXiv e-prints},
arXiv:1905.00391.
\doiurl{https://doi.org/10.48550/arXiv.1905.00391}.
\adsurl{2019arXiv190500391L}.
\end{botherref}
\endbibitem

\bibitem[\protect\citeauthoryear{{Li} et~al.}{2022}]{2022ApJ...936...37L}
\begin{barticle}
\bauthor{\bsnm{{Li}}, \binits{Q.}},
\bauthor{\bsnm{{Zhang}}, \binits{L.}},
\bauthor{\bsnm{{Yan}}, \binits{X.}},
\bauthor{\bsnm{{Norton}}, \binits{A.A.}},
\bauthor{\bsnm{{Wang}}, \binits{J.}},
\bauthor{\bsnm{{Yang}}, \binits{L.}},
\bauthor{\bsnm{{Xue}}, \binits{Z.}},
\bauthor{\bsnm{{Kong}}, \binits{D.}}:
\byear{2022},
\batitle{{Dependence of the Continuum Intensities on the Magnetic Fields at
  Different Evolution Phases of Sunspots}}.
\bjtitle{\apj}
\bvolume{936},
\bfpage{37}.
\doiurl{https://doi.org/10.3847/1538-4357/ac83b3}.
\adsurl{2022ApJ...936...37L}.
\end{barticle}
\endbibitem

\bibitem[\protect\citeauthoryear{{Lin}}{2012}]{2012SPIE.8446E..1DL}
\begin{bchapter}
\bauthor{\bsnm{{Lin}}, \binits{H.}}:
\byear{2012},
\bctitle{{SPIES: the spectropolarimetric imager for the energetic sun}}.
In: \beditor{\bsnm{{McLean}}, \binits{I.S.}},
\beditor{\bsnm{{Ramsay}}, \binits{S.K.}},
\beditor{\bsnm{{Takami}}, \binits{H.}} (eds.)
\bbtitle{Ground-based and Airborne Instrumentation for Astronomy IV},
\bsertitle{Society of Photo-Optical Instrumentation Engineers (SPIE) Conference
  Series}
\bseriesno{8446},
\bfpage{84461D}.
\doiurl{https://doi.org/10.1117/12.926830}.
\adsurl{2012SPIE.8446E..1DL}.
\end{bchapter}
\endbibitem

\bibitem[\protect\citeauthoryear{{Lites} et~al.}{1991}]{1991sopo.work....3L}
\begin{bchapter}
\bauthor{\bsnm{{Lites}}, \binits{B.W.}},
\bauthor{\bsnm{{Elmore}}, \binits{D.}},
\bauthor{\bsnm{{Murphy}}, \binits{G.}},
\bauthor{\bsnm{{Skumanich}}, \binits{A.}},
\bauthor{\bsnm{{Tomczyk}}, \binits{S.}},
\bauthor{\bsnm{{Dunn}}, \binits{R.B.}}:
\byear{1991},
\bctitle{{Preliminary results from the HAO/NSO Advanced Stokes Polarimeter
  prototype observing run.}}
In: \beditor{\bsnm{{November}}, \binits{L.J.}} (ed.)
\bbtitle{Solar Polarimetry},
\bfpage{3}.
\adsurl{1991sopo.work....3L}.
\end{bchapter}
\endbibitem

\bibitem[\protect\citeauthoryear{{Liu} et~al.}{2014}]{2014RAA....14..705L}
\begin{barticle}
\bauthor{\bsnm{{Liu}}, \binits{Z.}},
\bauthor{\bsnm{{Xu}}, \binits{J.}},
\bauthor{\bsnm{{Gu}}, \binits{B.-Z.}},
\bauthor{\bsnm{{Wang}}, \binits{S.}},
\bauthor{\bsnm{{You}}, \binits{J.-Q.}},
\bauthor{\bsnm{{Shen}}, \binits{L.-X.}},
\bauthor{\bsnm{{Lu}}, \binits{R.-W.}},
\bauthor{\bsnm{{Jin}}, \binits{Z.-Y.}},
\bauthor{\bsnm{{Chen}}, \binits{L.-F.}},
\bauthor{\bsnm{{Lou}}, \binits{K.}},
\bauthor{\bsnm{{Li}}, \binits{Z.}},
\bauthor{\bsnm{{Liu}}, \binits{G.-Q.}},
\bauthor{\bsnm{{Xu}}, \binits{Z.}},
\bauthor{\bsnm{{Rao}}, \binits{C.-H.}},
\bauthor{\bsnm{{Hu}}, \binits{Q.-Q.}},
\bauthor{\bsnm{{Li}}, \binits{R.-F.}},
\bauthor{\bsnm{{Fu}}, \binits{H.-W.}},
\bauthor{\bsnm{{Wang}}, \binits{F.}},
\bauthor{\bsnm{{Bao}}, \binits{M.-X.}},
\bauthor{\bsnm{{Wu}}, \binits{M.-C.}},
\bauthor{\bsnm{{Zhang}}, \binits{B.-R.}}:
\byear{2014},
\batitle{{New vacuum solar telescope and observations with high resolution}}.
\bjtitle{Research in Astronomy and Astrophysics}
\bvolume{14},
\bfpage{705}.
\doiurl{https://doi.org/10.1088/1674-4527/14/6/009}.
\adsurl{2014RAA....14..705L}.
\end{barticle}
\endbibitem

\bibitem[\protect\citeauthoryear{{L{\"o}fdahl}
  et~al.}{2021}]{2021A&A...653A..68L}
\begin{barticle}
\bauthor{\bsnm{{L{\"o}fdahl}}, \binits{M.G.}},
\bauthor{\bsnm{{Hillberg}}, \binits{T.}},
\bauthor{\bsnm{{de la Cruz Rodr{\'\i}guez}}, \binits{J.}},
\bauthor{\bsnm{{Vissers}}, \binits{G.}},
\bauthor{\bsnm{{Andriienko}}, \binits{O.}},
\bauthor{\bsnm{{Scharmer}}, \binits{G.B.}},
\bauthor{\bsnm{{Haugan}}, \binits{S.V.H.}},
\bauthor{\bsnm{{Fredvik}}, \binits{T.}}:
\byear{2021},
\batitle{{SSTRED: Data- and metadata-processing pipeline for CHROMIS and
  CRISP}}.
\bjtitle{\aap}
\bvolume{653},
\bfpage{A68}.
\doiurl{https://doi.org/10.1051/0004-6361/202141326}.
\adsurl{2021A&A...653A..68L}.
\end{barticle}
\endbibitem

\bibitem[\protect\citeauthoryear{{MacBride} et~al.}{2021}]{2021RSPTA.37900171M}
\begin{barticle}
\bauthor{\bsnm{{MacBride}}, \binits{C.D.}},
\bauthor{\bsnm{{Jess}}, \binits{D.B.}},
\bauthor{\bsnm{{Grant}}, \binits{S.D.T.}},
\bauthor{\bsnm{{Khomenko}}, \binits{E.}},
\bauthor{\bsnm{{Keys}}, \binits{P.H.}},
\bauthor{\bsnm{{Stangalini}}, \binits{M.}}:
\byear{2021},
\batitle{{Accurately constraining velocity information from spectral imaging
  observations using machine learning techniques}}.
\bjtitle{Philosophical Transactions of the Royal Society of London Series A}
\bvolume{379},
\bfpage{20200171}.
\doiurl{https://doi.org/10.1098/rsta.2020.0171}.
\adsurl{2021RSPTA.37900171M}.
\end{barticle}
\endbibitem

\bibitem[\protect\citeauthoryear{{Martinez Pillet}
  et~al.}{1990}]{1990ApJ...361L..81M}
\begin{barticle}
\bauthor{\bsnm{{Martinez Pillet}}, \binits{V.}},
\bauthor{\bsnm{{Garcia Lopez}}, \binits{R.J.}},
\bauthor{\bsnm{{del Toro Iniesta}}, \binits{J.C.}},
\bauthor{\bsnm{{Rebolo}}, \binits{R.}},
\bauthor{\bsnm{{Vazquez}}, \binits{M.}},
\bauthor{\bsnm{{Beckman}}, \binits{J.E.}},
\bauthor{\bsnm{{Char}}, \binits{S.}}:
\byear{1990},
\batitle{{Circular Polarization of the CA II H and K Lines in Solar Quiet and
  Active Regions}}.
\bjtitle{\apjl}
\bvolume{361},
\bfpage{L81}.
\doiurl{https://doi.org/10.1086/185832}.
\adsurl{1990ApJ...361L..81M}.
\end{barticle}
\endbibitem

\bibitem[\protect\citeauthoryear{{Mathur} et~al.}{2022}]{2022A&A...668A.153M}
\begin{barticle}
\bauthor{\bsnm{{Mathur}}, \binits{H.}},
\bauthor{\bsnm{{Joshi}}, \binits{J.}},
\bauthor{\bsnm{{Nagaraju}}, \binits{K.}},
\bauthor{\bsnm{{Rouppe van der Voort}}, \binits{L.}},
\bauthor{\bsnm{{Bose}}, \binits{S.}}:
\byear{2022},
\batitle{{Properties of shock waves in the quiet-Sun chromosphere}}.
\bjtitle{\aap}
\bvolume{668},
\bfpage{A153}.
\doiurl{https://doi.org/10.1051/0004-6361/202244332}.
\adsurl{2022A&A...668A.153M}.
\end{barticle}
\endbibitem

\bibitem[\protect\citeauthoryear{{Mattig} and
  {Zerfass}}{1991}]{1991A&A...241..212M}
\begin{barticle}
\bauthor{\bsnm{{Mattig}}, \binits{W.}},
\bauthor{\bsnm{{Zerfass}}, \binits{M.}}:
\byear{1991},
\batitle{{Oscillations in sunspots near the solar limb}}.
\bjtitle{\aap}
\bvolume{241},
\bfpage{212}.
\adsurl{1991A&A...241..212M}.
\end{barticle}
\endbibitem

\bibitem[\protect\citeauthoryear{Mohammadi and Eslami}{2010}]{Mohammadi2010}
\begin{barticle}
\bauthor{\bsnm{Mohammadi}, \binits{H.}},
\bauthor{\bsnm{Eslami}, \binits{E.}}:
\byear{2010},
\batitle{Investigation of spectral resolution in a Czerny Turner spectrograph}.
\bjtitle{Instruments and Experimental Techniques}
\bvolume{53},
\bfpage{549}.
\bisbn{1608-3180}.
\doiurl{https://doi.org/10.1134/S0020441210040147}.
\burl{https://doi.org/10.1134/S0020441210040147}.
\end{barticle}
\endbibitem

\bibitem[\protect\citeauthoryear{{Monson} et~al.}{2021}]{2021ApJ...915...16M}
\begin{barticle}
\bauthor{\bsnm{{Monson}}, \binits{A.J.}},
\bauthor{\bsnm{{Mathioudakis}}, \binits{M.}},
\bauthor{\bsnm{{Reid}}, \binits{A.}},
\bauthor{\bsnm{{Milligan}}, \binits{R.}},
\bauthor{\bsnm{{Kuridze}}, \binits{D.}}:
\byear{2021},
\batitle{{Flare-induced Photospheric Velocity Diagnostics}}.
\bjtitle{\apj}
\bvolume{915},
\bfpage{16}.
\doiurl{https://doi.org/10.3847/1538-4357/abfda8}.
\adsurl{2021ApJ...915...16M}.
\end{barticle}
\endbibitem

\bibitem[\protect\citeauthoryear{{Morosin} et~al.}{2022}]{2022A&A...664A...8M}
\begin{barticle}
\bauthor{\bsnm{{Morosin}}, \binits{R.}},
\bauthor{\bsnm{{de la Cruz Rodr{\'\i}guez}}, \binits{J.}},
\bauthor{\bsnm{{D{\'\i}az Baso}}, \binits{C.J.}},
\bauthor{\bsnm{{Leenaarts}}, \binits{J.}}:
\byear{2022},
\batitle{{Spatio-temporal analysis of chromospheric heating in a plage
  region}}.
\bjtitle{\aap}
\bvolume{664},
\bfpage{A8}.
\doiurl{https://doi.org/10.1051/0004-6361/202243461}.
\adsurl{2022A&A...664A...8M}.
\end{barticle}
\endbibitem

\bibitem[\protect\citeauthoryear{{M{\"u}ller}
  et~al.}{2013}]{2013SoPh..285...25M}
\begin{barticle}
\bauthor{\bsnm{{M{\"u}ller}}, \binits{D.}},
\bauthor{\bsnm{{Marsden}}, \binits{R.G.}},
\bauthor{\bsnm{{St. Cyr}}, \binits{O.C.}},
\bauthor{\bsnm{{Gilbert}}, \binits{H.R.}}:
\byear{2013},
\batitle{{Solar Orbiter . Exploring the Sun-Heliosphere Connection}}.
\bjtitle{\solphys}
\bvolume{285},
\bfpage{25}.
\doiurl{https://doi.org/10.1007/s11207-012-0085-7}.
\adsurl{2013SoPh..285...25M}.
\end{barticle}
\endbibitem

\bibitem[\protect\citeauthoryear{{M{\"u}ller}
  et~al.}{2020}]{2020A&A...642A...1M}
\begin{barticle}
\bauthor{\bsnm{{M{\"u}ller}}, \binits{D.}},
\bauthor{\bsnm{{St. Cyr}}, \binits{O.C.}},
\bauthor{\bsnm{{Zouganelis}}, \binits{I.}},
\bauthor{\bsnm{{Gilbert}}, \binits{H.R.}},
\bauthor{\bsnm{{Marsden}}, \binits{R.}},
\bauthor{\bsnm{{Nieves-Chinchilla}}, \binits{T.}},
\bauthor{\bsnm{{Antonucci}}, \binits{E.}},
\bauthor{\bsnm{{Auch{\`e}re}}, \binits{F.}},
\bauthor{\bsnm{{Berghmans}}, \binits{D.}},
\bauthor{\bsnm{{Horbury}}, \binits{T.S.}},
\bauthor{\bsnm{{Howard}}, \binits{R.A.}},
\bauthor{\bsnm{{Krucker}}, \binits{S.}},
\bauthor{\bsnm{{Maksimovic}}, \binits{M.}},
\bauthor{\bsnm{{Owen}}, \binits{C.J.}},
\bauthor{\bsnm{{Rochus}}, \binits{P.}},
\bauthor{\bsnm{{Rodriguez-Pacheco}}, \binits{J.}},
\bauthor{\bsnm{{Romoli}}, \binits{M.}},
\bauthor{\bsnm{{Solanki}}, \binits{S.K.}},
\bauthor{\bsnm{{Bruno}}, \binits{R.}},
\bauthor{\bsnm{{Carlsson}}, \binits{M.}},
\bauthor{\bsnm{{Fludra}}, \binits{A.}},
\bauthor{\bsnm{{Harra}}, \binits{L.}},
\bauthor{\bsnm{{Hassler}}, \binits{D.M.}},
\bauthor{\bsnm{{Livi}}, \binits{S.}},
\bauthor{\bsnm{{Louarn}}, \binits{P.}},
\bauthor{\bsnm{{Peter}}, \binits{H.}},
\bauthor{\bsnm{{Sch{\"u}hle}}, \binits{U.}},
\bauthor{\bsnm{{Teriaca}}, \binits{L.}},
\bauthor{\bsnm{{del Toro Iniesta}}, \binits{J.C.}},
\bauthor{\bsnm{{Wimmer-Schweingruber}}, \binits{R.F.}},
\bauthor{\bsnm{{Marsch}}, \binits{E.}},
\bauthor{\bsnm{{Velli}}, \binits{M.}},
\bauthor{\bsnm{{De Groof}}, \binits{A.}},
\bauthor{\bsnm{{Walsh}}, \binits{A.}},
\bauthor{\bsnm{{Williams}}, \binits{D.}}:
\byear{2020},
\batitle{{The Solar Orbiter mission. Science overview}}.
\bjtitle{\aap}
\bvolume{642},
\bfpage{A1}.
\doiurl{https://doi.org/10.1051/0004-6361/202038467}.
\adsurl{2020A&A...642A...1M}.
\end{barticle}
\endbibitem

\bibitem[\protect\citeauthoryear{{Nagaraju} et~al.}{2008}]{2008BASI...36...99N}
\begin{barticle}
\bauthor{\bsnm{{Nagaraju}}, \binits{K.}},
\bauthor{\bsnm{{Sankarasubramanian}}, \binits{K.}},
\bauthor{\bsnm{{Rangarajan}}, \binits{K.E.}},
\bauthor{\bsnm{{Ramesh}}, \binits{K.B.}},
\bauthor{\bsnm{{Singh}}, \binits{J.}},
\bauthor{\bsnm{{Devendran}}, \binits{P.}},
\bauthor{\bsnm{{Hariharan}}}:
\byear{2008},
\batitle{{On the performance of a dual-beam polarimeter at Kodaikanal Tower
  Telescope}}.
\bjtitle{Bulletin of the Astronomical Society of India}
\bvolume{36},
\bfpage{99}.
\adsurl{2008BASI...36...99N}.
\end{barticle}
\endbibitem

\bibitem[\protect\citeauthoryear{Okamoto}{2006}]{okamoto2006optical}
\begin{botherref}
\oauthor{\bsnm{Okamoto}, \binits{K.}}:
2006,
Optical fibers.
\textit{Fundamentals of Optical Waveguides},
57.
\end{botherref}
\endbibitem

\bibitem[\protect\citeauthoryear{Oostra}{2017}]{10.1119/1.4975109}
\begin{barticle}
\bauthor{\bsnm{Oostra}, \binits{B.}}:
\byear{2017},
\batitle{{A method for measuring magnetic fields in sunspots using
  Zeeman-broadened absorption lines}}.
\bjtitle{American Journal of Physics}
\bvolume{85},
\bfpage{295}.
\doiurl{https://doi.org/10.1119/1.4975109}.
\burl{https://doi.org/10.1119/1.4975109}.
\end{barticle}
\endbibitem

\bibitem[\protect\citeauthoryear{Osborne and
  Fletcher}{2022}]{10.1093/mnras/stac2570}
\begin{barticle}
\bauthor{\bsnm{Osborne}, \binits{C.M.J.}},
\bauthor{\bsnm{Fletcher}, \binits{L.}}:
\byear{2022},
\batitle{{Flare kernels may be smaller than you think: modelling the radiative
  response of chromospheric plasma adjacent to a solar flare}}.
\bjtitle{Monthly Notices of the Royal Astronomical Society}
\bvolume{516},
\bfpage{6066}.
\doiurl{https://doi.org/10.1093/mnras/stac2570}.
\burl{https://doi.org/10.1093/mnras/stac2570}.
\end{barticle}
\endbibitem

\bibitem[\protect\citeauthoryear{{Pesnell}, {Thompson}, and
  {Chamberlin}}{2012}]{2012SoPh..275....3P}
\begin{barticle}
\bauthor{\bsnm{{Pesnell}}, \binits{W.D.}},
\bauthor{\bsnm{{Thompson}}, \binits{B.J.}},
\bauthor{\bsnm{{Chamberlin}}, \binits{P.C.}}:
\byear{2012},
\batitle{{The Solar Dynamics Observatory (SDO)}}.
\bjtitle{\solphys}
\bvolume{275},
\bfpage{3}.
\doiurl{https://doi.org/10.1007/s11207-011-9841-3}.
\adsurl{2012SoPh..275....3P}.
\end{barticle}
\endbibitem

\bibitem[\protect\citeauthoryear{{Petersburg}
  et~al.}{2018}]{2018ApJ...853..181P}
\begin{barticle}
\bauthor{\bsnm{{Petersburg}}, \binits{R.R.}},
\bauthor{\bsnm{{McCracken}}, \binits{T.M.}},
\bauthor{\bsnm{{Eggerman}}, \binits{D.}},
\bauthor{\bsnm{{Jurgenson}}, \binits{C.A.}},
\bauthor{\bsnm{{Sawyer}}, \binits{D.}},
\bauthor{\bsnm{{Szymkowiak}}, \binits{A.E.}},
\bauthor{\bsnm{{Fischer}}, \binits{D.A.}}:
\byear{2018},
\batitle{{Modal Noise Mitigation through Fiber Agitation for Fiber-fed Radial
  Velocity Spectrographs}}.
\bjtitle{\apj}
\bvolume{853},
\bfpage{181}.
\doiurl{https://doi.org/10.3847/1538-4357/aaa487}.
\adsurl{2018ApJ...853..181P}.
\end{barticle}
\endbibitem

\bibitem[\protect\citeauthoryear{{Plowman} et~al.}{2022}]{2022arXiv221116635P}
\begin{botherref}
\oauthor{\bsnm{{Plowman}}, \binits{J.E.}},
\oauthor{\bsnm{{Auch{\`e}re}}, \binits{F.}},
\oauthor{\bsnm{{Aznar Cuadrado}}, \binits{R.}},
\oauthor{\bsnm{{Fludra}}, \binits{A.}},
\oauthor{\bsnm{{Fredvik}}, \binits{T.}},
\oauthor{\bsnm{{Hassler}}, \binits{D.M.}},
\oauthor{\bsnm{{Mandal}}, \binits{S.}},
\oauthor{\bsnm{{Peter}}, \binits{H.}}:
2022,
{SPICE PSF Correction: General Framework and Capability Demonstration}.
\textit{arXiv e-prints},
arXiv:2211.16635.
\doiurl{https://doi.org/10.48550/arXiv.2211.16635}.
\adsurl{2022arXiv221116635P}.
\end{botherref}
\endbibitem

\bibitem[\protect\citeauthoryear{{Poduval} et~al.}{2013}]{2013ApJ...765..144P}
\begin{barticle}
\bauthor{\bsnm{{Poduval}}, \binits{B.}},
\bauthor{\bsnm{{DeForest}}, \binits{C.E.}},
\bauthor{\bsnm{{Schmelz}}, \binits{J.T.}},
\bauthor{\bsnm{{Pathak}}, \binits{S.}}:
\byear{2013},
\batitle{{Point-spread Functions for the Extreme-ultraviolet Channels of
  SDO/AIA Telescopes}}.
\bjtitle{\apj}
\bvolume{765},
\bfpage{144}.
\doiurl{https://doi.org/10.1088/0004-637X/765/2/144}.
\adsurl{2013ApJ...765..144P}.
\end{barticle}
\endbibitem

\bibitem[\protect\citeauthoryear{Povel}{1995}]{10.1117/12.200596}
\begin{barticle}
\bauthor{\bsnm{Povel}, \binits{H.-P.}}:
\byear{1995},
\batitle{{Imaging Stokes polarimetry with piezoelastic modulators and
  charge-coupled-device image sensors}}.
\bjtitle{Optical Engineering}
\bvolume{34},
\bfpage{1870 }.
\doiurl{https://doi.org/10.1117/12.200596}.
\burl{https://doi.org/10.1117/12.200596}.
\end{barticle}
\endbibitem

\bibitem[\protect\citeauthoryear{{Psaltis} et~al.}{1990}]{1990Natur.343..325P}
\begin{barticle}
\bauthor{\bsnm{{Psaltis}}, \binits{D.}},
\bauthor{\bsnm{{Brady}}, \binits{D.}},
\bauthor{\bsnm{{Gu}}, \binits{X.-G.}},
\bauthor{\bsnm{{Lin}}, \binits{S.}}:
\byear{1990},
\batitle{{Holography in artificial neural networks}}.
\bjtitle{\nat}
\bvolume{343},
\bfpage{325}.
\doiurl{https://doi.org/10.1038/343325a0}.
\adsurl{1990Natur.343..325P}.
\end{barticle}
\endbibitem

\bibitem[\protect\citeauthoryear{{Puschmann}
  et~al.}{2012}]{2012AN....333..880P}
\begin{barticle}
\bauthor{\bsnm{{Puschmann}}, \binits{K.G.}},
\bauthor{\bsnm{{Denker}}, \binits{C.}},
\bauthor{\bsnm{{Kneer}}, \binits{F.}},
\bauthor{\bsnm{{Al Erdogan}}, \binits{N.}},
\bauthor{\bsnm{{Balthasar}}, \binits{H.}},
\bauthor{\bsnm{{Bauer}}, \binits{S.M.}},
\bauthor{\bsnm{{Beck}}, \binits{C.}},
\bauthor{\bsnm{{Bello Gonz{\'a}lez}}, \binits{N.}},
\bauthor{\bsnm{{Collados}}, \binits{M.}},
\bauthor{\bsnm{{Hahn}}, \binits{T.}},
\bauthor{\bsnm{{Hirzberger}}, \binits{J.}},
\bauthor{\bsnm{{Hofmann}}, \binits{A.}},
\bauthor{\bsnm{{Louis}}, \binits{R.E.}},
\bauthor{\bsnm{{Nicklas}}, \binits{H.}},
\bauthor{\bsnm{{Okunev}}, \binits{O.}},
\bauthor{\bsnm{{Mart{\'\i}nez Pillet}}, \binits{V.}},
\bauthor{\bsnm{{Popow}}, \binits{E.}},
\bauthor{\bsnm{{Seelemann}}, \binits{T.}},
\bauthor{\bsnm{{Volkmer}}, \binits{R.}},
\bauthor{\bsnm{{Wittmann}}, \binits{A.D.}},
\bauthor{\bsnm{{Woche}}, \binits{M.}}:
\byear{2012},
\batitle{{The GREGOR Fabry-P{\'e}rot Interferometer}}.
\bjtitle{Astronomische Nachrichten}
\bvolume{333},
\bfpage{880}.
\doiurl{https://doi.org/10.1002/asna.201211734}.
\adsurl{2012AN....333..880P}.
\end{barticle}
\endbibitem

\bibitem[\protect\citeauthoryear{{Puschmann}
  et~al.}{2013}]{2013OptEn..52h1606P}
\begin{barticle}
\bauthor{\bsnm{{Puschmann}}, \binits{K.G.}},
\bauthor{\bsnm{{Denker}}, \binits{C.}},
\bauthor{\bsnm{{Balthasar}}, \binits{H.}},
\bauthor{\bsnm{{Louis}}, \binits{R.E.}},
\bauthor{\bsnm{{Popow}}, \binits{E.}},
\bauthor{\bsnm{{Woche}}, \binits{M.}},
\bauthor{\bsnm{{Beck}}, \binits{C.}},
\bauthor{\bsnm{{Seelemann}}, \binits{T.}},
\bauthor{\bsnm{{Volkmer}}, \binits{R.}}:
\byear{2013},
\batitle{{GREGOR Fabry-P{\'e}rot interferometer and its companion the blue
  imaging solar spectrometer}}.
\bjtitle{Optical Engineering}
\bvolume{52},
\bfpage{081606}.
\doiurl{https://doi.org/10.1117/1.OE.52.8.081606}.
\adsurl{2013OptEn..52h1606P}.
\end{barticle}
\endbibitem

\bibitem[\protect\citeauthoryear{{Qiu} et~al.}{2022}]{2022SCPMA..6589603Q}
\begin{barticle}
\bauthor{\bsnm{{Qiu}}, \binits{Y.}},
\bauthor{\bsnm{{Rao}}, \binits{S.}},
\bauthor{\bsnm{{Li}}, \binits{C.}},
\bauthor{\bsnm{{Fang}}, \binits{C.}},
\bauthor{\bsnm{{Ding}}, \binits{M.}},
\bauthor{\bsnm{{Li}}, \binits{Z.}},
\bauthor{\bsnm{{Ni}}, \binits{Y.}},
\bauthor{\bsnm{{Wang}}, \binits{W.}},
\bauthor{\bsnm{{Hong}}, \binits{J.}},
\bauthor{\bsnm{{Hao}}, \binits{Q.}},
\bauthor{\bsnm{{Dai}}, \binits{Y.}},
\bauthor{\bsnm{{Chen}}, \binits{P.}},
\bauthor{\bsnm{{Wan}}, \binits{X.}},
\bauthor{\bsnm{{Xu}}, \binits{Z.}},
\bauthor{\bsnm{{You}}, \binits{W.}},
\bauthor{\bsnm{{Yuan}}, \binits{Y.}},
\bauthor{\bsnm{{Tao}}, \binits{H.}},
\bauthor{\bsnm{{Li}}, \binits{X.}},
\bauthor{\bsnm{{He}}, \binits{Y.}},
\bauthor{\bsnm{{Liu}}, \binits{Q.}}:
\byear{2022},
\batitle{{Calibration procedures for the CHASE/HIS science data}}.
\bjtitle{Science China Physics, Mechanics, and Astronomy}
\bvolume{65},
\bfpage{289603}.
\doiurl{https://doi.org/10.1007/s11433-022-1900-5}.
\adsurl{2022SCPMA..6589603Q}.
\end{barticle}
\endbibitem

\bibitem[\protect\citeauthoryear{{Quintero Noda}
  et~al.}{2022}]{2022A&A...666A..21Q}
\begin{barticle}
\bauthor{\bsnm{{Quintero Noda}}, \binits{C.}},
\bauthor{\bsnm{{Schlichenmaier}}, \binits{R.}},
\bauthor{\bsnm{{Bellot Rubio}}, \binits{L.R.}},
\bauthor{\bsnm{{L{\"o}fdahl}}, \binits{M.G.}},
\bauthor{\bsnm{{Khomenko}}, \binits{E.}},
\bauthor{\bsnm{{Jur{\v{c}}{\'a}k}}, \binits{J.}},
\bauthor{\bsnm{{Leenaarts}}, \binits{J.}},
\bauthor{\bsnm{{Kuckein}}, \binits{C.}},
\bauthor{\bsnm{{Gonz{\'a}lez Manrique}}, \binits{S.J.}},
\bauthor{\bsnm{{Gun{\'a}r}}, \binits{S.}},
\bauthor{\bsnm{{Nelson}}, \binits{C.J.}},
\bauthor{\bsnm{{de la Cruz Rodr{\'\i}guez}}, \binits{J.}},
\bauthor{\bsnm{{Tziotziou}}, \binits{K.}},
\bauthor{\bsnm{{Tsiropoula}}, \binits{G.}},
\bauthor{\bsnm{{Aulanier}}, \binits{G.}},
\bauthor{\bsnm{{Aboudarham}}, \binits{J.}},
\bauthor{\bsnm{{Allegri}}, \binits{D.}},
\bauthor{\bsnm{{Alsina Ballester}}, \binits{E.}},
\bauthor{\bsnm{{Amans}}, \binits{J.P.}},
\bauthor{\bsnm{{Asensio Ramos}}, \binits{A.}},
\bauthor{\bsnm{{Bail{\'e}n}}, \binits{F.J.}},
\bauthor{\bsnm{{Balaguer}}, \binits{M.}},
\bauthor{\bsnm{{Baldini}}, \binits{V.}},
\bauthor{\bsnm{{Balthasar}}, \binits{H.}},
\bauthor{\bsnm{{Barata}}, \binits{T.}},
\bauthor{\bsnm{{Barczynski}}, \binits{K.}},
\bauthor{\bsnm{{Barreto Cabrera}}, \binits{M.}},
\bauthor{\bsnm{{Baur}}, \binits{A.}},
\bauthor{\bsnm{{B{\'e}chet}}, \binits{C.}},
\bauthor{\bsnm{{Beck}}, \binits{C.}},
\bauthor{\bsnm{{Bel{\'\i}o-As{\'\i}n}}, \binits{M.}},
\bauthor{\bsnm{{Bello-Gonz{\'a}lez}}, \binits{N.}},
\bauthor{\bsnm{{Belluzzi}}, \binits{L.}},
\bauthor{\bsnm{{Bentley}}, \binits{R.D.}},
\bauthor{\bsnm{{Berdyugina}}, \binits{S.V.}},
\bauthor{\bsnm{{Berghmans}}, \binits{D.}},
\bauthor{\bsnm{{Berlicki}}, \binits{A.}},
\bauthor{\bsnm{{Berrilli}}, \binits{F.}},
\bauthor{\bsnm{{Berkefeld}}, \binits{T.}},
\bauthor{\bsnm{{Bettonvil}}, \binits{F.}},
\bauthor{\bsnm{{Bianda}}, \binits{M.}},
\bauthor{\bsnm{{Bienes P{\'e}rez}}, \binits{J.}},
\bauthor{\bsnm{{Bonaque-Gonz{\'a}lez}}, \binits{S.}},
\bauthor{\bsnm{{Braj{\v{s}}a}}, \binits{R.}},
\bauthor{\bsnm{{Bommier}}, \binits{V.}},
\bauthor{\bsnm{{Bourdin}}, \binits{P.-A.}},
\bauthor{\bsnm{{Burgos Mart{\'\i}n}}, \binits{J.}},
\bauthor{\bsnm{{Calchetti}}, \binits{D.}},
\bauthor{\bsnm{{Calcines}}, \binits{A.}},
\bauthor{\bsnm{{Calvo Tovar}}, \binits{J.}},
\bauthor{\bsnm{{Campbell}}, \binits{R.J.}},
\bauthor{\bsnm{{Carballo-Mart{\'\i}n}}, \binits{Y.}},
\bauthor{\bsnm{{Carbone}}, \binits{V.}},
\bauthor{\bsnm{{Carlin}}, \binits{E.S.}},
\bauthor{\bsnm{{Carlsson}}, \binits{M.}},
\bauthor{\bsnm{{Castro L{\'o}pez}}, \binits{J.}},
\bauthor{\bsnm{{Cavaller}}, \binits{L.}},
\bauthor{\bsnm{{Cavallini}}, \binits{F.}},
\bauthor{\bsnm{{Cauzzi}}, \binits{G.}},
\bauthor{\bsnm{{Cecconi}}, \binits{M.}},
\bauthor{\bsnm{{Chulani}}, \binits{H.M.}},
\bauthor{\bsnm{{Cirami}}, \binits{R.}},
\bauthor{\bsnm{{Consolini}}, \binits{G.}},
\bauthor{\bsnm{{Coretti}}, \binits{I.}},
\bauthor{\bsnm{{Cosentino}}, \binits{R.}},
\bauthor{\bsnm{{C{\'o}zar-Castellano}}, \binits{J.}},
\bauthor{\bsnm{{Dalmasse}}, \binits{K.}},
\bauthor{\bsnm{{Danilovic}}, \binits{S.}},
\bauthor{\bsnm{{De Juan Ovelar}}, \binits{M.}},
\bauthor{\bsnm{{Del Moro}}, \binits{D.}},
\bauthor{\bsnm{{del Pino Alem{\'a}n}}, \binits{T.}},
\bauthor{\bsnm{{del Toro Iniesta}}, \binits{J.C.}},
\bauthor{\bsnm{{Denker}}, \binits{C.}},
\bauthor{\bsnm{{Dhara}}, \binits{S.K.}},
\bauthor{\bsnm{{Di Marcantonio}}, \binits{P.}},
\bauthor{\bsnm{{D{\'\i}az Baso}}, \binits{C.J.}},
\bauthor{\bsnm{{Diercke}}, \binits{A.}},
\bauthor{\bsnm{{Dineva}}, \binits{E.}},
\bauthor{\bsnm{{D{\'\i}az-Garc{\'\i}a}}, \binits{J.J.}},
\bauthor{\bsnm{{Doerr}}, \binits{H.-P.}},
\bauthor{\bsnm{{Doyle}}, \binits{G.}},
\bauthor{\bsnm{{Erdelyi}}, \binits{R.}},
\bauthor{\bsnm{{Ermolli}}, \binits{I.}},
\bauthor{\bsnm{{Escobar Rodr{\'\i}guez}}, \binits{A.}},
\bauthor{\bsnm{{Esteban Pozuelo}}, \binits{S.}},
\bauthor{\bsnm{{Faurobert}}, \binits{M.}},
\bauthor{\bsnm{{Felipe}}, \binits{T.}},
\bauthor{\bsnm{{Feller}}, \binits{A.}},
\bauthor{\bsnm{{Feijoo Amoedo}}, \binits{N.}},
\bauthor{\bsnm{{Femen{\'\i}a Castell{\'a}}}, \binits{B.}},
\bauthor{\bsnm{{Fernandes}}, \binits{J.}},
\bauthor{\bsnm{{Ferro Rodr{\'\i}guez}}, \binits{I.}},
\bauthor{\bsnm{{Figueroa}}, \binits{I.}},
\bauthor{\bsnm{{Fletcher}}, \binits{L.}},
\bauthor{\bsnm{{Franco Ordovas}}, \binits{A.}},
\bauthor{\bsnm{{Gafeira}}, \binits{R.}},
\bauthor{\bsnm{{Gardenghi}}, \binits{R.}},
\bauthor{\bsnm{{Gelly}}, \binits{B.}},
\bauthor{\bsnm{{Giorgi}}, \binits{F.}},
\bauthor{\bsnm{{Gisler}}, \binits{D.}},
\bauthor{\bsnm{{Giovannelli}}, \binits{L.}},
\bauthor{\bsnm{{Gonz{\'a}lez}}, \binits{F.}},
\bauthor{\bsnm{{Gonz{\'a}lez}}, \binits{J.B.}},
\bauthor{\bsnm{{Gonz{\'a}lez-Cava}}, \binits{J.M.}},
\bauthor{\bsnm{{Gonz{\'a}lez Garc{\'\i}a}}, \binits{M.}},
\bauthor{\bsnm{{G{\"o}m{\"o}ry}}, \binits{P.}},
\bauthor{\bsnm{{Gracia}}, \binits{F.}},
\bauthor{\bsnm{{Grauf}}, \binits{B.}},
\bauthor{\bsnm{{Greco}}, \binits{V.}},
\bauthor{\bsnm{{Grivel}}, \binits{C.}},
\bauthor{\bsnm{{Guerreiro}}, \binits{N.}},
\bauthor{\bsnm{{Guglielmino}}, \binits{S.L.}},
\bauthor{\bsnm{{Hammerschlag}}, \binits{R.}},
\bauthor{\bsnm{{Hanslmeier}}, \binits{A.}},
\bauthor{\bsnm{{Hansteen}}, \binits{V.}},
\bauthor{\bsnm{{Heinzel}}, \binits{P.}},
\bauthor{\bsnm{{Hern{\'a}ndez-Delgado}}, \binits{A.}},
\bauthor{\bsnm{{Hern{\'a}ndez Su{\'a}rez}}, \binits{E.}},
\bauthor{\bsnm{{Hidalgo}}, \binits{S.L.}},
\bauthor{\bsnm{{Hill}}, \binits{F.}},
\bauthor{\bsnm{{Hizberger}}, \binits{J.}},
\bauthor{\bsnm{{Hofmeister}}, \binits{S.}},
\bauthor{\bsnm{{J{\"a}gers}}, \binits{A.}},
\bauthor{\bsnm{{Janett}}, \binits{G.}},
\bauthor{\bsnm{{Jarolim}}, \binits{R.}},
\bauthor{\bsnm{{Jess}}, \binits{D.}},
\bauthor{\bsnm{{Jim{\'e}nez Mej{\'\i}as}}, \binits{D.}},
\bauthor{\bsnm{{Jolissaint}}, \binits{L.}},
\bauthor{\bsnm{{Kamlah}}, \binits{R.}},
\bauthor{\bsnm{{Kapit{\'a}n}}, \binits{J.}},
\bauthor{\bsnm{{Ka{\v{s}}parov{\'a}}}, \binits{J.}},
\bauthor{\bsnm{{Keller}}, \binits{C.U.}},
\bauthor{\bsnm{{Kentischer}}, \binits{T.}},
\bauthor{\bsnm{{Kiselman}}, \binits{D.}},
\bauthor{\bsnm{{Kleint}}, \binits{L.}},
\bauthor{\bsnm{{Klvana}}, \binits{M.}},
\bauthor{\bsnm{{Kontogiannis}}, \binits{I.}},
\bauthor{\bsnm{{Krishnappa}}, \binits{N.}},
\bauthor{\bsnm{{Ku{\v{c}}era}}, \binits{A.}},
\bauthor{\bsnm{{Labrosse}}, \binits{N.}},
\bauthor{\bsnm{{Lagg}}, \binits{A.}},
\bauthor{\bsnm{{Landi Degl'Innocenti}}, \binits{E.}},
\bauthor{\bsnm{{Langlois}}, \binits{M.}},
\bauthor{\bsnm{{Lafon}}, \binits{M.}},
\bauthor{\bsnm{{Laforgue}}, \binits{D.}},
\bauthor{\bsnm{{Le Men}}, \binits{C.}},
\bauthor{\bsnm{{Lepori}}, \binits{B.}},
\bauthor{\bsnm{{Lepreti}}, \binits{F.}},
\bauthor{\bsnm{{Lindberg}}, \binits{B.}},
\bauthor{\bsnm{{Lilje}}, \binits{P.B.}},
\bauthor{\bsnm{{L{\'o}pez Ariste}}, \binits{A.}},
\bauthor{\bsnm{{L{\'o}pez Fern{\'a}ndez}}, \binits{V.A.}},
\bauthor{\bsnm{{L{\'o}pez Jim{\'e}nez}}, \binits{A.C.}},
\bauthor{\bsnm{{L{\'o}pez L{\'o}pez}}, \binits{R.}},
\bauthor{\bsnm{{Manso Sainz}}, \binits{R.}},
\bauthor{\bsnm{{Marassi}}, \binits{A.}},
\bauthor{\bsnm{{Marco de la Rosa}}, \binits{J.}},
\bauthor{\bsnm{{Marino}}, \binits{J.}},
\bauthor{\bsnm{{Marrero}}, \binits{J.}},
\bauthor{\bsnm{{Mart{\'\i}n}}, \binits{A.}},
\bauthor{\bsnm{{Mart{\'\i}n G{\'a}lvez}}, \binits{A.}},
\bauthor{\bsnm{{Mart{\'\i}n Hernando}}, \binits{Y.}},
\bauthor{\bsnm{{Masciadri}}, \binits{E.}},
\bauthor{\bsnm{{Mart{\'\i}nez Gonz{\'a}lez}}, \binits{M.}},
\bauthor{\bsnm{{Matta-G{\'o}mez}}, \binits{A.}},
\bauthor{\bsnm{{Mato}}, \binits{A.}},
\bauthor{\bsnm{{Mathioudakis}}, \binits{M.}},
\bauthor{\bsnm{{Matthews}}, \binits{S.}},
\bauthor{\bsnm{{Mein}}, \binits{P.}},
\bauthor{\bsnm{{Merlos Garc{\'\i}a}}, \binits{F.}},
\bauthor{\bsnm{{Moity}}, \binits{J.}},
\bauthor{\bsnm{{Montilla}}, \binits{I.}},
\bauthor{\bsnm{{Molinaro}}, \binits{M.}},
\bauthor{\bsnm{{Molodij}}, \binits{G.}},
\bauthor{\bsnm{{Montoya}}, \binits{L.M.}},
\bauthor{\bsnm{{Munari}}, \binits{M.}},
\bauthor{\bsnm{{Murabito}}, \binits{M.}},
\bauthor{\bsnm{{N{\'u}{\~n}ez Cagigal}}, \binits{M.}},
\bauthor{\bsnm{{Oliviero}}, \binits{M.}},
\bauthor{\bsnm{{Orozco Su{\'a}rez}}, \binits{D.}},
\bauthor{\bsnm{{Ortiz}}, \binits{A.}},
\bauthor{\bsnm{{Padilla-Hern{\'a}ndez}}, \binits{C.}},
\bauthor{\bsnm{{Pa{\'e}z Ma{\~n}{\'a}}}, \binits{E.}},
\bauthor{\bsnm{{Paletou}}, \binits{F.}},
\bauthor{\bsnm{{Pancorbo}}, \binits{J.}},
\bauthor{\bsnm{{Pastor Ca{\~n}edo}}, \binits{A.}},
\bauthor{\bsnm{{Pastor Yabar}}, \binits{A.}},
\bauthor{\bsnm{{Peat}}, \binits{A.W.}},
\bauthor{\bsnm{{Pedichini}}, \binits{F.}},
\bauthor{\bsnm{{Peixinho}}, \binits{N.}},
\bauthor{\bsnm{{Pe{\~n}ate}}, \binits{J.}},
\bauthor{\bsnm{{P{\'e}rez de Taoro}}, \binits{A.}},
\bauthor{\bsnm{{Peter}}, \binits{H.}},
\bauthor{\bsnm{{Petrovay}}, \binits{K.}},
\bauthor{\bsnm{{Piazzesi}}, \binits{R.}},
\bauthor{\bsnm{{Pietropaolo}}, \binits{E.}},
\bauthor{\bsnm{{Pleier}}, \binits{O.}},
\bauthor{\bsnm{{Poedts}}, \binits{S.}},
\bauthor{\bsnm{{P{\"o}tzi}}, \binits{W.}},
\bauthor{\bsnm{{Podladchikova}}, \binits{T.}},
\bauthor{\bsnm{{Prieto}}, \binits{G.}},
\bauthor{\bsnm{{Quintero Nehrkorn}}, \binits{J.}},
\bauthor{\bsnm{{Ramelli}}, \binits{R.}},
\bauthor{\bsnm{{Ramos Sapena}}, \binits{Y.}},
\bauthor{\bsnm{{Rasilla}}, \binits{J.L.}},
\bauthor{\bsnm{{Reardon}}, \binits{K.}},
\bauthor{\bsnm{{Rebolo}}, \binits{R.}},
\bauthor{\bsnm{{Regalado Olivares}}, \binits{S.}},
\bauthor{\bsnm{{Reyes Garc{\'\i}a-Talavera}}, \binits{M.}},
\bauthor{\bsnm{{Riethm{\"u}ller}}, \binits{T.L.}},
\bauthor{\bsnm{{Rimmele}}, \binits{T.}},
\bauthor{\bsnm{{Rodr{\'\i}guez Delgado}}, \binits{H.}},
\bauthor{\bsnm{{Rodr{\'\i}guez Gonz{\'a}lez}}, \binits{N.}},
\bauthor{\bsnm{{Rodr{\'\i}guez-Losada}}, \binits{J.A.}},
\bauthor{\bsnm{{Rodr{\'\i}guez Ramos}}, \binits{L.F.}},
\bauthor{\bsnm{{Romano}}, \binits{P.}},
\bauthor{\bsnm{{Roth}}, \binits{M.}},
\bauthor{\bsnm{{Rouppe van der Voort}}, \binits{L.}},
\bauthor{\bsnm{{Rudawy}}, \binits{P.}},
\bauthor{\bsnm{{Ruiz de Galarreta}}, \binits{C.}},
\bauthor{\bsnm{{Ryb{\'a}k}}, \binits{J.}},
\bauthor{\bsnm{{Salvade}}, \binits{A.}},
\bauthor{\bsnm{{S{\'a}nchez-Capuchino}}, \binits{J.}},
\bauthor{\bsnm{{S{\'a}nchez Rodr{\'\i}guez}}, \binits{M.L.}},
\bauthor{\bsnm{{Sangiorgi}}, \binits{M.}},
\bauthor{\bsnm{{Say{\`e}de}}, \binits{F.}},
\bauthor{\bsnm{{Scharmer}}, \binits{G.}},
\bauthor{\bsnm{{Scheiffelen}}, \binits{T.}},
\bauthor{\bsnm{{Schmidt}}, \binits{W.}},
\bauthor{\bsnm{{Schmieder}}, \binits{B.}},
\bauthor{\bsnm{{Scir{\`e}}}, \binits{C.}},
\bauthor{\bsnm{{Scuderi}}, \binits{S.}},
\bauthor{\bsnm{{Siegel}}, \binits{B.}},
\bauthor{\bsnm{{Sigwarth}}, \binits{M.}},
\bauthor{\bsnm{{Sim{\~o}es}}, \binits{P.J.A.}},
\bauthor{\bsnm{{Snik}}, \binits{F.}},
\bauthor{\bsnm{{Sliepen}}, \binits{G.}},
\bauthor{\bsnm{{Sobotka}}, \binits{M.}},
\bauthor{\bsnm{{Socas-Navarro}}, \binits{H.}},
\bauthor{\bsnm{{Sola La Serna}}, \binits{P.}},
\bauthor{\bsnm{{Solanki}}, \binits{S.K.}},
\bauthor{\bsnm{{Soler Trujillo}}, \binits{M.}},
\bauthor{\bsnm{{Soltau}}, \binits{D.}},
\bauthor{\bsnm{{Sordini}}, \binits{A.}},
\bauthor{\bsnm{{Sosa M{\'e}ndez}}, \binits{A.}},
\bauthor{\bsnm{{Stangalini}}, \binits{M.}},
\bauthor{\bsnm{{Steiner}}, \binits{O.}},
\bauthor{\bsnm{{Stenflo}}, \binits{J.O.}},
\bauthor{\bsnm{{{\v{S}}t{\v{e}}p{\'a}n}}, \binits{J.}},
\bauthor{\bsnm{{Strassmeier}}, \binits{K.G.}},
\bauthor{\bsnm{{Sudar}}, \binits{D.}},
\bauthor{\bsnm{{Suematsu}}, \binits{Y.}},
\bauthor{\bsnm{{S{\"u}tterlin}}, \binits{P.}},
\bauthor{\bsnm{{Tallon}}, \binits{M.}},
\bauthor{\bsnm{{Temmer}}, \binits{M.}},
\bauthor{\bsnm{{Tenegi}}, \binits{F.}},
\bauthor{\bsnm{{Tritschler}}, \binits{A.}},
\bauthor{\bsnm{{Trujillo Bueno}}, \binits{J.}},
\bauthor{\bsnm{{Turchi}}, \binits{A.}},
\bauthor{\bsnm{{Utz}}, \binits{D.}},
\bauthor{\bsnm{{van Harten}}, \binits{G.}},
\bauthor{\bsnm{{van Noort}}, \binits{M.}},
\bauthor{\bsnm{{van Werkhoven}}, \binits{T.}},
\bauthor{\bsnm{{Vansintjan}}, \binits{R.}},
\bauthor{\bsnm{{Vaz Cedillo}}, \binits{J.J.}},
\bauthor{\bsnm{{Vega Reyes}}, \binits{N.}},
\bauthor{\bsnm{{Verma}}, \binits{M.}},
\bauthor{\bsnm{{Veronig}}, \binits{A.M.}},
\bauthor{\bsnm{{Viavattene}}, \binits{G.}},
\bauthor{\bsnm{{Vitas}}, \binits{N.}},
\bauthor{\bsnm{{V{\"o}gler}}, \binits{A.}},
\bauthor{\bsnm{{von der L{\"u}he}}, \binits{O.}},
\bauthor{\bsnm{{Volkmer}}, \binits{R.}},
\bauthor{\bsnm{{Waldmann}}, \binits{T.A.}},
\bauthor{\bsnm{{Walton}}, \binits{D.}},
\bauthor{\bsnm{{Wisniewska}}, \binits{A.}},
\bauthor{\bsnm{{Zeman}}, \binits{J.}},
\bauthor{\bsnm{{Zeuner}}, \binits{F.}},
\bauthor{\bsnm{{Zhang}}, \binits{L.Q.}},
\bauthor{\bsnm{{Zuccarello}}, \binits{F.}},
\bauthor{\bsnm{{Collados}}, \binits{M.}}:
\byear{2022},
\batitle{{The European Solar Telescope}}.
\bjtitle{\aap}
\bvolume{666},
\bfpage{A21}.
\doiurl{https://doi.org/10.1051/0004-6361/202243867}.
\adsurl{2022A&A...666A..21Q}.
\end{barticle}
\endbibitem

\bibitem[\protect\citeauthoryear{{Ramesh} et~al.}{2016}]{2016ExA....42..271R}
\begin{barticle}
\bauthor{\bsnm{{Ramesh}}, \binits{K.B.}},
\bauthor{\bsnm{{Vasantharaju}}, \binits{N.}},
\bauthor{\bsnm{{Pruthvi}}, \binits{H.}},
\bauthor{\bsnm{{Reardon}}, \binits{K.}}:
\byear{2016},
\batitle{{Solar dynamics imaging system a back-end instrument for the proposed
  NLST}}.
\bjtitle{Experimental Astronomy}
\bvolume{42},
\bfpage{271}.
\doiurl{https://doi.org/10.1007/s10686-016-9509-y}.
\adsurl{2016ExA....42..271R}.
\end{barticle}
\endbibitem

\bibitem[\protect\citeauthoryear{{Ranganathan}
  et~al.}{2018}]{2018ApJ...867...77R}
\begin{barticle}
\bauthor{\bsnm{{Ranganathan}}, \binits{M.}},
\bauthor{\bsnm{{Sankarasubramanian}}, \binits{K.}},
\bauthor{\bsnm{{Bayanna}}, \binits{A.R.}},
\bauthor{\bsnm{{Mathew}}, \binits{S.K.}}:
\byear{2018},
\batitle{{Lenslet Array and Fabry-P{\'e}rot Based Hyperspectral Imaging}}.
\bjtitle{\apj}
\bvolume{867},
\bfpage{77}.
\doiurl{https://doi.org/10.3847/1538-4357/aae30d}.
\adsurl{2018ApJ...867...77R}.
\end{barticle}
\endbibitem

\bibitem[\protect\citeauthoryear{{Rao} et~al.}{2015}]{2015JATIS...1b4001R}
\begin{barticle}
\bauthor{\bsnm{{Rao}}, \binits{C.}},
\bauthor{\bsnm{{Gu}}, \binits{N.}},
\bauthor{\bsnm{{Zhu}}, \binits{L.}},
\bauthor{\bsnm{{Huang}}, \binits{J.}},
\bauthor{\bsnm{{Li}}, \binits{C.}},
\bauthor{\bsnm{{Cheng}}, \binits{Y.}},
\bauthor{\bsnm{{Liu}}, \binits{Y.}},
\bauthor{\bsnm{{Cao}}, \binits{X.}},
\bauthor{\bsnm{{Zhang}}, \binits{M.}},
\bauthor{\bsnm{{Zhang}}, \binits{L.}},
\bauthor{\bsnm{{Liu}}, \binits{H.}},
\bauthor{\bsnm{{Wan}}, \binits{Y.}},
\bauthor{\bsnm{{Xian}}, \binits{H.}},
\bauthor{\bsnm{{Ma}}, \binits{W.}},
\bauthor{\bsnm{{Bao}}, \binits{H.}},
\bauthor{\bsnm{{Zhang}}, \binits{X.}},
\bauthor{\bsnm{{Guan}}, \binits{C.}},
\bauthor{\bsnm{{Chen}}, \binits{D.}},
\bauthor{\bsnm{{Li}}, \binits{M.}}:
\byear{2015},
\batitle{{1.8-m solar telescope in China: Chinese Large Solar Telescope}}.
\bjtitle{Journal of Astronomical Telescopes, Instruments, and Systems}
\bvolume{1},
\bfpage{024001}.
\doiurl{https://doi.org/10.1117/1.JATIS.1.2.024001}.
\adsurl{2015JATIS...1b4001R}.
\end{barticle}
\endbibitem

\bibitem[\protect\citeauthoryear{{Rao} et~al.}{2020}]{2020SCPMA..6309631R}
\begin{barticle}
\bauthor{\bsnm{{Rao}}, \binits{C.}},
\bauthor{\bsnm{{Gu}}, \binits{N.}},
\bauthor{\bsnm{{Rao}}, \binits{X.}},
\bauthor{\bsnm{{Li}}, \binits{C.}},
\bauthor{\bsnm{{Zhang}}, \binits{L.}},
\bauthor{\bsnm{{Huang}}, \binits{J.}},
\bauthor{\bsnm{{Kong}}, \binits{L.}},
\bauthor{\bsnm{{Zhang}}, \binits{M.}},
\bauthor{\bsnm{{Cheng}}, \binits{Y.}},
\bauthor{\bsnm{{Pu}}, \binits{Y.}},
\bauthor{\bsnm{{Bao}}, \binits{H.}},
\bauthor{\bsnm{{Guo}}, \binits{Y.}},
\bauthor{\bsnm{{Liu}}, \binits{Y.}},
\bauthor{\bsnm{{Yang}}, \binits{J.}},
\bauthor{\bsnm{{Zhong}}, \binits{L.}},
\bauthor{\bsnm{{Wang}}, \binits{C.}},
\bauthor{\bsnm{{Fang}}, \binits{K.}},
\bauthor{\bsnm{{Zhang}}, \binits{X.}},
\bauthor{\bsnm{{Chen}}, \binits{D.}},
\bauthor{\bsnm{{Wang}}, \binits{C.}},
\bauthor{\bsnm{{Fan}}, \binits{X.}},
\bauthor{\bsnm{{Yan}}, \binits{Z.}},
\bauthor{\bsnm{{Chen}}, \binits{K.}},
\bauthor{\bsnm{{Wei}}, \binits{X.}},
\bauthor{\bsnm{{Zhu}}, \binits{L.}},
\bauthor{\bsnm{{Liu}}, \binits{H.}},
\bauthor{\bsnm{{Wan}}, \binits{Y.}},
\bauthor{\bsnm{{Xian}}, \binits{H.}},
\bauthor{\bsnm{{Ma}}, \binits{W.}}:
\byear{2020},
\batitle{{First light of the 1.8-m solar telescope-CLST}}.
\bjtitle{Science China Physics, Mechanics, and Astronomy}
\bvolume{63},
\bfpage{109631}.
\doiurl{https://doi.org/10.1007/s11433-019-1557-3}.
\adsurl{2020SCPMA..6309631R}.
\end{barticle}
\endbibitem

\bibitem[\protect\citeauthoryear{{Rast} et~al.}{2021}]{2021SoPh..296...70R}
\begin{barticle}
\bauthor{\bsnm{{Rast}}, \binits{M.P.}},
\bauthor{\bsnm{{Bello Gonz{\'a}lez}}, \binits{N.}},
\bauthor{\bsnm{{Bellot Rubio}}, \binits{L.}},
\bauthor{\bsnm{{Cao}}, \binits{W.}},
\bauthor{\bsnm{{Cauzzi}}, \binits{G.}},
\bauthor{\bsnm{{Deluca}}, \binits{E.}},
\bauthor{\bsnm{{de Pontieu}}, \binits{B.}},
\bauthor{\bsnm{{Fletcher}}, \binits{L.}},
\bauthor{\bsnm{{Gibson}}, \binits{S.E.}},
\bauthor{\bsnm{{Judge}}, \binits{P.G.}},
\bauthor{\bsnm{{Katsukawa}}, \binits{Y.}},
\bauthor{\bsnm{{Kazachenko}}, \binits{M.D.}},
\bauthor{\bsnm{{Khomenko}}, \binits{E.}},
\bauthor{\bsnm{{Landi}}, \binits{E.}},
\bauthor{\bsnm{{Mart{\'\i}nez Pillet}}, \binits{V.}},
\bauthor{\bsnm{{Petrie}}, \binits{G.J.D.}},
\bauthor{\bsnm{{Qiu}}, \binits{J.}},
\bauthor{\bsnm{{Rachmeler}}, \binits{L.A.}},
\bauthor{\bsnm{{Rempel}}, \binits{M.}},
\bauthor{\bsnm{{Schmidt}}, \binits{W.}},
\bauthor{\bsnm{{Scullion}}, \binits{E.}},
\bauthor{\bsnm{{Sun}}, \binits{X.}},
\bauthor{\bsnm{{Welsch}}, \binits{B.T.}},
\bauthor{\bsnm{{Andretta}}, \binits{V.}},
\bauthor{\bsnm{{Antolin}}, \binits{P.}},
\bauthor{\bsnm{{Ayres}}, \binits{T.R.}},
\bauthor{\bsnm{{Balasubramaniam}}, \binits{K.S.}},
\bauthor{\bsnm{{Ballai}}, \binits{I.}},
\bauthor{\bsnm{{Berger}}, \binits{T.E.}},
\bauthor{\bsnm{{Bradshaw}}, \binits{S.J.}},
\bauthor{\bsnm{{Campbell}}, \binits{R.J.}},
\bauthor{\bsnm{{Carlsson}}, \binits{M.}},
\bauthor{\bsnm{{Casini}}, \binits{R.}},
\bauthor{\bsnm{{Centeno}}, \binits{R.}},
\bauthor{\bsnm{{Cranmer}}, \binits{S.R.}},
\bauthor{\bsnm{{Criscuoli}}, \binits{S.}},
\bauthor{\bsnm{{Deforest}}, \binits{C.}},
\bauthor{\bsnm{{Deng}}, \binits{Y.}},
\bauthor{\bsnm{{Erd{\'e}lyi}}, \binits{R.}},
\bauthor{\bsnm{{Fedun}}, \binits{V.}},
\bauthor{\bsnm{{Fischer}}, \binits{C.E.}},
\bauthor{\bsnm{{Gonz{\'a}lez Manrique}}, \binits{S.J.}},
\bauthor{\bsnm{{Hahn}}, \binits{M.}},
\bauthor{\bsnm{{Harra}}, \binits{L.}},
\bauthor{\bsnm{{Henriques}}, \binits{V.M.J.}},
\bauthor{\bsnm{{Hurlburt}}, \binits{N.E.}},
\bauthor{\bsnm{{Jaeggli}}, \binits{S.}},
\bauthor{\bsnm{{Jafarzadeh}}, \binits{S.}},
\bauthor{\bsnm{{Jain}}, \binits{R.}},
\bauthor{\bsnm{{Jefferies}}, \binits{S.M.}},
\bauthor{\bsnm{{Keys}}, \binits{P.H.}},
\bauthor{\bsnm{{Kowalski}}, \binits{A.F.}},
\bauthor{\bsnm{{Kuckein}}, \binits{C.}},
\bauthor{\bsnm{{Kuhn}}, \binits{J.R.}},
\bauthor{\bsnm{{Kuridze}}, \binits{D.}},
\bauthor{\bsnm{{Liu}}, \binits{J.}},
\bauthor{\bsnm{{Liu}}, \binits{W.}},
\bauthor{\bsnm{{Longcope}}, \binits{D.}},
\bauthor{\bsnm{{Mathioudakis}}, \binits{M.}},
\bauthor{\bsnm{{McAteer}}, \binits{R.T.J.}},
\bauthor{\bsnm{{McIntosh}}, \binits{S.W.}},
\bauthor{\bsnm{{McKenzie}}, \binits{D.E.}},
\bauthor{\bsnm{{Miralles}}, \binits{M.P.}},
\bauthor{\bsnm{{Morton}}, \binits{R.J.}},
\bauthor{\bsnm{{Muglach}}, \binits{K.}},
\bauthor{\bsnm{{Nelson}}, \binits{C.J.}},
\bauthor{\bsnm{{Panesar}}, \binits{N.K.}},
\bauthor{\bsnm{{Parenti}}, \binits{S.}},
\bauthor{\bsnm{{Parnell}}, \binits{C.E.}},
\bauthor{\bsnm{{Poduval}}, \binits{B.}},
\bauthor{\bsnm{{Reardon}}, \binits{K.P.}},
\bauthor{\bsnm{{Reep}}, \binits{J.W.}},
\bauthor{\bsnm{{Schad}}, \binits{T.A.}},
\bauthor{\bsnm{{Schmit}}, \binits{D.}},
\bauthor{\bsnm{{Sharma}}, \binits{R.}},
\bauthor{\bsnm{{Socas-Navarro}}, \binits{H.}},
\bauthor{\bsnm{{Srivastava}}, \binits{A.K.}},
\bauthor{\bsnm{{Sterling}}, \binits{A.C.}},
\bauthor{\bsnm{{Suematsu}}, \binits{Y.}},
\bauthor{\bsnm{{Tarr}}, \binits{L.A.}},
\bauthor{\bsnm{{Tiwari}}, \binits{S.}},
\bauthor{\bsnm{{Tritschler}}, \binits{A.}},
\bauthor{\bsnm{{Verth}}, \binits{G.}},
\bauthor{\bsnm{{Vourlidas}}, \binits{A.}},
\bauthor{\bsnm{{Wang}}, \binits{H.}},
\bauthor{\bsnm{{Wang}}, \binits{Y.-M.}},
\bauthor{\bsnm{{NSO and DKIST Project}}},
\bauthor{\bsnm{{DKIST Instrument Scientists}}},
\bauthor{\bsnm{{DKIST Science Working Group}}},
\bauthor{\bsnm{{DKIST Critical Science Plan Community}}}:
\byear{2021},
\batitle{{Critical Science Plan for the Daniel K. Inouye Solar Telescope
  (DKIST)}}.
\bjtitle{\solphys}
\bvolume{296},
\bfpage{70}.
\doiurl{https://doi.org/10.1007/s11207-021-01789-2}.
\adsurl{2021SoPh..296...70R}.
\end{barticle}
\endbibitem

\bibitem[\protect\citeauthoryear{{Rayleigh}}{1879}]{doi:10.1080/14786447908639684}
\begin{barticle}
\bauthor{\bsnm{{Rayleigh}}, \binits{L.}}:
\byear{1879},
\batitle{XXXI. Investigations in optics, with special reference to the
  spectroscope}.
\bjtitle{The London, Edinburgh, and Dublin Philosophical Magazine and Journal
  of Science}
\bvolume{8},
\bfpage{261}.
\doiurl{https://doi.org/10.1080/14786447908639684}.
\burl{https://doi.org/10.1080/14786447908639684}.
\end{barticle}
\endbibitem

\bibitem[\protect\citeauthoryear{{Reardon} and
  {Cavallini}}{2008}]{2008A&A...481..897R}
\begin{barticle}
\bauthor{\bsnm{{Reardon}}, \binits{K.P.}},
\bauthor{\bsnm{{Cavallini}}, \binits{F.}}:
\byear{2008},
\batitle{{Characterization of Fabry-Perot interferometers and multi-etalon
  transmission profiles. The IBIS instrumental profile}}.
\bjtitle{\aap}
\bvolume{481},
\bfpage{897}.
\doiurl{https://doi.org/10.1051/0004-6361:20078473}.
\adsurl{2008A&A...481..897R}.
\end{barticle}
\endbibitem

\bibitem[\protect\citeauthoryear{{Regalado Olivares}
  et~al.}{2022}]{2022SPIE12188E..57R}
\begin{bchapter}
\bauthor{\bsnm{{Regalado Olivares}}, \binits{S.}},
\bauthor{\bsnm{{L{\'o}pez L{\'o}pez}}, \binits{R.}},
\bauthor{\bsnm{{Collados}}, \binits{M.}},
\bauthor{\bsnm{{Quintero Noda}}, \binits{C.}},
\bauthor{\bsnm{{Mart{\'\i}nez Gonz{\'a}lez}}, \binits{M.J.}},
\bauthor{\bsnm{{Esteban Pozuelo}}, \binits{S.}},
\bauthor{\bsnm{{Gelly}}, \binits{B.}}:
\byear{2022},
\bctitle{{Optical design of an image slicer-based integral field unit for the
  THEMIS solar telescope}}.
In: \bbtitle{Society of Photo-Optical Instrumentation Engineers (SPIE)
  Conference Series},
\bsertitle{Society of Photo-Optical Instrumentation Engineers (SPIE) Conference
  Series}
\bseriesno{12188},
\bfpage{1218857}.
\doiurl{https://doi.org/10.1117/12.2629265}.
\adsurl{2022SPIE12188E..57R}.
\end{bchapter}
\endbibitem

\bibitem[\protect\citeauthoryear{Ren and Wang}{2020}]{10.1093/pasj/psaa006}
\begin{barticle}
\bauthor{\bsnm{Ren}, \binits{D.}},
\bauthor{\bsnm{Wang}, \binits{G.}}:
\byear{2020},
\batitle{{A low-cost and duplicable portable solar adaptive optics system based
  on LabVIEW hybrid programming}}.
\bjtitle{Publications of the Astronomical Society of Japan}
\bvolume{72},
\bfpage{30}.
\doiurl{https://doi.org/10.1093/pasj/psaa006}.
\burl{https://doi.org/10.1093/pasj/psaa006}.
\end{barticle}
\endbibitem

\bibitem[\protect\citeauthoryear{Resisi, Popoff, and
  Bromberg}{2021}]{lpor.202000553}
\begin{barticle}
\bauthor{\bsnm{Resisi}, \binits{S.}},
\bauthor{\bsnm{Popoff}, \binits{S.M.}},
\bauthor{\bsnm{Bromberg}, \binits{Y.}}:
\byear{2021},
\batitle{Image Transmission Through a Dynamically Perturbed Multimode Fiber by
  Deep Learning}.
\bjtitle{Laser \& Photonics Reviews}
\bvolume{15},
\bfpage{2000553}.
\doiurl{https://doi.org/10.1002/lpor.202000553}.
\burl{https://onlinelibrary.wiley.com/doi/abs/10.1002/lpor.202000553}.
\end{barticle}
\endbibitem

\bibitem[\protect\citeauthoryear{{Rezaei}, {Beck}, and
  {Schmidt}}{2012}]{2012A&A...541A..60R}
\begin{barticle}
\bauthor{\bsnm{{Rezaei}}, \binits{R.}},
\bauthor{\bsnm{{Beck}}, \binits{C.}},
\bauthor{\bsnm{{Schmidt}}, \binits{W.}}:
\byear{2012},
\batitle{{Variation in sunspot properties between 1999 and 2011 as observed
  with the Tenerife Infrared Polarimeter}}.
\bjtitle{\aap}
\bvolume{541},
\bfpage{A60}.
\doiurl{https://doi.org/10.1051/0004-6361/201118635}.
\adsurl{2012A&A...541A..60R}.
\end{barticle}
\endbibitem

\bibitem[\protect\citeauthoryear{{Rimmele}}{2004}]{2004SPIE.5490...34R}
\begin{bchapter}
\bauthor{\bsnm{{Rimmele}}, \binits{T.R.}}:
\byear{2004},
\bctitle{{Recent advances in solar adaptive optics}}.
In: \beditor{\bsnm{{Bonaccini Calia}}, \binits{D.}},
\beditor{\bsnm{{Ellerbroek}}, \binits{B.L.}},
\beditor{\bsnm{{Ragazzoni}}, \binits{R.}} (eds.)
\bbtitle{Advancements in Adaptive Optics},
\bsertitle{Society of Photo-Optical Instrumentation Engineers (SPIE) Conference
  Series}
\bseriesno{5490},
\bfpage{34}.
\doiurl{https://doi.org/10.1117/12.551764}.
\adsurl{2004SPIE.5490...34R}.
\end{bchapter}
\endbibitem

\bibitem[\protect\citeauthoryear{{Rimmele} et~al.}{2020}]{2020SoPh..295..172R}
\begin{barticle}
\bauthor{\bsnm{{Rimmele}}, \binits{T.R.}},
\bauthor{\bsnm{{Warner}}, \binits{M.}},
\bauthor{\bsnm{{Keil}}, \binits{S.L.}},
\bauthor{\bsnm{{Goode}}, \binits{P.R.}},
\bauthor{\bsnm{{Kn{\"o}lker}}, \binits{M.}},
\bauthor{\bsnm{{Kuhn}}, \binits{J.R.}},
\bauthor{\bsnm{{Rosner}}, \binits{R.R.}},
\bauthor{\bsnm{{McMullin}}, \binits{J.P.}},
\bauthor{\bsnm{{Casini}}, \binits{R.}},
\bauthor{\bsnm{{Lin}}, \binits{H.}},
\bauthor{\bsnm{{W{\"o}ger}}, \binits{F.}},
\bauthor{\bsnm{{von der L{\"u}he}}, \binits{O.}},
\bauthor{\bsnm{{Tritschler}}, \binits{A.}},
\bauthor{\bsnm{{Davey}}, \binits{A.}},
\bauthor{\bsnm{{de Wijn}}, \binits{A.}},
\bauthor{\bsnm{{Elmore}}, \binits{D.F.}},
\bauthor{\bsnm{{Fehlmann}}, \binits{A.}},
\bauthor{\bsnm{{Harrington}}, \binits{D.M.}},
\bauthor{\bsnm{{Jaeggli}}, \binits{S.A.}},
\bauthor{\bsnm{{Rast}}, \binits{M.P.}},
\bauthor{\bsnm{{Schad}}, \binits{T.A.}},
\bauthor{\bsnm{{Schmidt}}, \binits{W.}},
\bauthor{\bsnm{{Mathioudakis}}, \binits{M.}},
\bauthor{\bsnm{{Mickey}}, \binits{D.L.}},
\bauthor{\bsnm{{Anan}}, \binits{T.}},
\bauthor{\bsnm{{Beck}}, \binits{C.}},
\bauthor{\bsnm{{Marshall}}, \binits{H.K.}},
\bauthor{\bsnm{{Jeffers}}, \binits{P.F.}},
\bauthor{\bsnm{{Oschmann}}, \binits{J.M.}},
\bauthor{\bsnm{{Beard}}, \binits{A.}},
\bauthor{\bsnm{{Berst}}, \binits{D.C.}},
\bauthor{\bsnm{{Cowan}}, \binits{B.A.}},
\bauthor{\bsnm{{Craig}}, \binits{S.C.}},
\bauthor{\bsnm{{Cross}}, \binits{E.}},
\bauthor{\bsnm{{Cummings}}, \binits{B.K.}},
\bauthor{\bsnm{{Donnelly}}, \binits{C.}},
\bauthor{\bsnm{{de Vanssay}}, \binits{J.-B.}},
\bauthor{\bsnm{{Eigenbrot}}, \binits{A.D.}},
\bauthor{\bsnm{{Ferayorni}}, \binits{A.}},
\bauthor{\bsnm{{Foster}}, \binits{C.}},
\bauthor{\bsnm{{Galapon}}, \binits{C.A.}},
\bauthor{\bsnm{{Gedrites}}, \binits{C.}},
\bauthor{\bsnm{{Gonzales}}, \binits{K.}},
\bauthor{\bsnm{{Goodrich}}, \binits{B.D.}},
\bauthor{\bsnm{{Gregory}}, \binits{B.S.}},
\bauthor{\bsnm{{Guzman}}, \binits{S.S.}},
\bauthor{\bsnm{{Guzzo}}, \binits{S.}},
\bauthor{\bsnm{{Hegwer}}, \binits{S.}},
\bauthor{\bsnm{{Hubbard}}, \binits{R.P.}},
\bauthor{\bsnm{{Hubbard}}, \binits{J.R.}},
\bauthor{\bsnm{{Johansson}}, \binits{E.M.}},
\bauthor{\bsnm{{Johnson}}, \binits{L.C.}},
\bauthor{\bsnm{{Liang}}, \binits{C.}},
\bauthor{\bsnm{{Liang}}, \binits{M.}},
\bauthor{\bsnm{{McQuillen}}, \binits{I.}},
\bauthor{\bsnm{{Mayer}}, \binits{C.}},
\bauthor{\bsnm{{Newman}}, \binits{K.}},
\bauthor{\bsnm{{Onodera}}, \binits{B.}},
\bauthor{\bsnm{{Phelps}}, \binits{L.}},
\bauthor{\bsnm{{Puentes}}, \binits{M.M.}},
\bauthor{\bsnm{{Richards}}, \binits{C.}},
\bauthor{\bsnm{{Rimmele}}, \binits{L.M.}},
\bauthor{\bsnm{{Sekulic}}, \binits{P.}},
\bauthor{\bsnm{{Shimko}}, \binits{S.R.}},
\bauthor{\bsnm{{Simison}}, \binits{B.E.}},
\bauthor{\bsnm{{Smith}}, \binits{B.}},
\bauthor{\bsnm{{Starman}}, \binits{E.}},
\bauthor{\bsnm{{Sueoka}}, \binits{S.R.}},
\bauthor{\bsnm{{Summers}}, \binits{R.T.}},
\bauthor{\bsnm{{Szabo}}, \binits{A.}},
\bauthor{\bsnm{{Szabo}}, \binits{L.}},
\bauthor{\bsnm{{Wampler}}, \binits{S.B.}},
\bauthor{\bsnm{{Williams}}, \binits{T.R.}},
\bauthor{\bsnm{{White}}, \binits{C.}}:
\byear{2020},
\batitle{{The Daniel K. Inouye Solar Telescope - Observatory Overview}}.
\bjtitle{\solphys}
\bvolume{295},
\bfpage{172}.
\doiurl{https://doi.org/10.1007/s11207-020-01736-7}.
\adsurl{2020SoPh..295..172R}.
\end{barticle}
\endbibitem

\bibitem[\protect\citeauthoryear{{Robertson}}{2013}]{2013PASA...30...48R}
\begin{barticle}
\bauthor{\bsnm{{Robertson}}, \binits{J.G.}}:
\byear{2013},
\batitle{{Quantifying Resolving Power in Astronomical Spectra}}.
\bjtitle{\pasa}
\bvolume{30},
\bfpage{e048}.
\doiurl{https://doi.org/10.1017/pasa.2013.26}.
\adsurl{2013PASA...30...48R}.
\end{barticle}
\endbibitem

\bibitem[\protect\citeauthoryear{{Rouppe van der Voort}
  et~al.}{2003}]{2003A&A...403..277R}
\begin{barticle}
\bauthor{\bsnm{{Rouppe van der Voort}}, \binits{L.H.M.}},
\bauthor{\bsnm{{Rutten}}, \binits{R.J.}},
\bauthor{\bsnm{{S{\"u}tterlin}}, \binits{P.}},
\bauthor{\bsnm{{Sloover}}, \binits{P.J.}},
\bauthor{\bsnm{{Krijger}}, \binits{J.M.}}:
\byear{2003},
\batitle{{La Palma observations of umbral flashes}}.
\bjtitle{\aap}
\bvolume{403},
\bfpage{277}.
\doiurl{https://doi.org/10.1051/0004-6361:20030237}.
\adsurl{2003A&A...403..277R}.
\end{barticle}
\endbibitem

\bibitem[\protect\citeauthoryear{{Rouppe van der Voort}
  et~al.}{2021}]{2021A&A...648A..54R}
\begin{barticle}
\bauthor{\bsnm{{Rouppe van der Voort}}, \binits{L.H.M.}},
\bauthor{\bsnm{{Joshi}}, \binits{J.}},
\bauthor{\bsnm{{Henriques}}, \binits{V.M.J.}},
\bauthor{\bsnm{{Bose}}, \binits{S.}}:
\byear{2021},
\batitle{{Signatures of ubiquitous magnetic reconnection in the deep atmosphere
  of sunspot penumbrae}}.
\bjtitle{\aap}
\bvolume{648},
\bfpage{A54}.
\doiurl{https://doi.org/10.1051/0004-6361/202040171}.
\adsurl{2021A&A...648A..54R}.
\end{barticle}
\endbibitem

\bibitem[\protect\citeauthoryear{{Rouppe van der Voort}
  et~al.}{2017}]{2017ApJ...851L...6R}
\begin{barticle}
\bauthor{\bsnm{{Rouppe van der Voort}}, \binits{L.}},
\bauthor{\bsnm{{De Pontieu}}, \binits{B.}},
\bauthor{\bsnm{{Scharmer}}, \binits{G.B.}},
\bauthor{\bsnm{{de la Cruz Rodr{\'\i}guez}}, \binits{J.}},
\bauthor{\bsnm{{Mart{\'\i}nez-Sykora}}, \binits{J.}},
\bauthor{\bsnm{{N{\'o}brega-Siverio}}, \binits{D.}},
\bauthor{\bsnm{{Guo}}, \binits{L.J.}},
\bauthor{\bsnm{{Jafarzadeh}}, \binits{S.}},
\bauthor{\bsnm{{Pereira}}, \binits{T.M.D.}},
\bauthor{\bsnm{{Hansteen}}, \binits{V.H.}},
\bauthor{\bsnm{{Carlsson}}, \binits{M.}},
\bauthor{\bsnm{{Vissers}}, \binits{G.}}:
\byear{2017},
\batitle{{Intermittent Reconnection and Plasmoids in UV Bursts in the Low Solar
  Atmosphere}}.
\bjtitle{\apjl}
\bvolume{851},
\bfpage{L6}.
\doiurl{https://doi.org/10.3847/2041-8213/aa99dd}.
\adsurl{2017ApJ...851L...6R}.
\end{barticle}
\endbibitem

\bibitem[\protect\citeauthoryear{{Sankarasubramanian}, {Hasan}, and
  {Rangarajan}}{2010}]{2010ASSP...19..156S}
\begin{bchapter}
\bauthor{\bsnm{{Sankarasubramanian}}, \binits{K.}},
\bauthor{\bsnm{{Hasan}}, \binits{S.S.}},
\bauthor{\bsnm{{Rangarajan}}, \binits{K.E.}}:
\byear{2010},
\bctitle{{Spectropolarimetry with the NLST}}.
In: \bbtitle{Magnetic Coupling between the Interior and Atmosphere of the Sun},
\bsertitle{Astrophysics and Space Science Proceedings}
\bseriesno{19},
\bfpage{156}.
\doiurl{https://doi.org/10.1007/978-3-642-02859-5_12}.
\adsurl{2010ASSP...19..156S}.
\end{bchapter}
\endbibitem

\bibitem[\protect\citeauthoryear{{Scharmer}}{2017}]{2017psio.confE..85S}
\begin{bchapter}
\bauthor{\bsnm{{Scharmer}}, \binits{G.}}:
\byear{2017},
\bctitle{{SST/CHROMIS: a new window to the solar chromosphere}}.
In: \bbtitle{SOLARNET IV: The Physics of the Sun from the Interior to the Outer
  Atmosphere},
\bfpage{85}.
\adsurl{2017psio.confE..85S}.
\end{bchapter}
\endbibitem

\bibitem[\protect\citeauthoryear{{Scharmer}}{2006}]{2006A&A...447.1111S}
\begin{barticle}
\bauthor{\bsnm{{Scharmer}}, \binits{G.B.}}:
\byear{2006},
\batitle{{Comments on the optimization of high resolution Fabry-P{\'e}rot
  filtergraphs}}.
\bjtitle{\aap}
\bvolume{447},
\bfpage{1111}.
\doiurl{https://doi.org/10.1051/0004-6361:20052981}.
\adsurl{2006A&A...447.1111S}.
\end{barticle}
\endbibitem

\bibitem[\protect\citeauthoryear{{Scharmer} et~al.}{2003}]{2003SPIE.4853..341S}
\begin{bchapter}
\bauthor{\bsnm{{Scharmer}}, \binits{G.B.}},
\bauthor{\bsnm{{Bjelksjo}}, \binits{K.}},
\bauthor{\bsnm{{Korhonen}}, \binits{T.K.}},
\bauthor{\bsnm{{Lindberg}}, \binits{B.}},
\bauthor{\bsnm{{Petterson}}, \binits{B.}}:
\byear{2003},
\bctitle{{The 1-meter Swedish solar telescope}}.
In: \beditor{\bsnm{{Keil}}, \binits{S.L.}},
\beditor{\bsnm{{Avakyan}}, \binits{S.V.}} (eds.)
\bbtitle{\procspie},
\bsertitle{Society of Photo-Optical Instrumentation Engineers (SPIE) Conference
  Series}
\bseriesno{4853},
\bfpage{341}.
\doiurl{https://doi.org/10.1117/12.460377}.
\adsurl{2003SPIE.4853..341S}.
\end{bchapter}
\endbibitem

\bibitem[\protect\citeauthoryear{{Scharmer} et~al.}{2008}]{2008ApJ...689L..69S}
\begin{barticle}
\bauthor{\bsnm{{Scharmer}}, \binits{G.B.}},
\bauthor{\bsnm{{Narayan}}, \binits{G.}},
\bauthor{\bsnm{{Hillberg}}, \binits{T.}},
\bauthor{\bsnm{{de la Cruz Rodriguez}}, \binits{J.}},
\bauthor{\bsnm{{L{\"o}fdahl}}, \binits{M.G.}},
\bauthor{\bsnm{{Kiselman}}, \binits{D.}},
\bauthor{\bsnm{{S{\"u}tterlin}}, \binits{P.}},
\bauthor{\bsnm{{van Noort}}, \binits{M.}},
\bauthor{\bsnm{{Lagg}}, \binits{A.}}:
\byear{2008},
\batitle{{CRISP Spectropolarimetric Imaging of Penumbral Fine Structure}}.
\bjtitle{\apjl}
\bvolume{689},
\bfpage{L69}.
\doiurl{https://doi.org/10.1086/595744}.
\adsurl{2008ApJ...689L..69S}.
\end{barticle}
\endbibitem

\bibitem[\protect\citeauthoryear{{Schlichenmaier}
  et~al.}{2023}]{2023A&A...669A..78S}
\begin{barticle}
\bauthor{\bsnm{{Schlichenmaier}}, \binits{R.}},
\bauthor{\bsnm{{Pitters}}, \binits{D.}},
\bauthor{\bsnm{{Borrero}}, \binits{J.M.}},
\bauthor{\bsnm{{Schubert}}, \binits{M.}}:
\byear{2023},
\batitle{{Effects of solar evolution on finite acquisition time of Fabry-Perot
  interferometers in high resolution solar physics}}.
\bjtitle{\aap}
\bvolume{669},
\bfpage{A78}.
\doiurl{https://doi.org/10.1051/0004-6361/202244640}.
\adsurl{2023A&A...669A..78S}.
\end{barticle}
\endbibitem

\bibitem[\protect\citeauthoryear{{Schmidt} et~al.}{2012}]{2012AN....333..796S}
\begin{barticle}
\bauthor{\bsnm{{Schmidt}}, \binits{W.}},
\bauthor{\bsnm{{von der L{\"u}he}}, \binits{O.}},
\bauthor{\bsnm{{Volkmer}}, \binits{R.}},
\bauthor{\bsnm{{Denker}}, \binits{C.}},
\bauthor{\bsnm{{Solanki}}, \binits{S.K.}},
\bauthor{\bsnm{{Balthasar}}, \binits{H.}},
\bauthor{\bsnm{{Bello Gonzalez}}, \binits{N.}},
\bauthor{\bsnm{{Berkefeld}}, \binits{T.}},
\bauthor{\bsnm{{Collados}}, \binits{M.}},
\bauthor{\bsnm{{Fischer}}, \binits{A.}},
\bauthor{\bsnm{{Halbgewachs}}, \binits{C.}},
\bauthor{\bsnm{{Heidecke}}, \binits{F.}},
\bauthor{\bsnm{{Hofmann}}, \binits{A.}},
\bauthor{\bsnm{{Kneer}}, \binits{F.}},
\bauthor{\bsnm{{Lagg}}, \binits{A.}},
\bauthor{\bsnm{{Nicklas}}, \binits{H.}},
\bauthor{\bsnm{{Popow}}, \binits{E.}},
\bauthor{\bsnm{{Puschmann}}, \binits{K.G.}},
\bauthor{\bsnm{{Schmidt}}, \binits{D.}},
\bauthor{\bsnm{{Sigwarth}}, \binits{M.}},
\bauthor{\bsnm{{Sobotka}}, \binits{M.}},
\bauthor{\bsnm{{Soltau}}, \binits{D.}},
\bauthor{\bsnm{{Staude}}, \binits{J.}},
\bauthor{\bsnm{{Strassmeier}}, \binits{K.G.}},
\bauthor{\bsnm{{Waldmann }}, \binits{T.A.}}:
\byear{2012},
\batitle{{The 1.5 meter solar telescope GREGOR}}.
\bjtitle{Astronomische Nachrichten}
\bvolume{333},
\bfpage{796}.
\doiurl{https://doi.org/10.1002/asna.201211725}.
\adsurl{2012AN....333..796S}.
\end{barticle}
\endbibitem

\bibitem[\protect\citeauthoryear{{Schou} et~al.}{2012}]{2012SoPh..275..229S}
\begin{barticle}
\bauthor{\bsnm{{Schou}}, \binits{J.}},
\bauthor{\bsnm{{Scherrer}}, \binits{P.H.}},
\bauthor{\bsnm{{Bush}}, \binits{R.I.}},
\bauthor{\bsnm{{Wachter}}, \binits{R.}},
\bauthor{\bsnm{{Couvidat}}, \binits{S.}},
\bauthor{\bsnm{{Rabello-Soares}}, \binits{M.C.}},
\bauthor{\bsnm{{Bogart}}, \binits{R.S.}},
\bauthor{\bsnm{{Hoeksema}}, \binits{J.T.}},
\bauthor{\bsnm{{Liu}}, \binits{Y.}},
\bauthor{\bsnm{{Duvall}}, \binits{T.L.}},
\bauthor{\bsnm{{Akin}}, \binits{D.J.}},
\bauthor{\bsnm{{Allard}}, \binits{B.A.}},
\bauthor{\bsnm{{Miles}}, \binits{J.W.}},
\bauthor{\bsnm{{Rairden}}, \binits{R.}},
\bauthor{\bsnm{{Shine}}, \binits{R.A.}},
\bauthor{\bsnm{{Tarbell}}, \binits{T.D.}},
\bauthor{\bsnm{{Title}}, \binits{A.M.}},
\bauthor{\bsnm{{Wolfson}}, \binits{C.J.}},
\bauthor{\bsnm{{Elmore}}, \binits{D.F.}},
\bauthor{\bsnm{{Norton}}, \binits{A.A.}},
\bauthor{\bsnm{{Tomczyk}}, \binits{S.}}:
\byear{2012},
\batitle{{Design and Ground Calibration of the Helioseismic and Magnetic Imager
  (HMI) Instrument on the Solar Dynamics Observatory (SDO)}}.
\bjtitle{\solphys}
\bvolume{275},
\bfpage{229}.
\doiurl{https://doi.org/10.1007/s11207-011-9842-2}.
\adsurl{2012SoPh..275..229S}.
\end{barticle}
\endbibitem

\bibitem[\protect\citeauthoryear{{Singh}}{2008}]{2008JApA...29..345S}
\begin{barticle}
\bauthor{\bsnm{{Singh}}, \binits{J.}}:
\byear{2008},
\batitle{{Proposed national large solar telescope}}.
\bjtitle{Journal of Astrophysics and Astronomy}
\bvolume{29},
\bfpage{345}.
\doiurl{https://doi.org/10.1007/s12036-008-0045-7}.
\adsurl{2008JApA...29..345S}.
\end{barticle}
\endbibitem

\bibitem[\protect\citeauthoryear{Sinha et~al.}{2004}]{10.1117/1.1775230}
\begin{barticle}
\bauthor{\bsnm{Sinha}, \binits{A.}},
\bauthor{\bsnm{Barbastathis}, \binits{G.}},
\bauthor{\bsnm{Liu}, \binits{W.}},
\bauthor{\bsnm{Psaltis}, \binits{D.}}:
\byear{2004},
\batitle{{Imaging using volume holograms}}.
\bjtitle{Optical Engineering}
\bvolume{43},
\bfpage{1959 }.
\doiurl{https://doi.org/10.1117/1.1775230}.
\burl{https://doi.org/10.1117/1.1775230}.
\end{barticle}
\endbibitem

\bibitem[\protect\citeauthoryear{{Socas-Navarro}, {Trujillo Bueno}, and {Ruiz
  Cobo}}{2000}]{2000Sci...288.1396S}
\begin{barticle}
\bauthor{\bsnm{{Socas-Navarro}}, \binits{H.}},
\bauthor{\bsnm{{Trujillo Bueno}}, \binits{J.}},
\bauthor{\bsnm{{Ruiz Cobo}}, \binits{B.}}:
\byear{2000},
\batitle{{Anomalous Polarization Profiles in Sunspots: Possible Origin of
  Umbral Flashes}}.
\bjtitle{Science}
\bvolume{288},
\bfpage{1396}.
\doiurl{https://doi.org/10.1126/science.288.5470.1396}.
\adsurl{2000Sci...288.1396S}.
\end{barticle}
\endbibitem

\bibitem[\protect\citeauthoryear{{Socas-Navarro}
  et~al.}{2006}]{2006SoPh..235...55S}
\begin{barticle}
\bauthor{\bsnm{{Socas-Navarro}}, \binits{H.}},
\bauthor{\bsnm{{Elmore}}, \binits{D.}},
\bauthor{\bsnm{{Pietarila}}, \binits{A.}},
\bauthor{\bsnm{{Darnell}}, \binits{A.}},
\bauthor{\bsnm{{Lites}}, \binits{B.W.}},
\bauthor{\bsnm{{Tomczyk}}, \binits{S.}},
\bauthor{\bsnm{{Hegwer}}, \binits{S.}}:
\byear{2006},
\batitle{{Spinor: Visible and Infrared Spectro-Polarimetry at the National
  Solar Observatory}}.
\bjtitle{\solphys}
\bvolume{235},
\bfpage{55}.
\doiurl{https://doi.org/10.1007/s11207-006-0020-x}.
\adsurl{2006SoPh..235...55S}.
\end{barticle}
\endbibitem

\bibitem[\protect\citeauthoryear{{Solanki} et~al.}{2010}]{2010ApJ...723L.127S}
\begin{barticle}
\bauthor{\bsnm{{Solanki}}, \binits{S.K.}},
\bauthor{\bsnm{{Barthol}}, \binits{P.}},
\bauthor{\bsnm{{Danilovic}}, \binits{S.}},
\bauthor{\bsnm{{Feller}}, \binits{A.}},
\bauthor{\bsnm{{Gandorfer}}, \binits{A.}},
\bauthor{\bsnm{{Hirzberger}}, \binits{J.}},
\bauthor{\bsnm{{Riethm{\"u}ller}}, \binits{T.L.}},
\bauthor{\bsnm{{Sch{\"u}ssler}}, \binits{M.}},
\bauthor{\bsnm{{Bonet}}, \binits{J.A.}},
\bauthor{\bsnm{{Mart{\'\i}nez Pillet}}, \binits{V.}},
\bauthor{\bsnm{{del Toro Iniesta}}, \binits{J.C.}},
\bauthor{\bsnm{{Domingo}}, \binits{V.}},
\bauthor{\bsnm{{Palacios}}, \binits{J.}},
\bauthor{\bsnm{{Kn{\"o}lker}}, \binits{M.}},
\bauthor{\bsnm{{Bello Gonz{\'a}lez}}, \binits{N.}},
\bauthor{\bsnm{{Berkefeld}}, \binits{T.}},
\bauthor{\bsnm{{Franz}}, \binits{M.}},
\bauthor{\bsnm{{Schmidt}}, \binits{W.}},
\bauthor{\bsnm{{Title}}, \binits{A.M.}}:
\byear{2010},
\batitle{{{\sc Sunrise}: Instrument, Mission, Data, and First Results}}.
\bjtitle{\apjl}
\bvolume{723},
\bfpage{L127}.
\doiurl{https://doi.org/10.1088/2041-8205/723/2/L127}.
\adsurl{2010ApJ...723L.127S}.
\end{barticle}
\endbibitem

\bibitem[\protect\citeauthoryear{{Sparrow}}{1916}]{1916ApJ....44...76S}
\begin{barticle}
\bauthor{\bsnm{{Sparrow}}, \binits{C.M.}}:
\byear{1916},
\batitle{{On Spectroscopic Resolving Power}}.
\bjtitle{\apj}
\bvolume{44},
\bfpage{76}.
\doiurl{https://doi.org/10.1086/142271}.
\adsurl{1916ApJ....44...76S}.
\end{barticle}
\endbibitem

\bibitem[\protect\citeauthoryear{{SPICE Consortium}
  et~al.}{2020}]{2020A&A...642A..14S}
\begin{barticle}
\bauthor{\bsnm{{SPICE Consortium}}},
\bauthor{\bsnm{{Anderson}}, \binits{M.}},
\bauthor{\bsnm{{Appourchaux}}, \binits{T.}},
\bauthor{\bsnm{{Auch{\`e}re}}, \binits{F.}},
\bauthor{\bsnm{{Aznar Cuadrado}}, \binits{R.}},
\bauthor{\bsnm{{Barbay}}, \binits{J.}},
\bauthor{\bsnm{{Baudin}}, \binits{F.}},
\bauthor{\bsnm{{Beardsley}}, \binits{S.}},
\bauthor{\bsnm{{Bocchialini}}, \binits{K.}},
\bauthor{\bsnm{{Borgo}}, \binits{B.}},
\bauthor{\bsnm{{Bruzzi}}, \binits{D.}},
\bauthor{\bsnm{{Buchlin}}, \binits{E.}},
\bauthor{\bsnm{{Burton}}, \binits{G.}},
\bauthor{\bsnm{{B{\"u}chel}}, \binits{V.}},
\bauthor{\bsnm{{Caldwell}}, \binits{M.}},
\bauthor{\bsnm{{Caminade}}, \binits{S.}},
\bauthor{\bsnm{{Carlsson}}, \binits{M.}},
\bauthor{\bsnm{{Curdt}}, \binits{W.}},
\bauthor{\bsnm{{Davenne}}, \binits{J.}},
\bauthor{\bsnm{{Davila}}, \binits{J.}},
\bauthor{\bsnm{{Deforest}}, \binits{C.E.}},
\bauthor{\bsnm{{Del Zanna}}, \binits{G.}},
\bauthor{\bsnm{{Drummond}}, \binits{D.}},
\bauthor{\bsnm{{Dubau}}, \binits{J.}},
\bauthor{\bsnm{{Dumesnil}}, \binits{C.}},
\bauthor{\bsnm{{Dunn}}, \binits{G.}},
\bauthor{\bsnm{{Eccleston}}, \binits{P.}},
\bauthor{\bsnm{{Fludra}}, \binits{A.}},
\bauthor{\bsnm{{Fredvik}}, \binits{T.}},
\bauthor{\bsnm{{Gabriel}}, \binits{A.}},
\bauthor{\bsnm{{Giunta}}, \binits{A.}},
\bauthor{\bsnm{{Gottwald}}, \binits{A.}},
\bauthor{\bsnm{{Griffin}}, \binits{D.}},
\bauthor{\bsnm{{Grundy}}, \binits{T.}},
\bauthor{\bsnm{{Guest}}, \binits{S.}},
\bauthor{\bsnm{{Gyo}}, \binits{M.}},
\bauthor{\bsnm{{Haberreiter}}, \binits{M.}},
\bauthor{\bsnm{{Hansteen}}, \binits{V.}},
\bauthor{\bsnm{{Harrison}}, \binits{R.}},
\bauthor{\bsnm{{Hassler}}, \binits{D.M.}},
\bauthor{\bsnm{{Haugan}}, \binits{S.V.H.}},
\bauthor{\bsnm{{Howe}}, \binits{C.}},
\bauthor{\bsnm{{Janvier}}, \binits{M.}},
\bauthor{\bsnm{{Klein}}, \binits{R.}},
\bauthor{\bsnm{{Koller}}, \binits{S.}},
\bauthor{\bsnm{{Kucera}}, \binits{T.A.}},
\bauthor{\bsnm{{Kouliche}}, \binits{D.}},
\bauthor{\bsnm{{Marsch}}, \binits{E.}},
\bauthor{\bsnm{{Marshall}}, \binits{A.}},
\bauthor{\bsnm{{Marshall}}, \binits{G.}},
\bauthor{\bsnm{{Matthews}}, \binits{S.A.}},
\bauthor{\bsnm{{McQuirk}}, \binits{C.}},
\bauthor{\bsnm{{Meining}}, \binits{S.}},
\bauthor{\bsnm{{Mercier}}, \binits{C.}},
\bauthor{\bsnm{{Morris}}, \binits{N.}},
\bauthor{\bsnm{{Morse}}, \binits{T.}},
\bauthor{\bsnm{{Munro}}, \binits{G.}},
\bauthor{\bsnm{{Parenti}}, \binits{S.}},
\bauthor{\bsnm{{Pastor-Santos}}, \binits{C.}},
\bauthor{\bsnm{{Peter}}, \binits{H.}},
\bauthor{\bsnm{{Pfiffner}}, \binits{D.}},
\bauthor{\bsnm{{Phelan}}, \binits{P.}},
\bauthor{\bsnm{{Philippon}}, \binits{A.}},
\bauthor{\bsnm{{Richards}}, \binits{A.}},
\bauthor{\bsnm{{Rogers}}, \binits{K.}},
\bauthor{\bsnm{{Sawyer}}, \binits{C.}},
\bauthor{\bsnm{{Schlatter}}, \binits{P.}},
\bauthor{\bsnm{{Schmutz}}, \binits{W.}},
\bauthor{\bsnm{{Sch{\"u}hle}}, \binits{U.}},
\bauthor{\bsnm{{Shaughnessy}}, \binits{B.}},
\bauthor{\bsnm{{Sidher}}, \binits{S.}},
\bauthor{\bsnm{{Solanki}}, \binits{S.K.}},
\bauthor{\bsnm{{Speight}}, \binits{R.}},
\bauthor{\bsnm{{Spescha}}, \binits{M.}},
\bauthor{\bsnm{{Szwec}}, \binits{N.}},
\bauthor{\bsnm{{Tamiatto}}, \binits{C.}},
\bauthor{\bsnm{{Teriaca}}, \binits{L.}},
\bauthor{\bsnm{{Thompson}}, \binits{W.}},
\bauthor{\bsnm{{Tosh}}, \binits{I.}},
\bauthor{\bsnm{{Tustain}}, \binits{S.}},
\bauthor{\bsnm{{Vial}}, \binits{J.-C.}},
\bauthor{\bsnm{{Walls}}, \binits{B.}},
\bauthor{\bsnm{{Waltham}}, \binits{N.}},
\bauthor{\bsnm{{Wimmer-Schweingruber}}, \binits{R.}},
\bauthor{\bsnm{{Woodward}}, \binits{S.}},
\bauthor{\bsnm{{Young}}, \binits{P.}},
\bauthor{\bsnm{{de Groof}}, \binits{A.}},
\bauthor{\bsnm{{Pacros}}, \binits{A.}},
\bauthor{\bsnm{{Williams}}, \binits{D.}},
\bauthor{\bsnm{{M{\"u}ller}}, \binits{D.}}:
\byear{2020},
\batitle{{The Solar Orbiter SPICE instrument. An extreme UV imaging
  spectrometer}}.
\bjtitle{\aap}
\bvolume{642},
\bfpage{A14}.
\doiurl{https://doi.org/10.1051/0004-6361/201935574}.
\adsurl{2020A&A...642A..14S}.
\end{barticle}
\endbibitem

\bibitem[\protect\citeauthoryear{{Steiner} et~al.}{2013}]{2013SPIE.8870E..0HS}
\begin{bchapter}
\bauthor{\bsnm{{Steiner}}, \binits{R.}},
\bauthor{\bsnm{{Pesch}}, \binits{A.}},
\bauthor{\bsnm{{Erdmann}}, \binits{L.H.}},
\bauthor{\bsnm{{Burkhardt}}, \binits{M.}},
\bauthor{\bsnm{{Gatto}}, \binits{A.}},
\bauthor{\bsnm{{Wipf}}, \binits{R.}},
\bauthor{\bsnm{{Diehl}}, \binits{T.}},
\bauthor{\bsnm{{Vink}}, \binits{H.J.P.}},
\bauthor{\bsnm{{van den Bosch}}, \binits{B.G.}}:
\byear{2013},
\bctitle{{Fabrication of low straylight holographic gratings for space
  applications}}.
In: \beditor{\bsnm{{Mouroulis}}, \binits{P.}},
\beditor{\bsnm{{Pagano}}, \binits{T.S.}} (eds.)
\bbtitle{Imaging Spectrometry XVIII},
\bsertitle{Society of Photo-Optical Instrumentation Engineers (SPIE) Conference
  Series}
\bseriesno{8870},
\bfpage{88700H}.
\doiurl{https://doi.org/10.1117/12.2025269}.
\adsurl{2013SPIE.8870E..0HS}.
\end{bchapter}
\endbibitem

\bibitem[\protect\citeauthoryear{{Stenflo}, {Baur}, and
  {Elmore}}{1980}]{1980A&A....84...60S}
\begin{barticle}
\bauthor{\bsnm{{Stenflo}}, \binits{J.O.}},
\bauthor{\bsnm{{Baur}}, \binits{T.G.}},
\bauthor{\bsnm{{Elmore}}, \binits{D.F.}}:
\byear{1980},
\batitle{{Resonance-line polarization: IV. Observations of non-magnetic line
  polarization and its center-to-limb variations.}}
\bjtitle{\aap}
\bvolume{84},
\bfpage{60}.
\adsurl{1980A&A....84...60S}.
\end{barticle}
\endbibitem

\bibitem[\protect\citeauthoryear{Tomczyk
  et~al.}{2016}]{https://doi.org/10.1002/2016JA022871}
\begin{barticle}
\bauthor{\bsnm{Tomczyk}, \binits{S.}},
\bauthor{\bsnm{Landi}, \binits{E.}},
\bauthor{\bsnm{Burkepile}, \binits{J.T.}},
\bauthor{\bsnm{Casini}, \binits{R.}},
\bauthor{\bsnm{DeLuca}, \binits{E.E.}},
\bauthor{\bsnm{Fan}, \binits{Y.}},
\bauthor{\bsnm{Gibson}, \binits{S.E.}},
\bauthor{\bsnm{Lin}, \binits{H.}},
\bauthor{\bsnm{McIntosh}, \binits{S.W.}},
\bauthor{\bsnm{Solomon}, \binits{S.C.}},
\bauthor{\bparticle{de} \bsnm{Toma}, \binits{G.}},
\bauthor{\bparticle{de} \bsnm{Wijn}, \binits{A.G.}},
\bauthor{\bsnm{Zhang}, \binits{J.}}:
\byear{2016},
\batitle{Scientific objectives and capabilities of the Coronal Solar Magnetism
  Observatory}.
\bjtitle{Journal of Geophysical Research: Space Physics}
\bvolume{121},
\bfpage{7470}.
\doiurl{https://doi.org/10.1002/2016JA022871}.
\burl{https://agupubs.onlinelibrary.wiley.com/doi/abs/10.1002/2016JA022871}.
\end{barticle}
\endbibitem

\bibitem[\protect\citeauthoryear{{Tritschler}
  et~al.}{2016}]{2016AN....337.1064T}
\begin{barticle}
\bauthor{\bsnm{{Tritschler}}, \binits{A.}},
\bauthor{\bsnm{{Rimmele}}, \binits{T.R.}},
\bauthor{\bsnm{{Berukoff}}, \binits{S.}},
\bauthor{\bsnm{{Casini}}, \binits{R.}},
\bauthor{\bsnm{{Kuhn}}, \binits{J.R.}},
\bauthor{\bsnm{{Lin}}, \binits{H.}},
\bauthor{\bsnm{{Rast}}, \binits{M.P.}},
\bauthor{\bsnm{{McMullin}}, \binits{J.P.}},
\bauthor{\bsnm{{Schmidt}}, \binits{W.}},
\bauthor{\bsnm{{W{\"o}ger}}, \binits{F.}},
\bauthor{\bsnm{{DKIST Team}}}:
\byear{2016},
\batitle{{Daniel K. Inouye Solar Telescope: High-resolution observing of the
  dynamic Sun}}.
\bjtitle{Astronomische Nachrichten}
\bvolume{337},
\bfpage{1064}.
\doiurl{https://doi.org/10.1002/asna.201612434}.
\adsurl{2016AN....337.1064T}.
\end{barticle}
\endbibitem

\bibitem[\protect\citeauthoryear{Tu et~al.}{2021}]{Tu2021}
\begin{barticle}
\bauthor{\bsnm{Tu}, \binits{H.-T.}},
\bauthor{\bsnm{Jiang}, \binits{A.-Q.}},
\bauthor{\bsnm{Chen}, \binits{J.-K.}},
\bauthor{\bsnm{Lu}, \binits{W.-J.}},
\bauthor{\bsnm{Zang}, \binits{K.-Y.}},
\bauthor{\bsnm{Tang}, \binits{H.-Q.}},
\bauthor{\bsnm{Yoshie}, \binits{O.}},
\bauthor{\bsnm{Xiang}, \binits{X.-D.}},
\bauthor{\bsnm{Lee}, \binits{Y.-P.}},
\bauthor{\bsnm{Zhao}, \binits{H.-B.}},
\bauthor{\bsnm{Zheng}, \binits{Y.-X.}},
\bauthor{\bsnm{Wang}, \binits{S.-Y.}},
\bauthor{\bsnm{Guo}, \binits{J.}},
\bauthor{\bsnm{Zhang}, \binits{R.-J.}},
\bauthor{\bsnm{Li}, \binits{J.}},
\bauthor{\bsnm{Yang}, \binits{Y.-M.}},
\bauthor{\bsnm{Lynch}, \binits{W.D.}},
\bauthor{\bsnm{Chen}, \binits{L.-Y.}}:
\byear{2021},
\batitle{A coma-free super-high resolution optical spectrometer using 44 high
  dispersion sub-gratings}.
\bjtitle{Scientific Reports}
\bvolume{11},
\bfpage{1093}.
\bisbn{2045-2322}.
\doiurl{https://doi.org/10.1038/s41598-020-80307-z}.
\burl{https://doi.org/10.1038/s41598-020-80307-z}.
\end{barticle}
\endbibitem

\bibitem[\protect\citeauthoryear{{Uexkuell}, {Kneer}, and
  {Mattig}}{1983}]{1983A&A...123..263U}
\begin{barticle}
\bauthor{\bsnm{{Uexkuell}}, \binits{M.V.}},
\bauthor{\bsnm{{Kneer}}, \binits{F.}},
\bauthor{\bsnm{{Mattig}}, \binits{W.}}:
\byear{1983},
\batitle{{The chromosphere above sunspot umbrae. IV - Frequency analysis of
  umbral oscillations}}.
\bjtitle{\aap}
\bvolume{123},
\bfpage{263}.
\adsurl{1983A&A...123..263U}.
\end{barticle}
\endbibitem

\bibitem[\protect\citeauthoryear{{van Noort}
  et~al.}{2022}]{2022A&A...668A.149V}
\begin{barticle}
\bauthor{\bsnm{{van Noort}}, \binits{M.}},
\bauthor{\bsnm{{Bischoff}}, \binits{J.}},
\bauthor{\bsnm{{Kramer}}, \binits{A.}},
\bauthor{\bsnm{{Solanki}}, \binits{S.K.}},
\bauthor{\bsnm{{Kiselman}}, \binits{D.}}:
\byear{2022},
\batitle{{A prototype of a microlensed hyperspectral imager for solar
  observations}}.
\bjtitle{\aap}
\bvolume{668},
\bfpage{A149}.
\doiurl{https://doi.org/10.1051/0004-6361/202243464}.
\adsurl{2022A&A...668A.149V}.
\end{barticle}
\endbibitem

\bibitem[\protect\citeauthoryear{{Vieira} et~al.}{2019}]{2019AGUFMSH13C3447A}
\begin{bchapter}
\bauthor{\bsnm{{Vieira}}, \binits{L.E.A.}},
\bauthor{\bsnm{{Lago}}, \binits{A.}},
\bauthor{\bsnm{{Rockenbach}}, \binits{M.}},
\bauthor{\bsnm{{Guarnieri}}, \binits{F.L.}},
\bauthor{\bsnm{{Da Silva}}, \binits{L.A.}},
\bauthor{\bsnm{{Carlesso}}, \binits{F.}},
\bauthor{\bsnm{{Alves}}, \binits{L.R.}},
\bauthor{\bsnm{{Souza}}, \binits{V.M.C.E.S.}},
\bauthor{\bsnm{{Jauer}}, \binits{P.R.}}:
\byear{2019},
\bctitle{{Status of the Galileo Solar Space Telescope Mission (GSST)
  proposal}}.
In: \bbtitle{AGU Fall Meeting Abstracts}
\bseriesno{2019},
\bfpage{SH13C}.
\adsurl{2019AGUFMSH13C3447A}.
\end{bchapter}
\endbibitem

\bibitem[\protect\citeauthoryear{{Wang} et~al.}{2013}]{2013RAA....13.1240W}
\begin{barticle}
\bauthor{\bsnm{{Wang}}, \binits{R.}},
\bauthor{\bsnm{{Xu}}, \binits{Z.}},
\bauthor{\bsnm{{Jin}}, \binits{Z.-Y.}},
\bauthor{\bsnm{{Li}}, \binits{Z.}},
\bauthor{\bsnm{{Fu}}, \binits{Y.}},
\bauthor{\bsnm{{Liu}}, \binits{Z.}}:
\byear{2013},
\batitle{{The first observation and data reduction of the Multi-wavelength
  Spectrometer on the New Vacuum Solar Telescope}}.
\bjtitle{Research in Astronomy and Astrophysics}
\bvolume{13},
\bfpage{1240}.
\doiurl{https://doi.org/10.1088/1674-4527/13/10/012}.
\adsurl{2013RAA....13.1240W}.
\end{barticle}
\endbibitem

\bibitem[\protect\citeauthoryear{Wang, Yokoyama, and Iijima}{2021}]{Wang_2021}
\begin{barticle}
\bauthor{\bsnm{Wang}, \binits{Y.}},
\bauthor{\bsnm{Yokoyama}, \binits{T.}},
\bauthor{\bsnm{Iijima}, \binits{H.}}:
\byear{2021},
\batitle{Fast Magnetic Wave Could Heat the Solar Low-beta Chromosphere}.
\bjtitle{The Astrophysical Journal Letters}
\bvolume{916},
\bfpage{L10}.
\doiurl{https://doi.org/10.3847/2041-8213/ac10c7}.
\burl{https://dx.doi.org/10.3847/2041-8213/ac10c7}.
\end{barticle}
\endbibitem

\bibitem[\protect\citeauthoryear{{Wells} et~al.}{2015}]{2015PASP..127..646W}
\begin{barticle}
\bauthor{\bsnm{{Wells}}, \binits{M.}},
\bauthor{\bsnm{{Pel}}, \binits{J.-W.}},
\bauthor{\bsnm{{Glasse}}, \binits{A.}},
\bauthor{\bsnm{{Wright}}, \binits{G.S.}},
\bauthor{\bsnm{{Aitink-Kroes}}, \binits{G.}},
\bauthor{\bsnm{{Azzollini}}, \binits{R.}},
\bauthor{\bsnm{{Beard}}, \binits{S.}},
\bauthor{\bsnm{{Brandl}}, \binits{B.R.}},
\bauthor{\bsnm{{Gallie}}, \binits{A.}},
\bauthor{\bsnm{{Geers}}, \binits{V.C.}},
\bauthor{\bsnm{{Glauser}}, \binits{A.M.}},
\bauthor{\bsnm{{Hastings}}, \binits{P.}},
\bauthor{\bsnm{{Henning}}, \binits{T.}},
\bauthor{\bsnm{{Jager}}, \binits{R.}},
\bauthor{\bsnm{{Justtanont}}, \binits{K.}},
\bauthor{\bsnm{{Kruizinga}}, \binits{B.}},
\bauthor{\bsnm{{Lahuis}}, \binits{F.}},
\bauthor{\bsnm{{Lee}}, \binits{D.}},
\bauthor{\bsnm{{Martinez-Delgado}}, \binits{I.}},
\bauthor{\bsnm{{Mart{\'\i}nez-Galarza}}, \binits{J.R.}},
\bauthor{\bsnm{{Meijers}}, \binits{M.}},
\bauthor{\bsnm{{Morrison}}, \binits{J.E.}},
\bauthor{\bsnm{{M{\"u}ller}}, \binits{F.}},
\bauthor{\bsnm{{Nakos}}, \binits{T.}},
\bauthor{\bsnm{{O'Sullivan}}, \binits{B.}},
\bauthor{\bsnm{{Oudenhuysen}}, \binits{A.}},
\bauthor{\bsnm{{Parr-Burman}}, \binits{P.}},
\bauthor{\bsnm{{Pauwels}}, \binits{E.}},
\bauthor{\bsnm{{Rohloff}}, \binits{R.-R.}},
\bauthor{\bsnm{{Schmalzl}}, \binits{E.}},
\bauthor{\bsnm{{Sykes}}, \binits{J.}},
\bauthor{\bsnm{{Thelen}}, \binits{M.P.}},
\bauthor{\bsnm{{van Dishoeck}}, \binits{E.F.}},
\bauthor{\bsnm{{Vandenbussche}}, \binits{B.}},
\bauthor{\bsnm{{Venema}}, \binits{L.B.}},
\bauthor{\bsnm{{Visser}}, \binits{H.}},
\bauthor{\bsnm{{Waters}}, \binits{L.B.F.M.}},
\bauthor{\bsnm{{Wright}}, \binits{D.}}:
\byear{2015},
\batitle{{The Mid-Infrared Instrument for the James Webb Space Telescope, VI:
  The Medium Resolution Spectrometer}}.
\bjtitle{\pasp}
\bvolume{127},
\bfpage{646}.
\doiurl{https://doi.org/10.1086/682281}.
\adsurl{2015PASP..127..646W}.
\end{barticle}
\endbibitem

\bibitem[\protect\citeauthoryear{{Wilson}}{1968}]{1968ApJ...153..221W}
\begin{barticle}
\bauthor{\bsnm{{Wilson}}, \binits{O.C.}}:
\byear{1968},
\batitle{{Flux Measurements at the Centers of Stellar H- and K-Lines}}.
\bjtitle{\apj}
\bvolume{153},
\bfpage{221}.
\doiurl{https://doi.org/10.1086/149652}.
\adsurl{1968ApJ...153..221W}.
\end{barticle}
\endbibitem

\bibitem[\protect\citeauthoryear{{W{\"o}ger}, {von der L{\"u}he}, and
  {Reardon}}{2008}]{2008A&A...488..375W}
\begin{barticle}
\bauthor{\bsnm{{W{\"o}ger}}, \binits{F.}},
\bauthor{\bsnm{{von der L{\"u}he}}, \binits{O.}},
\bauthor{\bsnm{{Reardon}}, \binits{K.}}:
\byear{2008},
\batitle{{Speckle interferometry with adaptive optics corrected solar data}}.
\bjtitle{\aap}
\bvolume{488},
\bfpage{375}.
\doiurl{https://doi.org/10.1051/0004-6361:200809894}.
\adsurl{2008A&A...488..375W}.
\end{barticle}
\endbibitem

\bibitem[\protect\citeauthoryear{{Wu} et~al.}{2016}]{2016RScI...87kD616W}
\begin{barticle}
\bauthor{\bsnm{{Wu}}, \binits{C.R.}},
\bauthor{\bsnm{{Huang}}, \binits{J.}},
\bauthor{\bsnm{{Gao}}, \binits{W.}},
\bauthor{\bsnm{{Gao}}, \binits{W.}},
\bauthor{\bsnm{{Xu}}, \binits{Z.}},
\bauthor{\bsnm{{Chang}}, \binits{J.F.}},
\bauthor{\bsnm{{Hou}}, \binits{Y.M.}},
\bauthor{\bsnm{{Jin}}, \binits{Z.}},
\bauthor{\bsnm{{Xu}}, \binits{J.C.}},
\bauthor{\bsnm{{Duan}}, \binits{Y.M.}},
\bauthor{\bsnm{{Zhang}}, \binits{P.F.}},
\bauthor{\bsnm{{Chen}}, \binits{Y.J.}},
\bauthor{\bsnm{{Zhang}}, \binits{L.}},
\bauthor{\bsnm{{Wu}}, \binits{Z.W.}},
\bauthor{\bsnm{{Li}}, \binits{J.G.}}:
\byear{2016},
\batitle{{Measurement of the deuterium Balmer series line emission on EAST}}.
\bjtitle{Review of Scientific Instruments}
\bvolume{87},
\bfpage{11D616}.
\doiurl{https://doi.org/10.1063/1.4961293}.
\adsurl{2016RScI...87kD616W}.
\end{barticle}
\endbibitem

\bibitem[\protect\citeauthoryear{{Xuan} et~al.}{1998}]{1998A&AS..129..553X}
\begin{barticle}
\bauthor{\bsnm{{Xuan}}, \binits{J.Y.}},
\bauthor{\bsnm{{Gu}}, \binits{X.M.}},
\bauthor{\bsnm{{Lin}}, \binits{J.}},
\bauthor{\bsnm{{Jiang}}, \binits{Y.C.}},
\bauthor{\bsnm{{Zhong}}, \binits{S.H.}},
\bauthor{\bsnm{{Li}}, \binits{Y.S.}}:
\byear{1998},
\batitle{{Time evolution and morphological characteristics of white light flare
  on 18 January 1989}}.
\bjtitle{\aaps}
\bvolume{129},
\bfpage{553}.
\doiurl{https://doi.org/10.1051/aas:1998203}.
\adsurl{1998A&AS..129..553X}.
\end{barticle}
\endbibitem

\bibitem[\protect\citeauthoryear{{Yurchyshyn}
  et~al.}{2020}]{2020ApJ...896..150Y}
\begin{barticle}
\bauthor{\bsnm{{Yurchyshyn}}, \binits{V.}},
\bauthor{\bsnm{{Kilcik}}, \binits{A.}},
\bauthor{\bsnm{{{\c{S}}ahin}}, \binits{S.}},
\bauthor{\bsnm{{Abramenko}}, \binits{V.}},
\bauthor{\bsnm{{Lim}}, \binits{E.-K.}}:
\byear{2020},
\batitle{{Spatial Distribution of the Origin of Umbral Waves in a Sunspot
  Umbra}}.
\bjtitle{\apj}
\bvolume{896},
\bfpage{150}.
\doiurl{https://doi.org/10.3847/1538-4357/ab91b8}.
\adsurl{2020ApJ...896..150Y}.
\end{barticle}
\endbibitem

\bibitem[\protect\citeauthoryear{{Zeeman}}{1897}]{1897ApJ.....5..332Z}
\begin{barticle}
\bauthor{\bsnm{{Zeeman}}, \binits{P.}}:
\byear{1897},
\batitle{{On the Influence of Magnetism on the Nature of the Light Emitted by a
  Substance.}}
\bjtitle{\apj}
\bvolume{5},
\bfpage{332}.
\doiurl{https://doi.org/10.1086/140355}.
\adsurl{1897ApJ.....5..332Z}.
\end{barticle}
\endbibitem

\bibitem[\protect\citeauthoryear{Zhang}{2020}]{Zhang2020}
\begin{barticle}
\bauthor{\bsnm{Zhang}, \binits{H.}}:
\byear{2020},
\batitle{Diagnostic of spectral lines in magnetized solar atmosphere: Formation
  of the H$\beta$ line in sunspots}.
\bjtitle{Science China Physics, Mechanics \& Astronomy}
\bvolume{63},
\bfpage{119611}.
\bisbn{1869-1927}.
\doiurl{https://doi.org/10.1007/s11433-020-1584-9}.
\burl{https://doi.org/10.1007/s11433-020-1584-9}.
\end{barticle}
\endbibitem

\bibitem[\protect\citeauthoryear{Zhao}{2003}]{doi:10.1366/000370203322554527}
\begin{barticle}
\bauthor{\bsnm{Zhao}, \binits{J.}}:
\byear{2003},
\batitle{Image Curvature Correction and Cosmic Removal for High-Throughput
  Dispersive Raman Spectroscopy}.
\bjtitle{Applied Spectroscopy}
\bvolume{57},
\bfpage{1368}.
\bcomment{PMID: 14658150}.
\doiurl{https://doi.org/10.1366/000370203322554527}.
\burl{https://doi.org/10.1366/000370203322554527}.
\end{barticle}
\endbibitem

\end{thebibliography}
%
%
%
%

\end{article} 
\end{document}